\documentclass[10pt,a4paper]{report}
\usepackage[utf8x]{inputenc}
\usepackage{amsfonts}
\usepackage{latexsym}
\usepackage{amsmath}
\usepackage{amssymb}
\usepackage{amsthm}
\usepackage{parskip}
\let\listoftables\relax

\ifx\paragraph\undefined\else
\let\oldparagraph\paragraph
\renewcommand{\paragraph}[1]{\oldparagraph{#1}\mbox{}}
\fi
\ifx\subparagraph\undefined\else
\let\oldsubparagraph\subparagraph
\renewcommand{\subparagraph}[1]{\oldsubparagraph{#1}\mbox{}}
\fi

\usepackage{fancyhdr}
\usepackage{wallpaper}

\definecolor{gray75}{gray}{0.75}

\usepackage{sectsty} 
\usepackage[normalem]{ulem} 
\sectionfont{\rmfamily\mdseries\Large} 
\subsectionfont{\rmfamily\mdseries\scshape\large} 
\subsubsectionfont{\rmfamily\bfseries\upshape\large} 
\usepackage[font={footnotesize}]{caption}

\usepackage{siunitx}

\usepackage{setspace}
\usepackage[top=20.5mm,bottom=20.5mm,left=40.5mm,right=20.5mm]{geometry}
\usepackage{hyperref}
\hypersetup{colorlinks=true, linkcolor=black}

\usepackage{booktabs}
\usepackage{threeparttable}
\usepackage{array}
\newcolumntype{x}[1]{%
	>{\centering\arraybackslash}m{#1}}%

\date{}

\usepackage{slashed}
\usepackage{upgreek}
\usepackage{mathrsfs}

\usepackage{relsize}
\usepackage{graphicx}

\usepackage{graphicx}
\usepackage{verbatim}
\usepackage{latexsym}
\usepackage{mathchars}
\usepackage{setspace}
\usepackage{dirtytalk}
\usepackage{appendix}
\usepackage{dsfont}

\newsavebox{\uuunit}
\input{blocked.sty}
\input{uhead.sty}
\input{boxit.sty}
\input{icthesis.sty}

\newcommand{\NN}{{\sf I\kern-0.14emN}}   % Natural numbers
\newcommand{\ZZ}{{\sf Z\kern-0.45emZ}}   % Integers
\newcommand{\QQQ}{{\sf C\kern-0.48emQ}}   % Rational numbers
\newcommand{\RR}{{\sf I\kern-0.14emR}}   % Real numbers

\newcommand{\be}{\begin{eqnarray}}
\newcommand{\bea}{\begin{eqnarray}}
\newcommand{\ee}{\end{eqnarray}}
\newcommand{\eea}{\begin{eqnarray}}
\def \la {\label}

\def\l{\lambda}

\def\e{\epsilon}

\def\G{\Gamma}

\def\bbe{{\bf{e}}}
\font\mybb=msbm10 at 11pt

\def\bb#1{\hbox{\mybb#1}}

\def\bR {\bb{R}}

\def\bC {\bb{C}}

\def\tn {{\tilde{\nabla}}}
\def\bI {\bb{I}}
\def\bS {\bb{S}}

\newcommand{\R}{{\mathbb R}}

\newcommand{\ben}{\begin{eqnarray*}}
	\newcommand{\een}{\end{eqnarray*}}

\newcommand{\bem}{\begin{pmatrix}}
	\newcommand{\eem}{\end{pmatrix}}
\newcommand{\bl}{\begin{align}}
\newcommand{\el}{\end{align}}

\def\e{\epsilon}
\def\k{\kappa}             % Also, \varkappa (see below)
\def\l{\lambda}
\def\m{\mu}
\def\n{\nu}

\def\p{\pi}
\def\pa{\partial}

\def\r{\rho}                                     %     \varrho
\def\F{\Phi}
\def\G{\Gamma}

\newcommand{\ba}{\begin{array}}
	\newcommand{\ea}{\end{array}}

\newenvironment{frcseries}{\fontfamily{frc}\selectfont}{}

\def \la {\label}

\def\l{\lambda}

\def\e{\epsilon}

\def\G{\Gamma}

\def\bbe{{\bf{e}}}
\font\mybb=msbm10 at 11pt

\def\bb#1{\hbox{\mybb#1}}

\def\bR {\bb{R}}

\def\bC {\bb{C}}

\def\tn {{\tilde{\nabla}}}
\def\bI {\bb{I}}
\def\bS {\bb{S}}

\newcommand{\normallinespacing}{\renewcommand{\baselinestretch}{1.5} \normalsize}

\newcommand{\narrowlinespacing}{\renewcommand{\baselinestretch}{1.0} \normalsize}

\newcommand{\syncc}{~\stackrel{\textstyle \rhd\kern-0.57em\lhd}{\scriptstyle L}~}

\newtheorem{definition}{Definition}[chapter]

\begin{document}
	
	\begin{titlepage}
		\begin{center}
			\ThisULCornerWallPaper{.3}{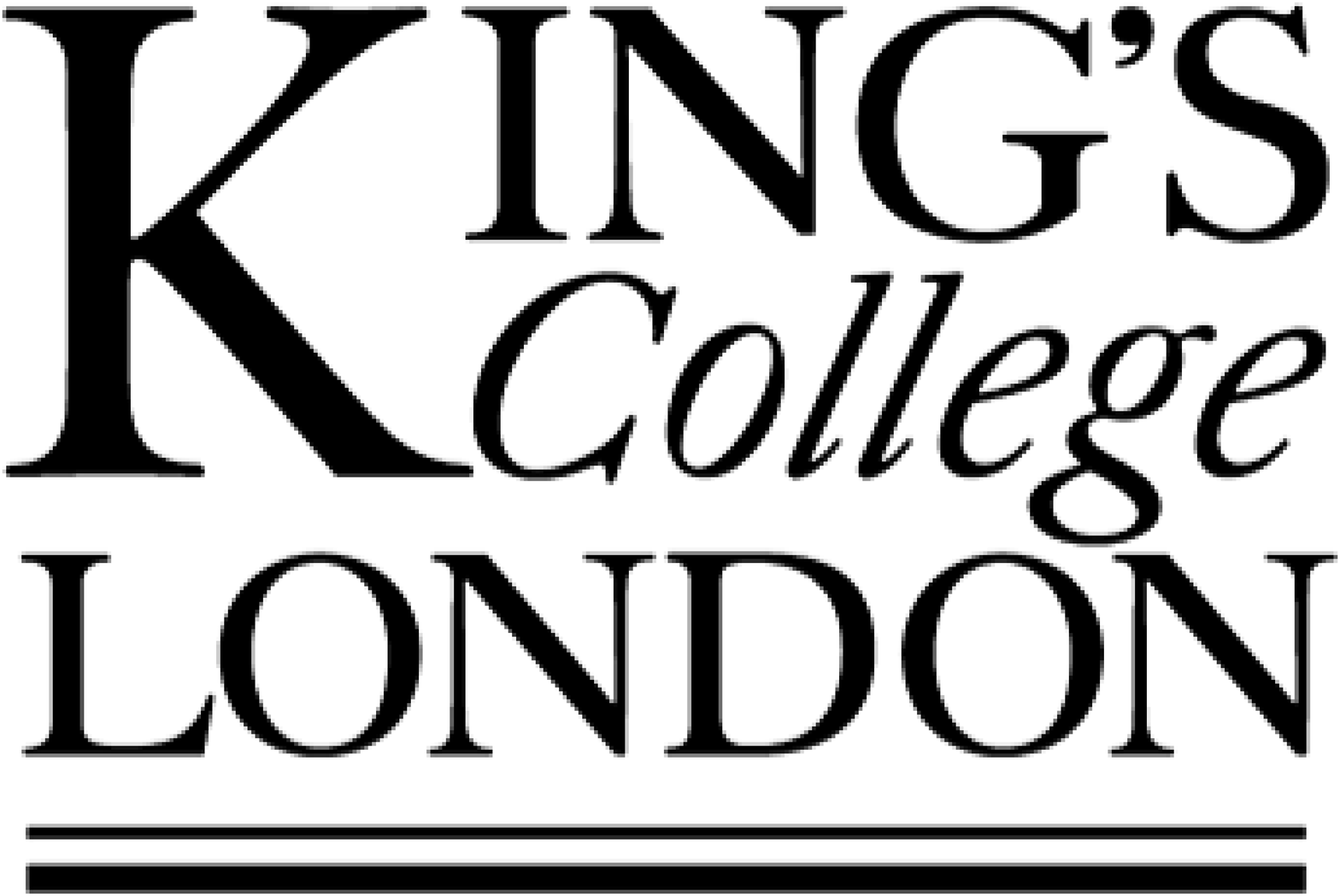}
			
			\vspace*{2cm}
			\huge
			Dynamical supersymmetry enhancement of black hole horizons
			
			\vspace{1.5cm}
			
			\Large
			Usman Kayani
			
			\vspace{1.5cm}
			
			\normalsize
			A thesis presented for the degree of\\
			Doctor of Philosophy
			
			\vfill
			
			\normalsize
			Supervised by:\\
			Dr. Jan Gutowski\\
			
			\vspace{0.8cm}
			
			% Uncomment the following line
			% to add a centered university logo
			%\includegraphics[width=0.4\textwidth]{style/univ_logo.e}
			
			\normalsize
			Department of Mathematics\\
			King's College London, UK\\
			November 2018
			
			% Except where otherwise noted, content in this thesis is licensed under a Creative Commons Attribution 4.0 License (http://creativecommons.org/licenses/by/4.0), which permits unrestricted use, distribution, and reproduction in any medium, provided the original work is properly cited. Copyright 2015, Author.
			
		\end{center}
	\end{titlepage}
	
	\preface
	\addcontentsline{toc}{chapter}{Abstract}

\begin{abstract}

This thesis is devoted to the study of dynamical symmetry enhancement of black hole horizons in string theory. In particular, we consider supersymmetric horizons in the low energy limit of string theory known as supergravity and we prove the {\it horizon conjecture} for a number of supergravity theories. We first give important examples of symmetry enhancement in $D=4$ and the mathematical preliminaries required for the analysis. Type IIA supergravity is the low energy limit of $D=10$ IIA string theory, but also the dimensional reduction of $D=11$ supergravity which itself the low energy limit of M-theory.  We prove that Killing horizons in IIA supergravity with compact spatial sections preserve an even number of supersymmetries. By analyzing the global properties of the Killing spinors, we prove that the near-horizon geometries undergo a supersymmetry enhancement. This follows from a set of generalized Lichnerowicz-type theorems we establish, together with an index theory argument. We also show that the symmetry algebra of horizons with non-trivial fluxes includes an $\mathfrak{sl}(2, \mathbb{R})$ subalgebra. As an intermediate step in the proof, we also demonstrate new Lichnerowicz type theorems for spin bundle connections whose holonomy is contained in a general linear group. We prove the same result for Roman's Massive IIA supergravity. We also consider the near-horizon geometry of supersymmetric extremal black holes in un-gauged and gauged 5-dimensional supergravity, coupled to abelian vector multiplets. We consider important examples in $D=5$ such as the BMPV and supersymmetric black ring solution, and investigate the near-horizon geometry to show the enhancement of the symmetry algebra of the Killing vectors. We repeat a similar analysis as above to prove the horizon conjecture. We also investigate the conditions on the geometry of the spatial horizon section $\cal{S}$.

\end{abstract}
	\cleardoublepage

\addcontentsline{toc}{chapter}{Acknowledgements}

\begin{acknowledgements}
	
Firstly, I would like to express my gratitude to Jan Gutowski for his guidance in my research, and for his patience and understanding.

I would also like to thank my PhD mentor, Geoffrey Cantor, who has guided me through the trials and tribulations and Jean Alexandre for all his support in helping me cross the finish line.

Finally, I would like to thank my family and friends, especially my parents, for all their support over the years and throughout my education and life.

\end{acknowledgements}
	\clearpage

\narrowlinespacing

\vspace*{4mm}

\say{Black holes provide theoreticians with an important theoretical laboratory to test ideas. Conditions within a black hole are so extreme, that by analyzing aspects of black holes we see space and time in an exotic environment, one that has shed important, and sometimes perplexing, new light on their fundamental nature.} \\
\\
\emph{Brian Greene}

\normallinespacing
	
	\body
	\chapter{Introduction}
\section{Black holes}
A black hole is a region of spacetime that has such strong gravitational effects that nothing, not even light can escape. This literally makes the region black as no photons or massive particles are able to escape this region, and hence we cannot observe it. In other words, it is the region where light rays cannot escape to infinity. They can be formed from the gravitational collapse of massive stars but are also of interest in their own right, independently of how they were formed. Far from being hungry beasts devouring everything in their vicinity as shown in popular culture, we now believe certain (supermassive) black holes actually drive the evolution of galaxies \cite{richstone}.

In 2016, the Laser Interferometer Gravitational-Wave Observatory (LIGO) made the first detection of gravitational waves from a binary black hole merger \cite{abbott} which was a milestone in physics and astronomy. It confirmed a major prediction of Einstein's general theory of relativity and marked the beginning of the new field of gravitational-wave astronomy. Gravitational waves carry information about their origins and about the nature of gravity that cannot otherwise be obtained. The gravitational waves were produced in the final moments before the merger of two black holes, 14 and 8 times the mass of the sun to produce a single black hole 21 times the mass of the sun. During the merger, which occurred approximately 1.4 billion years ago, a quantity of energy roughly equivalent to the mass of the sun was converted into gravitational waves. The detected signal comes from the last 27 orbits of the black holes before their merger. In 2017, another observation by LIGO and Virgo gravitational-wave detectors of a binary neutron star inspiral signal, called GW170817 \cite{ligo1} and associated to the event a gamma-ray burst \cite{ligo2} was independently observed \cite{ligo3}. These led to new bounds and constraints for fundamental physics, such as the stringent constraint on the difference between the speed of gravity and the speed of light, a new bound on local Lorentz invariance violations and a constraint on the Shapiro delay between gravitational and electromagnetic radiation which enabled a new test of the equivalence principle \cite{ligo2}. As a result, these bounds also constrain the allowed parameter space of the alternative theories of gravity that offer gravitational explanations for the origin of dark energy or dark matter.

Black hole research is at the forefront of modern physics because much about black holes is still unknown. Currently, the best two theories we have that describe the known universe are the Standard Model of particle physics (a quantum field theory) and general relativity (a classical field theory). Most of the time these two theories do not talk to each other, with the exception of two arenas: the singularity before the Big Bang and the singularity formed within a black hole. The singularity is a point of infinite density and thus the physical description necessarily requires quantum gravity. Indeed Roger Penrose and Stephen Hawking also showed 48 years ago that, according to general relativity, any object that collapses to form a black hole will go on to collapse to a singularity inside the black hole \cite{penrose}. This means that there are strong gravitational effects on arbitrarily short distance scales inside a black hole and such short distance scales, we certainly need to use a quantum theory to describe the collapsing matter.  

General relativity is not capable of describing what happens near a singularity and if one tries to quantize gravity \textit{naively}, we find divergences that we can't cancel because gravity is non-renormalizable\footnote{In particular, the number of counterterms in the Lagrangian, required to cancel the divergences is infinite and thus the process of renormalization fails.} \cite{goroff}. String theory is a broad and varied subject that attempts to remedy this and to address a number of deep questions of fundamental physics. It has been applied to a variety of problems in black hole physics, early universe cosmology, nuclear physics, and condensed matter physics, and it has stimulated a number of major developments in pure mathematics. Because string theory potentially provides a unified description of gravity and particle physics, it is a candidate for a theory of everything, a self-contained mathematical model that describes all fundamental forces and forms of matter.

\begin{comment}
In 1964, Roger Penrose proves that an imploding star will necessarily produce a singularity once it has formed an event horizon. The presence of a singularity in the classical theory also means that once we go sufficiently far into the black hole, we can no longer predict what will happen. It is hoped that this failure of the classical theory can be cured by quantizing gravity as well.
\end{comment}

In the currently accepted models of stellar evolution, black holes are thought to arise when massive stars undergo gravitational collapse, and many galaxies are thought to contain supermassive black holes at their centers. Black holes are also important for theoretical reasons, as they present profound challenges for theorists attempting to understand the quantum aspects of gravity. String theory has proved to be an important tool for investigating the theoretical properties of black holes because it provides a framework in which theorists can study their thermodynamics, aided in particular by properties such as supersymmetry.

When we talk about the masses of black holes, we usually compare them with the mass of our Sun (also known as a solar mass denoted by $M_{\odot}$). Based upon the length of the year, the distance from Earth to the Sun (an astronomical unit or AU), and the gravitational constant (G), one solar mass is given by:
\bea
M_{\odot}={\frac {4\pi ^{2}\times (1\,\mathrm {AU} )^{3}}{G\times (1\,\mathrm {yr} )^{2}}} = (1.98855 \pm 0.00025) \times 10^{30} kg \ .
\ee
There are three types of black holes that can arise:
\begin{itemize}
	\item Stellar black holes are formed from the gravitational collapse of a massive star. They have masses ranging from about 5-99$M_{\odot}$
	\item Supermassive black holes are the largest type, on the order of $10^2-10^{12} M_{\odot}$
	\item Miniature black holes, also known as quantum mechanical black holes are hypothetical tiny black holes, for which quantum mechanical effects play a role. These will have masses about the same as Mount Everest.
\end{itemize}

The first person to come up with the idea of a black hole was John Mitchell in 1783,
followed (independently) by Pierre-Simon Laplace in 1796. These prototypical black objects, called 'dark stars', were considered from the point of view of Newton's law of motion and gravitation. In particular, Mitchell calculated that when the escape velocity at the surface of a star was equal to or greater than lightspeed, the generated light would be gravitationally trapped, so that the star would not be visible to a distant astronomer.  The event horizon is the boundary of this region, also known as the `point of no return' because once you go past this point it is impossible to turn back. To see this, consider an object of mass $m$ and a spherically symmetric body of mass $M$ and radius $R$. The total energy is,
\begin{eqnarray}
\label{te}
E = \frac{1}{2}m v^2 - \frac{G m M}{r} \ ,
\end{eqnarray}
where $r$ is the distance from the centre of mass of the body to the object. We can also write this as $r = R + h$, where $h$ is the distance from the surface of the body to the object, but the region we will consider is at $r \geq R$ or $h \geq 0$. An object that makes it to $r \rightarrow \infty$ will have $E \geq 0$ as the kinetic energy will dominate for large $r$. This implies,
\begin{eqnarray}
v \geq v_e(r) \equiv \sqrt{\frac{2GM}{r}} \ ,
\end{eqnarray}
where $v_e(r)$ is the escape velocity, which is the lowest velocity which an object must have in order to escape the gravitational attraction from a spherical body of mass $M$ at a given distance $r$. By energy conservation, we have at the surface of the body ($r=R$),
\begin{eqnarray}
v \geq  v_{e}(R) = \sqrt{\frac{2GM}{R}} \ .
\end{eqnarray}
This is independent of the mass of the escaping object, by the equivalence of inertial and gravitational masses. Now the escape velocity at the surface is greater than the speed of light, $v_e (R) > c$ if
\begin{eqnarray}
R < r_s \equiv \frac{2GM}{c^2} \ .
\end{eqnarray}
The radius $r_s$ is now understood in terms of the Schwarzschild radius. The surface $r = r_s $ acts as an ``event horizon" if the body fits inside this radius. Of course, there is nothing particularly special about the speed of light $c$ in the Newtonian (non-relativistic) gravity. Objects in principle could move faster than $c$ which means that they may always escape the would-be black hole. Moreover,
a photon emitted from such an object does leave the object, although it would eventually fall back in. Thus there are no real black holes (a body from which nothing can escape) in Newton's gravity. 

As we now understand photons to be massless, so the naive analysis described above is not physically realistic.
However, it is still interesting to consider such objects as a starting point.
The consideration of such ``dark stars" raised the intriguing possibility that
the universe may contain a large number of massive objects which cannot be
observed directly. This principle turns out to be remarkably close to our current understanding of cosmology. However, the accurate description of a black hole has to be general relativistic.
\begin{figure}[h!]
	\begin{center}
		\includegraphics[width=200pt]{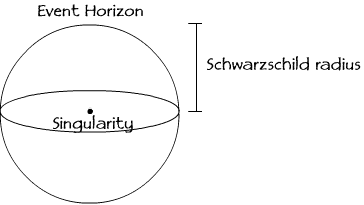}
	\end{center}
	\caption{Event horizon of a black hole enclosing a singularity (credit: \url{https://goo.gl/CAXqfD})}
	\label{fig:sradius}
\end{figure}
Isaac Newton was to be overturned by Albert Einstein with his special theory of relativity in 1905, which showed that the speed of light was constant in any reference frame and then his general theory of relativity in 1915 which describes gravity as the curvature of space-time \cite{einsteingr};
\bea
R_{\mu \nu} - \frac{1}{2}R g_{\mu \nu} = \frac{8 \pi G}{c^4} T_{\mu \nu} \ .
\ee
The theory relates the metric tensor $g_{\mu \nu}$, which defines a spacetime geometry\footnote{This can also be written as a line element with $ds^2 = g_{\mu \nu}dx^{\mu} dx^{\nu}$ and $\mu=0,1,2,3$}, with a source of gravity that is associated with mass or energy and $R_{\mu \nu}$ is the Ricci tensor.
Black holes arise in general relativity as a consequence of the solution 
to the Einstein field equations found by  Karl Schwarzschild in 1916.
Although the Schwarzschild solution was originally formulated in order
to describe the gravitational field of the solar system, and to
understand the motion of objects passing through this field, 
the idea of a black hole remained as an intriguing possibility.
Einstein himself thought that these solutions were simply a mathematical curiosity and not physically relevant. In fact, in 1939 Einstein tried to show that stars cannot collapse under gravity by assuming matter cannot be compressed beyond a certain point. 

The Schwarzschild radius can also be realized by considering the Schwarzschild solution in general relativity i.e any non-rotating, non-charged spherically-symmetric body that is smaller than its Schwarzschild radius forms a black hole. This idea was first promoted by David Finkelstein in 1958 who theorised that the Schwarzschild radius of a black hole is a causality barrier: an event horizon. The Schwarzschild solution is an exact solution and was found within only a few months of the publication of Einstein’s field equations. Instead of dealing only with weak-field corrections to Newtonian gravity, full nonlinear features of the theory could be studied, most notably gravitational collapse and singularity formation. The Schwarzschild metric written in Schwarzschild Coordinates $(t, r, \theta, \phi)$ is given by,\footnote{$d\Omega_2^2 = d\theta^2 + \sin^2{\theta}d\phi^2$ is the metric of the unit 2-sphere $S^2$} 
\bea
ds^2 = -f(r) c^2 dt^2 + f(r)^{-1} dr^2 + r^2 d\Omega_2^2 \ ,
\ee
where $f(r) = 1-r_s / r$. In this form, the metric has a coordinate singularity at $r=r_s$ which is removable upon an appropriate change of coordinates. The Kretschmann scalar $K$ for this metric elucidates the fact that the singularity at $r=0$ cannot be removed by a coordinate transformation,
\bea
K \equiv R_{\mu \nu, \rho \sigma}R^{\mu \nu, \rho \sigma} = \frac{48 G^2M^2}{c^4 r^6} \ ,
\ee
where $R_{\mu \nu, \rho \sigma}$ is the Riemann tensor. A more accurate general relativistic description of the event horizon is that within this horizon all light-like paths and hence all paths in the forward light cones of particles within the horizon are warped as to fall farther into the hole.
\begin{figure}[h!]
	\begin{center}
		\includegraphics[width=200pt]{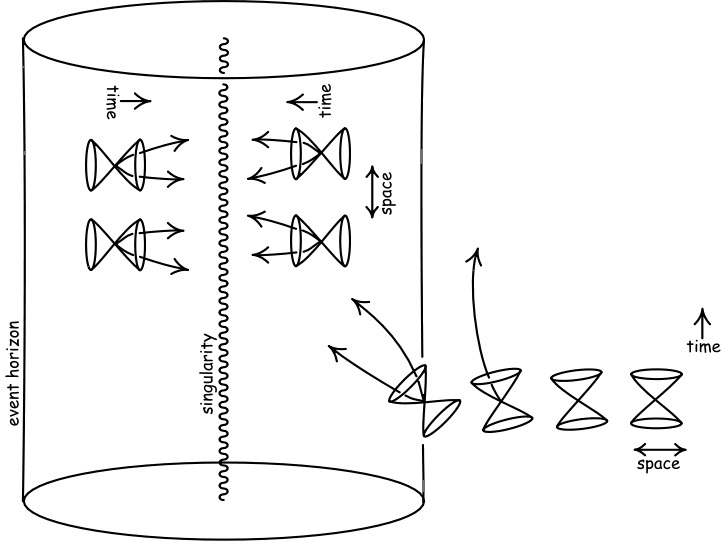}
	\end{center}
	\caption{Light cones tipping near a black hole (credit: \url{https://goo.gl/riTdG1})}
	\label{fig:tipping}
\end{figure}
In order to define the black hole region and event horizon in a concrete way, we will need to consider some definitions. Let $({\cal M},g)$ be an asymptotically flat manifold.
\theoremstyle{definition}
\begin{definition}{The {\bf causal past} of $U \subset {\cal M}$ is}
	\begin{eqnarray}
	\label{cp}
	J^{-}(U) = \big\{p\in {\cal M} : \exists \textrm{ a future directed causal curve from }p \textrm{ to some } q\in U \big\} \nonumber \ .
	\end{eqnarray}\end{definition}Let us denote $\mathscr{I}^+$ as {\bf future null infinity}, the limit points of future directed null rays in ${\cal M}$.
\theoremstyle{definition}
\begin{definition}{The {\bf black hole region} $B$ is the region of spacetime in ${\cal M}$ that is not contained in the Causal past of future null infinity $J^{-}(\mathscr{I}^+)$, $B = {\cal M} \setminus J^{-}(\mathscr{I}^+)$ i.e}
	\begin{eqnarray}
	\label{bhr}
	B &=& \big\{p\in {\cal M} : \nexists \textrm{ a future directed casual curve from }p\textrm{ to }\mathscr{I}^+ \big\} \nonumber \ .
	\end{eqnarray}
\end{definition}
\theoremstyle{definition}
\begin{definition}{The {\bf event horizon} $\cal{H}$ is the boundary of $B$ in ${\cal M}$ i.e the boundary of the causal past of future null infinity $J^{-}(\mathscr{I}^+)$ in ${\cal M}$, ${\cal{H}} = \partial B = \partial J^{-}(\mathscr{I}^+)$}
\end{definition}
\begin{figure}[h!]
	\begin{center}
		\includegraphics[width=180pt]{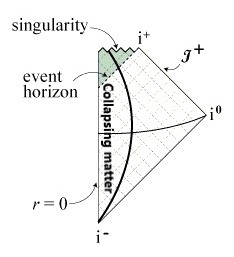}    
	\end{center}
	\caption{Penrose diagram of a collapsing star where $i^+$, $i^-$ denote future and past timelike infinity respectively and $i^0$ is spacelike infinity}
	\label{fig:penrose}
\end{figure}
\theoremstyle{definition}
\begin{definition}{A hypersurface $\cal{H}$ defined by a smooth function $S(x^{\mu}) = 0$, where $S \in  C^{\infty}({\cal M})$ is called a {\bf null hypersurface} if the normal vector field $\xi^{\mu} \sim g^{\mu \nu}\partial_{\nu}S$ is null, $g(\xi, \xi)=g_{\mu \nu}\xi^{\mu}\xi^{\nu} = 0$ on $\cal{H}$.}
\end{definition}
For such surfaces one can show that
\begin{eqnarray}
\xi^{\nu}\nabla_{\nu}\xi^{\mu} = \kappa \xi^{\mu} \ ,
\end{eqnarray}
where $\kappa$ is a measure to the extent to which the parametrization is not affine. In the context of Killing horizons, $\kappa$ can physically be interpreted as the surface gravity, which we will see in the next section. If we denote $l$ to be the normal to $\cal{H}$ which corresponds to an affine parametrization, i.e $l^{\nu}\nabla_{\nu}l^{\mu} = 0$ and $\xi = f(x)l$ for some $f(x)$ then $\kappa = \xi^{\mu}\partial_{\mu}\ln |f|$.

One can also show that the event horizon $\cal{H}$ is a null hypersurface since all its normals are null on $\cal{H}$. The advantage of the event horizon $\cal{H}$ is that the boundary of a past set is always a null hypersurface and is ruled by null geodesics which stay on the boundary. This allows one to prove general properties of the horizon. The disadvantage is that the event horizon cannot be determined locally and one needs the entire future of the spacetime to know where the event horizon is. For practical reasons, one might a consider a more local definition of a horizon.

\subsection{Other solutions: Reisser-Nordstr\"om metric (1918)}

The existence of the Schwarzschild solution set in motion a search for other
exact solutions and in 1918, Reissner and Nordstr\"om solved
the Einstein-Maxwell field equations for charged spherically-symmetric non-rotating
systems. Gravity coupled to the electromagnetic field is described by the Einstein-Maxwell action
\bea
S = \frac{1}{16 \pi G} \int \sqrt{-g}(R - F_{\mu \nu}F^{\mu \nu})d^4 x \ ,
\ee
where $F_{\mu \nu} = \nabla_{\mu}A_{\nu} - \nabla_{\nu}A_{\mu}$ and $A_{\mu}$ is the electromagnetic (four-)potential. The normalisation of the Maxwell term is such that the Coulomb force between two charges $Q_1$ and $Q_2$ separated by a sufficiently large distance $r$ is
\bea
F_{Coulomb} = G \frac{|Q_1 Q_2|}{r^2} \ .
\ee
This corresponds to geometrised units of charge.
The equations of motion derived from the variation of the Einstein-Maxwell action are
\bea
R_{\mu \nu} - \frac{1}{2} R g_{\mu \nu} &=& 2\bigg(F_{\mu \lambda}F_{\nu}{}^{\lambda} - \frac{1}{4}g_{\mu \nu}F_{\rho \sigma}F^{\rho \sigma}\bigg) \ ,
\nonumber \\
\nabla_{\mu}F^{\mu \nu} &=& 0 \ .
\ee
They admit the spherically symmetric solution,
\bea
ds^2 &=& -\bigg(1 - \frac{r_s}{r} + \frac{r_Q^2}{r^2}\bigg) c^2 dt^2 + \bigg(1 - \frac{r_s}{r} + \frac{r_Q^2}{r^2}\bigg)^{-1}dr^2 + r^2 d\Omega^2_2
\cr
&=& -\bigg(1-\frac{r_+}{r}\bigg)\bigg(1-\frac{r_-}{r}\bigg) c^2 dt^2 +\bigg(1-\frac{r_+}{r}\bigg)^{-1}\bigg(1-\frac{r_-}{r}\bigg)^{-1}dr^2 + r^2 d\Omega^2_2 \ ,
\label{rnsolution}
\ee
where we define the length-scale corresponding to the electric charge $Q$ as,
\bea
r^2_{Q} = \frac{\kappa_e Q^2 G}{c^4} \ ,
\ee
where $\kappa_e$ is the Coulomb constant. This is known as the Reissner-Nordstr\"om solution and the two horizons given by $r_{\pm} = \frac{r_s}{2} \pm \frac{1}{2}\sqrt{r_s^2 -4r_Q^2}$ and the extremal limit where the horizons coincide corresponds to $Q=M$ in natural units\footnote{$G = c = \hbar = 1$}. The electric potential is $A_{t} = \frac{Q}{r}$ and the other components of $A_{\mu}$ vanish. We therefore interpret $Q$ as the electric charge of the black hole and $M$ as its mass. Without loss of generality we assume that $Q > 0$. By a theorem analogous to Birkhoff's theorem, the Reissner-Nordstrom solution is the unique (stationary and asymptotically flat) spherically symmetric solution to the 
Einstein-Maxwell equations with cosmological constant $\Lambda = 0$.

Gravitatational theories coupled to Maxwell but also {\it dilaton} fields $\Phi$ emerge from several more fundemental theories such with Kaluza-Klein reduction and compactification. The addition of the dilaton field and the dynamics of the black hole in this theory display interesting properties when compared to the standard Reissner-Nordstr\"om black hole. For this reason, we briefly consider an Einstein-Maxwell gravity coupled to a dilaton field $\Phi$ with dilaton coupling constant $\alpha$. The action is \cite{Garfinkle},
\bea
S = \frac{1}{16 \pi G} \int \sqrt{-g}(R -2\nabla^{\mu}\Phi \nabla_{\mu}\Phi - e^{-2\alpha \Phi}F_{\mu \nu}F^{\mu \nu})d^4 x \ .
\ee
The equations of motion derived from this variation are
\bea
R_{\mu \nu} - \frac{1}{2} R g_{\mu \nu} &=& 2\bigg(F_{\mu \lambda}F_{\nu}{}^{\lambda} - \frac{1}{4}g_{\mu \nu}F_{\rho \sigma}F^{\rho \sigma}\bigg) + 2\nabla_{\mu}\Phi \nabla_{\nu} \Phi - g_{\mu \nu}\nabla^{\lambda}\Phi \nabla_{\lambda}\Phi \ ,
\nonumber \\
\nabla_{\mu}(e^{-2\alpha \Phi}F^{\mu \nu}) &=& 0 \ ,
\nonumber \\
\nabla^{\mu}\nabla_{\mu}\Phi &=& \frac{1}{2}\alpha e^{-2\alpha \Phi}F_{\rho \sigma}F^{\rho \sigma} \ .
\ee
The parameter $\alpha$ is a dimensionless constant and the behaviour of the theory shows non-trivial dependence on $\alpha$. The spherically symmetric black hole solutions of this action are given by \cite{Garfinkle},
\bea
ds^2 &=& -\bigg(1-\frac{r_+}{r}\bigg)\bigg(1-\frac{r_-}{r}\bigg)^{\frac{1-\alpha^2}{1+ \alpha^2}}dt^2 +\bigg(1-\frac{r_+}{r}\bigg)^{-1}\bigg(1-\frac{r_-}{r}\bigg)^{\frac{\alpha^2 - 1}{1+ \alpha^2}}dr^2 
\nonumber \\
&+& r^2\bigg(1-\frac{r_-}{r}\bigg)^{\frac{2\alpha^2}{1+\alpha^2}} d\Omega^2_2 \ ,
\ee
where the two inner and outer horizons are located at (in natural units),
\bea
r_+ = M + \sqrt{M^2 - (1-\alpha^2)Q^2},~~ r_{-} = \bigg(\frac{1+\alpha^2}{1-\alpha^2}\bigg)(M - \sqrt{M^2 - (1-\alpha^2)Q^2}) \ .
\ee
The extremal limit where the two horizons coincide corresponds to $Q = M\sqrt{1+\alpha^2}$ which is singular for generic $\alpha$. The Maxwell and dilaton fields are,
\bea
A_{t} = \frac{Q}{r},~~ e^{2 \alpha \Phi}=
\bigg(1-\frac{r_-}{r}\bigg)^{\frac{2\alpha^2}{1+\alpha^2}} \ .
\ee
For $\alpha < 1$ in order to preserve reality one must have $|\frac{Q}{M}| \leq \frac{1}{\sqrt{1-\alpha^2}}$ but for $\alpha > 1$ we do not have this restriction. 

\subsection{Kerr metric (1963)}
Already in 1918, Lense and Thirring had found the exterior field of a rotating sphere to the first order in the angular momentum, but many were after a simple exact solution that was physically relevant.
Astrophysically, we know that stars (and for that matter planets) rotate, and from the weak-field approximation to the Einstein equations we even know the approximate form of the metric at sufficiently large distances from a stationary isolated body of mass $M$ and angular momentum $J$, given by (in natural units),
\bea
\label{weak}
ds^2 &=& -\bigg[1 - \frac{2M}{r} + O\bigg(\frac{1}{r^2}\bigg)\bigg]dt^2 
- \bigg[\frac{4 J \sin^2 \theta}{r} + O\bigg(\frac{1}{r^2}\bigg)\bigg]d \phi dt
\nonumber \\
&+& \bigg[1 + \frac{2M}{r} + O\bigg(\frac{1}{r^2}\bigg)\bigg] \bigg(dr^2 + r^2 d\Omega^2_2\bigg) \ .
\ee
This metric is perfectly adequate for almost all solar system
tests of general relativity, but there  are well-known astrophysical
situations for which this approximation is inadequate and so a ``strong field" solution required physically. Furthermore, if a rotating star were to undergo gravitational collapse, then the resulting black hole would be expected to retain some portion of its initial
angular momentum. Thus suggesting on physical grounds that there should be an extension of the Schwarzschild geometry to the situation where the central body carries angular momentum.
From the weak-field metric (\ref{weak}), we can clearly see that angular
momentum destroys the spherical symmetry and this lack of spherical symmetry
makes the calculations more difficult. It took another 45 years to find another exact solution and in 1963, Kerr solved the Einstein vacuum field equations for uncharged symmetric rotating systems, deriving the Kerr metric. The Kerr metric describes the geometry of a spacetime for a rotating body with mass $M$ and angular momentum $J$ and is given by,
\bea
ds^{2}&=&-\bigg(1-{\frac {r_{s}r}{\Sigma }}\bigg)c^{2}dt^{2}+{\frac {\Sigma }{\Delta }}dr^{2}+\Sigma d\theta ^{2}
\nonumber \\
&+&(r^{2}+a^{2}+{\frac {r_{s}ra^{2}}{\Sigma }}\sin ^{2}\theta )\sin ^{2}\theta \ d\phi ^{2}-{\frac {2r_{s}ra\sin ^{2}\theta }{\Sigma }}\,c\,dt\,d\phi \ ,
\label{kerrsol}
\ee
where the coordinates  $r,\theta ,\phi$  are standard spherical coordinate system, which are equivalent to the cartesian coordinates,
\bea
x &=& \sqrt {r^2 + a^2} \sin\theta\cos\phi \ ,
\nonumber \\
y &=& \sqrt {r^2 + a^2} \sin\theta\sin\phi \ ,
\nonumber \\
z &=& r\cos \theta \ ,
\ee
and $r_s$ is the Schwarzschild radius, and where $a, \Sigma$ and $\Delta$ are given by,
\bea
a &=&\frac {J}{Mc} \ ,
\nonumber \\
\Sigma &=& r^{2}+a^{2}\cos ^{2}\theta \ ,
\nonumber \\
\Delta &=& r^{2}-r_{s}r+a^{2} \ .
\ee
The horizons are located at $r_{\pm} = \frac{r_s}{2} \pm \frac{1}{2}\sqrt{r_s^2 -4a^2}$ and the extremal limit corresponds to taking $a=M$ in natural units. 

Astrophysical black holes have a null charge $Q$ and they all
belong to the Kerr family (when $a = 0$ it reduces to the Schwarzschild solution). In contrast to the Kerr solution, the Reissner-Nordstr\"om solution of Einstein's equation is spherically symmetric
which makes the analysis much simpler. Now that another 55 years have elapsed, we can see the impact of this exact solution. It has significantly influenced our understanding of general relativity and in astrophysics the discovery of rotating black holes together with a simple way to treat their properties has revolutionized the subject.

\subsection{The Kerr-Newman geometry (1965)} 

The Kerr-Newmann solution is the most general metric that describes the geometry of spacetime for a rotating charged black hole with mass mass $M$, charge $Q$ and angular momentum $J$. The metric is
\bea
ds^2 =-\bigg({\frac {dr^{2}}{\Delta }}+d\theta ^{2}\bigg)\Sigma+(c\,dt-a\sin ^{2}\theta \,d\phi )^{2}{\frac {\Delta }{\Sigma}} -((r^{2}+a^{2})d\phi -ac\,dt)^{2}{\frac {\sin ^{2}\theta }{\Sigma}} \ ,
\label{kerrnewsol}
\ee
where the coordinates $(r, \theta, \phi)$ are standard spherical coordinate system, and where,
\bea
\Delta &=& r^{2}-r_{s}r+a^{2}+r_{Q}^{2} \ ,
\nonumber \\
\Sigma &=& r^{2}+a^{2}\cos ^{2}\theta \ .
\ee
The horizons are located at $r_{\pm} = \frac{r_s}{2} \pm \frac{1}{2}\sqrt{r_s^2 - 4a^2 -4r_Q^2}$ and the extremal limit is $Q^2 = M^2 -a^2$ in natural units. In this metric, $a=0$ is a special case of the Reissner-Nordstr\"om solution, while $Q=0$ corresponds to the Kerr solution. 

\subsection{The no-hair theorem}
In 1967, Werner Israel \cite{israel1} presented the proof of the no-hair theorem at King's College London. After the first version of the no-hair theorem for the uniqueness of the Schwarzschild metric in 1967, the result was quickly generalized to the cases of charged or spinning black holes \cite{israel2, carter}. In $D=4$ these imply that the Einstein equations admit a unique class of asymptotically flat black hole solutions, parametrized by only three externally observable classical parameters $(M, Q, J)$. A key step is to establish the horizon topology theorem, which proves that the event horizon of a stationary black hole must have $S^2$ topology \cite{Htopology1}, which was shown in 1972. All other information about the matter which formed a black hole or is falling into it, ``disappears" behind the black-hole event horizon and is therefore permanently inaccessible to external observers.

\subsubsection{Higher dimensional black holes}

The proof the no-hair theorem relies on the Gauss-Bonnet theorem applied to the 2-manifold spatial horizon section and therefore does not generalize to higher dimensions. Thus for dimensions $D>4$, uniqueness theorems for asymptotically flat black holes lose
their validity. Indeed, the first example of how the classical uniqueness theorems break down in higher dimensions is given by the five-dimensional black ring solution
\cite{Emparan:2001wn, BR1}. There exist (vacuum) black rings with the same asymptotic conserved charges as the Myers-Perry black hole\footnote{This is a vacuum solution in $D=5$ \cite{myers} analogous to the Kerr solution in $D=4$} but with a different horizon topology.  Even more exotic solutions in five dimensions are now known to exist, such as the solutions obtained in \cite{Horowitz:2017fyg}, describing asymptotically flat black holes which possess a non-trivial topological structure outside the event horizon, but whose
near-horizon geometry is the same as that of the BMPV solution\footnote{The BMPV solution in $D = 5$ is a stationary, non-static, non-rotating black hole with angular momentum $J$
	and electric charge $Q$ \cite{BMPV1}.}.

The known black hole solutions in four dimensions have also been generalised to higher
dimensions \cite{Emparan:2008eg}. For instance, there exists a solution to the Einstein equations in any dimension $D$,
which is a generalization of the Schwarzschild metric which was discovered by Tangherlini \cite{Tangherlini}. In $D$ dimensions, a generalisation of the Reissner-Nordstr\"om black hole has the metric,
\bea
ds^2 = -\bigg(1-\frac{2\mu}{r^{D-3}} + \frac{q}{r^{2(D-3)}}\bigg)dt^2 + \bigg(1-\frac{2\mu}{r^{D-3}} + \frac{q}{r^{2(D-3)}}\bigg)^{-1}dr^2 + r^2 d\Omega_{D-2}^2 \ ,
\ee
where $d\Omega_{D-2}^2$ is the line element on the unit $S^{D-2}$ sphere and,
\bea
\mu = \frac{8 \pi G M}{(D-2) A_{D-2}},~~ q = \frac{8 \pi G Q^2}{(D-3)(D-2)},~~ A_{N-1} = \frac{2\pi^{\tfrac{N}{2}}}{\Gamma(\tfrac{N}{2})} \ ,
\ee
where $A_{N-1}$ is the area of the sphere $S^{N-1}$ and the horizons are located at $r_{\pm} = (\mu \pm \sqrt{\mu^2 - q})^{\tfrac{1}{D-3}}$.
Furthermore, the generalisation of the Kerr metric to higher dimensions was found by Myers and Perry \cite{myers}. The Myers-Perry
solution is specified by the mass $M$ and a set of angular momenta $J_r$, where $r = 1, \dots,rank[SO(D − 1)]$, and the horizon topology is $S^{D−2}$. In $D = 5$, for solutions which have
only one non-zero angular momentum $J_1 = J \neq 0$, they found the bound 
$J^2 \leq 32GM^3/(27\pi)$,
which is a generalisation of the known four dimensional Kerr bound $J \leq GM^2$. However for
$D > 5$, the momentum is unbounded, and the black hole can be ultra-spinning. Different black hole solutions can also have the same near-horizon geometry, for example in $D = 5$ the extremal self-dual Myers-Perry black hole and the extremal $J = 0$ Kaluza-Klein black hole.

\subsection{The laws of black hole mechanics and entropy}
In 1972 Stephen Hawking proved that the area of a classical black hole's event horizon cannot decrease and along with James Bardeen, Brandon Carter \cite{bhmech}, they derived four laws of black hole mechanics in analogy with the laws of thermodynamics.

\begin{description} 
	\item[Zeroth Law:]
	\begin{description}
		The surface gravity $\kappa$
		is constant over the event horizon of a stationary black hole.
	\end{description}
	\item[First Law:]
	\begin{description}
		\[
		d M = \frac{\kappa}{8\pi}  d A + \Omega
		d J + \Phi d Q \ ,
		\]
		where $M$ is the total black hole mass, $A$ the surface area of
		the horizon, $\Omega$ is the angular velocity, $J$ is the angular momentum, $\Phi$ is the electrostatic potential and $Q$ is the electric charge. For the Reissner-Nordstr\"om black hole one has,
		\bea
		\kappa = \frac{r_+ - r_-}{2r_+^2}, ~\Phi = \frac{Q}{r_+},~ A = 4\pi r_+^2 \ ,
		\ee
	\end{description}
	\item[Second Law:]
	\begin{description}
		$d A \ge 0$
	\end{description}
	\item[Third Law:] 
	\begin{description}
		$\kappa = 0$ is not
		achievable by any physical process.
	\end{description}
\end{description}

Also in 1972, Jacob Bekenstein \cite{bekenstein} conjectured that black holes have an entropy proportional to their surface area due to information loss effects. Stephen Hawking was able to prove this conjecture in 1974, when he applied quantum field theory to black hole spacetimes and showed that black holes will radiate particles with a black-body spectrum, causing the black hole to evaporate \cite{hradiation}. We can describe the interaction of some quantum matter with gravity by quantising the matter on a fixed, classical gravitational background. That is, we can try quantising the matter, but not gravity. This will work only if the gravitational field is weak, outside a large black hole, but not near the singularity. Using this approach, Hawking showed that by studying quantum matter fields on a classical black hole background, we find that, when the matter fields are initially in the vacuum, there is a steady stream of outgoing radiation, which has a temperature determined by its mass and charge, known as Hawking radiation. This decreases the mass of the black holes, so eventually, the black hole will disappear and the temperature increases as the black hole shrinks, which implies that the black hole will disappear abruptly, in a final flash of radiation. The black hole emits a blackbody radiation with temperature,
\bea
T_{BH} = \frac{\kappa\hbar}{2\pi} = \frac{\hbar c^3}{8\pi G M k_{B}} \ ,
\ee
where $\hbar$ is the reduced Planck constant, $c$ is the speed of light, $k_B$ is the Boltzmann constant, $G$ is the gravitational constant, and $M$ is the mass of the black hole. For the Reisnner-Nordstr\"om metric we have,
\bea
T_{RN} =  \frac{\hbar c^3\sqrt{G^2 M^2 - \kappa_e G Q^2}}{2\pi \kappa_B(2GM(GM+ \sqrt{G^2M^2 - \kappa_e G Q^2}) - \kappa_e G Q^2)} \ .
\ee
where $\kappa_e$ is the Coulomb constant. There is also an analogy between the classical laws governing black holes, and the laws of thermodynamics. Thermodynamics is just an approximate description of the behaviour of large groups of particles, which works because the particles obey statistical mechanics.  Since black holes have a non-zero temperature, the classical laws of black holes can be interpreted in terms of 
the laws of thermodynamics applied to black holes. We expect there to be
some more fundamental (quantum) description of black holes, which in particular would give some understanding of black hole microstates and whose statistical properties
give rise to the classical laws governing black holes in terms of statistical mechanics.\footnote{String theory and the AdS/CFT correspondence provide a way to understand black hole entropy \cite{micro1} by matching the Bekenstein-Hawking entropy with counting black hole microstates $\Omega$, $S_{BH} = S_{micro} \equiv \log \Omega$} The entropy of the black hole can be computed in terms of the area as;
\bea
S_{BH} = \frac{k_{B} A}{4 \ell^2_P} \ ,
\ee 
where $A = 4\pi r_+^2$ is the area of the event horizon and $\ell_P = \sqrt{G\hbar/c^3}$ is the Planck length.  For the Reisnner-Nordstr\"om metric we have,
\bea
S_{RN} = \frac{\pi \kappa_B}{G \hbar c} (GM + \sqrt{G^2 M^2 -\kappa_e G Q^2})^2 \ .
\ee

\section{Symmetry in physics}
At the centre of fundamental physics stands the concept of symmetry,
often implemented using the mathematics of group theory and Lie algebras.
A symmetry transformation is a mathematical transformation which leaves
all measurable quantities intact.  Spacetime symmetries correspond to transformations on a field theory acting explicitly on the spacetime coordinates
\bea
x^{\mu} \rightarrow x'^{\mu}(x^\nu),~~ \mu,\nu=0,1,2,\dots,d-1 \ .
\ee
A Killing vector $X$ defined by,
\bea
(\mathcal{L}_X g)_{\mu \nu} = \nabla_{\mu}X_{\nu} + \nabla_{\nu}X_{\mu} = 0 \,
\ee
is a coordinate
independent way of describing spacetime symmetries. There is a certain maximal
amount of symmetry that the geometry can have in a given number of
dimension and the number of linearly independent Killing vectors $N \leq \frac{d(d+1)}{2}$ tells you what amount of symmetry you have. The space of all Killing vector fields form the Lie algebra ${\mathfrak{g}} = Lie(G)$ of the isometry group $G = Isom({\cal{M}})$ of a (semi) Riemannian manifold $\cal{M}$. The group of isometries of such a connected smooth manifold is always a Lie group. However, a Lie group can also include subgroups of discrete isometries that, cannot be represented by continuous isometries and thus they have no associated Killing vectors. For example, $\mathbb{R}^3$ equipped with the standard metric has a Lie group of isometries which is the semidirect product of  rotations $O(3)$ and space translations $\mathbb{R}^3$ around a fixed point. The first subgroup of isometries, $O(3)$, admits a discrete subgroup given by $\{I,−I\}$, but the spatial inversion $−I$ cannot be associated with any Killing field. 

Internal symmetries correspond to transformations of the different fields of the field theory,
\bea
\Phi^a (x) \rightarrow M^{a}{}_{b} \Phi^b (x) \ .
\ee
The indices $a, b$ label the corresponding fields. If $M^{a}{}_{b}$ is constant then then we have a global symmetry, if $M^{a}{}_{b}(x)$ is spacetime dependent then we have a local symmetry. The process of making a global symmetry into a local symmetry with $M^{a}{}_{b} \rightarrow M^{a}{}_{b}(x)$ is known as gauging. If the Lagrangian (or the action) of the theory is invariant under the symmetry
transformation we say that the theory has the symmetry. This doesn’t necessarily mean that the
solution has that symmetry - there are a variety of ways to break a
symmetry. Suffice to say, that the solution
has a symmetry if it is invariant under the symmetry transformation. Consider as an example the massless complex scalar field $\phi$ with Lagrangian,
\bea
\label{lagra1}
{\cal{L}} = \partial_{\mu}\phi \partial^{\mu}{\bar{\phi}} \ .
\ee
This theory has a $U(1)$ symmetry, acting as constant phase shift on the field $\phi$ as,
\bea
\label{sym}
\phi(x) \rightarrow e^{i \Lambda}\phi(x)
\ee
To construct a theory with local symmetry, we promote $\Lambda \rightarrow \Lambda(x)$ which allows the phase to depend on the space time coordinate. (\ref{sym}) is no longer a symmetry of the Lagrangian (\ref{lagra1}), in order to repair the symmetry we need to introduce the covariant derivative of a gauge field over the spacetime manifold,
\bea
D_{\mu}\phi(x) \equiv \partial_{\mu}\phi(x) -i A_{\mu} \phi(x) \ ,
\ee
where $A_{\mu}$ transforms under a local $U(1)$ gauge transformation as,
\bea
A_{\mu} \rightarrow A_{\mu} + \partial_{\mu} \Lambda(x) \ ,
\ee
and we can thus replace the ordinary derivative with this covariant derivative in the Lagrangian (\ref{lagra1}) to get,
\bea
{\cal{L}} = D_{\mu}\phi D^{\mu}{\bar{\phi}} \ .
\ee

\subsection{Noether's theorem}
Noether's theorem, which was proven by Emmy Noether in 1915, relates the global symmetries  of the action of a physical system to the conservation laws. The actions used in the original work can have any number of derivatives and still under certain conditions one finds a conservation of the Noether current, once the Euler-Lagrange field equations are satisfied. This is true even when the relevant actions have an infinite number of derivatives \cite{townsend2}. Noether transformations laws acting on the fields can be linear or nonlinear, or some combination of these. As a consequence of Noether's theorem, symmetries label and classify particles according to the different conserved quantum numbers identified by the spacetime and internal symmetries (mass, spin, charge, colour, etc.). In this regard symmetries actually, define an elementary particle according to the behaviour of the corresponding field with respect to the corresponding symmetry. This property was used to classify particles not only as fermions and bosons but also to group them in multiplets.

\begin{comment}
\subsection{Killing vectors}
\subsection{Lie derivative}
\subsection{Lie algebra}
\end{comment}

\subsection{Poincar\'e symmetry}
The Poincare group corresponds to the basic symmetries of special relativity, it acts on spacetime coordinates $x^\mu$ as follows:
\bea
x^{\mu} \rightarrow x'^{\mu} = \Lambda^{\mu}{}_{\nu} x^{\nu} + a^\mu \ ,
\ee
where $\Lambda^{\mu}{}_{\nu}$ correspond to the Lorentz transformations, and $a^{\mu}$ the translations. The Poincaré algebra is the Lie algebra of the Poincaré group. More specifically, the proper ($\det \Lambda =1$), orthochronous ($\Lambda^{0}{}_{0} \geq 1$) part of the Lorentz subgroup, $SO^+(1, 3)$, connected to the identity. Generators for the Poincaré group are $M_{\mu \nu}$ and $P_{\mu}$ and the Poincare algebra is given by the commutation relations:
\bea
[P_\mu, P_\nu] &=& 0 \ ,
\cr
[M_{\mu \nu}, P_{\rho}] &=& i(\eta_{\mu \rho}P_{\nu} - \eta_{\nu \rho}P_{\mu}) \ ,
\cr
[M_{\mu \nu}, M_{\rho \sigma}] &=& i(\eta_{\mu \rho}M_{\nu \sigma} - \eta_{\mu \sigma}M_{\nu \rho} - \eta_{\nu \rho}M_{\mu \sigma} + \eta_{\nu \sigma}M_{\mu \rho}) \ .
\ee

To include new symmetries we extend the Poincar\'e group via new anticommuting
symmetry generators, and postulate their anticommutation relations. To
ensure that these unobserved symmetries do not conflict with experimental results
obtained thus far we must assume that supersymmetry is spontaneously broken,
allowing the superpartners to be more massive than the energy scales probed thus
far.

\section{Quantum gravity}
\begin{figure}[h!]
	\begin{center}
		\includegraphics[width=350pt]{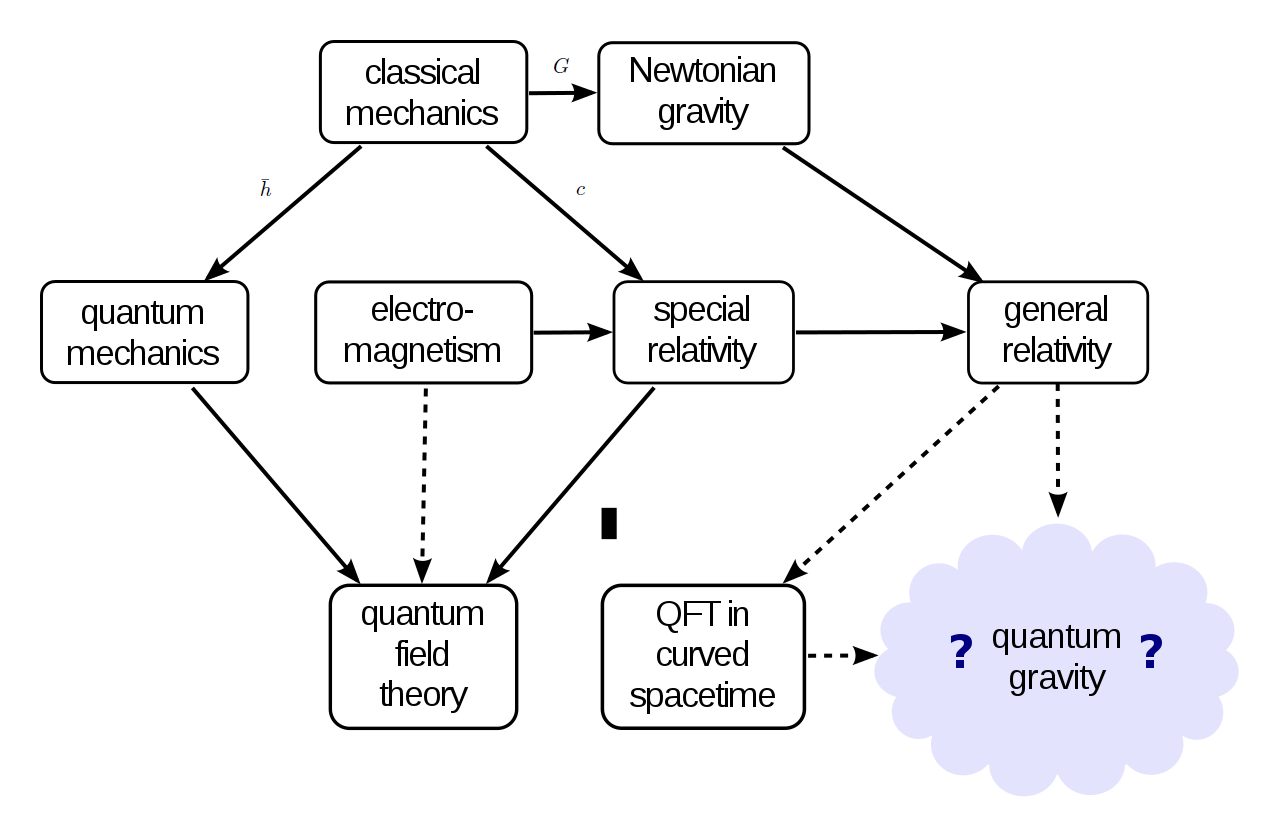}
	\end{center}
	\caption{Diagram showing where quantum gravity sits in the hierarchy of physical theories (credit: \url{https://goo.gl/9RmT3Z})}
	\label{fig:qg}
\end{figure}
Quantising matter fields on a black hole background teaches us a lot about black holes. However, we need a quantum theory of gravity to understand the fundamental principles underlying black hole thermodynamics. On short distance scales, such as near the singularity, we certainly need to use a quantum theory to describe the collapsing matter as general relativity provides no basis for working out what happens next as the equations no longer make sense or have any predictive power. It is hoped that this failure of the classical theory can be cured by quantising gravity. A complete quantum theory of gravity would also be able
to explain the nature of the end-point of black 
hole evaporation. The same problem crops up when trying to explain the big bang, which is thought to have started with a singularity. 
\begin{figure}[h!]
	\begin{center}
		\includegraphics[width=430pt]{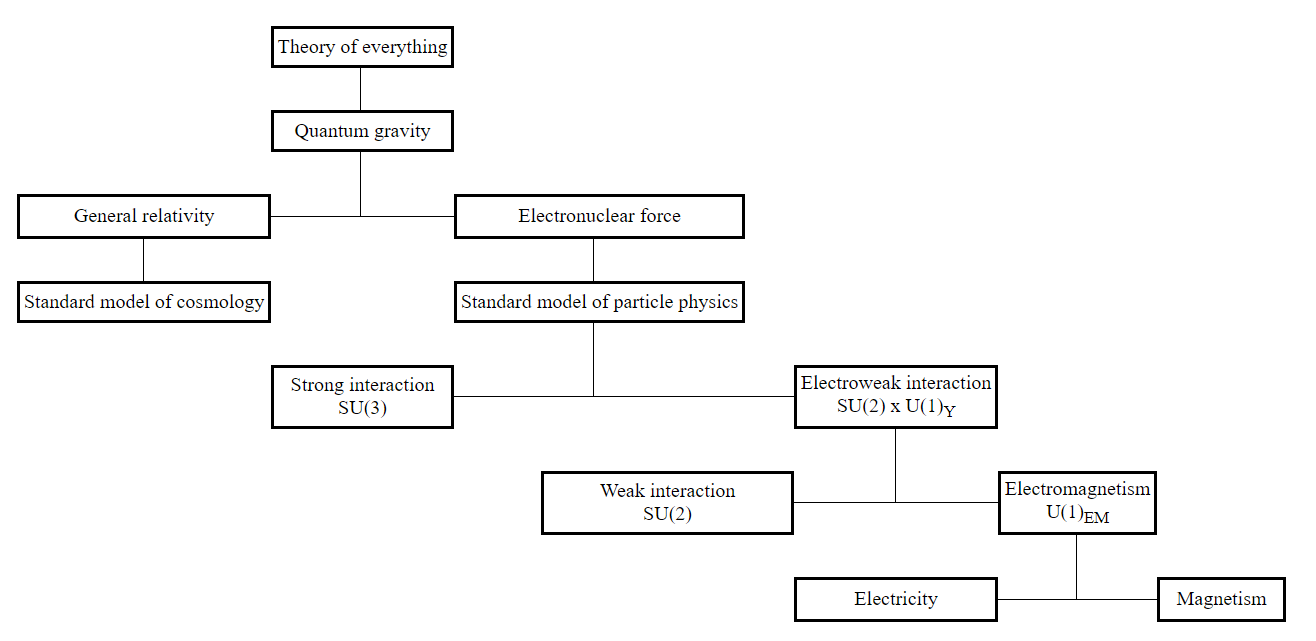}
	\end{center}
	\caption{Theory Of Everything}
	\label{fig:toe}
\end{figure}

Another motivation for quantum gravity would be for the unification of the laws of nature. The standard model describes strong, weak and electromagnetic force with gauge group $SU(3) \times SU(2) \times U(1)$ and the experimental verification of quantum electrodynamics has been the most stringently tested theories in physics. General relativity which describes gravity is another theory which has been rigorously tested in the very strong field limit, observing to date no deviations from the theory. Naive attempts at reconciling these theories have failed, yet a true unified theory of everything must include all the forces of nature. In Figure \ref{fig:toe}, each unification step leads one level up and electroweak unification occurs at around 100 GeV, grand unification is predicted to occur at $10^{16}$ GeV, and unification of the GUT force with gravity is expected at the Planck energy, roughly $10^{19}$ GeV.

Formulating a theory of quantum gravity has turned out to be one of the hardest problems of theoretical physics. Attempts to incorporate gravity into the quantum framework as if it were just another force like electromagnetism or the nuclear forces have failed; such models are riddled with infinities since gravity is non-renormalizable in quantum field theory \cite{goroff} and do not make any sense physically. Until now, the question of how a proper quantum theory of gravity should look has not found a complete answer.  One theory which offers some hope, particularly for understanding black holes\footnote{It turns out that electrically charged extreme $D = 4$ black holes
	are approximate descriptions of higher dimensional ($D = 5$) fundamental
	string backgrounds \cite{exstr1}.}, is string theory. 

%Four fundamental forces of nature: Strong, Weak, Electromagnetism and Gravity. Standard model gauge group $SU(3) \times SU(2) \times U(1)$.

\subsection{Kaluza-Klein unification and compactification}
\begin{figure}[h!]
	\begin{center}
		\includegraphics[width=150pt]{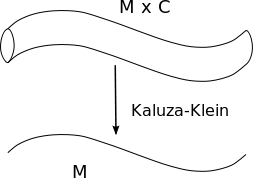}
	\end{center}
	\caption{Kaluza-Klein compactification (credit: \url{https://goo.gl/cXymk1})}
	\label{fig:KK}
\end{figure}
The Kaluza-Klein theory is a {\it classical} unified field theory of gravitation and electromagnetism. The crucial idea of compactification is built around the idea of a fifth dimension beyond the usual four of space and time is considered an important precursor to string theory and supergravity. Compactification is one way of modifying the number of dimensions in a physical theory, where some of the extra dimensions are assumed to ``close up" on themselves to form circles. In the limit where these curled up dimensions become very small, one obtains a theory in which spacetime effectively has a lower number of dimensions. 

The original idea came from Theodor Kaluza in 1919 which included extension of general relativity to five dimensions \cite{kaluza}. The metric for this theory ${\hat g}_{A B}$ has 15 components, ten components are identified with the four-dimensional spacetime metric $g_{\mu \nu}$, four components with the electromagnetic vector potential $A_{\mu}$, and one component with an unidentified scalar field $\Phi$ sometimes called the ``dilaton". More precisely we have,
\bea
{\hat g}_{\mu \nu} = g_{\mu \nu} + \Phi^2 A_{\mu} A_{\nu},~~ {\hat g}_{5 \nu} = {\hat g}_{\nu 5} = \Phi^2 A_{\nu},~~ {\hat g}_{5 5} = \Phi^2 \ .
\ee
Kaluza also introduced the ``cylinder condition", which states that no component of the five-dimensional metric depends on the fifth dimension. Without this assumption, the field equations of five-dimensional relativity become intractable as they grow in complexity. The five-dimensional (vacuum) Einstein equations yield the four-dimensional Einstein field equations, the Maxwell equations for the electromagnetic field $A_{\mu}$, and an equation for the scalar field $\Phi$ given by,
\bea
R_{\mu \nu} - \frac{1}{2}g_{\mu \nu}R &=& \frac{1}{2}\Phi^2 \bigg(F_{\mu \alpha}F_{\nu}{}^{\alpha} - \frac{1}{4}F_{\alpha \beta}F^{\alpha \beta}\bigg) + \Phi^{-1}\bigg(\nabla_{\mu}\nabla_{\nu} \Phi - g_{\mu \nu}\nabla^{\lambda}\nabla_{\lambda}\Phi \bigg) \ ,
\cr
\nabla^{\beta}( \Phi^3 F_{\alpha \beta} )&=& 0 \ ,
\cr
\nabla^{\mu}\nabla_{\mu} \Phi  &=& \frac{1}{4}\Phi^3 F_{\alpha \beta}F^{\alpha \beta} \ .
\ee
The Einstein equation and field equation for $A_{\mu}$ has the form of the vacuum Maxwell equations if the scalar field is constant. It turns out that the scalar field cannot be set to a constant without constraining the electromagnetic field $A_{\mu}$. The earlier treatments by Kaluza and Klein did not have an adequate description of the scalar field $\Phi$, and did not realize the implied constraint.

\subsection{Supersymmetry}
\begin{figure}[h!]
	\begin{center}
		\includegraphics[width=250pt]{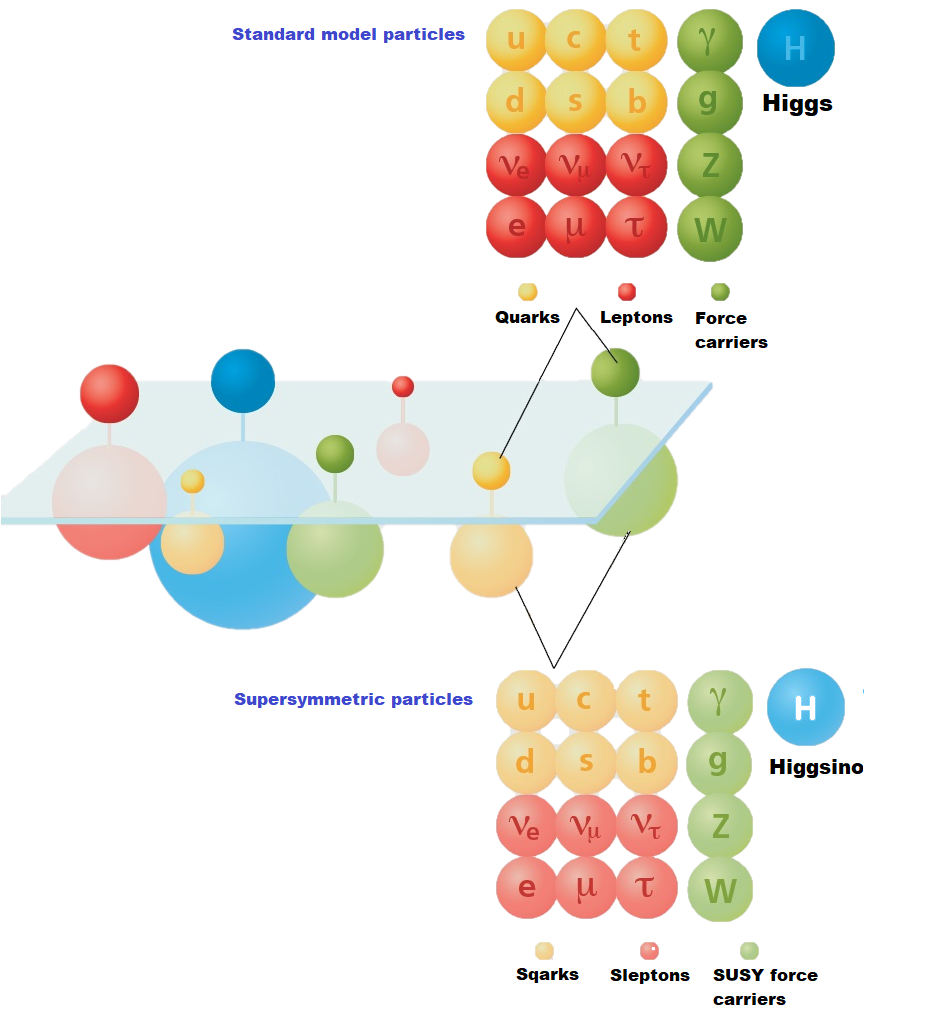}
	\end{center}
	\caption{The Standard Model and Supersymmetry (credit: \url{https://goo.gl/QcqXih})}
	\label{fig:susy}
\end{figure}
Supersymmetry introduces a symmetry between fermions and bosons and is an attractive solution to the dark matter problem \cite{jungman}. As
a consequence of supersymmetry, all fermion particles get their own boson
superpartner, and all boson particles get their own fermion superpartner.
It turns out that it is possible to have symmetry
transformations where the symmetry parameter is not a phase shift or some Lorentz scalar, but actually a spinor. The matter fields of the fermions are spinors, so when we take the symmetry parameter to be a spinor, the symmetry transformation necessarily relates the bosons to fermions, and vice versa. Schematically this can be represented as,
\bea
\delta_1 B \sim \bar{\epsilon_1}F,~~
\delta_2 F \sim \partial B \epsilon_2 \ ,
\ee
where $F$ and $B$ denote the fermionic and bosonic fields respectively. The anticommuting parameter $\epsilon$ known as a spinor must have dimension $[\epsilon] = -\frac{1}{2}$ in mass units since $[B] =1$ and $[F] = \frac{3}{2}$, which implies the presence of the derivative operator in the second transformation to ensure dimensional consistency. If we consider the effect of successive supersymmetry variations,
\bea
\{\delta_1, \delta_2 \} \sim a^{\mu}\partial_{\mu}B,~~ a^{\mu} = \epsilon_2 \gamma^{\mu} {\bar{\epsilon}}_1
\ee
Supersymmetry is not just a symmetry between bosons and fermions, it is also an extension of the Poincar\'e symmetry, which is the symmetry of Minkowski spacetime. When talking about supersymmetry, one frequently considers rigid supersymmetry which is a global symmetry, whose symmetry parameters do not
depend on the point in spacetime. If we gauge it and make the global symmetry into a local one with $\epsilon \rightarrow \epsilon(x)$, it is possible to supersymmetrize general relativity by combining the standard bosonic metric with a gravitino, as well as more general bosonic and fermionic matter terms.

For this reason, rather than talking about local supersymmetry or gauged supersymmetry, the established term is supergravity.  In supergravity theories, the vanishing of the supersymmetry variations when we set the fermions to zero are known as the Killing spinor equations. Similarly
to symmetry, there is a maximum amount of possible supersymmetry, and the number of linearly independent Killing spinors determines how much of that supersymmetry is realized for a given bosonic solution.

\subsection{Super-poincar\'e symmetry}
In order to have a supersymmetric extension of the Poincar\'e algebra, we first introduce graded algebras. Let ${\cal O}_a$ be operators of a Lie algebra, then
\bea
{\cal O}_a {\cal O}_b = (-1)^{\eta_a \eta_b} {\cal O}_b {\cal O}_a = iC^{e}{}_{a b}{\cal O}_e \ ,
\ee
where the gradings take the values $\eta_a = 0$ for a bosonic generator ${\cal O}_a$ and $\eta_a = 1$ for a fermionic generator ${\cal O}_a$. For supersymmetry, generators are the Poincar\'e generators $M_{\mu \nu}$, $P_{\mu}$ and the spinor generators $Q^{A}_{\alpha}$, ${\bar{Q}}^{A}_{{\dot \alpha}}$ where $A=1,\dots,\cal{N}$.
The simplest supersymmetric extension with ${\cal N} = 1$ is defined by the following commutation relations,
\bea
[Q_{\alpha}, P^{\mu}] &=& 0 \ ,
\cr
\{Q_{\alpha}, {\bar{Q}}_{{\dot \beta}} \} &=& 2 (\sigma^\mu)_{\alpha {\dot \beta}} P_{\mu} \ ,
\ee
where $P_{\mu}$ are the generators of translation as before and $\sigma^\mu$ are Pauli matrices. As before $\mu$ is a spacetime index, but $\alpha$ is a spinor index and has $\alpha = 1,2$ where a dot over the index transforms according to an inequivalent conjugate spinor representation. Combined with the Poincar\'e algebra, this is a closed algebra since all the super-Jacobi identities are satisfied. 
For the case of ${\cal N} > 1$, it is known as extended supersymmetry where each of the generators will be labelled by the index $A$ and will contain additional (anti)-commutation relations with central charges $Z^{A B}$ given by,
\bea
\{Q_{\alpha}^{A}, {\bar{Q}}_{{\dot \beta}}^B \} &=& 2 (\sigma^\mu)_{\alpha {\dot \beta}} P_{\mu} \delta^{A B} \ ,
\cr
\{Q_{\alpha}^{A},Q_{\beta}^B \} &=& \epsilon_{\alpha \beta}Z^{A B} \ .
\ee
The central charges are anti-symmetric $Z^{A B} = -Z^{B A}$ and commute with all the generators. For model building, it has been assumed that almost all the supersymmetries would be broken in nature leaving just ${\cal{N}}=1$ supersymmetry.

\subsection{Superstring theory}

String theory is a theoretical framework in which the point-like particles are replaced by one-dimensional objects called strings. It describes the dynamics of these strings; how they propagate through space and interact with each other. On distance scales larger than the string scale which on the order of the Planck length ($\sim 10^{−35}$ meters), are where effects of quantum gravity become significant. A string looks just like an ordinary particle, with its mass, charge, and other properties determined by the vibrational modes of the string. One of the many vibrational states of the string corresponds to the graviton, a quantum mechanical particle that carries gravitational force, which makes it a perfect candidate for a theory of quantum gravity. String theory was first studied in 1969-1970 in the context of the strong nuclear force, however, this description made many predictions that directly contradicted experimental findings and it was abandoned in favour of quantum chromodynamics. 

The earliest version of string theory was known as bosonic string theory which was discovered by Schwarz and Scherk \cite{schwarz}, and independently Yoneya \cite{yoneya} in 1974. It was realized that the very properties that made string theory unsuitable to describe nuclear physics, as it contained a bosonic field of spin-2, made it a promising candidate for a quantum theory of gravity. They studied the boson-like patterns of string vibration and found that their properties exactly matched those of the graviton. 
In order to detail the action, we recall that the dynamics of a point-like particle of mass $m$ moving in Minkowski spacetime is described by the action,
\bea
S = -m \int ds \ .
\ee
Consider now a string of length $\ell_s$ moving in Minkowski spacetime. The dynamics are governed by the Nambu-Goto action,
\bea
S = -T \int d{\cal A} \ ,
\ee
where $d{\cal A}$ is the infinitesimal area element of the string world-sheet. The Polyakov action which is classically equivalent describes a two-dimensional sigma model given by,
\bea
S = -T\int d^2\sigma \sqrt{-\gamma} \gamma^{a b}g_{\mu \nu}\nabla_{a}X^{\mu} \nabla_{b}X^{\nu}
\ee
where $\sigma^a$ are coordinates on the string world-sheet with metric $\gamma_{a b}$ and spacetime metric $g_{\mu \nu}$. The requirement that unphysical states with negative norm disappear implies that the dimension of spacetime is 26, a feature that was originally discovered by Claud Lovelace in 1971 \cite{lovelace} but the open string spectrum still contained a tachyon as a ground state. Furthermore, any unified theory of physics should also contain fermions and it turns out including fermions provides a way to eliminate the tachyon from the spectrum. Investigating how a string theory may include fermions in its spectrum led to the invention of supersymmetry \cite{gervais}. It was later developed into superstring theory and its action looks like,
\bea
S = -T\int d^2\sigma \sqrt{-\gamma}\bigg(\gamma^{a b}g_{\mu \nu}\nabla_{a}X^{\mu} \nabla_{b}X^{\nu} - i g_{\mu \nu}{\bar \psi}^{\mu}\Gamma^{a}\nabla_{a} \psi^{\nu} \bigg) \ ,
\ee
where $\psi^{\mu}$ is a fermion, and $\Gamma^{a}$ the gamma matrices in the 2-dimensional world-sheet. Quantum mechanical consistency now requires that the dimension of spacetime be 10 for superstring theory, which had been originally discovered by John H. Schwarz in 1972 \cite{schwarz2}. 

Consider a closed bosonic string in a more general background consisting of massless states $(\Phi, h_{\mu \nu}, B_{\mu \nu})$ generated by closed strings in the bulk \cite{callan}. The resulting action is called the non-linear sigma model,
\bea
S = -T\int d^2\sigma \bigg[\bigg(\sqrt{-\gamma} \gamma^{a b}g_{\mu \nu} - \epsilon^{a b}B_{\mu \nu}\bigg)\nabla_{a}X^{\mu} \nabla_{b}X^{\nu} - \alpha' \sqrt{-\gamma} \Phi {\cal {R}}^{(2)}\bigg] \ ,
\ee
where the background metric is given by $g_{\mu \nu} = \eta_{\mu \nu} + h_{\mu \nu}$ and ${\cal {R}}^{(2)}$ is the Ricci-scalar of the worldsheet metric $\gamma$. The $\beta$-functionals associated to the ``coupling constants" $\Phi$, $h_{\mu \nu}$ and $B_{\mu \nu}$ vanish and in the lowest order in $\alpha'$ are given by \cite{callan},
\bea
\beta^{(\Phi)} &=& R + \frac{1}{12}H_{\mu \nu \rho}H^{\mu \nu \rho} - 4\nabla^{\rho}\nabla_{\rho}\Phi + 4 \nabla^{\rho}\Phi \nabla_{\rho}\Phi + O(\alpha')= 0 \ ,
\nonumber \\
\beta^{(h)}_{\mu \nu} &=& R_{\mu \nu} - \frac{1}{4}H_{\mu \rho \sigma}H_{\nu}{}^{\rho \sigma} + 2 \nabla_{\mu}\nabla_{\nu} \Phi + O(\alpha') = 0 \ ,
\nonumber \\
\beta^{(B)}_{\mu \nu} &=& \frac{1}{2}\nabla^{\rho}H_{\rho \mu \nu} - H_{\mu \nu \rho}\nabla^{\rho}\Phi + O(\alpha') = 0 \ ,
\ee
where $R_{\mu \nu}$ is the Ricci tensor of the space-time, and $H_{\mu \nu \rho} = 3\nabla_{[\mu}B_{\nu \rho]}$. The form of these equations suggests they can be interpreted as equations of motion for the background fields. Indeed they can also be obtained from the following low energy effective action,
\bea
\label{laction}
S = \int d^{26} x \sqrt{-g} e^{-2\Phi}\bigg(R + 4\nabla^{\mu}\Phi \nabla_{\mu}\Phi - \frac{1}{12}H_{\mu \nu \rho}H^{\mu \nu \rho} + O(\alpha')\bigg) \ .
\ee
So far this is only for bosonic string theory and in $D=26$, but this can be extended to the supersymmetric case in $D=10$. It turns out that the low energy effective descriptions of superstring theories in $D=10$, except type I, have one part in common known as the `common sector', which is the Neveu-Schwarz (NS-NS) sector given by the ten dimensional analogue of (\ref{laction}). Five consistent versions of superstring theory were developed before it was conjectured that they were all different limiting cases of a single theory in eleven dimensions known as M-theory. The low energy limits of these superstring theories coincide with supergravity.

\subsubsection{The first superstring revolution}

The first superstring revolution began in 1984 with the discovery of anomaly cancellation in type I string theory via the Green-Schwarz mechanism \cite{green} and the subsequent discovery of the heterotic string was made by David Gross, Jeffrey Harvey, Emil Martinec, and Ryan Rohm in 1985 \cite{gross}. Also in the same year, it was realized by Philip Candelas, Gary Horowitz, Andrew Strominger, and Edward Witten that to obtain $N=1$ supersymmetry, the six extra dimensions need to be compactified on a Calabi-Yau manifold \cite{candelas}. By then, five separate superstring theories had been described: type I, type II (IIA and IIB) \cite{green2}, and heterotic ($SO(32)$ and $E_8 \times E_8$) \cite{gross}.  Eric Bergshoeff, Ergin Sezgin, and Paul Townsend in 1987 also showed the existence of supermembranes instead of superstrings in eleven dimensions \cite{bergshoeff}.

\subsubsection{The second superstring revolution}

\begin{figure}[h!]
	\begin{center}
		\includegraphics[width=180pt]{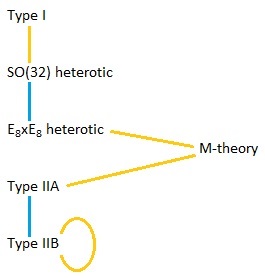}
	\end{center}
	\caption{A diagram of string theory dualities. Yellow lines indicate S-duality. Blue lines indicate T-duality. (credit: \url{https://goo.gl/j771v7})}
	\label{fig:dual}
\end{figure}

In the early 90s, Edward Witten and others found strong evidence that the different superstring theories that were discovered a decade earlier were different limits of an 11-dimensional theory that became known as M-theory \cite{witten2, duff}. The different versions of superstring theory were unified, by new equivalences from dualities and symmetries. These are known as the S-duality, T-duality, U-duality, mirror symmetry, and conifold transitions. In particular, the S-duality shows the relationship between type I superstring theory with heterotic $SO(32)$ superstring theory, and type IIB theory with itself. The T-duality also relates type I superstring theory to both type IIA and type IIB superstring theories with certain boundary conditions and the U-duality is a combination of the S-duality and T-duality transformations. The mirror symmetry also shows that IIA and IIB string theory can be compactified on different Calabi-Yau manifolds giving rise to the same physics and the conifold transitions details the connection between all possible Calabi-Yau manifolds.

\begin{figure}[h!]
	\begin{center}
		\includegraphics[width=250pt]{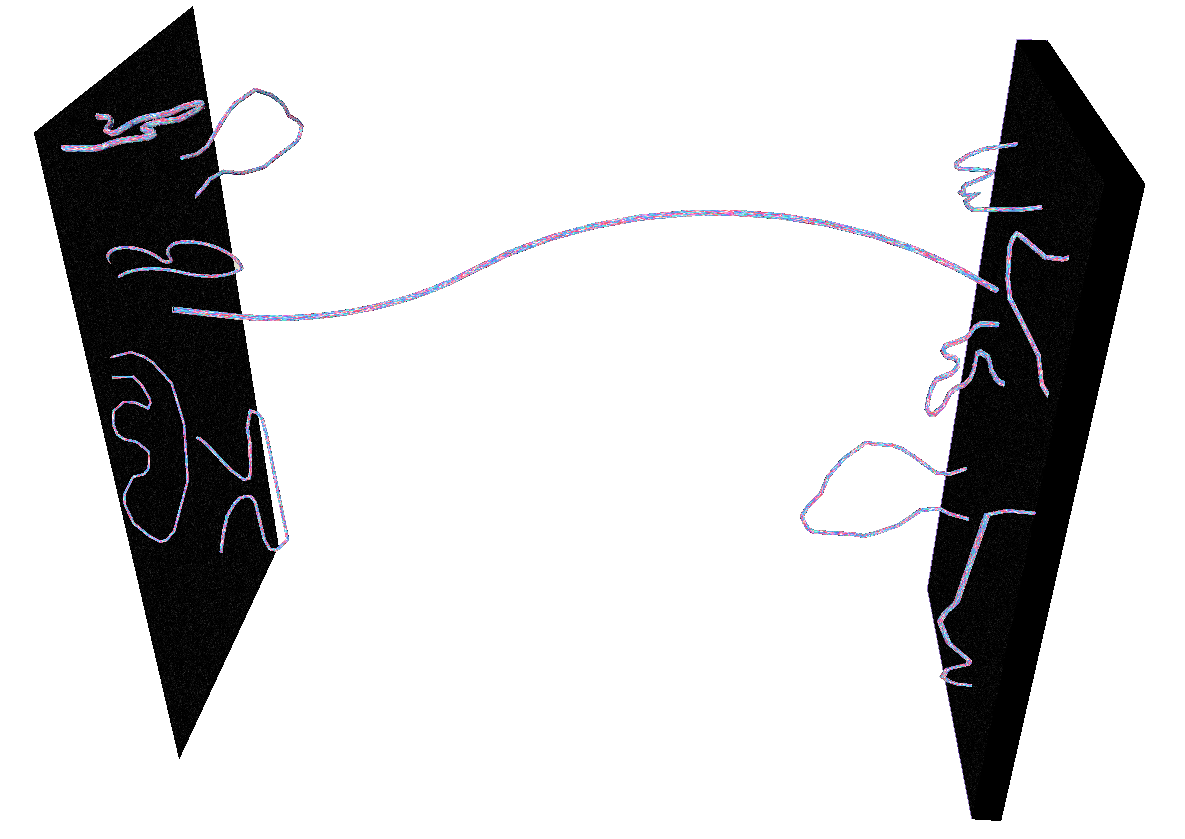}
	\end{center}
	\caption{Open strings attached to a pair of D-branes (credit: \url{https://goo.gl/nvdmgM})}
	\label{fig:dbrane}
\end{figure}

In 1995, Joseph Polchinski discovered that the theory requires the inclusion of higher-dimensional objects, called D-branes\footnote{Where ``D" in D-brane refers to a Dirichlet boundary condition on the system}\cite{polchinski}. These are the sources of electric and magnetic Ramond–Ramond (R-R) fields that are required by the string duality. A brane is a physical object that generalizes the notion of a point particle to higher dimensions. A point particle can be viewed as a brane of dimension zero or a 0-brane, while a string is a brane of dimension one or a 1-brane. It is also possible to consider higher-dimensional branes and in dimension $p$, they are known as $p$-branes. The word brane comes from the word ``membrane" which refers to a two-dimensional brane or 2-brane. Branes are dynamical objects which can propagate through spacetime according to the rules of quantum mechanics, and they have mass and can have other attributes such as charge. A $p$-brane sweeps out a $(p+1)$-dimensional volume in spacetime called its world volume. D-branes are an important class of branes that arise when one considers open strings, and it when propagates through spacetime, it's endpoints are required to lie on a D-brane.  

Another important discovery, known as the AdS/CFT correspondence, relates string theory to certain quantum field theories and has led to many insights in pure mathematics. It was first conjectured in 1997-1998 by Juan Maldacena \cite{maldacena, maldacena2}. In particular, he conjectured a duality between type IIB string theory on $AdS_5 \times S^5$ and $N = 4$ supersymmetric Yang-Mills theory, a gauge theory in four-dimensional Minkowski spacetime. An interesting property of the AdS/CFT correspondence is the duality between strong and weak coupling; both theories describe the same physics through a dictionary that relates quantities in one theory to quantities in the other, so both can be used to calculate
the same physical quantities e.g calculating quantities in a strongly coupled regime of one theory by doing a calculation of the desired dual quantity in the weakly coupled regime of the other theory. It is also a realization of the holographic principle\footnote{Describing a $(d + 1)$-dimensional gravity theory in terms of a lower $d$-dimensional system is reminiscent of an optical hologram that stores a three-dimensional image on a two dimensional photographic plate} which is believed to provide a resolution of the black hole information paradox \cite{maldacena3} and has helped elucidate the mysteries of black holes suggested by Stephen Hawking.

\begin{comment}
\begin{figure}[h!]
\includegraphics[width=250pt]{worldsheet.png}
\caption{Wiki: World line, worldsheet, and world volume, as they are derived from particles, strings, and branes.}
\label{fig:worldsheet}
\end{figure}
\end{comment}

\section{Supergravity}
Supergravity is often called a ``square root" of general relativity. Indeed,
a supersymmetric extension of the Poincar\'e algebra is reminiscent of Dirac's procedure of
obtaining a spin-$\tfrac{1}{2}$ wave equation from the scalar wave equation. Supergravity has played an important role in theoretical physics, merging the theory of general relativity and supersymmetry. In fact, supergravity arises naturally when we promote supersymmetry to a local (gauge) symmetry \cite{wess}. A bosonic sector of extended supergravities, apart from the graviton, contains scalar and vector fields. In ${\cal N} = 2,  D=4$ supergravity, the bosonic sector is the usual Einstein-Maxwell theory. Supergravity is also understood to be the low-energy limit of string theory, that is if we truncate to the massless modes. In supergravity theories, supersymmetry transformations are generated by a set of spinors $\epsilon_{I}(x)$ for $I = 1,\dots\,{\cal{N}}$, where ${\cal{N}}$ is the number of supersymmetries of the given supergravity. In $D=4$, the spinors $\epsilon_{I}(x)$ can be taken to be Weyl or Majorana, and we have $2{\cal{N}}$ complex or $4{\cal{N}}$ real associated charges. Such transformations schematically have the form \cite{wess2, volkov, akulov},
\bea
\delta B = \bar{\epsilon}(B + \bar{F}F)F,~~ \delta F = \partial \epsilon + (B + \bar{F}F) \epsilon F \ ,
\ee
where $F$ and $B$ denote the fermionic and bosonic fields respectively. Since all Supergravity theories will contain at least a graviton $g$, which is a bosonic field, it will also have a corresponding gravitino $\psi$, which is a fermionic field. Such theories will have the transformations,
\bea
\delta e^{a}_{M} = \alpha \bar{\epsilon} \Gamma^a\psi_{M},~~ \delta \psi_{M} = \nabla_{M} \epsilon + \Psi_{M}\epsilon,~~ \alpha \in \mathbb{C} \ ,
\ee
and possibly additional transformations for the other bosonic and fermionic fields of the supergravity theory in question, where $\alpha$ depends on the particular supergravity and $\Psi_M$ depends on the fluxes of that theory. When looking for classical supergravity solutions, we will not get any constraints from the variation of the bosons since these correspond to vanishing fermion fields and become trivial, but the variation of fermions gives us differential and possibly algebraic equations for the spinor $\epsilon$ which are called Killing spinor equations (KSEs) and take the form,
\bea
\delta \psi_M \equiv {\cal{D}}_{M}\epsilon = 0,~~ \delta \lambda_I \equiv {\cal{A}}_I \epsilon = 0 \ ,
\ee
where ${\cal{D}}_{M}$ is known as the supercovariant derivative and $\delta \lambda_I$ collectively denote the variations of the additional fermions in theory pertaining to the algebraic conditions. The integrability conditions of the KSEs of a particular Supergravity theory can also be written in terms of the field equations and Bianchi identities of that theory. Also, if $\epsilon$ is a Killing spinor, one can show for all supergravity theories using a $Spin$-invariant inner product that,
\bea
K =  \langle \epsilon, \Gamma^{\mu} \epsilon \rangle \partial_{\mu} \ ,
\ee
is a Killing vector. If we take the inner product given by $\langle \epsilon, \Gamma_{\mu} \epsilon \rangle = \bar{\epsilon}\Gamma_{0}\Gamma_{\mu}\epsilon$ with $\epsilon \neq 0$, then the vector $K$ is not identically zero. As an example, let us consider these for Heterotic supergravity; the bosonic fields are the metric $g$, a dilaton field $\Phi$, a 3-form $H$ and a non-abelian 2-form field $F$ and the fermionic fields are the gravitino $\psi_{\mu}$, the dilatino $\lambda$ and gaugino $\chi$ which we set to zero. 
Let us first consider the integrability conditions for the KSEs. The field equations and Bianchi identities are,
\bea
E_{\mu \nu} &=& R_{\mu \nu} + 2 \nabla_\mu \nabla_\nu \Phi
-{1 \over 4} H_{\mu \lambda_1 \lambda_2} H_\nu{}^{\lambda_1 \lambda_2} = 0 \ ,
\nonumber \\
F\Phi &=& \nabla^\mu \nabla_\mu \Phi
- 2 \nabla_\lambda \Phi \nabla^\lambda \Phi
+{1 \over 12} H_{\lambda_1 \lambda_2 \lambda_3}H^{\lambda_1 \lambda_2 \lambda_3} = 0 \ ,
\nonumber \\
FH_{\mu \nu} &=& \nabla^{\rho}(e^{-2\Phi}H_{\mu \nu \rho}) = 0 \ ,
\nonumber \\
FF_{\mu} &=& \nabla^{\nu}(e^{-2\Phi}F_{\mu \nu}) - \frac{1}{2}e^{-2\Phi}H_{\mu \nu \rho}F^{\nu \rho} = 0 \ ,
\nonumber \\
BH_{\mu \nu \rho \sigma} &=& \nabla_{[\mu}H_{\nu \rho \sigma]} = 0 \ ,
\nonumber \\
BF_{\mu \nu \rho} &=& \nabla_{[\mu}F_{\nu \rho]} = 0 \ .
\ee
and the KSEs are,
\bea
{\cal D}_{\mu}\epsilon &=& \nabla_{\mu}\epsilon - \frac{1}{8}H_{\mu \nu \rho}\Gamma^{\nu \rho}\epsilon = 0 \ ,
\nonumber \\
{\cal{A}}\epsilon &=& \nabla_{\mu}{\Phi}\Gamma^{\mu}\epsilon - \frac{1}{12}H_{\mu \nu \rho}\Gamma^{\mu \nu \rho}\epsilon = 0 \ ,
\nonumber \\
{\cal {F}}\epsilon &=& F_{\mu \nu}\Gamma^{\mu \nu}\epsilon = 0 \ ,
\ee
One can also show that the dilaton field equation $F\Phi$ is implied by the other field equations and Bianchi identities by establishing,
\bea
\nabla_{\nu}{(F\Phi)} =  - 2 E_{\nu \lambda}\nabla^{\lambda}{\Phi} 
+ \nabla^{\mu}{(E_{\mu \nu})} - \frac{1}{2}\nabla_{\nu}{(E^{\mu}{}_{\mu})}
- \frac{1}{3}BH_{\nu}{}^{\lambda_1 \lambda_2 \lambda_3} H_{\lambda_1 \lambda_2 \lambda_3} + \frac{1}{4}FH_{\lambda_1 \lambda_2}H_{\nu}{}^{\lambda_1 \lambda_2} \ .
\nonumber \\
\ee
The integrability conditions of the KSEs can be expressed in terms of the field equations and Bianchi identities as follows,
\bea
\Gamma^{\nu}[{\cal{D}}_{\mu},{\cal{D}}_{\nu}]\epsilon - [{\cal{D}}_{\mu},{\cal{A}}]\epsilon &=& \bigg(-\frac{1}{2}E_{\mu \nu}\Gamma^{\nu}  - \frac{1}{4}e^{2\Phi}FH_{\mu \nu}\Gamma^{\nu} - \frac{1}{6}BH_{\mu \nu \rho \lambda}\Gamma^{\nu \rho \lambda} \bigg)\epsilon \ ,
\nonumber \\
\Gamma^{\mu}[{\cal{D}}_{\mu},{\cal{A}}]\epsilon - 2{\cal{A}}^2 \epsilon &=& \bigg(F\Phi - \frac{1}{4}e^{2\Phi}FH_{\mu \nu}\Gamma^{\mu \nu} - \frac{1}{12}BH_{\mu \nu \rho \lambda}\Gamma^{\mu \nu \rho \lambda}\bigg)\epsilon \ ,
\nonumber \\
\Gamma^{\mu}[{\cal{D}}_{\mu}, {\cal{F}}]\epsilon + [{\cal{F}},{\cal{A}}]\epsilon &=& \bigg(-2e^{2\Phi}FF_{\mu}\Gamma^{\mu} + BF_{\mu \nu \rho}\Gamma^{\mu \nu \rho}\bigg)\epsilon \ .
\ee
To show that $K$ is a Killing vector in heterotic supergravity, we consider the supercovariant derivative for a 1-form $K_{\mu}$ is given by,
\bea
{\cal D}_{\mu}K_{\nu} = \nabla_{\mu}K_{\nu} + \frac{1}{2}H_{\mu \nu}{}^{\rho}K_{\rho} \ .
\ee
If we compute the supercovariant derivative directly from the inner product we have,
\bea
{\cal D}_{\mu}K_{\nu} = {\cal D}_{\mu} \langle \epsilon, \Gamma_{\nu} \epsilon \rangle = \langle {\cal D}_{\mu}\epsilon, \Gamma_{\nu} \epsilon \rangle + \langle \epsilon, \Gamma_{\nu} {\cal D}_{\mu}\epsilon \rangle = 0 \ ,
\ee
since $\epsilon$ is a Killing spinor, this implies,
\bea
\nabla_{\mu}K_{\nu} = -\frac{1}{2}H_{\mu \nu}{}^{\rho}K_{\rho} \ ,
\ee
and hence $K$ is a Killing vector since $H$ is skew-symmetric and we have,
\bea
\nabla_{(\mu}K_{\nu)} = 0 \ .
\ee
Investigations into supergravity first began in the mid-1970's through the work
of Freedman, Van Nieuwenhuizen and Ferrara, at around the time that the implications
of supersymmetry for quantum field theory were first beginning to be
understood \cite{freedman}. 
\begin{figure}[h!]
	\begin{center}
		\includegraphics[width=300pt]{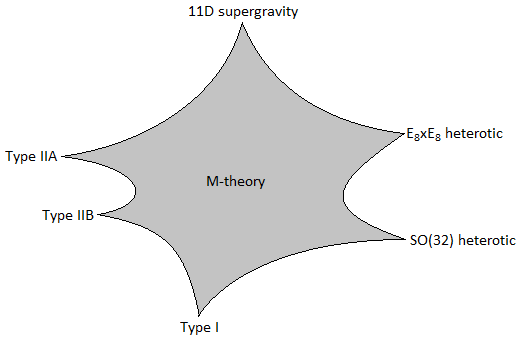}
	\end{center}
	\caption{A schematic illustration of the relationship between M-theory, the five superstring theories, and eleven-dimensional supergravity. (credit: \url{https://goo.gl/5Q4Aw4})}
	\label{fig:sugras}
\end{figure}
Supergravity means that we have a symmetry whose symmetry parameter is a spinor that depends on the position in the spacetime. We are looking for supersymmetric geometries, and we get them by insisting
that the solution is invariant under the supersymmetry transformation and if it is we call the supersymmetry parameter a Killing spinor. On a bosonic background, the Killing spinor is a solution to the KSEs, which are given by demanding that the fermions vanish. In the limit where quantum gravity effects are small, the superstring theories which are related by certain limits and dualities, give rise to different types of supergravity.
The focus of this work has been type IIA, massive IIA and $D=5$ (gauged and ungauged) supergravity coupled to an arbitrary number of vector multiplets. Type IIA supergravity is
a ten-dimensional theory which can be obtained either by taking a certain limit in type IIA string theory or by doing a dimensional reduction of $D=11$ supergravity on $S^1$. $D=5$ ungauged supergravity can be obtained from reducing $D=11$ supergravity on a $T^6$, or more generally a Calabi-Yau compactification \cite{cadavid} and $D=5$ gauged supergravity can be obtained from reducing IIB supergravity on $S^5$ \cite{gutperle}.

\subsection{$D=4, N=8$ supergravity}
In $D=4$, the $N=8$ Supergravity is the most symmetric quantum field theory which involves gravity and a finite number of fields. It can be obtained from $D=11$ supergravity compactified on a particular Calabi Yau manifold. The theory was found to predict rather than assume the correct charges for fundamental particles, and potentially offered to replicate much of the content of the standard model \cite{vannie} and Stephen Hawking once speculated that this theory could be the theory of everything \cite{hawkingd4n8}. This optimism, however, proved to be short-lived, and in
particular, it was not long before a number of gauge and gravitational anomalies
were discovered; seemingly fatal flaws which would render the theory inconsistent \cite{green}. In later years this was initially abandoned in favour of String Theory. There has been renewed interest in the 21st century with the possibility that the theory may be finite. 

\subsection{$D=11, N=1$ supergravity}
$D=11$ supergravity generated considerable excitement as the first potential candidate for the theory of everything. In 1977, Werner Nahm was able to show that $11$ dimensions are the largest number of dimensions consistent with a single graviton, and more dimensions will manifest  particles with spins greater than 2 \cite{nahm} and are thus unphysical. In 1981 Edward Witten also showed $D=11$ as the smallest number of dimensions big enough to contain the gauge groups of the Standard Model \cite{witten}. Many techniques exist to embed the standard model gauge group in supergravity in any number of dimensions like the obligatory gauge symmetry in type I and heterotic string theories, and obtained in type II string theory by compactification on certain Calabi–Yau manifolds. The D-branes engineer gauge symmetries too. 

In 1978 Cremmer, Julia and Scherk (CJS) found the classical action for an 11-dimensional supergravity theory \cite{cremmer}. This remains today the only known classical 11-dimensional theory with local supersymmetry and no fields of spin higher than two. Other 11-dimensional theories which are known and quantum-mechanically inequivalent reduce to the CJS theory when one imposes the classical equations of motion.  In 1980 Peter Freund and M. A. Rubin showed that compactification from $D=11$ dimensions preserving all the SUSY generators could occur in two ways, leaving only 4 or 7 macroscopic dimensions \cite{rubin}. There are many possible compactifications, but the Freund-Rubin compactification's invariance under all of the supersymmetry transformations preserves the action. Finally, the first two results to establish the uniqueness of the theory in $D=11$ dimensions, while the third result appeared to specify the theory, and the last explained why the observed universe appears to be four-dimensional. For $D=11$ supergravity, the field content has the graviton, the gravitino and a 3-form potential $A^{(3)}$.

Initial excitement about the 10-dimensional theories and the string theories that provide their quantum completion diminished by the end of the 1980s. After the second superstring revolution occurred, Joseph Polchinski realized that D-branes, which he discovered six years earlier, corresponds to string versions of the $p$-branes known in supergravity theories. 

In ten dimensions a class of solutions of IIA/IIB string theory, satisfying Dirichelet
boundary conditions in certain directions. These solutions are called $Dp$-branes (or just D-branes) and the charge of the $Dp$-brane is carried by a RR
gauge field. $Dp$-branes exist for all values $0 \leq p \leq 9$ and are all related by T duality where $p$ is even for IIA and odd for IIB. The $Dp$-brane solution in type II theories in $D=10$ is given by \cite{brane1} \cite{brane2},
\bea
ds^2 &=& H(r)^{-\frac{1}{2}}\eta_{\mu \nu}dx^\mu dx^\mu + H(r)^{\frac{1}{2}}\delta_{i j}dy^i dy^j
\nonumber \\
F_{(p+2)} &=& -d(H^{-1}) \wedge \omega_{(1,p)},~~ e^{\Phi} = H^{\frac{3-p}{4}}
\nonumber \\
H(r) &=& 1 + \frac{\Lambda_{Dp}}{r^{7-p}},~~\Lambda_{Dp} = (2\sqrt{\pi})^{5-p}\Gamma\bigg(\frac{7-p}{2}\bigg)g_{s} \alpha'^{\frac{1}{2}(7-p)} N \ ,
\ee
where $x^\mu$ and $y^i$ are the parallel and transverse dimensions with $\mu =0,\cdots,p$, $i=1,\cdots, 9-p$ and $r$ is the radial coordinate defined by $r^2 = y_i y^i$. 

String theory perturbation didn't restrict these $p$-branes and thanks to supersymmetry, supergravity $p$-branes were understood to play an important role in our understanding of string theory and using this Edward Witten and many others could show all of the perturbative string theories as descriptions of different states in a single theory that Edward Witten named M-theory \cite{witten2}, \cite{duff}. Furthermore, he argued that M-theory's low energy limit is described by $D=11$ supergravity. Certain extended
objects, the $M2$ and $M5$-branes, which arise in the context $D=11$ supergravity,
but whose relationship with the ten-dimensional branes was unclear, was then
understood to play an important role in M-theory related by various dualities.

Both the M-brane and the D-brane solutions are characterized by a harmonic function $H$ which depends
only on the coordinates transverse to the brane and is harmonic on this transverse
space. This suggests that the M-brane and D-brane solutions are related, and indeed one
finds that direct dimensional reduction of $M2$-brane and a double-dimensional reduction of the $M5$-brane in $D = 11$ leads to branes in IIA supergravity e.g Fundemental strings ($F1$-branes) and $D2$-branes are simply $M2$-branes wrapped or not wrapped on the eleventh dimension and similarly the $D4$-branes and $NS5$-branes both correspond to $M5$-branes.

It is a general feature that in the presence of branes, the flat space supersymmetry algebras are modified beyond the super Poincar\'e algebra by including terms which contain topological charges of the branes.
For individual brane configurations these take the form,
\bea
\frac{1}{p!}(C \Gamma_{\mu_1 \dots \mu_p})_{\alpha \beta}Z^{\mu_1 \dots \mu_p} \ ,
\ee
where $C$ is the charge conjugation matrix, $X^{\mu}$ are spacetime coordinates and $\Gamma_{\mu_1 \dots \mu_p}$ an antisymmetric combination of $p$ Dirac Gamma matrices in a particular supergravity theory. For example in $D=11$ supergravity, the theory contains the $M2$ and $M5$-Brane which have the 2-form $Z_{\mu \nu}$ and 5-form $Z_{\mu_1 \dots \mu_5}$ charges respectively and the SUSY algebra therefore takes the form,
\bea
\{Q_{\alpha}, Q_{\beta}\} = (C\Gamma^{\mu})_{\alpha \beta}P_{\mu} + \frac{1}{2!}(C\Gamma_{\mu_1 \mu_2})_{\alpha \beta}Z^{\mu_1 \mu_2} + \frac{1}{5!}(C\Gamma_{\mu_1 \dots \mu_5})_{\alpha \beta}Z^{\mu_1 \dots \mu_5} \ .
\ee
Therefore, supergravity comes full circle and uses a common framework in understanding features of string theories, M-theory, and their compactifications to lower spacetime dimensions.

\begin{figure}[h!]
	\begin{center}
		\includegraphics[width=250pt]{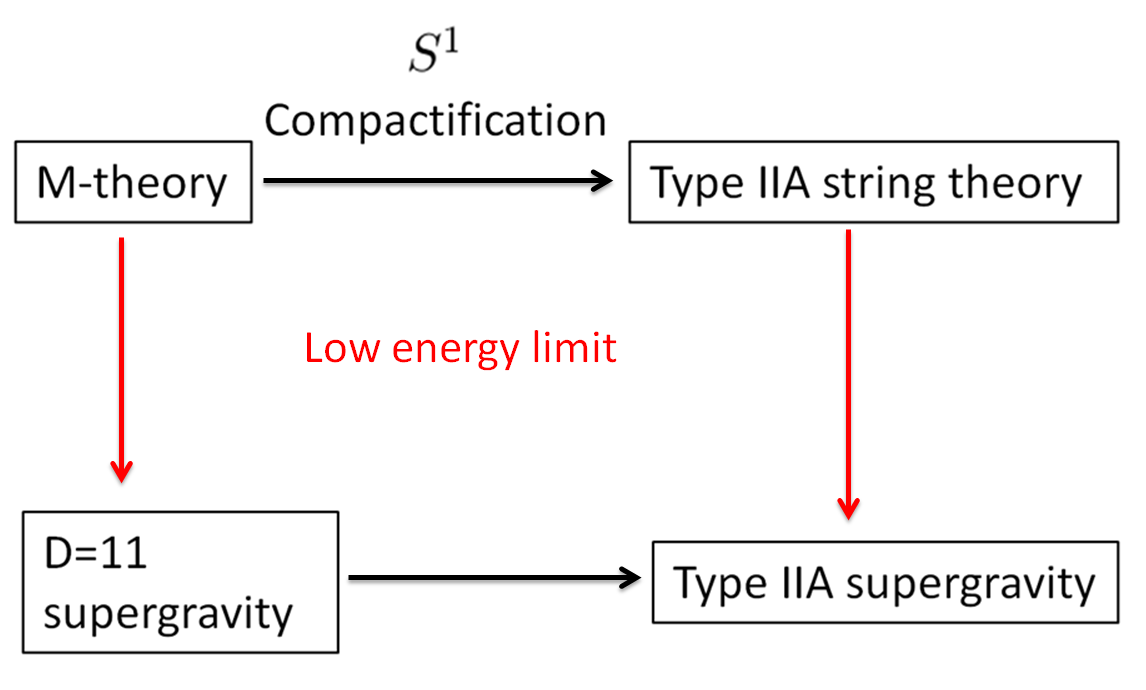}
	\end{center}
	\caption{Compactification/Low energy limit}
	\label{fig:msugraiia}
\end{figure}

In string theory, black holes can be constructed as systems of intersecting branes. In particular, solutions of five-dimensional ungauged supergravity can be uplifted to solutions of $D=11$ supergravity  \cite{townsend}, while solutions in $D = 5$ gauged supergravity can be uplifted to solutions in $D=10$ type IIB
supergravity \cite{chamblin} e.g the supersymmetric
black ring becomes a black supertube when uplifted to higher dimensions \cite{elvang} and is reproduced in M-theory as a system of intersecting $M2$- and $M5$-branes. Another way to construct a black hole solution is to construct a configuration of {\it wrapped} branes which upon dimensional reduction yields a black hole spacetime \cite{ddp1} \cite{ddp2} e.g type IIB theory on $T^5$ 
with a $D1$-$D5$-$PP$ system such that $N_{D5}$ $D5$-branes are wrapped on the whole of $T^5$; $N_{D1}$ $D1$-branes wrapped on $S^1$ of length $2\pi R$ and momentum $N= N_{W}/R$ carried along the $S^1$. A $D=5$ solution is obtained if five of the dimensions of the $D=10$ IIB theory are sufficiently small upon compactification on $S^1 \times T^4$ and a supergravity analysis can be used if the black holes are sufficiently large \cite{ddp1}.

\section{Summary of Research}
It has been known for some time that the black hole solutions of 4-dimensional theories with physical matter
couplings have event horizons with spherical topology \cite{Htopology1}. As a result, black holes have a spherical shape and they describe
the gravitational field of stars after they have undergone gravitational collapse. In addition in four dimensions, several black hole uniqueness theorems \cite{israel2, carter, 4dunique} have been shown which essentially state that black holes in four dimensions are determined by their mass, angular momentum and charge. In recent years most of the proposed unified theories of all four fundamental forces of nature, like string theory or supergravity, are defined in more than four dimensions. Many such theories have black hole solutions \cite{Emparan:2008eg} and as a result, the shape, as well as the uniqueness theorems for black holes as we have mentioned earlier, need re-examining. It is known that in five dimensions that apart from the spherical black holes, there are solutions
for which the event horizon has ring topology. Moreover, there are known examples of five-dimensional black objects with spherical
horizon topology \cite{lucietti}, which nevertheless has the same asymptotic charges with spherical horizon cross sections such as the BMPV solution \cite{BMPV1}, \cite{gauntlett}. As a result, the uniqueness theorems of black holes in four do not extend to five and possibly higher dimensions \cite{dunique}.

Supergravity theories are gravitational theories coupled to appropriate matter, including fermionic fields, and describe the dynamics of string theory at low energies. One way to investigate the geometry of black holes is to assume that the solutions have a sufficiently large number of commuting rotational isometries. 
However, in this thesis, we shall instead consider supersymmetric black holes.
This means  that in addition to field equations the black holes solve a set of first-order non-linear
differential equations which arise from the vanishing of the supersymmetry transformations of the fermions
of supergravity theories. These are the so-called Killing spinor equations (KSEs).
The understanding of supersymmetric black holes that it is proposed is facilitated
by the recent progress that has been made towards understanding the geometry of all supersymmetric backgrounds of supergravity theories. In particular, we shall exploit the fact that an extremal black hole has a well-defined near-horizon limit which solves the same field equations as the full black hole solution. Classifying near-horizon geometries is important as it reveals the geometry and hence the topology and isometries of the horizon of the full black hole solution. It also becomes simpler to solve as we can reduce it to $(D-2)$-dimensional problem on a compact spatial manifold. The classification of near-horizon geometries in a particular theory provides important information about which black hole solutions are possible.

The enhancement of supersymmetry near to brane and black hole horizons has been known
for some time. In the context of branes, many solutions are known which exhibit supersymmetry
enhancement near to the brane \cite{gwgpkt, Schwarz:1983qr}. For example, the geometry of D3-branes doubles its supersymmetry to become the maximally supersymmetric $AdS_5 \times S^5$
solution. The bosonic symmetry of the {\it near-horizon region} is given by the product of the symmetry groups of the constituent factors of the metric, an $SO(2,4)$-factor for $AdS_5$ and a factor $SO(6)$ for the $S^5$. Similarly for the $M2$- and $M5$ branes, we have an $AdS_4 \times S^7$ near-horizon geometry for the $M2$ and a $AdS_7 \times S^4$ for the $M5$. Both these near-horizon geometries have enhanced supersymmetry and allow for $32$ real supercharges. The $M2$-brane had a bosonic symmetry group $SO(2,3) \times SO(8)$ while the $M5$-branes near-horizon geometry has a symmetry $SO(2,6) \times SO(5)$.

This phenomenon played a crucial role in the early development of the
AdS/CFT correspondence \cite{maldacena}. Black hole solutions are also known to exhibit
supersymmetry enhancement; for example in the case of the five-dimensional BMPV black hole \cite{BMPV1, Chamseddine:1996pi, Chamseddine:1999qs}. Further recent interest in the geometry of black hole horizons has arisen in the context of
the Bondi-Metzner-Sachs (BMS)-type symmetries associated with black holes, following 
\cite{Hawking:2016msc, Hawking:2016sgy, Averin:2016ybl, Donnay:2015abr}.
In particular, the analysis of the asymptotic symmetry group of Killing horizons was undertaken in
\cite{Akhmedov:2017ftb}. In that case, an infinite dimensional symmetry group is obtained,
analogous to the BMS symmetry group of asymptotically flat solutions. 

Another important observation in the study of black holes is the attractor mechanism \cite{attract1}.
This states that the entropy is obtained by extremizing an entropy function which depends only on
the near-horizon parameters and conserved charges, and if this admits a unique extremum then the entropy is independent of the asymptotic values of the moduli.
In the case of 4-dimensional solutions the analysis of \cite{astef} implies that if
the solution admits $SO(2,1) \times U(1)$ symmetry, and the horizon has spherical topology, then such a mechanism holds. 
In $D=4, 5$ it is known that all known asymptotically flat black hole solutions exhibit attractor mechanism behaviour which follow from near-horizon symmetry theorems \cite{nhsymmetrythm1} for any Einstein-Maxwell-scalar-CS theory.
In particular, a generalization of the analysis
of \cite{astef} to five dimensions requires the existence of
a $SO(2,1) \times U(1)^2$ symmetry, where all the possibilities have been classified for $D=5$ minimal ungauged supergravity \cite{u1proof}. Near-horizon geometries of asymptotically $AdS_5$ 
supersymmetric black holes admitting a $SO(2,1) \times U(1)^2$ symmetry have been 
classified in \cite{Kunduri:2006ek, Kunduri:2007qy}. 

It remains to be determined
if all supersymmetric near-horizon geometries fall into this class. There is no general proof of an attractor mechanism for higher dimensional black holes ($D>5$) as it depends largely on the properties of the geometry of the horizon
section e.g for $D = 10$ heterotic, it remains undetermined
if there are near-horizon geometries with non-constant dilaton $\Phi$.

We shall consider the following conjecture concerning the properties
of supersymmetric regular near-horizon geometries:
\vspace{2mm}
\newtheorem{thm}{Theorem}
\begin{thm}[{\it The Horizon Conjecture}]
	Assuming all fields are smooth
	and the spatial cross section of the event horizon, ${\cal{S}}$, is smooth and compact without boundary,
	\begin{itemize}
\item The number of Killing spinors $N$, $N\not=0$,  of Killing horizons $\cal{H}$ in supergravity\footnote{Generated by a Killing vector $X$ with $X^{\mu}X^{\nu}g_{\mu \nu} = 0$ on $\cal{H}$ as introduced in the following section} is given by
	\begin{eqnarray}
		N = 2N_{-} + \mathrm{Index}(D_{E})~,
		\label{indexcon}
		\label{index}
		\end{eqnarray}
		where $N_{-}\in {\mathbb{N}}_{>0}$ and $D_{E}$ is a Dirac operator twisted by a vector bundle $E$, defined on the spatial horizon section $\mathcal{S}$, which depends on the gauge symmetries of the supergravity theory in question,
		\item Horizons with non-trivial fluxes and $N_{-} \neq 0$ admit an $\mathfrak{sl}(2,{\mathbb{R}})$ symmetry subalgebra.
	\end{itemize} 
\end{thm}
One simple motivational example of (super)symmetry enhancement which we will consider in the next chapter is the $\mathbb{R} \times SO(3)$ isometry group of the Reissner-Nordstr\"om black hole which in the extremal near-horizon limit enhances to $SL(2, \mathbb{R}) × SO(3)$ with near-horizon geometry $AdS_2 \times S^2$. In addition, viewing the extreme Reissner-Nordstrom black hole as a solution of the ${\cal{N}} = 2, D = 4$ minimal supergravity, the $N = 4$ supersymmetry of the solution also
enhances to $N = 8$ near the horizon.

The main focus of this thesis is to prove the horizon conjecture for supersymmetric black hole horizons of IIA, massive IIA and $D=5$ (gauged and ungauged) supergravity with vector multiplets. In particular, the first part of
the horizon conjecture as applied to these theories will establish that
there is supersymmetry enhancement, which gives rise to symmetry enhancement
in the form of the $\mathfrak{sl}(2,{\mathbb{R}})$ symmetry, as mentioned
in the second part of the horizon conjecture. Such symmetry enhancement
also produces additional conditions on the geometry of the solution.
The methodology used to investigate these problems involves techniques in differential geometry, differential equations on compact manifolds, and requires some knowledge on general relativity and supergravity. Algebraic and differential topology are also essential in the analysis. 

The proofs that we establish in this thesis for (super)symmetry enhancement rely on establishing Lichnerowicz-type theorems and an index theory argument. A similar proof has been given for supergravity horizons in $D=11$, IIB, $D=5$ minimal gauged and $D=4$ gauged \cite{11index, iibindex, 5dindex, 4dindex}. 
We shall also prove that the near-horizon geometries admit a $\mathfrak{sl}(2,\mathbb{R})$
symmetry algebra. In general, we find that the orbits of the generators of $\mathfrak{sl}(2,\mathbb{R})$ are
3-dimensional, though in some special cases they are 2-dimensional. In these special cases,
the geometry is a warped product $AdS_2 \times_w {\cal{S}}$.
The properties of $AdS_2$ and their relationship to black hole entropy
have been examined in \cite{strominger, sen}.
Our results, together with those of our previous calculations, implies 
that the $\mathfrak{sl}(2,\mathbb{R})$ symmetry is a universal property of supersymmetric black holes. This has previously been observed for generic non-supersymmetric extremal horizons \cite{genextrsl}. 

Unlike most previous investigations of near horizon geometries, e.g \cite{gutbh, reall, bilin}, we do not assume the vector bilinear matching condition, which is the identification of the stationary Killing vector field of a black hole with the vector Killing spinor bilinear; in fact we prove this is the case for the theories under consideration. In particular, we find that the emergence of an isometry generated by the spinor, from the solution of the KSEs, is proportional to Killing vector which generates the Killing horizon. Thus previous results which assumed the bilinear condition automatically follow for the theories that we consider e.g for $D=5$ ungauged supergravity, the analysis of \cite{gutbh} classifies the possible near-horizon geometries which we also revisit in Chapter 6 from the conditions that we establish. This also allows us to investigate the properties of the near-horizon geometry in $D=5$ gauged supergravity and eliminate certain solutions.

The new Lichnerowicz type theorems established in this thesis are of interest
because they have certain free parameters appearing in the definition
of various connections and Dirac operators on ${\cal{S}}$. Such freedom
to construct more general types of Dirac operators in this way is related to
the fact that the minimal set of Killing spinor equations consists not
only of parallel conditions on the spinors but also certain algebraic conditions. These algebraic conditions do
not arise in the case of $D=11$ supergravity. Remarkably, the Lichnerowicz type
theorems imply not only the parallel transport conditions but also
the algebraic ones as well. The solution of the KSEs is essential to the investigation of geometries
of supersymmetric horizons. We show that the enhancement of the supersymmetry
produces a corresponding symmetry enhancement, and describe the resulting conditions
on the geometry. The only
assumptions we make are that the fields are smooth (or at least $C^2$ differentiable)
and the spatial horizon section ${\cal S}$ is compact without boundary.\footnote{This
	is an assumption which can be relaxed. To extend the proof to horizons with non-compact ${\cal S}$, one has to impose boundary conditions on the fields.}

\subsection{Plan of Thesis}

In {\bf Chapter 2}, we describe the properties of Killing Horizons and introduce Gaussian Null Coordinates (GNC), and we explicitly state important examples in $D=4$ in these coordinates, paying particular attention to extremal horizons. We also introduce the Near-Horizon Limit (NHL) and give explicit examples for the extremal Reissner-Nordstr\"om, Kerr, and Kerr-Newman to demonstrate the symmetry enhancement with the ${\mathfrak{sl}}(2,{\mathbb{R}})$ Lie algebra of the Killing vectors. We then give an overview of the curvature of the near-horizon geometry, field strengths in the near-horizon limit, the supercovariant derivative, the maximum principle, and the classical Lichnerowicz theorem.

In {\bf Chapter 3}, we summarize the various types of supergravity theories which we will consider in this thesis. We begin with an overview of $D=11$ supergravity stating the action, the supersymmetry variations and the field equations, which we will do for every supergravity theory that we consider. We also give the details for dimensional reduction on ${\cal M}_{10} \times S^1$ which gives $D=10$ IIA supergravity on ${\cal{M}}_{10}$. We also give the Romans Massive IIA which give extra terms that depend on the mass parameter $m$. Next, we consider the dimensional reduction of $D=11$ supergravity on ${\cal M}_{5} \times CY_3$ and give the details for $D=5$ (ungauged) supergravity coupled to an arbitrary number of vector multiplets on ${\cal M}_{5}$ and the Very Special Geometry associated with the Calabi Yau manifold. Finally, we summarize the details for $D=5$ gauged supergravity with vector multiplets, which give extra terms that depend 
on the gauging parameter $\chi$.

We begin the {\it general analysis} of supersymmetric near-horizon geometries by solving the KSEs along the lightcone direction, identifying the redundant conditions and stating the independent KSEs, field equations and Bianchi identities given as the restriction on the spatial horizon section. By an application of the Hopf maximum principle, we establish generalised Lichnerowicz theorems relating the zero modes of the horizon Dirac equation with the KSEs. Using the index theorem, we then establish the enhancement of supersymmetry and show that the number of supersymmetries must double. Finally, by identifying the isometries generated by the Killing vectors of the Killing spinor bilinear with the solution along the light cone, we show the enhancement of symmetry with the $\mathfrak{sl}(2,\mathbb{R})$ lie subalgebra.

In {\bf Chapter 4}, we give the details of the {\it general analysis} in proving the horizon conjecture for IIA supergravity. In particular, we give the near-horizon fields, the horizon Bianchi identities and field equations. We then solve the KSEs of IIA supergravity along the lightcone and identify the independent conditions. We then prove the supersymmetry enhancement by identifying the horizon Dirac equation, establishing the Lichnerowizc theorems and using the index theory. We also give the details of the $\mathfrak{sl}(2,\mathbb{R})$ symmetry enhancement for IIA horizons. In {\bf Chapter 5} we repeat the {\it general analysis} with the addition of the mass parameter $m$.

In {\bf Chapter 6} we give a brief introduction into the near-horizon geometry of the BMPV and black ring solution in Gaussian Null Coordinates, and explicitly show how the symmetry enhancement in the near-horizon limit produces Killing vectors which satisfy the $\mathfrak{sl}(2,\mathbb{R})$ Lie algebra. We then repeat the same {\it general analysis} as previous chapters, for $D=5$ supergravity (gauged and ungauged) with vector multiplets. We also highlight differences in the gauged and ungauged theories, particularly in counting the number of supersymmetries and the conditions on the geometry of $\cal{S}$.

In the appendices, we give the calculations required for the analysis of the various supergravity theories we have considered. In particular, {\bf Appendix A} gives the details for the derivation of Gaussian null coordinates, other regular coordinate systems and we explicitly give the expressions used in computing the spin connection, Riemann curvature tensor and Ricci tensor for near-horizon geometries in terms of the frame basis. {\bf Appendix B} gives the details for the gamma matrices and Clifford algebra conventions for arbitrary spacetime dimensions. In {\bf Appendix C} we give the calculations for IIA supergravity; first giving the integrability conditions from the KSEs which give rise to the field equations.  We give a derivation of the Dilaton field equation which is implied by the other field equations. We then give a proof that the bilinears constructed from Killing spinors give rise to isometries and preserve all the fluxes. Next we give the details in establishing the conditions that we get from the solution along the lightcone in terms of the independent KSEs. We also give a proof of the Lichneorwicz principle using a maximum principle by calculating the Laplacian of the norm of the spinors. We also give an alternative derivation using a partial integration argument. In {\bf Appendix D} and {\bf Appendix E}, we repeat these details for Massive IIA, $D=5$ ungauged and gauged supergravity. Finally in {\bf Appendix F} we give the generic calculations that are required for the analysis of the $\mathfrak{sl}(2,\mathbb{R})$ symmetry and spinor bilinears.

\newpage

\section{Statement of Originality}

I hereby declare that except where specific reference is made to the work of others,
the contents of this dissertation are original and have not been submitted in whole
or in part for consideration for any other degree or qualification in this, or any other university. This dissertation is my own work and contains nothing which is the outcome of work done in collaboration with others, except as specified in the text and references. This thesis contains fewer than 100,000 words excluding the bibliography, footnotes, and equations.

\section{Publications}
U.~Kayani, G.~Papadopoulos, J.~Gutowski, and U.~Gran,
``Dynamical symmetry enhancement near IIA horizons,''
JHEP {\bf 1506}, 139 (2015)
doi:10.1007/JHEP06(2015)139,

%%CITATION = doi:10.1007/JHEP06(2015)139;%%
U.~Kayani, G.~Papadopoulos, J.~Gutowski, and U.~Gran,
``Dynamical symmetry enhancement near massive IIA horizons,''
Class.\ Quant.\ Grav.\  {\bf 32}, no. 23, 235004 (2015)
doi:10.1088/0264-9381/32/23/235004

U.~Kayani, ``Symmetry enhancement of extremal horizons in $D = 5$ supergravity,''
%``Symmetry enhancement of extremal horizons in D  =  5 supergravity,''
Class.\ Quant.\ Grav.\  {\bf 35}, no. 12, 125013 (2018)
doi:10.1088/1361-6382/aac30c
%%CITATION = doi:10.1088/1361-6382/aac30c;%%
	\chapter{Killing Horizons and Near-Horizon Geometry}

\label{ch:background}
In this section we describe the properties of Killing Horizons and introduce Gaussian Null Coordinates which are particularly well adapted for such geometries, illustrated by certain important examples in $D=4$.\footnote{We use natural units with $G = c = \hbar = 1$} As the purpose of this thesis is to investigate the geometric properties of supersymmetric near-horizon geometries, it will be particularly advantageous to work in a co-ordinate system 
which is specially adapted to describe Killing Horizons. 

In what follows, we will assume that the black hole event horizon is a Killing horizon. Rigidity theorems imply that the black hole horizon is Killing
for both non-extremal and extremal black holes, under certain assumptions, have been constructed, e.g. \cite{rigidity1, gnull, axi1, axi2}. The assumption that the event horizon is Killing
enables the introduction of Gaussian Null co-ordinates \cite{isen, gnull} in a neighbourhood
of the horizon. The analysis of the near-horizon geometry is significantly simpler than that of the
full black hole solution, as the near-horizon limit reduces the system to a set of
equations on a co-dimension 2 surface, ${\cal{S}}$, which is the spatial section of
the event horizon. 

\section{Killing horizons}
\theoremstyle{definition}
\begin{definition}{A null hypersurface $\cal{H}$ is a {\bf Killing Horizon} of a Killing vector field $\xi$ if it is normal to $\cal{H}$ i.e $\exists$ a Killing vector field $\xi$ everywhere on the spacetime $\cal{M}$ which becomes null only on the horizon $\cal{H}$.}
\end{definition}
If $\xi$ is a Killing vector then the Killing horizon $\cal{H}$ can be identified with the surface given by $g(\xi,\xi) = 0$. A Killing horizon is a more local description of a horizon since it can be formulated in terms of local coordinates.
For an example, consider the Schwarzchild metric,
\bea
ds^2 = -\bigg(1-\frac{2M}{r}\bigg)dt^2 + \bigg(1-\frac{2M}{r}\bigg)^{-1}dr^2 + r^2 (d\theta^2 +  \sin^2{\theta} d\phi^2) \ .
\label{scwarzsol}
\ee
The Killing horizon is generated by the timelike Killing vector $\xi = \partial_{t}$ which becomes null on the horizon $r=2M$ since $g(\xi, \xi) = -(1-\frac{2M}{r})$. 

Associated to a Killing horizon is a geometrical quantity known as the surface gravity $\kappa$. If the surface gravity vanishes, then the Killing horizon is said to be extreme or degenerate. The surface gravity $\kappa$ is defined as,
\begin{eqnarray}
\xi^{\nu}\nabla_{\nu}\xi^{\mu}\big|_{\cal{H}} = \kappa \xi^{\mu} \ .
\end{eqnarray}
This can be rewritten as
\begin{eqnarray}
\nabla_{\mu}(\xi^2)\big|_{\cal{H}} = -2\kappa \xi_{\mu} \ .
\end{eqnarray}
By Frobenius theorem, a vector $\xi^\mu$ is hyperspace orthogonal if,
\begin{eqnarray}
\xi_{[\mu}\nabla_{\nu}\xi_{\rho]} = 0 \ .
\end{eqnarray}
Since $\xi$ is Killing, we can rewrite this as,
\begin{eqnarray}
\xi_{\rho} \nabla_{\mu}{\xi_{\nu}} = - 2\xi_{[\mu}\nabla_{\nu]}{\xi_{\rho}} \ .
\end{eqnarray}
By contracting with $\nabla^{\mu}{\xi^{\nu}}$ and evaluating on $\cal{H}$ we get,
\begin{eqnarray}
\xi_{\rho} (\nabla_{\mu}{\xi_{\nu}})(\nabla^{\mu}{\xi^{\nu}}) = -2 (\nabla^{\mu}{\xi^{\nu}})(\xi_{[\mu}\nabla_{\nu]}\xi_{\rho}) = - 2\kappa \xi^{\mu}\nabla_{\mu}{\xi_{\rho}} = -2\kappa^2 \xi_{\rho} \ .
\end{eqnarray}
Thus we can write,
\begin{eqnarray}
\kappa^2 = -\frac{1}{2}(\nabla_{\mu}\xi_{\nu})(\nabla^{\mu}\xi^{\nu})\big|_{\cal{H}} \ .
\end{eqnarray}
The surface gravity of a static Killing horizon can be interpreted as the acceleration, as exerted at infinity, needed to keep an object on the horizon. For the Schwarzchild metric the surface gravity is $\kappa = \frac{1}{4M}$ which is non-vanishing.

\section{Gaussian null coordinates}
In order to study near-horizon geometries we need to introduce a coordinate system which is regular and adapted to the horizon. We will consider a $D$-dimensional stationary black hole metric,
for which the horizon is a Killing horizon, and the metric is regular at the horizon. A set of Gaussian Null coordinates \cite{isen, gnull} $\{u, r, y^{I}\}$
will be used to describe the metric, where $r$ denotes the coordinate transverse to the horizon as the radial distance away from the event horizon which is located at $r=0$ and $y^I,~ I=1, \dots, D-2$ are local co-ordinates on ${\cal S}$. The metric components have no dependence on $u$, and the timelike isometry $\xi = {\partial \over \partial u}$ is null on the horizon at $r=0$. As shown in \cite{gnull} (see Appendix A) the black hole metric in a patch containing the horizon is given by,
\bea
\label{gncmetric}
ds^2 = 2du dr + 2r h_I(y, r) du dy^I - r f(y, r) du^2 + ds_{\cal S}^2 \ .
\ee
The spatial horizon section ${\cal S}$ is given by $u=const,~ r=0$ with the metric 
\bea
ds_{\cal S}^2  = \gamma_{I J}(y, r)dy^I dy^J \ .
\ee
where $\gamma_{I J}$ is the metric on the spatial horizon section $\cal{S}$ and $f, h_{I}$ and $\gamma_{I J}$ are smooth functions of $(r, y^I)$ so that the spacetime is smooth.
We assume that ${\cal{S}}$, when restricted to $r=const.$ for sufficiently small values of $r$, is compact and without boundary. The
1-form $h$, scalar $\Delta$ and metric $\gamma$ are functions of $r$ and $y^{I}$; they are smooth in $r$ and regular at the horizon.
The surface gravity associated with the Killing vector $\xi$ can be computed from this metic, to obtain $\kappa = \frac{1}{2}f(y,0)$. 

It is instructive to consider a number of important 4-dimensional examples. In each
case the co-ordinate transformation used to write the metric in regular coordinates around the horizon, GNC and Kerr coordinates, which removes the co-ordinate singularity at the horizon.
\vspace{2mm}
\newtheorem{eg}{Example}
\begin{eg}
	Consider the Schwarzschild solution ({\ref{scwarzsol}}) and
	make the change of coordinates $(t, r, \theta, \phi) \rightarrow (u, r, \theta, \phi)$ with $t \rightarrow u + \lambda(r)$ and
	\bea
	\lambda(r) = -r-2M\ln(r-2M) \ .
	\ee
	Thus in GNC the metric can be written as,
	\bea
	ds^2 = -r(r+2M)^{-1}du^2 + 2du dr + (r+2M)^2 (d\theta^2 +  \sin^2{\theta} d\phi^2) \ .
	\ee
	where we have also made shift $r \rightarrow 2M + r$ so that the horizon is now located at $r=0$. We remark that the derivation of the Gaussian null co-ordinates for the Schwarzschild solution is identical to that of the standard Eddington-Finkelstein co-ordinates.
\end{eg}
\vspace{2mm}
\begin{eg}It is also straightforward to consider Reisser-Nordstr\"om solution 
({\ref{rnsolution}}). We make the same co-ordinate transformation as for the Schwarzschild analysis, 
but take
\bea
\lambda(r) = -r - M\ln(r^2 -2Mr + Q^2) + \bigg(\frac{Q^2 - 2M^2}{\sqrt{Q^2 - M^2}}\bigg)\arctan{\bigg(\frac{r-M}{\sqrt{Q^2 - M^2}}\bigg)} \ .
\ee
This produces the following metric
\bea
\label{rngncoord}
ds^2 = -\bigg(1-\frac{2M}{r} + \frac{Q^2}{r^2}\bigg)du^2 + 2du dr + r^2 (d\theta^2 +  \sin^2{\theta} d\phi^2) \ .
\ee
The event horizon is located at the outer horizon $r = r_+ \equiv M + \sqrt{M^2 - Q^2}$. We can also make the shift  $r \rightarrow r_+ + r$ so that the horizon is now located at $r=0$.
\end{eg}
\vspace{2mm}
\begin{eg}For the Kerr metric, given in ({\ref{kerrsol}}), we make the change of co-ordinates $(t, r, \theta, \phi) \rightarrow (u, r, \theta, \tilde{\phi})$ with $t \rightarrow u + \lambda_1(r),~ \phi \rightarrow \tilde{\phi} + \lambda_2(r)$ and
take
\bea
\lambda_1(r) &=&
-r - M\ln(r^2 -2Mr + a^2) - \bigg(\frac{2M^2}{\sqrt{a^2 - m^2}}\bigg)\arctan{\bigg(\frac{r-M}{\sqrt{a^2 - M^2}}\bigg)} \ ,
\nonumber \\
\lambda_2(r) &=& - \bigg(\frac{a}{\sqrt{a^2 - m^2}}\bigg)\arctan{\bigg(\frac{r-M}{\sqrt{a^2 - M^2}}\bigg)} \ .
\ee
which produces the metric
\bea
ds^2 &=& -\bigg(\frac{r^2 - 2Mr + a^2\cos^2 \theta}{r^2 + a^2 \cos^2 \theta}\bigg)du^2 + 2du dr - \underbrace{a\sin^2 \theta dr d\phi}_{(*)} - \bigg(\frac{2aMr \sin^2 \theta}{r^2 + a^2\cos^2 \theta}\bigg) du d\phi
\nonumber \\
&+& (r^2+a^2 \cos^2 \theta)d\theta^2 + \bigg(\frac{\sin^2 \theta (a^2 (a^2 - 2Mr + r^2) \cos^2 \theta + (2Mr + r^2)a^2 + r^4)}{r^2 + a^2 \cos^2 \theta} \bigg)d\phi^2 \ ,
\nonumber \\
\ee
where the tildes have been dropped.
The event horizon is located at the outer horizon $r = r_+ \equiv M + \sqrt{M^2 - a^2}$. We can also make the shift  $r \rightarrow r_+ + r$ so that the horizon is now located at $r=0$. 
\end{eg}
\vspace{2mm}
\begin{eg}
	For the Kerr-Newman metric, given in ({\ref{kerrnewsol}}), on making the change of coordinates $(t, r, \theta, \phi) \rightarrow (u, r, \theta, \tilde{\phi})$ with $t \rightarrow u + \lambda_1(r),~ \phi \rightarrow \tilde{\phi} + \lambda_2(r)$ and taking
\bea
\lambda_1(r) &=& 
-r - M\ln(r^2 -2Mr + Q^2 + a^2) + \bigg(\frac{Q^2 - 2M^2}{\sqrt{Q^2 - M^2 + a^2}}\bigg)\arctan{\bigg(\frac{r-M}{\sqrt{Q^2 - M^2 + a^2}}\bigg)} \ ,
\nonumber \\
\lambda_2(r) &=& - \bigg(\frac{a}{\sqrt{Q^2 - M^2 + a^2}}\bigg)\arctan{\bigg(\frac{r-M}{\sqrt{Q^2 - M^2 + a^2}}\bigg)} \ ,
\ee
the following metric is found
\bea
ds^2 &=& -\bigg(\frac{r^2 - 2Mr + a^2\cos^2 \theta +Q^2}{r^2 + a^2 \cos^2 \theta}\bigg)du^2 + 2du dr - \underbrace{a\sin^2 \theta dr d\phi}_{(*)} 
\nonumber \\
&+& \bigg(\frac{a \sin^2 \theta(Q^2 - 2Mr)}{r^2 + a^2\cos^2 \theta}\bigg) du d\phi
+ (r^2+a^2 \cos^2 \theta)d\theta^2 
\nonumber \\
&+& \bigg(\frac{\sin^2 \theta (a^2 (a^2 + Q^2 - 2Mr + r^2) \cos^2 \theta + (2Mr + r^2 -Q^2)a^2 + r^4)}{r^2 + a^2 \cos^2 \theta} \bigg)d\phi^2 \ ,
\label{kngncoords}
\ee
where the tildes have been dropped.
The event horizon is located at the outer horizon $r = r_+ \equiv M + \sqrt{M^2 - Q^2 -a^2}$. We can also make the shift  $r \rightarrow r_+ + r$ so that the horizon is now located at $r=0$.
\end{eg}
The resulting metric for both Kerr and Kerr-Newman are expressed in terms of regular coordinates around the horizon, known as Kerr coordinates. These are evidently different from the usual coordinates in GNC since it contains a non-zero $dr d\phi$ term (*) and the Killing vector $\partial_u$ is not null on the horizon. Nonetheless, this extra term will disappear in the near-horizon limit for extreme horizons as we shall see (see Appendix A).

\subsection{Extremal horizons}

Since the near-horizon geometry is only well defined for extremal black holes, with vanishing surface gravity, it will be useful to consider some examples. The two examples of particular interest are the extremal Reisser-Nordstr\"om solution and the extremal Kerr-Newman solution.
\vspace{2mm}
\begin{eg}
For the case of Reisser-Nordstr\"om, the extremal solution is obtained by setting
$Q=M$. On taking the metric given in ({\ref{rngncoord}}) and setting $Q=M$, and also
shifting $r \rightarrow M + r$ so that the horizon is now located at $r=0$, we find the metric in GNC,
\bea
ds^2 = - r^2 \big(M+r\big)^{-2} du^2 + 2 du dr + \big(M+r\big)^2 (d\theta^2 +  \sin^2{\theta} d\phi^2) \ .
\ee
\end{eg}
\vspace{2mm}
\begin{eg}
For the case of the Kerr-Newman, the extremal solution is obtained by setting $Q^2 = M^2 - a^2$ in the metric ({\ref{kngncoords}}), and also shifting $r \rightarrow M + r$ so that the horizon is now located at $r=0$, we find the metric the following metric in Kerr coordinates,
\bea
ds^2 &=& -\bigg(\frac{r^2 -a^2 + a^2\cos^2 \theta}{(r+M)^2 + a^2 \cos^2 \theta}\bigg)du^2 + 2du dr - a\sin^2 \theta dr d\phi 
\nonumber \\
&-& \bigg(\frac{a\sin^2 \theta (a^2 +M^2 + 2Mr)}{(r+M)^2 + a^2\cos^2 \theta}\bigg) du d\phi
+ ((r+M)^2+a^2 \cos^2 \theta)d\theta^2 
\nonumber \\
&+& \bigg(\frac{\sin^2 \theta (a^2 r^2 \cos^2 \theta + a^4 + (r^2+4Mr + 2M^2 )a^2 + (r+M)^4)}{(r+M)^2 + a^2 \cos^2 \theta}\bigg) d\phi^2 \ .
\ee
\end{eg}
The extremal Kerr ($Q=0,~ a=M$) and the extreme RN is obtained by the extreme Kerr-Newman by setting $a=M$ and $a=0$ respectively. 

\section{The near-horizon limit}

Having constructed the Gaussian null co-ordinates, we shall consider a particular type of limit which exists for extremal solutions, called the near-horizon limit \cite{reall}. This limit can be thought of as a decoupling limit in which 
the asymptotic data at infinity is scaled away, however the geometric structure
in a neighbourhood very close to the horizon is retained.

We begin by considering the Gaussian null co-ordinates adapted to a Killing horizon
$\cal{H}$ associated with the Killing vector $\xi = \partial_u$, identified with the hypersurface given by $r=0$. The Killing vector becomes null on the horizon, since $g(\xi, \xi) = -r f(y, r)$. 
\begin{eqnarray}
ds^2 = 2(dr + r h_{I}(y,r)dy^{I} - \frac{1}{2}r f(y,r) du) du + \gamma_{I J}(y,r)dy^{I}dy^{J} \ .
\end{eqnarray}
As we have mentioned earlier, the surface gravity associated to the Killing vector $\xi$ is given by $\kappa =  \frac{1}{2}f(y,0)$. To take the near-horizon limit we first make the rescalings
\begin{eqnarray}
r\rightarrow \epsilon \hat{r}, \, u\rightarrow \epsilon^{-1} \hat{u}, \, y^{I}\rightarrow y^{I} \ ,
\end{eqnarray}
which produces the metric (after dropping the hats),
\begin{eqnarray}
ds^2 = 2(dr + r h_{I}(y,\epsilon r) dy^{I} - \frac{1}{2}r \epsilon^{-1} f(y, \epsilon r) du) du + \gamma_{I J}(y,\epsilon r)dy^{I}dy^{J} \ .
\end{eqnarray}
Since $f$ is analytic in $r$ we have an expansion
\begin{eqnarray}
f(y,r) = \sum_{n=0}^{\infty}\frac{r^n}{n!}\partial^{n}_{r}f\big|_{r=0} \ ,
\end{eqnarray}
and a similar expansion for $h_I$ and $\gamma_{I J}$. Therefore,
\begin{eqnarray}
\epsilon^{-1}f(y,\epsilon r) &=& \sum_{n=0}^{\infty} \epsilon^{n-1}\frac{r^n}{n!}\partial^{n}_{r}f\big|_{r=0} \nonumber \\
&=& \frac{f(y,0)}{\epsilon} + r \, \partial_{r}f\big|_{r=0} + \sum_{n=2}^{\infty} \epsilon^{n-1}\frac{r^n}{n!}\partial^{n}_{r}f\big|_{r=0} \ .
\end{eqnarray}
The near-horizon limit then corresponds to taking the limit $\epsilon \rightarrow 0$.
This limit is clearly only well-defined when $f(y,0)=0$, 
corresponding to vanishing surface gravity. Hence the near-horizon limit is only well defined for extreme black holes. Thus, for extremal black holes, after taking the near-horizon limit we have the metric,
\begin{eqnarray}
\label{nhmetricf}
ds_{NH}^2 = 2(dr + r h_{I}dy^{I} - \frac{1}{2}r^2 \Delta du) du + \gamma_{I J}dy^{I}dy^{J} \ ,
\label{nhmmx}
\end{eqnarray}
where we have defined $\Delta = \partial_{r}f\big|_{r=0}$ and $h_I, \gamma_{I J}$ are evaluated at $r=0$ so that the $r$-dependence is fixed on $\cal{H}$. $\{\Delta, h_I, \gamma_{I J}\}$ are collectively known as the near-horizon data and depend only on the coordinates $y^I$. In Appendix A consider an arbitrary metric written in the coordinates $(u,r,y^I)$ which is regular around the horizon $r=0$ generated by a Killing vector $\partial_u$. We consider the conditions on the metric components for the near-horizon limit to be well defined and show that the metric under a certain condition can be written as (\ref{nhmetricf}) upon identification of the near-horizon data. 

The near-horizon metric (\ref{nhmmx}) also has a new scale symmetry, $r \rightarrow \lambda r,~ u \rightarrow \lambda^{-1}u$ generated by the Killing vector $L=u\partial_{u} - r\partial_{r}$. This, together with the Killing vector $V=\partial_u$ satisfy the algebra $[V, L] = V$ and they form a 2-dimensional non-abelian symmetry group ${\cal{G}}_2$. We shall show that for a very large class of supersymmetric near-horizon geometries, this further enhances into a larger symmetry algebra, which will include a $\mathfrak{sl}(2,\mathbb{R})$ subalgebra. This has previously been shown for non-supersymmetric extremal black hole horizons \cite{genextrsl}.

Supersymmetric black holes in four and five dimensions are necessarily extreme. To see why this is to be expected, we recall that Killing spinors are the parameters of preserved supersymmetry of a solution, so a supersymmetric solution to any supergravity theory necessarily admits a Killing spinor $\epsilon$. The bilinear $K^{\mu} = \bar{\epsilon}\Gamma^{\mu}\epsilon$ is
a non-spacelike Killing vector field i.e. $K^2 \leq 0$.  Suppose a supersymmetric Killing horizon $\cal{H}$ is invariant under the action of $K$, then $K$ must be null and $dK^2 = -2\kappa K$ on the horizon. It follows that $K^2$ attains a maximum
on the horizon, and therefore $dK^2=0$ which implies that the horizon is extremal. It is also known in five dimensions that there exists a real scalar spinor bilinear $f$, with the property that $K^2=-f^2$. Assuming that the Killing spinor is analytic in $r$ in a neighbourhood of the horizon, this implies that $K^2 \sim -r^2$ in a neighbourhood of the horizon and this also implies that the horizon is extremal. A similar argument holds in four dimensions.

A near-extremal black hole is a black hole which is not far from the extremality. The calculations of the properties of near-extremal black holes are usually performed using perturbation theory around the extremal black hole; the expansion parameter known as non-extremality \cite{nextr1,nextr2}. In supersymmetric theories, near-extremal black holes are often small perturbations of supersymmetric black holes. Such black holes have a very small surface gravity and Hawking temperature, which consequently emit a small amount of Hawking radiation. Their black hole entropy can often be calculated in string theory, much like in the case of extremal black holes, at least to the first order in non-extremality.

To extend the horizon into the bulk away from the near-horizon limit, one has to consider the full $r$-dependence of the near-horizon data \cite{tdef, moduli}, which are evaluated at $r=0$ and thus depend only the coordinates $y^I$ of the spatial horizon section ${\cal S}$ in the near-horizon decoupling limit. We thus extend the data $\{\Delta(y), h_I(y), \gamma_{I J}(y)\} \rightarrow \{{\hat{\Delta}}(y,r), {\hat{h}_I}(y,r), {\hat{\gamma}_{I J}(y,r)}\}$, taylor expand around $r=0$ and consider the first order deformation of the horizon fields, where the usual near-horizon data is given by, 
\bea
{\hat{\Delta}}(y,0) =  \Delta,~ {\hat{h}}_{I}(y,0) = h_I,~ {\hat{\gamma}_{I J}(y,0)} = \gamma_{I J} \ .
\ee

\subsection{Examples of near-horizon geometries}
Now we will give examples of near-horizon geometries for the extremal Reisser-Nordstr\"om, Kerr and Kerr-Newman solution to illustrate the emergence of an extra isometry which forms the ${\mathfrak{sl}}(2,\bR)$ algebra \cite{genextrsl},
\vspace{2mm}
\begin{eg}
It is instructive to consider the case of the extremal Reisser-Nordstr\"om solution with metric written in Gaussian null co-ordinates as:
\bea
ds^2 = - r^2 \big(M+r\big)^{-2} du^2 + 2 du dr + \big(M+r\big)^2 (d\theta^2 +  \sin^2{\theta} d\phi^2) \ .
\ee
On taking the near-horizon limit as described previously, the metric becomes
\begin{eqnarray}
ds^2 = 2(dr  - \frac{1}{2}r^2 \Delta du) du + \gamma_{1 1}d\theta^2 + \gamma_{2 2}d\phi^2
\end{eqnarray}
with the near-horizon data,
\bea
\Delta = \frac{1}{M^2},~~ \gamma_{1 1} = M^2, ~~\gamma_{2 2} = M^2\sin^2 \theta \ ,
\ee
which is the metric of $AdS_2 \times S^2$. 
The isometries of $AdS_2$, denoted by $\{ K_1, K_2, K_3 \}$ are given by
\bea
K_1 = \partial_{u},~K_2 = -u\partial_{u} + r\partial_{r},~ K_3 = -\frac{u^2}{2}\partial_{u} + (M^2 + u r)\partial_{r} \ ,
\ee
which satisfy the $\mathfrak{sl}(2,\mathbb{R})$ algebra
\bea
[K_1, K_2] = -K_1,~[K_1, K_3] = K_2,~[K_2,K_3] = -K_3 \ ,
\ee
and the isometries of the $S^2$ are given by $\{ K_4, K_5, K_6 \}$, with
\bea
K_4 &=& \partial_{\phi},~~ K_5 =\sin{\phi}\partial_{\theta} + \cos{\phi}\cot{\theta}\partial_{\phi},~~
K_6 = \cos{\phi}\partial_{\theta} - \sin{\phi}\cot{\theta}\partial_{\phi} \ ,
\ee
which satisfy the Lie algebra ${\mathfrak{so}}(3)$,
\bea
[K_4, K_5] = K_6,~ [K_4, K_6] = -K_5,~ [K_5, K_6] = K_4 \ .
\ee
\end{eg}
\vspace{2mm}
\begin{eg}
Now let us consider the extremal Kerr metric. In the usual NHL we first take the extremal limit ($a=M$) in Kerr coordinates,
\bea
ds^2 &=& -\bigg(\frac{r^2 -M^2 +M^2\cos^2 \theta}{(r+M)^2 + M^2 \cos^2 \theta}\bigg)du^2 + 2du dr - M\sin^2 \theta dr d\phi 
\nonumber \\
&-& \bigg(\frac{2M^2(r+M)\sin^2 \theta }{(r+M)^2 + M^2\cos^2 \theta}\bigg) du d\phi
+ ((r+M)^2+M^2 \cos^2 \theta)d\theta^2 
\nonumber \\
&+& \bigg(\frac{(M^2 r^2 \cos^2 \theta + 4M^4 + 8M^3 r + 7M^2 r^2 + 4M r^3 + r^4)\sin^2 \theta }{(r+M)^2 + M^2 \cos^2 \theta}\bigg) d\phi^2 \ ,
\ee
and then the near-horizon limit 
\bea
r\rightarrow \epsilon \hat{r},~ u\rightarrow \epsilon^{-1} \hat{u},~ \phi \rightarrow \hat{\phi} + \frac{\hat{u}}{2M}\epsilon^{-1},~ \epsilon \rightarrow 0 \ ,
\ee
and subsequently drop the hats and repeat this after we make the change,
\bea
r \rightarrow \bigg(\frac{2}{\cos^2 \theta + 1}\bigg) \hat{r} \ ,
\ee
to get the metric into the form,
\begin{eqnarray}
ds^2 = 2(dr + r h_1 d\theta + r h_2 d\phi - \frac{1}{2}r^2 \Delta du) du + \gamma_{1 1}d\theta^2 + \gamma_{2 2}d\phi^2 \ ,
\end{eqnarray}
and the near-horizon data given by,
\bea
\Delta &=& \frac{(\cos^4  \theta + 6\cos^2  \theta - 3)}{M^2 (\cos^2  \theta + 1)^3 } \ ,
\nonumber \\
h_1 &=& \frac{2\cos \theta\sin \theta}{\cos^2  \theta + 1},~~ h_2 = \frac{4 \sin^2  \theta}{(\cos^2  \theta + 1)^2} \ ,
\nonumber \\
\gamma_{1 1} &=& M^2(\cos^2  \theta + 1),~~ \gamma_{2 2} = \frac{4M^2 \sin^2  \theta}{\cos^2  \theta + 1} \ .
\ee
The Killing vectors ${K_1, K_2, K_3, K_4}$ of this near-horizon metric are given by,
\bea
K_1 &=& \partial_{u},~ K_2 = -u\partial_{u} + r\partial_{r} - \partial_\phi \ ,
\nonumber \\
K_3 &=& -\frac{u^2}{2}\partial_{u} + (2M^2 + u r)\partial_{r} - u\partial_{\phi} \ ,
\nonumber \\
K_4 &=& \partial_{\phi} \ ,
\ee
with the Lie algebra $\mathfrak{sl}(2,\mathbb{R}) \times \mathfrak{u}(1)$,
\bea
[K_1,K_2]=-K_1,~ [K_1,K_3]= K_2,~[K_2,K_3]=-K_3 \ .
\ee
\end{eg}
\vspace{2mm}
\begin{eg}
Finally, we consider the Kerr-Newman metric in Kerr coordinates. We take the extremal limit ($Q^2 = M^2 - a^2$) 
\bea
ds^2 &=& -\bigg(\frac{r^2 -a^2 + a^2\cos^2 \theta}{(r+M)^2 + a^2 \cos^2 \theta}\bigg)du^2 + 2du dr - a\sin^2 \theta dr d\phi 
\nonumber \\
&-& \bigg(\frac{a (a^2 +M^2 + 2Mr)\sin^2 \theta}{(r+M)^2 + a^2\cos^2 \theta}\bigg) du d\phi
+ ((r+M)^2+a^2 \cos^2 \theta)d\theta^2 
\nonumber \\
&+& \bigg(\frac{ (a^2 r^2 \cos^2 \theta + a^4 + (r^2+4Mr + 2M^2 )a^2 + (r+M)^4)\sin^2 \theta}{(r+M)^2 + a^2 \cos^2 \theta}\bigg) d\phi^2 \ ,
\ee
and then the near-horizon limit,
\bea
r\rightarrow \epsilon \hat{r},~ u\rightarrow \epsilon^{-1} \hat{u},~ \phi \rightarrow \hat{\phi} + \frac{a \hat{u}}{(a^2 + M^2)}\epsilon^{-1},~ \epsilon \rightarrow 0 \ ,
\ee
after which we also make the change 
\bea
r \rightarrow \bigg(\frac{(a^2 + M^2)}{a^2\cos^2 \theta + M^2}\bigg)\hat{r} \ ,
\ee
and dropping the hats after each coordinate transformation to get the metric into the form,
\begin{eqnarray}
ds^2 = 2(dr + r h_1 d\theta + r h_2 d\phi - \frac{1}{2}r^2 \Delta du) du + \gamma_{1 1}d\theta^2 + \gamma_{2 2}d\phi^2 \ ,
\end{eqnarray}
with the near-horizon data,
\bea
\Delta &=& \frac{(a^4 \cos^4  \theta + 6a^2 M^2\cos^2  \theta - 4a^2 M^2 + M^4)}{ (a^2\cos^2  \theta + M^2)^3 } \ ,
\nonumber \\
h_1 &=& \frac{2a^2\cos \theta\sin \theta}{a^2\cos^2  \theta + M^2},~~ h_2 = \frac{2 a M(a^2 + M^2) \sin^2  \theta}{(a^2\cos^2  \theta + M^2)^2} \ ,
\nonumber \\
\gamma_{1 1} &=& a^2\cos^2  \theta + M^2,~~ \gamma_{2 2} = \frac{(a^2 +M^2)^2 \sin^2  \theta}{a^2\cos^2  \theta + M^2} \ .
\ee
The Killing vectors ${K_1, K_2, K_3, K_4}$ of this near-horizon metric are given by,
\bea
K_1 &=& \partial_{u},~K_{2} = -u\partial_{u} + r\partial_{r}- \bigg(\frac{2aM}{a^2 + M^2}\bigg) \partial_\phi \ ,
\nonumber \\
K_3 &=& -\frac{u^2}{2}\partial_{u} + (a^2 + M^2 + u r)\partial_{r} - \bigg(\frac{2aM u}{a^2 + M^2}\bigg)\partial_{\phi} \ ,
\nonumber \\
K_4 &=& \partial_{\phi} \ ,
\ee
with the Lie algebra $\mathfrak{sl}(2,\mathbb{R}) \times \mathfrak{u}(1)$,
\bea
[K_1,K_2]=-K_1,~ [K_1,K_3]= K_2,~[K_2,K_3]=-K_3 \ .
\ee
\end{eg}
As we have previously remarked, the isometries $K_1$ and $K_2$ are generic for all
near-horizon geometries. In these cases, an additional isometry $K_3$ is present; which also follow from known near-horizon symmetry theorems \cite{genextrsl} for non-supersymmetric extremal horizons.
We shall show that the emergence of such an extra isometry, in the near-horizon limit, which forms the $\mathfrak{sl}(2,\mathbb{R})$ algebra is generic for {\it supersymmetric} black holes.

\subsection{Curvature of the near-horizon geometry}

As we will see, geometric equations (such as Einstein's equations) for a near-horizon geometry can be equivalently written as geometric equations defined purely on a $(D-2)$-dimensional spatial cross section manifold ${\cal S}$ of the horizon. It is convenient to introduce a null-orthonormal frame for the near-horizon metric, denoted by $(\bbe^A)$, where $A=(+,-, i)$, $i=1, \dots, D-2$ and
\be
\label{basis1}
\bbe^+ = du, \qquad \bbe^- = dr + rh -{1 \over 2} r^2 \Delta du, \qquad \bbe^i = \bbe^i{}_I dy^I \ ,
\ee
so that $ds^2 = g_{A B} \bbe^A \bbe^B = 2 \bbe^+ \bbe^- + \delta_{ij} \bbe^i \bbe^j~,$ where $\bbe^i$ are vielbeins for the horizon metric $\delta_{i j}$. The dual basis vectors are frame derivatives which are expressed in terms of co-ordinate derivatives  as
\begin{eqnarray}
\label{frco}
\bbe_+ = \partial_+  = \partial_u  +{1 \over 2} r^2 \Delta \partial_r ~,~~
\bbe_- = \partial_-  = \partial_r ~,~~
\bbe_i = \partial_i  = {\tilde{\partial}}_i  -r h_i \partial_r  \ .
\end{eqnarray} 
The spin-connection 1-forms satisfy $d\bbe^A= -\Omega^A_{\phantom{A}B} \wedge \bbe^B$ and are given by
\begin{eqnarray}
\Omega_{+-} &=& -r\Delta \bbe^+ + \frac{1}{2}h_i \bbe^i \,, 
\nonumber \\
\Omega_{+i} &=& -\frac{1}{2}r^2(\partial_i \Delta-\Delta h_i)\bbe^+-\frac{1}{2}h_i\bbe^- +{1 \over 2} r dh_{ij} \bbe^j \,,~~~ 
\nonumber \\
\Omega_{-i} &=& -\frac{1}{2}h_i \bbe^+ \,,   \quad \Omega_{i j} = \tilde{\Omega}_{i j}- {1 \over 2} r dh_{ij} \bbe^+ \ ,
\end{eqnarray}
where $\tilde{\Omega}_{i j}$ are the spin-connection 1-forms of the $(D-2)$-manifold ${{\cal{S}}}$ with metric $\delta_{i j}$ and basis ${\bf{e}}^i$. Here we have made use of the following identities:
\bea
d{\bf{e^+}} = 0, \; \; \; d{\bf{e^-}} = {\bf{e^-}} \wedge h + r dh + \frac{1}{2}{\bf{e^+}}\wedge(-r^2\Delta h + r^2 d\Delta + 2r\Delta{\bf{e^-}}) \ .
\ee
The non-vanishing components of the spin connection are
\begin{eqnarray}
\label{spin}
&&\Omega_{-,+i} = -{1 \over 2} h_i~,~~~
\Omega_{+,+-} = -r \Delta, \quad \Omega_{+,+i} ={1 \over 2} r^2(  \Delta h_i - \partial_i \Delta),
\cr
&&\Omega_{+,-i} = -{1 \over 2} h_i, \quad \Omega_{+,ij} = -{1 \over 2} r dh_{ij}~,~~~
\Omega_{i,+-} = {1 \over 2} h_i, \quad \Omega_{i,+j} = -{1 \over 2} r dh_{ij},
\cr
&&\Omega_{i,jk}= \tilde\Omega_{i,jk} \ .
\end{eqnarray}
The curvature two-forms defined by $\rho_{AB}= d\Omega_{AB}+ \Omega_{AC} \wedge \Omega^{C}_{\phantom{C}B}$ give the Riemann tensor in this basis using $\rho_{AB}= \frac{1}{2} R_{ABCD} \bbe^C \wedge \bbe^D$ and are given in Appendix A.
The non-vanishing components of the Ricci tensor with respect to the basis ({\ref{basis1}}) are
\bea
R_{+-} &=& {1 \over 2} \tilde{\nabla}^i h_i - \Delta -{1 \over 2} h^2~,~~~
R_{ij} = {\hat{R}}_{ij} + \tilde{\nabla}_{(i} h_{j)} -{1 \over 2} h_i h_j \ ,
\nonumber \\
R_{++} &=& r^2 \bigg( {1 \over 2} \tilde{\nabla}^2 \Delta -{3 \over 2} h^i \tilde{\nabla}_i \Delta -{1 \over 2} \Delta \tilde{\nabla}^i h_i + \Delta h^2
+{1 \over 4} (dh)_{ij} (dh)^{ij} \bigg) \ ,
\nonumber \\
R_{+i} &=& r \bigg( {1 \over 2} \tilde{\nabla}^j (dh)_{ij} - (dh)_{ij} h^j - \tilde{\nabla}_i \Delta + \Delta h_i \bigg) \ ,
\ee
where ${\hat{R}}$ is the Ricci tensor of the metric $\delta_{i j}$ on the horizon section ${\cal S}$ in the $\bbe^i$ frame. The spacetime contracted Bianchi identity implies the following identities \cite{genextrsl} on ${\cal S}$:
\begin{eqnarray}
\label{Sid}
R_{++} &=& - \frac{1}{2} r (\tilde{\nabla}^i-2h^i) R_{+ i} \,,
\cr
R_{+ i} &=& r\bigg(-\tilde{\nabla}^j [R_{j i} - \frac{1}{2}\delta_{j i}( R^k_{~k}+2R_{+-})] + h^jR_{j i} - h_i R_{+-}\bigg) \ ,
\end{eqnarray}
which may also be verified by computing this directly from the above expressions. 

\subsection{The supercovariant derivative}
We can also decompose the supercovariant derivative of the spinor $\epsilon$ given by\footnote{We use the Clifford algebra conventions with mostly positive signature and $\{ \Gamma_{\mu}, \Gamma_{\nu} \} = 2g_{\mu \nu}$},
\bea
\label{covepsilon}
\nabla_{\mu} \epsilon = \partial_{\mu} \epsilon + \frac{1}{4}\Omega_{\mu, \nu \rho}\Gamma^{\nu \rho} \epsilon \ ,
\ee
with respect to the basis ({\ref{basis1}}), 
which will be useful later for the analysis of KSEs. After expanding each term and evaluating the components of the spin connection with ({\ref{spin}}) and the frame derivatives with ({\ref{frco}}) we have,
\bea
\nabla_{+} \epsilon &=& \partial_u \epsilon + \frac{1}{2}r^2 \Delta \partial_r \epsilon + \frac{1}{4}r^2(\Delta h_i - \partial_i \Delta)\Gamma^{+i}\epsilon - \frac{1}{4}h_i\Gamma^{-i}\epsilon 
- \frac{1}{2}r\Delta\Gamma^{+-}\epsilon - \frac{1}{8}r(dh)_{ij}\Gamma^{ij}\epsilon \ ,
\cr
\nabla_{-} \epsilon &=& \partial_{r} \epsilon - \frac{1}{4}h_i \Gamma^{+i}\epsilon \ ,
\cr
\nabla_{i} \epsilon &=& \tilde{\nabla}_{i}\epsilon - r\partial_{r} \epsilon h_i - \frac{1}{4}r(dh)_{ij}\Gamma^{+j}\epsilon + \frac{1}{4}h_i \Gamma^{+-}\epsilon \ .
\ee
The integrability condition for ({\ref{covepsilon}) can be written in terms of the Riemann and Ricci tensor as,
\bea
[\nabla_{\mu}, \nabla_{\nu}]\epsilon &=& \frac{1}{4} R_{\mu \nu, \rho \sigma} \Gamma^{\rho \sigma} \epsilon,~\Gamma^{\nu}[\nabla_{\mu}, \nabla_{\nu}]\epsilon  = -\frac{1}{2}R_{\mu \sigma} \Gamma^{\sigma}\epsilon \ .
\ee
Similarly, the covariant derivative of a vector $\xi^{\rho}$ can be written in terms of the spin connection as,
\bea
\nabla_{\mu}{\xi^{\rho}} = \partial_{\mu}{\xi^{\rho}} + \Omega_{\mu,}{}^{\rho}{}_{\lambda} \xi^{\lambda} \ ,
\ee
and for a Killing vector $\xi$ we can write the integrability condition associated with the covariant derivative in terms of the Riemann and Ricci tensor as,
\bea
[\nabla_{\mu}, \nabla_{\nu}]\xi^{\rho} = R^{\rho}{}_{\lambda, \mu \nu}\xi^{\lambda},~\nabla_{\mu}\nabla_{\nu} \xi^{\mu} = R_{\rho \nu}\xi^{\rho} \ .
\ee
\section{Field strengths}

Consider a $p$-form field strength, $F_{(p)}$. Suppose that
the components of this field strength, when written in the 
Gaussian null co-ordinates are independent of $u$ and smooth (or at least $C^2$) in $r$,
and furthermore that it admits a well-defined near-horizon limit. Such a field strength, after taking the near-horizon limit, can always be decomposed  with respect to the basis ({\ref{basis1}) as follows:
\bea
F_{(p)} = {\bf{e^+}} \wedge {\bf{e^-}} \wedge L_{(p-2)} + r{\bf{e^+}}\wedge M_{(p-1)} + N_{(p)},~~ p>1 \ .
\ee
where $L_{(p-2)}$, $M_{(p-1)}$ and $N_{(p)}$ are $p-2$, $p-1$ and $p$-forms
on the horizon spatial cross-section which are independent of $u$ and $r$.
On taking the exterior derivative one finds\footnote{$d_h \alpha = d\alpha - h \wedge \alpha$}
\bea 
dF_{(p)} = {\bf{e}^+} \wedge {\bf{e}^-} \wedge (d_h L_{(p-2)} - M_{(p-1)}) + r{\bf{e}^+}\wedge (-d_h M_{(p-1)} - dh \wedge L_{(p-2)}) + dN_{(p)} \ .
\ee
If $F_{(p)} = dA_{(p-1)}$ with gauge potential $A_{(p-1)}$ then $dF_{(p)} = 0$ as with the common Bianchi identities, we get the following conditions;
\bea
M_{(p-1)} = d_h L_{(p-2)},~~d_h M_{(p-1)} = -dh \wedge L_{(p-2)},~~dN_{(p)} = 0 \ .
\ee 
The third implies $N_{(p)}$ is a closed form on the spatial section $\cal{S}$. The second condition is not independent as it is implied by the first. 

We will now give a reminder of the maximum principle and the classical Lichnerozicz theorem, which are crucial in establishing the results of (super)symmetry enhancement.

\section{The maximum principle}

In the analysis of the global properties of near-horizon geometries, we shall
obtain various equations involving the Laplacian of a non-negative scalar $f$.
Typically $f$ will be associated with the modulus of a particular spinor.
Such equations will be analysed either by application of integration by parts, or
by the Hopf maximum principle. The background manifold ${\cal{N}}$
is assumed to be smooth and compact without boundary, and all tensors are also
assumed to be smooth.

In the former case, we shall obtain second order PDEs on $\cal{N}$ given by,
\bea
\nabla^i \nabla_i f + \lambda^i \nabla_i f + \nabla_{i}(\lambda^i)f = \alpha^2 \ ,
\ee
where $\lambda^i$ is a smooth vector and $\alpha \in \bR$. This can be rewritten as,
\bea
\nabla^{i} V_i = \alpha^2 \ ,
\ee
with $V_i = \nabla_i f + \lambda_i f$. By partial integration over $\cal{N}$, the LHS vanishes since it is a total derivative and we have,
\bea
\alpha = 0,~~ \nabla^i V_i = 0 \ .
\ee
In the latter case, we shall obtain PDEs of the form
\bea
\nabla^{i}\nabla_{i}f + \lambda^i \nabla_{i}f = \alpha^2 \ ,
\ee
and an application of the Hopf maximum principle, which states that if $f \geq 0$ is a $C^2$-function which attains a maximum value in $\cal{N}$ then,
\bea
f = const,~~ \alpha = 0,  \ .
\ee

\section{The classical Lichnerowicz theorem}

A particularly important aspect of the analysis of the Killing spinor equations associated with the near-horizon geometries of black holes is the proof of certain types of generalized Lichnerowicz theorems. These state that if a spinor is a zero mode of a certain class of near-horizon Dirac operators, then it is also parallel with respect to a particular class of supercovariant derivatives, and also satisfies various algebraic conditions. These Dirac operators and supercovariant connections depend linearly on certain types of $p$-form fluxes which appear
in the supergravity theories under consideration. Before attempting to derive these results it is instructive to recall how the classical Lichnerowicz theorem arises, in the case when the fluxes are absent.

On any spin compact manifold ${\cal{N}}$, without boundary, one can establish the equality
\bea
\int_{\cal{N}} \langle \Gamma^i \nabla_i \epsilon, \Gamma^j \nabla_j \epsilon \rangle=  \int_{\cal{N}} \langle  \nabla_i \epsilon ,  \nabla^i \epsilon \rangle+\int_{\cal{N}} {R\over 4} \langle \epsilon , \epsilon \rangle \ .
\label{classlich}
\ee
To show this, we let
\bea
{\cal{I}} = \int_{\cal{N}} \langle \nabla_i \epsilon ,\nabla^i \epsilon \rangle
- \langle \Gamma^i \nabla_i \epsilon ,
 \Gamma^j \nabla_j \epsilon \rangle \ .
\ee
This can be rewritten as\footnote{The gamma matrices are Hermitian $(\Gamma^i)^{\dagger} = \Gamma^i$ with respect to this inner product.},
\bea
{\cal{I}} = \int_{\cal{N}} -\nabla_{i}\langle \epsilon, \Gamma^{i j}\nabla_{j}\epsilon \rangle + \int_{\cal{N}} \langle \epsilon, \Gamma^{i j}\nabla_i \nabla_j \epsilon\rangle \ .
\ee
The first term vanishes since the integrand is a total derivative and for the second term we use $\Gamma^{i j}\nabla_i \nabla_j \epsilon = - \frac{1}{4}R \epsilon$, thus we have
\bea
{\cal{I}} = -\int_{\cal{N}} {R\over 4} \langle \epsilon , \epsilon \rangle \ ,
\ee
where $\nabla$ is the Levi-Civita connection, $\langle \cdot, \cdot\rangle$ is the real and positive definite $Spin$-invariant Dirac inner product (see Appendix B) identified with the standard Hermitian inner product and $R$ is the Ricci scalar. 
On considering the identity ({\ref{classlich}}), it is clear that if $R>0$ then
the Dirac operator has no zero modes. Moreover, if $R=0$, then the zero modes of the Dirac operator are parallel with respect to the Levi-Civita connection.

An alternative derivation of this result can be obtained by noting that if $\epsilon$ satisfies the Dirac equation $\Gamma^{i}\nabla_{i}\epsilon = 0$, then
\bea
\nabla^{i}\nabla_{i}  \parallel \epsilon \parallel^2 = \frac{1}{2}R  \parallel \epsilon \parallel^2 + 2\langle \nabla_{i}\epsilon, \nabla^{i}\epsilon \rangle \ .
\label{classlich2}
\ee
This identity is obtained by writing
\bea
\nabla^i \nabla_i \parallel \epsilon \parallel^2 = 2\langle\epsilon ,\nabla^i \nabla_i\epsilon\rangle + 2 \langle\nabla^i \epsilon, \nabla_i \epsilon\rangle \ .
\ee
To evaluate this expression note that
\bea
\nabla^i \nabla_i \epsilon = \Gamma^{i}\nabla_{i}(\Gamma^{j}\nabla_j \epsilon) -\Gamma^{i j}\nabla_i \nabla_j \epsilon = \frac{1}{4}R\epsilon \ .
\ee
On considering the identity ({\ref{classlich2}}), if $R \geq 0$, then the RHS of 
({\ref{classlich2}}) is non-negative. An application of the Hopf maximum principle  then implies that $ \parallel \epsilon \parallel^2$ is constant, and moreover
that $R=0$ and $\nabla \epsilon=0$.

	\chapter{Supergravity}

In this section we summarize the properties of various types of supergravity theories, whose near-horizon geometries will be considered later.

\label{ch:papers}

\section{$D=11$ to IIA supergravity}
It will turn out to be a rewarding path to begin with the eleven dimensional supergravity theory. Supersymmetry ensures that this theory is unique. Furthermore the IIA ten-dimensional supergravity has to be the dimensional reduction of this higher-dimensional theory,  since the two theories have the same supersymmetry algebras. $D=11$ supergravity on the spacetime ${\cal{M}}_{10} \times S^1$ is equivalent to IIA supergravity on the 10-dimensional manifold ${\cal{M}}_{10}$ where masses proportional to the inverse radius of $S^1$ are eliminated.

The field content of this theory is rather simple: for the bosons there are just graviton $G_{M N}$ with \(\frac{9\times 10}{2} -1 =44 \) components and a three-form potential $A^{(3)}$ with \(\frac{9\times 8\times 7}{3!}=84\) components, in representation of the SO(9) little group of massless particles in eleven dimensions. 
 There is also the gravitino $\psi_M$ with its \(16 \times 8\) degrees of freedom, in representation of the covering group \(Spin(9)\). This is indeed the same number as the number of massless degrees of freedom of the type II string theory.
 
\index{eleven-dimensional supergravity}
The bosonic part of the action is
\be\label{S11}
(16 \p G_\text{\tiny N}^{\text{\tiny(11)}}) \, S^{\text{\tiny(11)}} &=& \int d^{11}x \sqrt{-G} R- \frac{1}{2} \int F^{(4)} \wedge \star_{11} F^{(4)}  -\frac{1}{3!} \int A^{(3)}\wedge F^{(4)}\wedge F^{(4)}\;,
\nonumber \\
&=& \int d^{11}x \sqrt{-G} \bigg(R- \frac{1}{48}F_{M_1 M_2 M_3 M_4}F^{M_1 M_2 M_3 M_4}\bigg)  -\frac{1}{6} \int A^{(3)}\wedge F^{(4)}\wedge F^{(4)} \ ,
\nonumber \\
\ee
where \(F^{(4)}\) is the $U(1)$ field strength of the three-form potential \(F^{(4)} =  dA^{(3)}\). This leads to the field equations
\bea
R_{MN} &=& {1 \over 12} F_{M L_1 L_2 L_3} F_N{}^{L_1 L_2 L_3}
-{1 \over 144} G_{MN} F_{L_1 L_2 L_3 L_4}F^{L_1 L_2 L_3 L_4}~,
\cr
d \star  F &=&{1 \over 2} F \wedge F \ .
\la{feqns}
\ee
The supersymmetry variations of 11-dimensional supergravity for the bosons\footnote{These are trivial when we consider a classical background and set the fermions to zero} are given by\footnote{The frame fields $e^{a}_{M}$ are defined as $g^{M N}e^{a}_{M} e^{b}_{N} = \eta^{a b}$ where $a$ labels the local spacetime and $\eta_{a b}$ is the Lorentz metric},
\bea
\delta e^{a}_{M} = i \bar{\epsilon} \Gamma^{a} \psi_M,~~~ \delta A_{M_1 M_2 M_3} = 3i \bar{\epsilon} \Gamma_{[M_1 M_2}\psi_{M_3]} \ ,
\ee
while for the fermions we have,
\bea
\delta \psi_M = \nabla_M \epsilon
+\bigg(-{1 \over 288} \Gamma_M{}^{L_1 L_2 L_3 L_4} F_{L_1 L_2 L_3 L_4}
+{1 \over 36} F_{M L_1 L_2 L_3} \Gamma^{L_1 L_2 L_3} \bigg) \epsilon \ .
\ee
To dimensionally reduce it, write the eleven-dimensional metric as
\be\label{reduce_metric}
G_{MN} = e^{-\frac{2}{3}\F} \bem  g_{\m\n}+e^{2\F}A_\m A_\n &  e^{2\F}A_\m \\
e^{2\F}A_\n & e^{2\F}
\eem \ ,
\ee
where we use \(M, N, \dotsi = 0,1,\dotsi,10\) to denote the eleven-dimensional and  \(\m, \n, \dotsi = 0,1,\dotsi,9\) the ten-dimensional directions. We also reduce the three-form potential \(A^{(3)}_{MNP}\) as \(C_{\m\n\r}\) when it has no ``leg" in the 11-th direction and as \(A^{(3)}_{MN10}=B_{\m\n}\) and \(H = dB \) when it does. 
Under this field redefinition, and truncating all the dependence on the eleventh direction, the action reduces to\footnote{We use the notation $|F_{(p)}|^2 = \frac{1}{p!}F_{\mu_1 \cdots \mu_p}F^{\mu_1 \cdots \mu_p}$}
\bea
S^{(\text{\tiny IIA})} &=& S_{\text{NS}} + S_{\text{R}}^{(\text{\tiny IIA})}
 + S_{\text{C-S}}^{(\text{\tiny IIA})} \ ,
\nonumber \\
2\k^2 S_{\text{NS}}  &=& \int d^{10} \sqrt{-g} \, e^{-2\F} \left( R+ 4\pa_\m\F\pa^\m\F -\frac{1}{2} |H|^2  \right) \ ,
\nonumber \\
2\k^2 S_{\text{R}}^{(\text{\tiny IIA})} & =& -\frac{1}{2} \int d^{10}x \left( |F|^2 +  |G|^2\right) \ ,
\nonumber \\
2\k^2S_{\text{C-S}}^{(\text{\tiny IIA})}&=& -\frac{1}{2!} \int B \wedge dC \wedge dC \ ,
\ee
where \(F\) is the field strength of the Kaluza-Klein gauge field $A$ and $G$ is the field strength modified by the Chern-Simons term. This is the bosonic action for the type IIA supergravity that we want to construct. The bosonic field content  of IIA supergravity are the spacetime metric  $g$, the dilaton $\Phi$, the 2-form NS-NS gauge potential $B$,
and the 1-form and the 3-form RR gauge potentials  $A$ and $C$, respectively. In addition,
the  theory has  non-chiral fermionic fields consisting of a Majorana gravitino and  a Majorana dilatino  but these are set to zero
in all the computations that follow. The bosonic field strengths of IIA supergravity in the conventions of \cite{howe} are
\bea
F=dA~,~~~H=dB~,~~~G=dC-H\wedge A \ .
\ee
These lead to the Bianchi identities
\bea
\label{iiabian}
dF=0~,~~~dH=0~,~~~dG=F\wedge H \ .
\ee
The bosonic part of the IIA action in the string frame is
\bea
S &=& \int \sqrt{-g} \bigg(e^{-2 \Phi} \big(R+4 \nabla_\mu \Phi\nabla^\mu \Phi -{1 \over 12} H_{\lambda_1 \lambda_2 \lambda_3}
H^{\lambda_1 \lambda_2 \lambda_3} \big)
\nonumber \\
&-&{1 \over 4} F_{\mu \nu} F^{\mu \nu}
-{1 \over 48} G_{\mu_1 \mu_2 \mu_3 \mu_4}
G^{\mu_1 \mu_2 \mu_3 \mu_4}\bigg) + {1 \over 2} dC \wedge dC \wedge B \ .
\ee
This leads to the Einstein equation
\bea
R_{\mu \nu}&=&-2 \nabla_\mu \nabla_\nu \Phi
+{1 \over 4} H_{\mu \lambda_1 \lambda_2} H_\nu{}^{\lambda_1 \lambda_2}
+{1 \over 2} e^{2 \Phi} F_{\mu \lambda} F_\nu{}^\lambda
+{1 \over 12} e^{2 \Phi} G_{\mu \lambda_1 \lambda_2 \lambda_3}
G_\nu{}^{\lambda_1 \lambda_2 \lambda_3}
\nonumber \\
&+& g_{\mu \nu} \bigg(-{1 \over 8}e^{2 \Phi}
F_{\lambda_1 \lambda_2}F^{\lambda_1 \lambda_2}
-{1 \over 96}e^{2 \Phi} G_{\lambda_1 \lambda_2 \lambda_3
\lambda_4} G^{\lambda_1 \lambda_2 \lambda_3
\lambda_4} \bigg) \ ,
\ee
the dilaton field equation
\bea
\label{iiadileq}
\nabla^\mu \nabla_\mu \Phi
&=& 2 \nabla_\lambda \Phi \nabla^\lambda \Phi
-{1 \over 12} H_{\lambda_1 \lambda_2 \lambda_3}
H^{\lambda_1 \lambda_2 \lambda_3}+{3 \over 8} e^{2 \Phi}
F_{\lambda_1 \lambda_2} F^{\lambda_1 \lambda_2}
\nonumber \\
&+&{1 \over 96} e^{2 \Phi} G_{\lambda_1 \lambda_2 \lambda_3
\lambda_4} G^{\lambda_1 \lambda_2 \lambda_3
\lambda_4} \ ,
\ee
the 2-form field equation
\bea
\label{iiageq1}
\nabla^\mu F_{\mu \nu} +{1 \over 6} H^{\lambda_1 \lambda_2 \lambda_3} G_{\lambda_1 \lambda_2 \lambda_3 \nu} =0 \ ,
\ee
the 3-form field equation
\bea
\label{iiabeq1}
\nabla_\lambda \bigg( e^{-2 \Phi} H^{\lambda \mu \nu}\bigg) &=& -{1 \over 1152} \epsilon^{\mu \nu \lambda_1 \lambda_2
	\lambda_3 \lambda_4 \lambda_5 \lambda_6 \lambda_7 \lambda_8}
G_{\lambda_1 \lambda_2 \lambda_3 \lambda_4}
G_{\lambda_5 \lambda_6 \lambda_7 \lambda_8}
+ {1 \over 2} G^{\mu \nu \lambda_1 \lambda_2} F_{\lambda_1 \lambda_2} \ ,
\nonumber \\
\ee
and the 4-form field equation
\bea
\label{iiaceq1}
\nabla_\mu G^{\mu \nu_1 \nu_2 \nu_3}
+{1 \over 144} \epsilon^{\nu_1 \nu_2 \nu_3
\lambda_1 \lambda_2 \lambda_3 \lambda_4 \lambda_5
\lambda_6 \lambda_7} G_{\lambda_1 \lambda_2 \lambda_3
\lambda_4} H_{\lambda_5 \lambda_6 \lambda_7}=0 \ .
\ee
This completes the description of the dynamics of the bosonic part of IIA supergravity. Now let us consider the supersymmetry variations of the fields. We denote $\Gamma_{11}$  the
chirality matrix, defined as
\bea
 \G_{\m_1 \dots \m_{10}} = - \e_{\m_1 \dots \m_{10}} \G_{11} \ .
\ee
The supersymmetry transformations of all  fields to lowest
order in the fermions are 
\bea
\delta e^a &=& \bar{\e}\, \G^a \psi\,, 
\nonumber \\
\delta B_{(2)} &=& 2 \bar{\e}\, \G_{11} \G_{(1)} \psi\,,
\nonumber \\
\delta \phi &=& \frac{1}{2} \bar{\e}\, \l\,,
\nonumber \\
\delta C_{(1)} &=& - e^{-\phi} \bar{\e}\, \G_{11} \psi  + \frac{1}{2}
  e^{-\phi} \bar{\e}\, \G_{11} \G_{(1)} \l\,, 
\nonumber \\
\delta C_{(3)} &=& -3 e^{-\phi} \bar{\e}\, \G_{(2)} \psi + \frac{1}{2}
  e^{-\phi} \bar{\e}\, \G_{(3)} \l + 3 C_{(1)} \delta B_{(2)} \ ,
  \ee
for the bosons, while the fermions transform according
to\footnote{In the case of the fermions, we leave the index
structure explicit since contractions are involved.}
  \bea
\delta \psi_\m &=& \nabla_\mu\e +{1\over8} H_{\mu\nu_1\nu_2} \Gamma^{\nu_1\nu_2}\Gamma_{11}\e+{1\over16} e^\Phi F_{\nu_1\nu_2} \Gamma^{\nu_1\nu_2} \Gamma_\mu \Gamma_{11} \e
\cr
&+& {1\over 8\cdot 4!}e^\Phi G_{\nu_1\nu_2\nu_3\nu_4} \Gamma^{\nu_1\nu_2\nu_3\nu_4} \Gamma_\mu\e\,, \\
\delta \lambda &=& \partial_\mu\Phi\, \Gamma^\mu \e+{1\over12} H_{\mu_1\mu_2\mu_3} \Gamma^{\mu_1\mu_2\mu_3} \Gamma_{11} \e+{3\over8} e^\Phi F_{\mu_1\mu_2} \Gamma^{\mu_1\mu_2} \Gamma_{11} \e
\cr
&+& {1\over 4\cdot 4!}e^\Phi\, G_{\mu_1\mu_2\mu_3\mu_4} \Gamma^{\mu_1\mu_2\mu_3\mu_4}\e=0 \ .
  \ee
The KSEs of IIA supergravity are the vanishing conditions of the
gravitino and dilatino supersymmetry variations evaluated at the locus where
all fermions vanish.
These can be expressed as
\bea
{\cal D}_\mu\e&\equiv& \delta \psi_\m = 0~,\label{iiaGKSE}\\
{\cal A}\e&\equiv&~ \delta \lambda = 0 \ .
\label{iiaAKSE}
\ee
where $\epsilon$ is the supersymmetry parameter which from now on is taken to be a Majorana, but not Weyl, commuting spinor of $Spin(9,1)$. We use the spinor conventions of \cite{spinorconv1, spinorconv2}, see Appendix B and D for the Clifford algebra. The Dirac spinors of $Spin(9,1)$ are identified with $\Lambda^{*}(\bC^5)$ and the Majorana spinors span a real 32-dimensional subspace after imposing an appropriate reality condition.

Suppose that a $D=11$ background has a symmetry generated by a vector field $X$ with closed orbits. The spinorial Lie derivative ${\cal L}_X$ associated with a Killing vector $X$ is given by,
\bea
\mathcal{L}_X \epsilon \equiv X^{\mu}\nabla_{\mu}\epsilon + \frac{1}{4}\nabla_{\mu}X_{\nu}\Gamma^{\mu \nu}\epsilon \ .
\ee
The $D=11$ spinors are related to IIA spinors after an appropriate rescaling with the dilaton. To our knowledge, a supersymmetry generated by a Killing spinor $\epsilon$ survives the reduction from $D=11$ to IIA  along  $X$ iff,
\bea
{\cal L}_X \epsilon=0 \ ,
\ee
\section{IIA to Roman's massive IIA}
\begin{figure}[h!]
	\begin{center}
		\includegraphics[width=300pt]{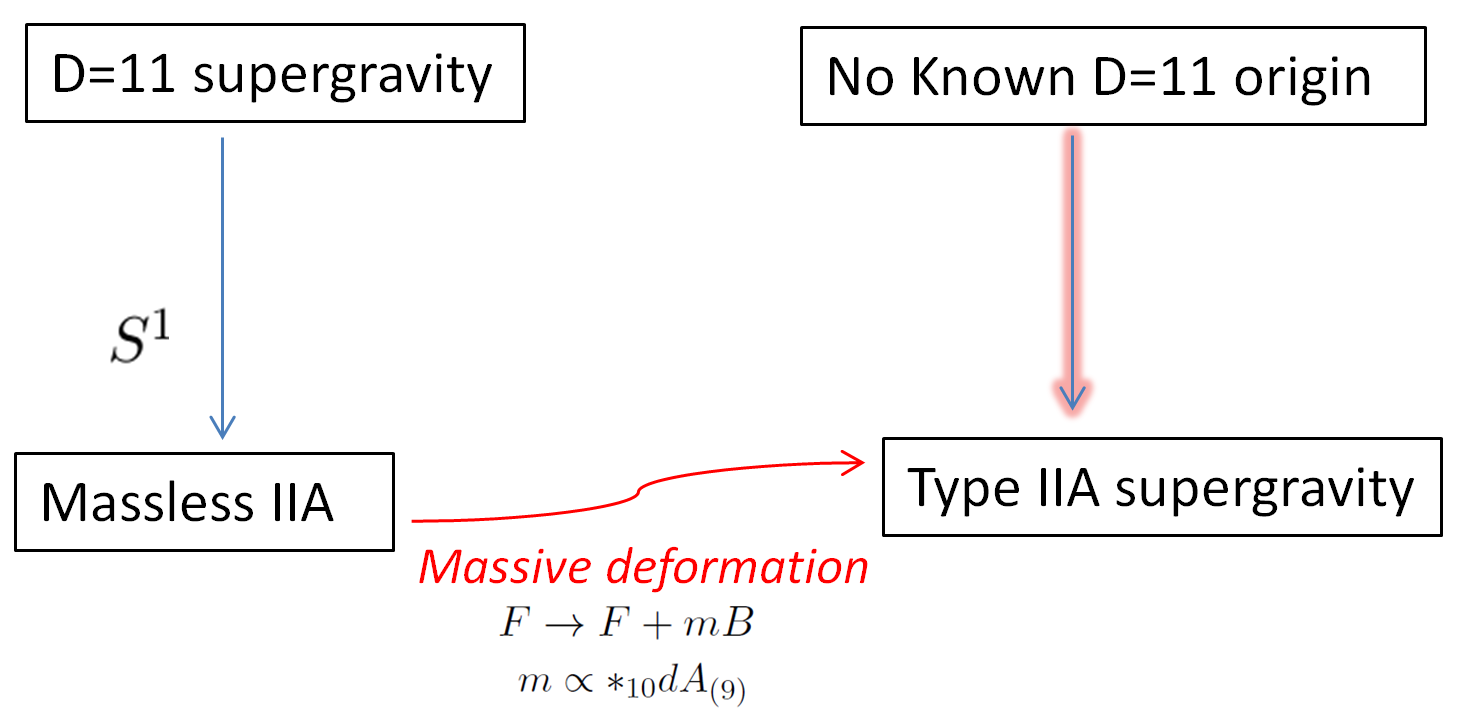}
	\end{center}
  \caption{Roman's massive IIA deformation}
  \label{fig:romans}
\end{figure}
The bosonic fields of massive IIA supergravity \cite{ romans} are the spacetime metric  $g$, the dilaton $\Phi$, the 2-form NS-NS gauge potential $B$, and the 1-form and the 3-form RR gauge potentials  $A$ and $C$, respectively. The theory also includes a mass parameter $m$ which induces a negative cosmological constant in the theory. In addition,
 fermionic fields of the theory are a Majorana gravitino $\psi_{\mu}$ and dilatino $\lambda$  which  are set to zero
in all the computations that follow.
The bosonic field strengths of massive IIA supergravity \cite{romans} in the conventions of \cite{roo} are
\bea
F=dA + mB ~,~~~H=dB~,~~~G=dC-H\wedge A + \frac{1}{2}m B \wedge B \ ,
\ee
implying the Bianchi identities
\bea
\label{miiabian}
dF=mH~,~~~dH=0~,~~~dG=F\wedge H \ ,
\ee
The bosonic part of the massive IIA action in the string frame is
\bea
S &=& \int \bigg[\sqrt{-g} \bigg(e^{-2 \Phi} \big(R+4 \nabla_\mu \Phi\nabla^\mu \Phi -{1 \over 12} H_{\lambda_1 \lambda_2 \lambda_3}
H^{\lambda_1 \lambda_2 \lambda_3} \big)
\nonumber \\
&-&{1 \over 4} F_{\mu \nu} F^{\mu \nu}
-{1 \over 48} G_{\mu_1 \mu_2 \mu_3 \mu_4}
G^{\mu_1 \mu_2 \mu_3 \mu_4} - \frac{1}{2}m^2 \bigg)
\nonumber \\
&+& {1 \over 2} dC \wedge dC \wedge B +{m \over 6} dC \wedge B \wedge B \wedge B+{m^2 \over 40} B \wedge B \wedge B \wedge B \wedge B \bigg] \ .
\ee
This leads to the Einstein equation
\bea
R_{\mu \nu}&=&-2 \nabla_\mu \nabla_\nu \Phi
+{1 \over 4} H_{\mu \lambda_1 \lambda_2} H_\nu{}^{\lambda_1 \lambda_2}
+{1 \over 2} e^{2 \Phi} F_{\mu \lambda} F_\nu{}^\lambda
+{1 \over 12} e^{2 \Phi} G_{\mu \lambda_1 \lambda_2 \lambda_3}
G_\nu{}^{\lambda_1 \lambda_2 \lambda_3}
\nonumber \\
&+& g_{\mu \nu} \bigg(-{1 \over 8}e^{2 \Phi}
F_{\lambda_1 \lambda_2}F^{\lambda_1 \lambda_2}
-{1 \over 96}e^{2 \Phi} G_{\lambda_1 \lambda_2 \lambda_3
\lambda_4} G^{\lambda_1 \lambda_2 \lambda_3
\lambda_4} - \frac{1}{4}e^{2\Phi}m^2 \bigg) \ ,
\ee
and the dilaton field equation
\bea
\label{miiadileq}
\nabla^\mu \nabla_\mu \Phi
&=& 2 \nabla_\lambda \Phi \nabla^\lambda \Phi
-{1 \over 12} H_{\lambda_1 \lambda_2 \lambda_3}
H^{\lambda_1 \lambda_2 \lambda_3}+{3 \over 8} e^{2 \Phi}
F_{\lambda_1 \lambda_2} F^{\lambda_1 \lambda_2}
\nonumber \\
&+&{1 \over 96} e^{2 \Phi} G_{\lambda_1 \lambda_2 \lambda_3
\lambda_4} G^{\lambda_1 \lambda_2 \lambda_3
\lambda_4} + \frac{5}{4}e^{2\Phi}m^2 \ ,
\ee
the 2-form field equation
\bea
\label{miiageq1}
\nabla^\mu F_{\mu \nu} +{1 \over 6} H^{\lambda_1 \lambda_2 \lambda_3} G_{\lambda_1 \lambda_2 \lambda_3 \nu} =0 \ ,
\ee
the 3-form field equation
\bea
\label{miiabeq1}
\nabla_\lambda \bigg( e^{-2 \Phi} H^{\lambda \mu \nu}\bigg) &=& mF^{\mu \nu} + {1 \over 2} G^{\mu \nu \lambda_1 \lambda_2} F_{\lambda_1 \lambda_2}- {1 \over 1152} \epsilon^{\mu \nu \lambda_1 \lambda_2
	\lambda_3 \lambda_4 \lambda_5 \lambda_6 \lambda_7 \lambda_8}
G_{\lambda_1 \lambda_2 \lambda_3 \lambda_4}
G_{\lambda_5 \lambda_6 \lambda_7 \lambda_8} \ ,
\nonumber \\
\ee
and the 4-form field equation
\bea
\label{miiaceq1}
\nabla_\mu G^{\mu \nu_1 \nu_2 \nu_3}
+{1 \over 144} \epsilon^{\nu_1 \nu_2 \nu_3
\lambda_1 \lambda_2 \lambda_3 \lambda_4 \lambda_5
\lambda_6 \lambda_7} G_{\lambda_1 \lambda_2 \lambda_3
\lambda_4} H_{\lambda_5 \lambda_6 \lambda_7}=0 \ ,
\ee
The supersymmetry transformations of all  fields to lowest
order in the fermions are 
\bea
\delta e^a &=& \bar{\e}\, \G^a \psi\,, 
\nonumber \\
\delta B_{(2)} &=& 2 \bar{\e}\, \G_{11} \G_{(1)} \psi\,,
\nonumber \\
\delta \phi &=& \frac{1}{2} \bar{\e}\, \l\,,
\nonumber \\
\delta C_{(1)} &=& - e^{-\phi} \bar{\e}\, \G_{11} \psi  + \frac{1}{2}
e^{-\phi} \bar{\e}\, \G_{11} \G_{(1)} \l\,, 
\nonumber \\
\delta C_{(3)} &=& -3 e^{-\phi} \bar{\e}\, \G_{(2)} \psi + \frac{1}{2}
e^{-\phi} \bar{\e}\, \G_{(3)} \l + 3 C_{(1)} \delta B_{(2)} \ ,
\ee
for the bosons, while the fermions transform according
to
  \bea
\delta \psi_\m &=& \nabla_\mu\e +{1\over8} H_{\mu\nu_1\nu_2} \Gamma^{\nu_1\nu_2}\Gamma_{11}\e+{1\over16} e^\Phi F_{\nu_1\nu_2} \Gamma^{\nu_1\nu_2} \Gamma_\mu \Gamma_{11} \e
\cr
&+&{1\over 8\cdot 4!}e^\Phi G_{\nu_1\nu_2\nu_3\nu_4} \Gamma^{\nu_1\nu_2\nu_3\nu_4} \Gamma_\mu\e + \frac{1}{8}e^\Phi m \Gamma_{\m} \e \,, \\
\delta \lambda &=& \partial_\mu\Phi\, \Gamma^\mu \e+{1\over12} H_{\mu_1\mu_2\mu_3} \Gamma^{\mu_1\mu_2\mu_3} \Gamma_{11} \e+{3\over8} e^\Phi F_{\mu_1\mu_2} \Gamma^{\mu_1\mu_2} \Gamma_{11} \e
\cr
&+& {1\over 4\cdot 4!}e^\Phi\, G_{\mu_1\mu_2\mu_3\mu_4} \Gamma^{\mu_1\mu_2\mu_3\mu_4}\e + \frac{5}{4}e^\Phi m \e=0 \quad \ .
  \ee
The KSEs of massive IIA supergravity are the vanishing conditions of the
gravitino and dilatino supersymmetry variations evaluated at the locus where
all fermions vanish.
These can be expressed as
\bea
{\cal D}_\mu\e&\equiv& \delta \psi_\m = 0~,\label{miiaGKSE}\\
{\cal A}\e&\equiv&~ \delta \lambda = 0 \ ,
\label{miiaAKSE}
\ee
where $\epsilon$ is the supersymmetry parameter which again is taken to be a Majorana, but not Weyl, commuting spinor of $Spin(9,1)$.

\section{$D=11$ to $D=5, N=2$ supergravity}
\subsection{Ungauged}
In this section, we briefly summarize some of the key properties of $D=5$, $N=2$ ungauged  supergravity \cite{dewit},
coupled to $k$ vector multiplets. We will also give the details of the compactification of
$D=11$ supergravity on a Calabi-Yau \cite{cadavid}. The bosonic part of the action is associated with a particular hypersurface $N$ of $\mathbb{R}^k$ defined by
\bea
V(X) = \frac{1}{6}C_{I J K}X^I X^J X^K = 1 \ .
\ee
For M-theory compactifications on a Calabi-Yau threefold $CY_3$ with Hodge numbers $h_{(1,1)}$,
$h_{(2,1)}$, and intersection numbers $C_{IJK}$, $V(X)$ represents the intersection
form of the Calabi-Yau threefold related to the overall volume of the Calabi-Yau threefold
and belongs to the so-called universal hypermultiplet. The scalars $X^I$
correspond
to the size of the 2-cycles of the Calabi-Yau threefold. The massless spectrum of the theory
contains $h_{(1,1)}$ − 1 vector multiplets with real scalar components defined by the moduli at
unit volume. Including the graviphoton, the theory has $h_{(1,1)}$ vector bosons. In addition
to the universal hypermultiplet present in any Calabi-Yau compactification, the theory
also contains $h_{(2,1)}$ hypermultiplets. 

To obtain the five-dimensional supergravity action, we first have to do a Kaluza-Klein reduction of eleven-dimensional supergravity, which is straightforward. We start with the eleven-dimensional supergravity action (\ref{S11}) which is completely fixed by $N=1$ supersymmetry in $D=11$ and $F_{4}=\mathrm{d}A_{3}$ is the field strength of the three-form gauge field. 

Recall that electric and magnetic charges are defined by integrating over the 4-sphere $S^4$ and 7-sphere $S^7$,
\bea
Q_{\mathrm{magnetic}}&=&Q_{\mathrm{M}5}\displaystyle \propto\int_{S^{4}}F_{4} \ ,
\nonumber \\
Q_{\mathrm{electric}}&=&Q_{\mathrm{M}2}\displaystyle \propto\int_{S^{7}}*_{11} F_4 \ ,
\ee
with $*_{11}$ being the eleven-dimensional Hodge star. To reduce on the compact manifold $\mathcal{M}_{6}$, we split the metric naturally into
\bea
\mathrm{d}s_{11}^{2}=\mathrm{d}s_{5}^{2}+\mathrm{d}s_{{\cal{M}}_6}^2 \ .
\ee
Besides the five-dimensional metric, we will find a number of gauge fields and scalar fields from the reduction. For the generic $N=2$ case, the gauge fields come from reduction of $A_{3}$, and the scalars come from the $CY_{3}$ moduli, of which there are two types,
\begin{itemize}
\item K\"ahler moduli, which combine into $D=5, N=2$ vector multiplets,
\item Complex structure moduli, which yield $D=5, N=2$ hypermultiplets.
\end{itemize}
The hypermultiplet scalars \cite{hyper1, hyper2} are dynamically decoupled for the purposes of investigating stationary solutions since they do not mix with the other fields (apart from the graviton) at the level of the equations of motion \cite{thy1, thy2}, and it is therefore consistent to set them to a constant value.\footnote{This is no longer true in gauged supergravities, where some of the hypermultiplets become charged under the vectors. \cite{gag1}} 

They are also normally neglected in black hole physics \cite{attract3}; at least in two-derivative gravity \cite{high1} they are just constants by no-hair theorems, and thus decouple effectively, a fact which is also elucidated by the attractor mechanism \cite{attract1, attract2}. We will thus neglect them in the following. One example is the overall size of ${\cal{M}}_{6}$, which we will just set to be a fixed constant, which we choose to be $\mathrm{V}\mathrm{o}\mathrm{l}({\cal{M}}_{6})=1$ in units of $\kappa_{11}^{2}=2\pi^{2}.$

To carry on, we expand the K\"ahler form $J$ on ${\cal{M}}_{6}$ in a complete basis of (1, 1)-forms $\{J_{I}\},$
\bea
J=\displaystyle \sum_{I}X^{I}J_{I},~~ I=1, \dots , h^{(1,1)} \ ,
\ee
$X^{I}$ are then the real K\"ahler moduli. The three-form gauge field can also be expanded using the following ansatz,
\bea
A_{3}=\displaystyle \sum_{I}A^{I}\wedge J_{I} \ ,
\ee
with $A^{I}$ being five-dimensional gauge fields. The four-form field strength thus decomposes as,
\bea
F_{4}=dA_{3} = \displaystyle \sum_{I}F^{I}\wedge J_{I} \ ,
\ee
yielding a collection of $U(1)$ gauge fields in five dimensions. The eleven-dimensional Chern- Simons term then reduces as
\bea
\int A_{3}\wedge F_{4}\wedge F_{4} &=& \int_{{\cal{M}}_{6}}J_{I}\wedge J_{J}\wedge J_{K}\int_{{\cal{M}}_5}A^{I}\wedge F^{J}\wedge F^{K}
\nonumber \\
&=& C_{IJK}\int_{{\cal{M}}_{5}}A^{I}\wedge F^{J}\wedge F^{K} \ ,
\ee
with $C_{IJK}$ being the 'triple intersection numbers'. Three two-cycles in a six-dimensional manifold generically intersect in a finite number of distinct points, which are counted by $C_{IJK}$. Also, since we set the volume of the internal manifold to one and the volume form is given by $J\wedge J\wedge J$, we enforced the `very special geometry' \cite{dewit} on the K\"ahler moduli,
\bea
V(X) = \displaystyle \mathrm{V}\mathrm{o}\mathrm{l}({\cal{M}}_{6})=\frac{1}{3!}\int_{{\cal{M}}_{6}}J\wedge J\wedge J=\frac{1}{3!}C_{IJK}X^{I}X^{J}X^{K}=1 \ .
\ee

It is straightforward to show that the kinetic term for the five-dimensional gauge fields is generated by $|F_{4}|^{2}$, and the kinetic terms for the K\"ahler moduli comes from $R^{(11)}$. Putting all the pieces together, one arrives at the $D=5,~ N=2$ action,
\bea
\label{5daction}
	S_5&=&-\frac{1}{4\pi^{2}}\int \mathrm{d}^{5}x\sqrt{-g}\bigg(R-Q_{IJ}\partial_{\mu}X^{I}\partial^{\mu}X^{J}-\frac{1}{2}Q_{IJ}F^{I}{}_{\mu \nu} F^{J \mu \nu}\bigg)
\nonumber \\
&+&\displaystyle \frac{C_{IJK}}{24\pi^{2}}\int_{\mathcal{M}_{6}}A^{I}\wedge F^{J}\wedge F^{K} \ ,
\ee
with,
\bea
Q_{IJ}=\frac{1}{2}\int_{\mathcal{M}_{6}}J_{I}\wedge*_6 J_{J}=  -\frac{1}{2}\frac{\partial}{\partial X^I}\frac{\partial}{\partial X^J}(\ln {V})|_{{V} =1} = -\frac{1}{2}C_{I J K}X^K + \frac{9}{2}X_{I}X_{J} \ .
\ee
We shall assume that the gauge coupling $Q_{IJ}$ is positive definite. (\ref{5daction}) is the action of $D=5, N=2$ supergravity coupled to an arbitrary number of abelian vector multiplets and the whole action can be derived as the supersymmetric completion of the Chern-Simons term. Note that the fact that the metric on the scalar manifold, $Q_{IJ},$ is the same metric contracting the kinetic terms for the gauge fields, is forced upon us by supersymmetry. All known asymptotically flat BPS black hole solutions can be embedded in this theory. The fields $\{ X^I = X^I(\phi) \, , I=0,\dots,k-1 \}$ are also standard coordinates on $\mathbb{R}^k$; and where $X_I$, the dual coordinate is defined by,
\bea
X_{I} = \frac{1}{6}C_{I J K}X^J X^K \ ,
\ee
and $C_{I J K}$ are constants which are symmetric in $IJK$. This allows us to express the hypersurface equation $V=1$ as $X^I X_I = 1$ and one can deduce that
\bea
\label{hyre}
\partial_a X_I &=& \frac{1}{3}C_{I J K}\partial_a X^J X^K 
\nonumber \\
X^I \partial_a X_I &=& X_I \partial_a X^I = 0 \ .
\ee
The bosonic part of the supergravity action can also be re-expressed as \cite{dewit},
\bea
S_{bos} &=&\int d^5 x \sqrt{-g} \bigg(R - \frac{1}{2}Q_{I J}(\phi)F^{I}{}_{\mu \nu} F^{J \mu \nu} - h_{a b}(\phi)\partial_{\mu} \phi^a \partial^{\mu} \phi^b  \bigg)
\nonumber \\
&+& \frac{1}{24}e^{\mu \nu \rho \sigma \tau}C_{I J K}F^{I}{}_{\mu \nu}F^{J}{}_{\rho \sigma}A^{K}{}_{\tau} \ ,
\ee
where $F^I = dA^{I}$, $I, J, K = 0, \dots, k-1$ are the 2-form Maxwell field strengths,
$\phi^a$ are scalars, $\mu, \nu, \rho, \sigma = 0, \dots, 4$, and $g$ is the metric of the five-dimensional spacetime.where $V_{I}$ are constants. The metric $h_{a b}$ on $N$ is given by,
\bea
\label{pbmetric}
h_{a b} = Q_{I J}\frac{\partial X^I}{\partial \phi^a}\frac{\partial X^J}{\partial \phi^b}|_{{V} =1} \ ,
\ee
where $\{ \phi^a \, , a=1,\dots,k-1 \}$ are local coordinates of $N$. 
We shall assume that the gauge coupling $Q_{IJ}$ is positive definite. In addition, the following relations also hold:
\bea
X_I &=& \frac{2}{3}Q_{I J}X^J
\nonumber \\
\partial_a X_I &=& - \frac{2}{3}Q_{I J}\partial_a X^J \ .
\ee
The Einstein equation is given by
\bea
\label{eins}
R_{\mu\nu}- Q_{I J}\left(  F^{I}{}_{\mu\lambda}F^{J}{}_{\nu}{}^{\lambda}
+\nabla_{\mu}X^{I}\nabla_{\nu}X^{J}-\frac{1}{6}g_{\mu\nu}F^{I}{}_{\rho\sigma
}F^{J\rho\sigma}\right)  = 0 \ .
\ee
The Maxwell gauge equations for $A^I$ are given by
\bea
\label{maxwell}
d(Q_{I J}\star_5 F^J) = \frac{1}{4}C_{I J K}F^J \wedge F^K \ ,
\ee
or equivalently, in components:
\bea
\label{maxeq}
\nabla_{\mu}(Q_{I J}F^{J \mu \nu}) = -\frac{1}{16}C_{I J K}e^{\nu \mu \rho \sigma \tau}F^J{}_{\mu \rho}F^K{}_{\sigma \tau} \ .
\ee
where $e^{\mu \nu \rho \sigma \kappa} = \sqrt{-g}\epsilon^{\mu \nu \rho \sigma \kappa}$. The scalar field equations for $\phi^a$ are $L_{I}\partial_{a}{X^{I}} = 0$ with
\bea
\label{scalar}
L_{I} = \nabla^{\mu}\nabla_{\mu}X_{I} + \bigg(-\frac{1}{6}C_{M N I} 
+ X_M X^P C_{N P I} \bigg)\bigg(\frac{1}{2}F^M{}_{\mu \nu}F^{N \mu \nu} + \nabla_{\mu}X^M \nabla^{\mu} X^N \bigg) \ .
\ee
We remark that if $L_{I}\partial_{a}{X^{I}} = 0$ 
for all $a=1, \dots , k-1$, then $L_{I} = f X_{I}$ where $f = X^{J}L_{J}$.
This result is established in Appendix E.
Using this, the scalar field equation can be rewritten as
\bea
\label{scalareq1}
&&\nabla^{\mu}\nabla_{\mu}{X_{I}} + \nabla_{\mu}{X^{M}} \nabla^{\mu}{X^{N}} \left( \frac{1}{2}C_{M N K} X_{I} X^{K} - \frac{1}{6}C_{I M N}\right) 
\nonumber \\
&+& \frac{1}{2}F^{M}{}_{\mu \nu} F^{N \mu \nu} \left(C_{I N P} X_{M} X^{P} - \frac{1}{6}C_{I M N}-6X_{I} X_{M} X_{N}+\frac{1}{6}C_{M N J} X_{I} X^{J}\right)  = 0 \ .
\ee
The KSEs are defined on a purely bosonic background, and are given as the vanishing of the supersymmetry transformations of the fermions at lowest order in fermions. The number of linearly independent Killing spinors determines how much supersymmetry is realised for a given solution. The KSEs \cite{klemm} can be expressed as,
\bea
{\cal D}_\mu\e&\equiv& \nabla_\mu \epsilon + \frac{i}{8}X_I\bigg(\Gamma_{\mu}{}^{\nu \rho} - 4\delta_{\mu}{}^{\nu}\Gamma^{\rho}\bigg)F^I{}_{\nu \rho} \epsilon   = 0
\label{Gkseo} \\
\label{Akseo}
{\cal A}^{I}\e &\equiv& \bigg[\bigg(\delta^{J}{}_{I} - X^{I}X_{J}\bigg)F^{J}{}_{\mu \nu}\Gamma^{\mu \nu} + 2i\Gamma^{\mu}\partial_{\mu}X^I\bigg]\e = 0 \ .
\ee
On decomposing $F^I$ as 
\bea
\label{decom}
F^I = FX^I + G^I \ ,
\ee
where
\bea
X_I F^I = F, \qquad X_I G^I = 0 \ .
\ee 
the KSEs can then be rewritten in terms of $F$ and $G^I$ as
\bea
{\cal D}_\mu\e&\equiv& \nabla_\mu \epsilon + \frac{i}{8}\bigg(\Gamma_{\mu}{}^{\nu \rho} - 4\delta_{\mu}{}^{\nu}\Gamma^{\rho} \bigg)F_{\nu \rho} \epsilon  = 0 \ , \label{5duGkse} 
\ee
and
\bea
{\cal A}^{I} \e &\equiv& \bigg[G^{I}{}_{\mu \nu}\Gamma^{\mu \nu} + 2i\Gamma^{\mu}\partial_{\mu}X^I\bigg] \e = 0  \ ,
\label{5d2Akse}
\ee
where $\epsilon$ is the supersymmetry parameter which is a Dirac spinor of $Spin(4, 1)$ and we use the spinor conventions of \cite{5dindex}, see Appendix B and E for the Clifford algebra. 
\subsection{Gauged}

The $D=5$ gauged supergravity with vector multiplets can be obtained by gauging the $U(1)$ subgroup of the $SU(2)$ automorphism group of the $N=2$ supersymmetry algebra, which breaks $SU(2)$ down to $U(1)$ \cite{gunaydin}, \cite{gunaydin2}. The gauging is achieved by introducing a linear combination of the abelian vector fields with $V_{I}A^{I}{}_{\mu}$ and coupling constant $\chi$. The $D=5$ gauged supergravity can also be obtained from type IIB supergravity compactified on $S^5$ \cite{gutperle}. In gauged supergravity theories, the action, field equations and supersymmetry transformations also get modified by $\chi$-dependent terms. In a bosonic background, these additional terms give rise to a scalar potential $U$ \cite{gunaydin}. In particular the terms which get modified are,
\bea
S_{bos} &=&\int d^5 x \sqrt{-g} \bigg(R - \frac{1}{2}Q_{I J}(\phi)F^{I}{}_{\mu \nu} F^{J \mu \nu} - h_{a b}(\phi)\partial_{\mu} \phi^a \partial^{\mu} \phi^b  + 2 \chi^2 U \bigg)
\nonumber \\
&&+ \frac{1}{24}e^{\mu \nu \rho \sigma \tau}C_{I J K}F^{I}{}_{\mu \nu}F^{J}{}_{\rho \sigma}A^{K}{}_{\tau} \ ,
\ee
where $U$ is the (gauge) scalar potential which can be expressed as,
\bea
U = 9V_I V_J\bigg(X^I X^J - \frac{1}{2}Q^{I J}\bigg) \ .
\ee
Where $V_{I}$ are constants. We shall again assume that the gauge coupling $Q_{IJ}$ is positive definite, and also
that the scalar potential is non-negative, $U \geq 0$.
The Einstein equation is given by
\bea
\label{eins2}
R_{\mu\nu}- Q_{I J}\left(  F^{I}{}_{\mu\lambda}F^{J}{}_{\nu}{}^{\lambda}
+\nabla_{\mu}X^{I}\nabla_{\nu}X^{J}-\frac{1}{6}g_{\mu\nu}F^{I}{}_{\rho\sigma
}F^{J\rho\sigma}\right) + \frac{2}{3}\chi^2 U g_{\mu \nu} = 0 \ .
\ee
The Maxwell gauge equations for $A^I$ are the same as the ungauged theory. The scalar field equations for $\phi^a$ become,
\bea
\label{scalar2}
&&\bigg[\nabla^{\mu}\nabla_{\mu}X_{I} + \bigg(-\frac{1}{6}C_{M N I} + X_M X^P C_{N P I} \bigg)\bigg(\frac{1}{2}F^M{}_{\mu \nu}F^{N \mu \nu} + \nabla_{\mu}X^M \nabla^{\mu} X^N \bigg) 
\nonumber \\
&&+ \frac{3}{2}\chi^2 C_{I J K}Q^{M J}Q^{N K}V_{M}V_{N}\bigg] \partial_a X^I = 0 \ ,
\ee
which as before implies,
\bea
\label{scalareq2}
&&\nabla^{\mu}\nabla_{\mu}{X_{I}} + \nabla_{\mu}{X^{M}} \nabla^{\mu}{X^{N}} \left( \frac{1}{2}C_{M N K} X_{I} X^{K} - \frac{1}{6}C_{I M N}\right) 
\nonumber \\
&+& \frac{1}{2}F^{M}{}_{\mu \nu} F^{N \mu \nu} \left(C_{I N P} X_{M} X^{P} - \frac{1}{6}C_{I M N}-6X_{I} X_{M} X_{N}+\frac{1}{6}C_{M N J} X_{I} X^{J}\right) 
\nonumber \\
&+& 3 \chi^2 V_{M} V_{N}\bigg(\frac{1}{2}C_{I J K}Q^{M J}Q^{N K} + X_{I}(Q^{M N} - 2 X^{M}X^{N})\bigg) = 0 \ .
\ee
The KSEs \cite{klemm} can be expressed as,
\bea
{\cal D}_\mu\e&\equiv& \nabla_\mu \epsilon + \frac{i}{8}X_I\bigg(\Gamma_{\mu}{}^{\nu \rho} - 4\delta_{\mu}{}^{\nu}\Gamma^{\rho}\bigg)F^I{}_{\nu \rho} \epsilon + \bigg(- \frac{3i}{2}\chi V_{I}A^{I}{}_{\mu} + \frac{1}{2}\chi V_I X^{I}\Gamma_\mu\bigg)\epsilon  = 0 \ ,
\label{Gkseo2} \\
\label{Akseo2}
{\cal A}^{I}\e &\equiv& \bigg[\bigg(\delta^{J}{}_{I} - X^{I}X_{J}\bigg)F^{J}{}_{\mu \nu}\Gamma^{\mu \nu} + 2i\Gamma^{\mu}\partial_{\mu}X^I - 6i\chi \bigg(Q^{I J} - \frac{2}{3}X^{I}X^{J}\bigg)V_J\bigg]\e = 0 \ .
\ee
On decomposing $F^I = FX^I + G^I$ as before with,
\bea
X_I F^I = F,~~~X_I G^I = 0 \ ,
\ee
the KSEs can then be rewritten in terms of $F$ and $G^I$ as
\bea
{\cal D}_\mu\e&\equiv& \nabla_\mu \epsilon + \frac{i}{8}\bigg(\Gamma_{\mu}{}^{\nu \rho} - 4\delta_{\mu}{}^{\nu}\Gamma^{\rho} \bigg)F_{\nu \rho} \epsilon + \bigg(- \frac{3i}{2}\chi V_{I}A^{I}{}_{\mu} + \frac{1}{2}\chi V_I X^{I}\Gamma_\mu\bigg)\epsilon = 0 \ ,
\label{5dgGkse} 
\\
{\cal A}^{I} \e &\equiv& \bigg[G^{I}{}_{\mu \nu}\Gamma^{\mu \nu} + 2i\Gamma^{\mu}\partial_{\mu}X^I - 6i\chi \bigg(Q^{I J} - \frac{2}{3}X^{I}X^{J}\bigg)V_J \bigg] \e = 0 \ .
\label{Akse2}
\ee

	\chapter{$D=10$ IIA Horizons}

In this chapter, we present the local, and global, analysis of the Killing spinor
equations for type IIA supergravity and investigate the resulting enhancement
of supersymmetry. This establishes the horizon conjecture for this theory.

The results presented for horizons in IIA supergravity do not follow from those that have been obtained for M-horizons in \cite{11index}. Although IIA supergravity is the dimensional
reduction of 11-dimensional supergravity, the reduction, after truncation of Kaluza-Klein
modes, does not always preserve all the supersymmetry of 11-dimensional solutions; for
a detailed analysis of these issues see \cite{bakas}, \cite{duff2}. As a result, for example, it does not directly follow
that IIA horizons preserve an even number of supersymmetries because M-horizons do as
shown in \cite{11index}. However, as we prove that both IIA and M-theory horizons preserve an even number of supersymmetries, one concludes that if the reduction process breaks some supersymmetry, then it always breaks an even number of supersymmetries.

\section{Horizon fields and KSEs}

\subsection{Near-horizon fields}
The description of the metric near extreme Killing horizons as expressed in Gaussian null coordinates \cite{isen, gnull}
can be adapted to include all IIA fields. In particular, one writes
\be
G &=& \bbe^+ \wedge \bbe^- \wedge X +r \bbe^+ \wedge Y + \tilde G~,~~
\cr
H &=&\bbe^+ \wedge \bbe^- \wedge L+ r \bbe^+ \wedge M + \tilde H~,~~~
\cr
F &=& \bbe^+ \wedge \bbe^- S + r \bbe^+ \wedge T+ \tilde F~,
\la{iiahormetr}
\ee
where we use the frame ({\ref{basis1}}),
and the dependence on the coordinates $u$ and $r$ is explicitly given. Moreover $\Phi$ and $\Delta$ are 0-forms, $h$, $L$ and $T$ are 1-forms, $X$, $M$ and $\tilde F$ are 2-forms,
$Y, \tilde H$ are 3-forms and $\tilde G$ is a 4-form on  the spatial horizon section ${\cal S}$, which is the co-dimension 2 submanifold given by the equation  $r=u=0$, i.e.~all these
components of the fields depend only on the coordinates of ${\cal S}$. It should be noted that one of our assumptions is that
all these forms on ${\cal S}$ are sufficiently differentiable, i.e.~we require at least $C^2$ differentiability
so that all the field equations and Bianchi identities are valid.
\subsection{Horizon Bianchi identities and field equations}

Substituting the fields (\ref{iiahormetr}) into the Bianchi identities  of IIA supergravity, one finds that
\bea
\label{iiabian2}
M&=&d_h L~,~~~T=d_h S~,~~~Y=d_h X-L\wedge \tilde F-S \tilde H~,~~~
\cr
d\tilde G&=&\tilde H\wedge \tilde F~,~~~d\tilde H=d\tilde F=0~,
\ee
where  $d_h \theta \equiv d \theta-
h \wedge \theta$ for any form $\theta$. These are the only independent Bianchi identities, see Appendix A.
Similarly, substituting the horizon fields into the field equations of IIA supergravity,  we find that
the 2-form field equation ({\ref{iiageq1}) gives
\bea
\label{iiafeq1}
\tn^i {\tilde{F}}_{ik}-h^i {\tilde{F}}_{ik}
+T_k -L^i X_{ik} +{1 \over 6} {\tilde{H}}^{\ell_1 \ell_2
\ell_3} {\tilde{G}}_{\ell_1 \ell_2 \ell_3 k}=0~,
\ee
the 3-form field equation  ({\ref{iiabeq1}}) gives
\bea
\label{iiafeq2}
\tn^i (e^{-2 \Phi}L_i) -{1 \over 2} {\tilde{F}}^{ij}
X_{ij} +{1 \over 1152}
\epsilon^{\ell_1 \ell_2 \ell_3 \ell_4 \ell_5 \ell_6 \ell_7
\ell_8} {\tilde{G}}_{\ell_1 \ell_2 \ell_3 \ell_4}
{\tilde{G}}_{\ell_5 \ell_6 \ell_7 \ell_8}=0
\ee
and
\bea
\label{iiafeq3}
&&\tn^i(e^{-2 \Phi} {\tilde{H}}_{imn})
-e^{-2 \Phi} h^i {\tilde{H}}_{imn}
+e^{-2 \Phi} M_{mn}+S X_{mn}-{1 \over 2}
{\tilde{F}}^{ij} {\tilde{G}}_{ijmn}
\nonumber \\
&&-{1 \over 48} \epsilon_{mn}{}^{\ell_1
\ell_2 \ell_3 \ell_4 \ell_5 \ell_6} X_{\ell_1 \ell_2}
{\tilde{G}}_{\ell_3 \ell_4 \ell_5 \ell_6} =0~,
\ee
and the 4-form field equation ({\ref{iiaceq1}}) gives
\bea
\label{iiafeq4}
\tn^i X_{ik} +{1 \over 144} \epsilon_k{}^{\ell_1 \ell_2
\ell_3 \ell_4 \ell_5 \ell_6 \ell_7}
{\tilde{G}}_{\ell_1 \ell_2 \ell_3 \ell_4} {\tilde{H}}_{\ell_5 \ell_6 \ell_7} =0
\ee
and
\bea
\label{iiafeq5}
&&\tn^i {\tilde{G}}_{ijkq}+Y_{jkq}-h^i {\tilde{G}}_{ijkq}
\nonumber \\
&&-{1 \over 12} \epsilon_{jkq}{}^{\ell_1 \ell_2 \ell_3
\ell_4 \ell_5} X_{\ell_1 \ell_2} {\tilde{H}}_{\ell_3 \ell_4 \ell_5}
-{1 \over 24}\epsilon_{jkq}{}^{\ell_1 \ell_2 \ell_3
\ell_4 \ell_5} {\tilde{G}}_{\ell_1 \ell_2 \ell_3 \ell_4}
L_{\ell_5} =0~,
\ee
where $\tilde \nabla$ is the Levi-Civita connection of the metric on ${\cal S}$.
In addition, the dilaton field equation ({\ref{iiadileq}}) becomes
\bea
\label{iiafeq6}
\tn^i \tn_i \Phi - h^i \tn_i \Phi &=&
2 \tn_i \Phi \tn^i \Phi +{1 \over 2} L_i L^i
-{1 \over 12} {\tilde{H}}_{\ell_1 \ell_2 \ell_3}
{\tilde{H}}^{\ell_1 \ell_2 \ell_3}-{3 \over 4} e^{2 \Phi}S^2
\nonumber \\
&+&{3 \over 8} e^{2 \Phi} {\tilde{F}}_{ij}
{\tilde{F}}^{ij} -{1 \over 8} e^{2 \Phi} X_{ij}X^{ij}
+{1 \over 96} e^{2 \Phi} {\tilde{G}}_{\ell_1 \ell_2 \ell_3
\ell_4} {\tilde{G}}^{\ell_1 \ell_2 \ell_3 \ell_4}~.
\ee
It remains to evaluate the Einstein field equation. This gives
\bea
\label{iiafeq7}
{1 \over 2} \tn^i h_i -\Delta -{1 \over 2}h^2
&=& h^i \tn_i \Phi -{1 \over 2} L_i L^i -{1 \over 4} e^{2 \Phi} S^2 -{1 \over 8} e^{2 \Phi} X_{ij} X^{ij}
\nonumber \\
&-&{1 \over 8} e^{2 \Phi} {\tilde{F}}_{ij} {\tilde{F}}^{ij}
-{1 \over 96} e^{2 \Phi} {\tilde{G}}_{\ell_1 \ell_2
\ell_3 \ell_4} {\tilde{G}}^{\ell_1 \ell_2 \ell_3 \ell_4}~,
\ee
and
\bea
\label{iiafeq8}
{\tilde{R}}_{ij} &=& -\tn_{(i} h_{j)}
+{1 \over 2} h_i h_j -2 \tn_i \tn_j \Phi
-{1 \over 2} L_i L_j +{1 \over 4} {\tilde{H}}_{i
\ell_1 \ell_2} {\tilde{H}}_j{}^{\ell_1 \ell_2}
\nonumber \\
&+&{1 \over 2} e^{2 \Phi} {\tilde{F}}_{i \ell}
{\tilde{F}}_j{}^\ell -{1 \over 2} e^{2 \Phi} X_{i \ell}
X_j{}^\ell+{1 \over 12} e^{2 \Phi} {\tilde{G}}_{i \ell_1 \ell_2 \ell_3} {\tilde{G}}_j{}^{\ell_1 \ell_2 \ell_3}
\nonumber \\
&+&\delta_{ij} \bigg({1 \over 4} e^{2 \Phi} S^2
-{1 \over 8} e^{2 \Phi} {\tilde{F}}_{\ell_1 \ell_2}
{\tilde{F}}^{\ell_1 \ell_2} +{1 \over 8} e^{2 \Phi}
X_{\ell_1 \ell_2}X^{\ell_1 \ell_2}
-{1 \over 96} e^{2 \Phi} {\tilde{G}}_{\ell_1 \ell_2
\ell_3 \ell_4} {\tilde{G}}^{\ell_1 \ell_2 \ell_3 \ell_4}
\bigg)~.
\nonumber \\
\ee
Above we have only stated the independent field equations. In fact, after substituting the near-horizon geometries into the IIA field equations, there are additional  equations that arise.
However, these are all implied from the above field equations and Bianchi identities. For completeness, these additional equations are given here.
We remark that there are a number of additional Bianchi identities, which are
\bea
dT + S dh + dS \wedge h &=&0~,
\nonumber \\
dM+L \wedge dh -h \wedge dL &=&0~,
\nonumber \\
dY + dh \wedge X - h \wedge dX + h \wedge (S {\tilde{H}}
+{\tilde{F}} \wedge L)+T \wedge {\tilde{H}}+{\tilde{F}} \wedge M &=&0~.
\ee
However, these Bianchi identities are implied by those in ({\ref{iiabian2}}).
There is also a number of additional field equations given by
\bea
\label{iiaauxeq1}
-\tn^i T_i + h^i T_i -{1 \over 2} dh^{ij} {\tilde{F}}_{ij}
-{1 \over 2} X_{ij}M^{ij} -{1 \over 6} Y_{ijk}{\tilde{H}}^{ijk}=0~,
\ee
\bea
\label{iiaauxeq2}
- \tn^i (e^{-2 \Phi}M_{ik}) + e^{-2 \Phi} h^i M_{ik}
-{1 \over 2} e^{-2 \Phi} dh^{ij} {\tilde{H}}_{ijk}
-T^i X_{ik} -{1 \over 2} {\tilde{F}}^{ij} Y_{ijk}
\nonumber \\
-{1 \over 144} \epsilon_k{}^{\ell_1
\ell_2 \ell_3 \ell_4 \ell_5 \ell_6 \ell_7}
Y_{\ell_1 \ell_2 \ell_3} {\tilde{G}}_{\ell_4 \ell_5 \ell_6 \ell_7} =0~,
\ee
\bea
\label{iiaauxeq3}
- \tn^i Y_{imn}+h^i Y_{imn}-{1 \over 2} dh^{ij}
{\tilde{G}}_{ijmn}
+{1 \over 36} \epsilon_{mn}{}^{\ell_1 \ell_2 \ell_3
\ell_4 \ell_5 \ell_6} Y_{\ell_1 \ell_2 \ell_3}
{\tilde{H}}_{\ell_4 \ell_5 \ell_6}
\nonumber \\
+{1 \over 48} \epsilon_{mn}{}^{\ell_1 \ell_2 \ell_3
\ell_4 \ell_5 \ell_6} {\tilde{G}}_{\ell_1 \ell_2 \ell_3
\ell_4} M_{\ell_5 \ell_6}=0~,
\ee
corresponding to equations obtained from
the $+$ component of ({\ref{iiageq1}}),
the $k$ component of ({\ref{iiabeq1}}) and
the $mn$ component of ({\ref{iiaceq1}}) respectively.
However, ({\ref{iiaauxeq1}}), ({\ref{iiaauxeq2}}) and
({\ref{iiaauxeq3}}) are implied by ({\ref{iiafeq1}})-
({\ref{iiafeq5}}) together with the Bianchi identities
({\ref{iiabian2}}).
Note also that the $++$ and $+i$ components of the
Einstein equation, which are
\bea
\label{iiaauxeq4}
{1 \over 2} \tn^i \tn_i \Delta -{3 \over 2} h^i \tn_i \Delta-{1 \over 2} \Delta \tn^i h_i
+ \Delta h^2 +{1 \over 4} dh_{ij} dh^{ij}
&=&(\tn^i \Delta - \Delta h^i)\tn_i \Phi 
\nonumber \\
&+&{1 \over 4} M_{ij}
M^{ij} +{1 \over 2} e^{2 \Phi} T_i T^i
\nonumber \\
&+&{1 \over 12} e^{2 \Phi} Y_{ijk} Y^{ijk} \ ,
\ee
and
\bea
\label{iiaauxeq5}
{1 \over 2} \tn^j dh_{ij}-dh_{ij} h^j - \tn_i \Delta + \Delta h_i
&=& dh_i{}^j \tn_j \Phi -{1 \over 2} M_i{}^j L_j
+{1 \over 4} M_{\ell_1 \ell_2} {\tilde{H}}_i{}^{\ell_1 \ell_2}
\nonumber \\
&-&{1 \over 2} e^{2 \Phi} S T_i +{1 \over 2} e^{2 \Phi} T^j {\tilde{F}}_{ij}
-{1 \over 4} e^{2 \Phi} Y_i{}^{\ell_1 \ell_2}
X_{\ell_1 \ell_2} 
\nonumber \\
&+&{1 \over 12} e^{2 \Phi}
Y_{\ell_1 \ell_2 \ell_3} {\tilde{G}}_i{}^{\ell_1 \ell_2
\ell_3} \ ,
\ee
are implied by ({\ref{iiafeq6}}), ({\ref{iiafeq7}}), ({\ref{iiafeq8}}),
together with ({\ref{iiafeq1}})-({\ref{iiafeq5}}),
and the Bianchi identities ({\ref{iiabian2}}).
To summarize, the independent Bianchi identities and field equations are given in
 ({\ref{iiabian2}})--({\ref{iiafeq8}}).
\subsection{Integration of KSEs along the lightcone}\label{ikse}
In what follows, we shall refer
to the ${\cal D}$ operator as the supercovariant connection.
Supersymmetric IIA horizons are those for which there exists an $\epsilon\not=0$ that is a solution of the KSEs. To find the conditions
on the fields required for such a solution to exist, we first integrate along the two lightcone directions, i.e.~we integrate the KSEs
along the $u$ and $r$ coordinates. To do this, we decompose $\epsilon$ as
\bea
\e=\e_++\e_-~,
\label{iiaksp1}
\ee
 where $\Gamma_\pm\epsilon_\pm=0$. To begin, we consider the $\mu = -$ component of the gravitino KSE (\ref{iiaGKSE}) which can be integrated to obtain, 
\bea\label{iialightconesol}
\e_+=\phi_+(u,y)~,~~~\e_-=\phi_-+r \Gamma_-\Theta_+ \phi_+~,
\ee
where $\partial_r \phi_{\pm} = 0$. Now we consider the $\mu=+$ component; on evaluating this component at $r=0$ we get,
\bea
\phi_-=\eta_-~,~~~\phi_+=\eta_++ u \Gamma_+ \Theta_-\eta_-~,
\ee
where $\partial_{u}\eta_{\pm} = \partial_r \eta_{\pm} = 0$ and,
\bea
\Theta_\pm&=&{1\over4} h_i\Gamma^i\mp{1\over4} \Gamma_{11} L_i \Gamma^i-{1\over16} e^{\Phi} \Gamma_{11} (\pm 2 S+\tilde F_{ij} \Gamma^{ij})
\nonumber \\
&-&{1\over8 \cdot 4!} e^{\Phi} (\pm 12 X_{ij} \Gamma^{ij}
+\tilde G_{ijkl} \Gamma^{ijkl})~,
\ee
and $\eta_\pm$ depend only on the coordinates of the spatial horizon section ${\cal S}$. As spinors on ${\cal S}$, $\eta_\pm$ are sections of the $Spin(8)$ bundle on ${\cal S}$
associated with the Majorana representation.  Equivalently, the $Spin(9,1)$ bundle $S$ on the spacetime when restricted to ${\cal S}$ decomposes
as $S=S_-\oplus S_+$ according to the lightcone projections $\Gamma_\pm$. Although $S_\pm$ are distinguished by the lightcone chirality, they are isomorphic
as $Spin(8)$ bundles over ${\cal S}$. We shall use this in the counting of supersymmetries of IIA horizons.

Substituting the solution of the KSEs along the lightcone directions (\ref{iialightconesol})  back into the gravitino KSE (\ref{iiaGKSE}) and appropriately expanding in the $r,u$ coordinates, we find that
for  the $\mu = \pm$ components, one obtains  the additional conditions
\bea
\label{iiaint1}
&&\bigg({1\over2}\Delta - {1\over8}(dh)_{ij}\Gamma^{ij} + {1\over8}M_{ij}\Gamma_{11}\Gamma^{ij} + 2\big( {1\over 4} h_i \Gamma^{i} - {1\over 4} L_{i}\Gamma_{11}\Gamma^{i} 
\nonumber \\
&&- {1\over 16} e^{\Phi}\Gamma_{11}(-2S + \tilde{F}_{i j}\Gamma^{i j}) 
- \frac{1}{8\cdot4!}e^{\Phi}(12X_{i j}\Gamma^{i j} - \tilde{G}_{i j k l}\Gamma^{i j k l})\big )\Theta_{+} \bigg)\phi_{+} = 0 \ ,
\ee
\bea
\label{iiaint2}
&&\bigg(\frac{1}{4}\Delta h_i \Gamma^{i} - \frac{1}{4}\partial_{i}\Delta \Gamma^{i} + \big(-\frac{1}{8}(dh)_{ij}\Gamma^{ij} - \frac{1}{8}M_{ij}\Gamma^{ij}\Gamma_{11} 
\cr
&&- \frac{1}{4}e^{\Phi}T_{i}\Gamma^{i}\Gamma_{11} + \frac{1}{24}e^{\Phi}Y_{i j k}\Gamma^{i j k} \big) \Theta_{+} \bigg) \phi_{+} = 0 \ ,
\ee
\bea
\label{iiaint3}
&&\bigg(-\frac{1}{2}\Delta - \frac{1}{8}(dh)_{ij}\Gamma^{ij} + \frac{1}{8}M_{ij}\Gamma^{ij}\Gamma_{11} - \frac{1}{4}e^{\Phi} T_{i}\Gamma^{i} \Gamma_{11} 
\cr
&&- \frac{1}{24}e^{\Phi}Y_{ijk}\Gamma^{ijk}
+ 2\big( -{1\over4} h_i\Gamma^i-{1\over4} \Gamma_{11} L_i \Gamma^i 
\cr
&&+{1\over16} e^\phi \Gamma_{11} (2 S+\tilde F_{ij} \Gamma^{ij})
-{1\over8 \cdot 4!} e^\phi (12 X_{ij} \Gamma^{ij}
+\tilde G_{ijkl} \Gamma^{ijkl}) \big) \Theta_{-} \bigg)\phi_{-} = 0 \ .
\ee
Similarly the $\mu=i$ component of the gravitino KSEs gives
\bea
\label{iiaint4}
&&\tilde \nabla_i \phi_\pm\mp {1\over 4} h_i \phi_\pm \mp {1\over4} \Gamma_{11} L_i \phi_\pm +{1\over8} \Gamma_{11} \tilde H_{ijk} \Gamma^{jk} \phi_\pm -{1\over16} e^\Phi \Gamma_{11} (\mp 2 S+ \tilde F_{kl} \Gamma^{kl}) \Gamma_i \phi_\pm
\cr
&&+{1\over 8\cdot 4!} e^\Phi (\mp 12 X_{kl} \Gamma^{kl}+ \tilde G_{j_1j_2j_3j_4} \Gamma^{j_1j_2j_3j_4}) \Gamma_i \phi_\pm=0 \ ,
\ee
and
\bea
\label{iiaint5}
&&\tilde \nabla_i \tau_{+} + \bigg( -\frac{3}{4}h_i - \frac{1}{16}e^{\Phi}X_{l_1 l_2}\Gamma^{l_1 l_2}\Gamma_{i} - \frac{1}{8\cdot4!}e^{\Phi} \tilde G_{l_1\cdots l_4}\Gamma^{l_1 \cdots l_4}\Gamma_{i}
\cr
&&- \Gamma_{11}(\frac{1}{4}L_i + \frac{1}{8}\tilde{H}_{i j k}\Gamma^{j k} + \frac{1}{8}e^{\Phi} S \Gamma_{i} +    \frac{1}{16}e^{\Phi}\tilde{F}_{l_1 l_2}\Gamma^{l_1 l_2}\Gamma_{i})\bigg )\tau_{+}
\cr
&&+ \bigg(-\frac{1}{4}(dh)_{ij}\Gamma^{j} - \frac{1}{4}M_{ij}\Gamma^{j}\Gamma_{11} + \frac{1}{8}e^{\Phi}T_{j}\Gamma^{j}\Gamma_{i}\Gamma_{11} 
+ \frac{1}{48}e^{\Phi}Y_{l_1 l_2 l_3}\Gamma^{l_1 l_2 l_3}\Gamma_{i} \bigg)\phi_{+} = 0 \ .
\ee
where we have set
\bea
\label{iiaint6}
\tau_{+} = \Theta_{+}\phi_{+} \ .
\ee
All the additional conditions above can be viewed as integrability conditions along the lightcone and mixed lightcone and ${\cal S}$ directions.
We shall demonstrate that upon using the field equations and the Bianchi identities, the only independent conditions are (\ref{iiacovr}).

Substituting the solution of the KSEs (\ref{iialightconesol})  into the dilatino KSE (\ref{iiaAKSE}) and expanding appropriately in the $r,u$ coordinates, one obtains the following additional conditions
\bea
\label{iiaint7}
&&\partial_i \Phi \Gamma^i \phi_{\pm} -{1\over12} \Gamma_{11} (\mp 6 L_i \Gamma^i+\tilde H_{ijk} \Gamma^{ijk}) \phi_\pm+{3\over8} e^\Phi \Gamma_{11} (\mp2 S+\tilde F_{ij} \Gamma^{ij})\phi_\pm
\cr
&&
+{1\over 4\cdot 4!}e^{\Phi} (\mp 12 X_{ij} \Gamma^{ij}+\tilde G_{j_1j_2j_3j_4} \Gamma^{j_1j_2j_3j_4}) \phi_\pm=0 \ ,
\ee
\be
\label{iiaint8}
&&-\bigg( \partial_{i}\Phi\Gamma^{i} + \frac{1}{12}\Gamma_{11} (6L_i \Gamma^{i} + \tilde{H}_{ijk}\Gamma^{ijk}) + \frac{3}{8}e^{\Phi}\Gamma_{11}(2S + \tilde{F}_{ij}\Gamma^{ij})
\cr
&&- \frac{1}{4\cdot 4!}e^{\Phi}(12X_{ij}\Gamma^{ij} + \tilde{G}_{ijkl}\Gamma^{ijkl})\bigg)\tau_{+}
\cr
&&+ \bigg(\frac{1}{4}M_{ij}\Gamma^{ij}\Gamma_{11} + \frac{3}{4}e^{\Phi}T_{i}\Gamma^{i}\Gamma_{11} + \frac{1}{24}e^{\Phi}Y_{ijk}\Gamma^{ijk}\bigg)\phi_{+}=0~.
\ee
We shall show that the only independent ones are those in (\ref{iiacovr}).

\subsection{Independent KSEs}
The substitution of the  spinor (\ref{iiaksp1}) into the KSEs produces a large number of additional conditions.  These can be seen
either as integrability conditions along the lightcone directions, as well as integrability conditions along the mixed lightcone and
${\cal S}$ directions, or as KSEs along ${\cal S}$. A detailed analysis, presented in Appendix C, of the formulae obtained reveals
that the independent KSEs are those that are obtained from the naive restriction of the IIA KSEs to ${\cal S}$.  In particular,
the independent KSEs are
\bea
\label{iiacovr}
\nabla_{i}^{(\pm)}\eta_{\pm}  = 0~,~~~\mathcal{A}^{(\pm)}\eta_{\pm} = 0~,
\ee
where
\bea
\nabla_{i}^{(\pm)}&=& \tilde{\nabla}_{i} + \Psi^{(\pm)}_{i}~,
\ee
with
\bea
\label{iiaalg1pm}
\Psi^{(\pm)}_{i} &=& \bigg( \mp \frac{1}{4}h_{i} \mp \frac{1}{16}e^{\Phi}X_{l_1 l_2}\Gamma^{l_1 l_2}\Gamma_{i} + \frac{1}{8.4!}e^{\Phi}{\tilde{G}}_{l_1 l_2 l_3 l_4}\Gamma^{l_1 l_2 l_3 l_4}\Gamma_{i}\bigg)
\cr
&+& \Gamma_{11}\bigg(\mp \frac{1}{4}L_{i} + \frac{1}{8}{\tilde{H}}_{i l_1 l_2}\Gamma^{l_1 l_2}
\pm \frac{1}{8}e^{\Phi}S\Gamma_{i} - \frac{1}{16}e^{\Phi}{\tilde{F}}_{l_1 l_2}\Gamma^{l_1 l_2}\Gamma_{i}\bigg)~,
\ee
and
\bea
\label{iiaalg2pm}
\mathcal{A}^{(\pm)} &=& \partial_i \Phi \Gamma^i  + \bigg(\mp \frac{1}{8}e^{\Phi}X_{l_1 l_2}\Gamma^{l_1 l_2} + \frac{1}{4.4!}e^{\Phi}{\tilde{G}}_{l_1 l_2 l_3 l_4}\Gamma^{l_1 l_2 l_3 l_4}\bigg)
\cr
&+& \Gamma_{11}\bigg(\pm \frac{1}{2}L_{i}\Gamma^{i} - \frac{1}{12}{\tilde{H}}_{i j k}\Gamma^{i j k} \mp \frac{3}{4}e^{\Phi}S + \frac{3}{8}e^{\Phi}{\tilde{F}}_{i j}\Gamma^{i j}\bigg)~.
\ee
Evidently, $\nabla^{(\pm)}$ arise from the supercovariant connection while $\mathcal{A}^{(\pm)}$ arise
from the dilatino KSE of IIA supergravity as restricted to ${\cal S}$ .
Furthermore, the analysis in Appendix C reveals that  if $\eta_{-}$ solves $(\ref{iiacovr})$ then
\bea
\eta_+ = \Gamma_{+}\Theta_{-}\eta_{-}~,
\label{iiaepfem}
\ee
 also solves $(\ref{iiacovr})$. This is the first indication that IIA horizons admit an even number of supersymmetries.  As we shall prove, the existence of the $\eta_+$ solution
 is also responsible for the $\mathfrak{sl}(2,\bR)$ symmetry of IIA horizons.
\section{Supersymmetry enhancement}
To prove that IIA horizons always admit an even number of supersymmetries, it suffices to prove that there are as many $\eta_+$ Killing spinors as there are $\eta_-$ Killing spinors,
i.e.~that the $\eta_+$ and $\eta_-$ Killing spinors come in pairs. For this, we shall identify the Killing spinors with the zero modes of  Dirac-like operators
which depend on the fluxes and then use the index theorem to count their modes.
\subsection{Horizon Dirac equations}
We define  horizon Dirac operators associated with the supercovariant derivatives following from the gravitino KSE as
\bea
{\cal D}^{(\pm)} \equiv \Gamma^{i}\nabla_{i}^{(\pm)} = \Gamma^{i}\tilde{\nabla}_{i} + \Psi^{(\pm)}~,
\ee
where
\bea
\label{iiaalg3pm}
\Psi^{(\pm)} \equiv \Gamma^{i}\Psi^{(\pm)}_{i} &=& \mp\frac{1}{4}h_{i}\Gamma^{i}
\mp\frac{1}{4}e^{\Phi}X_{i j}\Gamma^{i j}
\cr
&+& \Gamma_{11}\bigg(\pm \frac{1}{4}L_{i}\Gamma^{i} - \frac{1}{8}{\tilde{H}}_{i j k}\Gamma^{i j k} \mp e^{\Phi}S + \frac{1}{4}e^{\Phi}{\tilde{F}}_{i j}\Gamma^{i j} \bigg)~.
\ee
However, it turns out that
it is not possible to straightforwardly formulate Lichnerowicz theorems
to identify zero modes of these horizon Dirac operators with Killing spinors.
To proceed, we shall modify both the KSEs and the horizon Dirac operators.  For this first observe that an equivalent set of KSEs can be chosen
by redefining the supercovariant derivatives from the gravitino KSE as
\bea
\label{iiaredef1}
\hat{\nabla}_{i}^{(\pm)}=\nabla_{i}^{(\pm)}+ \kappa \Gamma_i {\cal A}^{(\pm)}~,
\ee
for some $\kappa\in \bR$, because
\bea
\hat{\nabla}_{i}^{(\pm)}\eta_\pm=0~,~~~{\cal A}^{(\pm)}\eta_\pm=0\Longleftrightarrow {\nabla}_{i}^{(\pm)}\eta_\pm=0~,~~~{\cal A}^{(\pm)}\eta_\pm=0~.
\ee
Similarly, one can modify the horizon Dirac operators as
\bea
\label{iiaredef2}
{\mathscr D}^{(\pm)}={\cal D}^{(\pm)} + q{\cal A}^{(\pm)}~,
\ee
for some $q\in \bR$. Clearly, if $q=8\kappa$, then ${\mathscr D}^{(\pm)}= \Gamma^i \hat{\nabla}_{i}^{(\pm)}$. However, we
shall not assume this in general. As we shall see, there is an appropriate choice of $q$ and appropriate choices of $\kappa$ such that
the Killing spinors can be identified with the zero modes of ${\mathscr D}^{(\pm)}$.
\subsection{Lichnerowicz type theorems for $\mathcal{D}^{(\pm)}$}
First let us establish that the $\eta_+$ Killing spinors can be identified with the zero modes of a ${\mathscr D}^{(+)}$.  It is straightforward to see
that if $\eta_+$ is a Killing spinor, then $\eta_+$ is a zero mode of ${\mathscr D}^{(+)}$. So it remains to demonstrate the converse.
For this assume that $\eta_+$ is a zero mode of ${\mathscr D}^{(+)}$, i.e.~${\mathscr D}^{(+)}\eta_+=0$. Then after some lengthy computation which utilizes the field equations and Bianchi identities, described
in Appendix C, one can establish the equality
\bea
{\tilde{\nabla}}^{i}{\tilde{\nabla}}_{i}\parallel\eta_+\parallel^2 - (2\tilde{\nabla}^i \Phi +  h^i) {\tilde{\nabla}}_{i}\parallel\eta_+\parallel^2 = 2\parallel{\hat\nabla^{(+)}}\eta_{+}\parallel^2 + (-4\kappa - 16 \kappa^2)\parallel\mathcal{A}^{(+)}\eta_+\parallel^2~,
\label{iiamaxprin}
\ee
provided that $q=-1$. $ \langle\cdot, \cdot \rangle$ is the Dirac inner product of $Spin(8)$, see Appendix B, which can be identified
with the standard Hermitian inner product on $\Lambda^*(\mathbb{C}^4)$ restricted on the real subspace of Majorana spinors and $\parallel \cdot \parallel$ is the associated norm. Therefore, $ \langle\cdot, \cdot \rangle$ is a real and positive definite. The $Spin(8)$ gamma matrices are Hermitian with respect to
$ \langle\cdot, \cdot \rangle$. It is clear that if the last term on the right-hand-side of the above identity is positive semi-definite, then one can apply the maximum principle on $\parallel\eta_+\parallel^2$
as the fields are assumed to be smooth, and ${\cal S}$ compact.
In particular,  if
\bea
-{1\over4} <\kappa<0~,
\label{rangek}
\ee
then the maximum principle implies that $\eta_+$ are Killing spinors and $\parallel\eta_+\parallel=\mathrm{const}$. Observe that if one takes ${\mathscr D}^{(+)}$ with $q=-1$, then ${\mathscr D}^{(+)}= \Gamma^i \hat{\nabla}_{i}^{(+)}$ provided that
$\kappa=-1/8$ which lies in the range (\ref{rangek}).
To summarize we have established that for $q=-1$ and $-{1 \over 4} < \kappa <0$,
\bea
\nabla_{i}^{(+)}\eta_+=0~,~~~{\cal A}^{(+)}\eta_+=0~\Longleftrightarrow~ {\mathscr D}^{(+)}\eta_+=0~.
\ee
Moreover $\parallel\eta_+\parallel^2$ is constant on ${\cal S}$.

Next we shall establish that the $\eta_-$ Killing spinors can also be identified with the zero modes of a modified horizon Dirac operator ${\mathscr D}^{(-)}$.
It is clear that all Killing spinors $\eta_-$ are zero modes of ${\mathscr D}^{(-)}$. To prove the converse,
suppose that $\eta_-$ satisfies ${\mathscr D}^{(-)} \eta_-=0$.
The proof proceeds by calculating the Laplacian of $\parallel \eta_-
\parallel^2$ as described in Appendix C, which requires the use of the
field equations and Bianchi identies.
One can then establish the formula
\bea
\label{iial2b}
{\tilde{\nabla}}^{i} \big( e^{-2 \Phi} V_i \big)
= -2 e^{-2 \Phi} \parallel{\hat\nabla^{(-)}}\eta_{-}\parallel^2 +   e^{-2 \Phi} (4 \kappa +16 \kappa^2) \parallel\mathcal{A}^{(-)}\eta_-\parallel^2~,
\ee
provided that $q=-1$, where
\bea
V=-d \parallel \eta_- \parallel^2 - \parallel \eta_- \parallel^2 h \ .
\ee
The last term on the RHS of ({\ref{iial2b}}) is negative semi-definite
if $
-{1 \over 4} < \kappa <0$.
Provided that this holds, on integrating ({\ref{iial2b}}) over
${\cal{S}}$ and assuming that ${\cal{S}}$ is compact and without boundary, one finds that ${\hat\nabla^{(-)}}\eta_{-}=0$ and $\mathcal{A}^{(-)}\eta_-=0$.
Therefore, we have shown that for $q=-1$ and $-{1 \over 4} < \kappa <0$,
\bea
\nabla_{i}^{(-)}\eta_-=0~,~~~{\cal A}^{(-)}\eta_-=0~\Longleftrightarrow~ {\mathscr D}^{(-)}\eta_-=0~ \ .
\ee
This concludes the relationship between Killing spinors and zero modes of modified horizon Dirac operators.
\subsection{Index theory and supersymmetry enhancement}
The analysis developed so far suffices to prove that IIA horizons preserve an even number of supersymmetries. Indeed, if $N_\pm$ is the number
of $\eta_\pm$ Killing spinors, then the number of supersymmetries of IIA horizon is $N=N_++N_-$. Utilizing the relation between the Killing spinors $\eta_\pm$
and the zero modes of the modified horizon Dirac operators ${\mathscr D}^{(\pm)}$ established in the previous two sections, we have that
\bea
N_\pm=\mathrm{dim}\,\mathrm{Ker}\, {\mathscr D}^{(\pm)}~.
\ee
Next let us focus on the index of the ${\mathscr D}^{(+)}$ operator. As we have mentioned, the spin bundle of the spacetime $S$ decomposes
on ${\cal S}$ as $S=S_+\oplus S_-$. Moreover, $S_+$ and $S_-$ are isomorphic as $Spin(8)$ bundles and are associated with the Majorana
non-Weyl ${\bf 16}$ representation. Furthermore ${\mathscr D}^{(+)}: \Gamma(S_+)\rightarrow \Gamma(S_+)$, where $ \Gamma(S_+)$ are the sections of $S_+$
and this action does not preserve the $Spin(8)$ chirality.  Since the principal symbol of ${\mathscr D}^{(+)}$ is the same as the principal symbol
of the standard Dirac operator acting on Majorana but not-Weyl spinors, the index vanishes\footnote{This should be contrasted to IIB horizons  where the horizon Dirac operators
 act on the Weyl spinors and map them to anti-Weyl ones.  As a result, the horizon Dirac operators have the same principal symbol as the standard Dirac operator acting on the Weyl spinors
 and so there is a non-trivial contribution from the index.} \cite{atiyah1}.  As a result, we conclude that
\bea
\mathrm{dim}\,\mathrm{Ker}\, {\mathscr D}^{(+)}= \mathrm{dim}\,\mathrm{Ker}\, ({\mathscr D}^{(+)})^\dagger~,
\ee
where $({\mathscr D}^{(+)})^\dagger$ is the adjoint of ${\mathscr D}^{(+)}$ with respect to the symmetric $Spin(8)$-invariant inner product $\langle \phantom{i},\phantom{i} \rangle$. Furthermore observe that
\bea
\big(e^{2 \Phi} \Gamma_-\big) \big({\mathscr D}^{(+)}\big)^\dagger
= {\mathscr D}^{(-)} \big(e^{2 \Phi} \Gamma_-\big), \qquad  ({\rm for} \ q=-1)~,
\ee
and so
\bea
N_-=\mathrm{dim}\,\mathrm{Ker}\, ({\mathscr D}^{(-)})=\mathrm{dim}\,\mathrm{Ker}\, ({\mathscr D}^{(+)})^\dagger~.
\ee
Therefore, we conclude that $N_+=N_-$ and so the number of supersymmetries of IIA horizons $N=N_++N_-=2 N_-$ is even. This proves
the first part of the conjecture ({\ref{index}}) for IIA horizons.

\section{The $\mathfrak{sl}(2,\bR)$ symmetry of IIA horizons}
\subsection{Construction of $\eta_+$ from $\eta_{-}$ Killing spinors}
In the investigation of the integrability conditions of the KSEs, we have demonstrated that if $\eta_-$ is a Killing spinor, then
$\eta_+ = \Gamma_+ \Theta_- \eta_-$ is also a Killing spinor, see (\ref{iiaepfem}). Since we know that the $\eta_+$ and $\eta_-$ Killing spinors
appear in pairs, the formula (\ref{iiaepfem}) provides a way to construct the $\eta_+$ Killing spinors from the $\eta_-$ ones.
However, this is the case provided that $\eta_+=\Gamma_+ \Theta_- \eta_-\not=0$. Here, we shall prove that for horizons with non-trivial fluxes
\bea
\mathrm{Ker}\, \Theta_-=\{0\}~,
\label{iiakerz}
\ee
 and so
the operator $\Gamma_+\Theta_-$ pairs the $\eta_-$ with the $\eta_+$ Killing spinors.
We shall prove  $\mathrm{Ker}\, \Theta_-=\{0\}$ using contradiction.  For this assume that $\Theta_-$ has a non-trivial kernel, i.e.~there is $\eta_-\not=0$
such that
\bea
\Theta_- \eta_-=0~.
\ee
If this is the case, then the last integrability condition in  (\ref{iiaint1}) gives that
\bea
\langle \eta_- ,  \bigg( -{1 \over 2} \Delta -{1 \over 8} dh_{ij} \Gamma^{ij} + \frac{1}{8}M_{i j}\Gamma^{i j}\Gamma_{11} 
- \frac{1}{4}e^{\Phi}T_{i}\Gamma^{i}\Gamma_{11} - \frac{1}{24}e^{\Phi}Y_{i j k}\Gamma^{i j k}\bigg)  \eta_- \rangle =0~.
\ee
This in turn implies that
\bea
\Delta \langle \eta_- , \eta_- \rangle =0~,
\ee
and hence
\bea
\Delta =0 \ ,
\ee
as $\eta_-$ is no-where vanishing.
Next the gravitino KSE $\nabla^{(-)}\eta_-=0$ implies that
\bea
{\tilde{\nabla}}_i \langle \eta_-, \eta_- \rangle &=& -{1 \over 2} h_i  \langle \eta_-, \eta_- \rangle
+ \langle \eta_{-}, \Gamma_{11}\bigg(-\frac{1}{2}L_{i}
+ \frac{1}{8}e^{\Phi}\tilde{F}_{\ell_1 \ell_2}\Gamma_i{}^{\ell_1 \ell_2} \bigg) \eta_{-} \rangle
\nonumber \\
&+& \langle \eta_- , \bigg(\frac{1}{4}e^{\Phi}X_{i \ell}\Gamma^{\ell} - \frac{1}{96}e^{\Phi}\tilde{G}_{\ell_1 \ell_2 \ell_3 \ell_4}\Gamma_i{}^{\ell_1 \ell_2 \ell_3 \ell_4}\bigg) \eta_- \rangle~,
\ee
which can be simplified further using
\bea
\langle \eta_{-}, \Gamma_{i}\Theta_{-}\eta_{-} \rangle &=& \frac{1}{4}h_{i} \langle \eta_-, \eta_- \rangle
+ \langle \eta_{-}, \Gamma_{11}\bigg(-\frac{1}{4}L_{i}
+ \frac{1}{16}e^{\Phi}\tilde{F}_{\ell_1 \ell_2}\Gamma_i{}^{\ell_1 \ell_2} \bigg) \eta_{-} \rangle
\nonumber \\
&+& \langle \eta_- , \bigg(\frac{1}{8}e^{\Phi}X_{i \ell}\Gamma^{\ell} - \frac{1}{192}e^{\Phi}\tilde{G}_{\ell_1 \ell_2 \ell_3 \ell_4}\Gamma_i{}^{\ell_1 \ell_2 \ell_3 \ell_4}\bigg) \eta_- \rangle~
 = 0~,
\ee
to yield,
\bea
\label{iianrm1a}
{\tilde{\nabla}}_i \parallel \eta_-\parallel^2 = - h_i  \parallel \eta_-\parallel^2~.
\ee
As $\eta_-$ is no-where zero, this implies that
\bea
dh=0~.
\ee
Substituting,  $\Delta=0$  and $dh=0$ into (\ref{iiaauxeq4}), we  find that
\bea
M=d_h L=0~,~~~T=d_h S=0~,~~~Y=d_h X-L\wedge \tilde F-S \tilde H=0~,
\ee
as well.
Returning to ({\ref{iianrm1a}}), on taking the divergence, and using ({\ref{iiafeq7}}) to
eliminate the ${\tilde{\nabla}}^i h_i$ term, one obtains
\bea
{\tilde{\nabla}}^i {\tilde{\nabla}}_i  \parallel \eta_-\parallel^2 &=& 2\tilde{\nabla}^{i}\Phi {\tilde{\nabla}}_i \parallel \eta_-\parallel^2+ \bigg(L^2+ \frac{1}{2}e^{2\Phi}S^2 
+ \frac{1}{4}e^{2\Phi}X^2 + \frac{1}{4}e^{2\Phi}\tilde{F}^2 + \frac{1}{48}e^{2\Phi}\tilde{G}^2\bigg)  \parallel \eta_-\parallel^2~.
\nonumber \\
\ee
Applying the maximum principle on $\parallel \eta_-\parallel^2$ we conclude that all the fluxes apart from the dilaton $\Phi$ and $\tilde H$ vanish and
$\parallel\eta_-\parallel$ is constant. The latter together with  (\ref{iianrm1a}) imply that $h=0$.
Next applying the maximum principle to the dilaton field equation (\ref{iiafeq6}), we conclude that the dilaton is constant and $\tilde H=0$.
Combining all the results so far, we conclude that all the fluxes vanish which is a contradiction to the assumption that not all of the fluxes vanish.  This establishes (\ref{iiakerz}).
Furthermore, the horizons for which $\Theta_- \eta_- =0$ ($\eta_- \neq 0$) are all local products  $\bR^{1,1}\times {\cal S}$,  where ${\cal S}$ up to a discrete identification is a product of Ricci flat Berger manifolds. Thus ${\cal S}$ has holonomy, $Spin(7)$ or $SU(4)$ or $Sp(2)$ as an  irreducible manifold,  and $G_2$ or $SU(3)$ or $Sp(1)\times Sp(1)$ or $Sp(1)$ or $\{1\}$ as a reducible one.

It remains to prove the second part of the conjecture that all IIA horizons with non-trivial fluxes admit an $\mathfrak{sl}(2,\bR)$ symmetry subalgebra.
As we shall demonstrate, this in fact is a consequence of our previous result that all IIA horizons admit an even number of supersymmetries. The  proof is
 very similar to that already given in the context of M-horizons in \cite{11index}.
\subsection{Killing vectors}
To begin, first note that the Killing spinor $\epsilon$ on the spacetime can be expressed in terms of $\eta_\pm$ as
\bea
\epsilon= \eta_++ u \Gamma_+\Theta_-\eta_-+ \eta_-+r \Gamma_-\Theta_+\eta_++ru \Gamma_-\Theta_+\Gamma_+\Theta_-\eta_-~,
\label{iiagensolkse}
\ee
which is derived after collecting the results of section \ref{ikse}.
Since the $\eta_-$ and $\eta_+$ Killing spinors appear in pairs for supersymmetric IIA horizons, let us choose a $\eta_-$ Killing spinor.  Then from the results
of the previous section, horizons with non-trivial fluxes also admit $\eta_+=\Gamma_+\Theta_-\eta_-$ as a Killing spinors. Using $\eta_-$ and $\eta_+=\Gamma_+\Theta_-\eta_-$,
one can construct two linearly independent Killing spinors on the  spacetime as
\bea
\epsilon_1=\eta_-+u\eta_++ru \Gamma_-\Theta_+\eta_+~,~~~\epsilon_2=\eta_++r\Gamma_-\Theta_+\eta_+~.
\ee
To continue, it is known from the general theory of supersymmetric IIA backgrounds that for any Killing spinors $\zeta_1$ and $\zeta_2$ the dual vector field of the 1-form
bilinear
\bea
K(\zeta_1, \zeta_2)=\langle(\Gamma_+-\Gamma_-) \zeta_1, \Gamma_a\zeta_2\rangle\, e^a~,
\ee
is a Killing vector and leaves invariant all the other fields of the theory.
Evaluating, the 1-form bilinears of the Killing spinor $\epsilon_1$ and $\epsilon_2$ and expanding with $a=(+,-,i)$, we find that
\bea
 K_1(\epsilon_1, \epsilon_2)&=& (2r \langle\Gamma_+\eta_-, \Theta_+\eta_+\rangle+  u^2 r \Delta \parallel \eta_+\parallel^2) \,{\bf{e}}^+-2u \parallel\eta_+\parallel^2\, {\bf{e}}^-+ V_i {\bf{e}}^i~,
 \cr
 K_2(\epsilon_2, \epsilon_2)&=& r^2 \Delta\parallel\eta_+\parallel^2 \,{\bf{e}}^+-2 \parallel\eta_+\parallel^2 {\bf{e}}^-~,
 \cr
 K_3(\epsilon_1, \epsilon_1)&=&(2\parallel\eta_-\parallel^2+4r u \langle\Gamma_+\eta_-, \Theta_+\eta_+\rangle+ r^2 u^2 \Delta \parallel\eta_+\parallel^2) {\bf{e}}^+ - 2u^2 \parallel\eta_+\parallel^2 {\bf{e}}^-+2u V_i {\bf{e}}^i~,
 \label{iiab1forms}
 \nonumber \\
 \ee
where we have set
\bea
\label{iiavii}
V_i =  \langle \Gamma_+ \eta_- , \Gamma_i \eta_+ \rangle~.
\ee
Moreover, we have used the identities
\bea
- \Delta\, \parallel\eta_+\parallel^2 +4  \parallel\Theta_+ \eta_+\parallel^2 =0~,~~~\langle \eta_+ , \Gamma_i \Theta_+ \eta_+ \rangle  =0~,
\ee
which follow from the first integrability condition in ({\ref{iiaint1}}),  $\parallel\eta_+\parallel=\mathrm{const}$ and the KSEs of $\eta_+$.

\subsection{$\mathfrak{sl}(2,\mathbb{R})$ symmetry of IIA-horizons}
To uncover the $\mathfrak{sl}(2,\mathbb{R})$ symmetry of IIA horizons it remains to compute the Lie bracket algebra of the vector fields associated to the 1-forms $K_1, K_2$ and $K_3$.
For this note that these vector fields can be expressed as
\bea
K_1&=&-2u \parallel\eta_+\parallel^2 \partial_u+ 2r \parallel\eta_+\parallel^2 \partial_r+ V^i \tilde \partial_i~,
\cr
K_2&=&-2 \parallel\eta_+\parallel^2 \partial_u~,
\cr
K_3&=&-2u^2 \parallel\eta_+\parallel^2 \partial_u +(2 \parallel\eta_-\parallel^2+ 4ru \parallel\eta_+\parallel^2)\partial_r+ 2u V^i \tilde \partial_i~,
\ee
where we have used the same symbol for the 1-forms and the associated vector fields. These expressions are
similar to those we have obtained for M-horizons in \cite{11index} apart form the range of the index $i$ which is different. Using the various identities we have obtained, a direct computation reveals
 that the Lie bracket algebra is
\bea
[K_1,K_2] =2 \parallel\eta_+\parallel^2 K_2~,~~ [K_2, K_3]=-4 \parallel\eta_+\parallel^2 K_1~,~~[K_3,K_1]=2 \parallel\eta_+\parallel^2 K_3~, \ \
\ee
which is isomorphic to $\mathfrak{sl}(2,\mathbb{R})$.
This proves the second part of the conjecture and completes the analysis.

\section{The geometry and isometries of ${\cal S}$}
First suppose that $V\not=0$. Then the conditions ${\cal L}_{K_a} g=0$ and ${\cal L}_{K_a} F=0$, $a=1,2,3$, where $F$ denotes collectively all the
fluxes of IIA supergravity, imply that
\bea
\tilde\nabla_{(i} V_{j)}=0~,~~~\tilde {\cal L}_Vh=\tilde {\cal L}_V\Delta=0~,~~~ \tilde {\cal L}_V \Phi=0~,
\nonumber \\
\tilde {\cal L}_V X=\tilde {\cal L}_V \tilde G=\tilde {\cal L}_V L=\tilde {\cal L}_V \tilde H=
\tilde {\cal L}_V S=\tilde {\cal L}_V \tilde F=0~,
\ee
i.e.~$V$ is an isometry of ${\cal S}$ and leaves all the fluxes on ${\cal S}$ invariant.
In addition, one also finds the useful identities
\bea
&&-2 \parallel\phi_+\parallel^2-h_i V^i+2 \langle\Gamma_+\phi_-, \Theta_+\phi_+\rangle=0~,~~~i_V (dh)+2 d \langle\Gamma_+\phi_-, \Theta_+\phi_+\rangle=0~,
\cr
&& 2 \langle\Gamma_+\phi_-, \Theta_+\phi_+\rangle-\Delta \parallel\phi_-\parallel^2=0~,~~~
V+ \parallel\phi_-\parallel^2 h+d \parallel\phi_-\parallel^2=0~,
\label{iiaconconx}
\ee
which imply that ${\cal L}_V\parallel\phi_-\parallel^2=0$. There are further restrictions on the geometry of ${\cal S}$ which will be explored elsewhere.
A special case arises for $V=0$ where the group action generated by $K_1, K_2$ and $K_3$ has only 2-dimensional orbits. A direct substitution of this condition in (\ref{iiaconconx}) reveals that
\bea
\Delta \parallel\phi_-\parallel^2=2 \parallel\phi_+\parallel^2~,~~~h=\Delta^{-1} d\Delta~.
\ee
Since $dh=0$ and $h$ is exact such horizons are static and a coordinate transformation $r\rightarrow \Delta r$ reveals that the horizon geometry is a warped product of $AdS_2$ with ${\cal S}$, $AdS_2\times_w {\cal S}$.

	\chapter{Roman's Massive IIA Supergravity}

In this chapter, we present a similar local and global analysis of the Killing spinor
equations for Roman's Massive type IIA supergravity and a proof for the horizon conjecture for this theory.

It was initially unclear if the horizon conjecture would work in the same way for massive
IIA supergravity, because of the presence of a negative cosmological constant. The proof of the conjecture relies on the application of the maximum principle to demonstrate certain Lichnerowicz type theorems. In turn the application of the maximum principle requires the positive semi-definiteness of certain terms which depend on the fluxes. The existence of a negative cosmological constant in the theory has the potential of invalidating these arguments as it can contribute with the opposite sign in the expressions required for the application of the maximum principle. We show that this is not the case and therefore the conjecture can be extended to massive IIA horizons. Additional delicate analysis was
required to establish this result, which we therefore consider as a separate case.

Nevertheless many of the steps in the proof of the conjecture for massive IIA horizons are similar to those presented for IIA horizons in the previous section. Because of this, we shall only state the key statements and formulae required for the proof of the conjecture.

\section{Horizon fields and KSEs}

\subsection{Horizon Bianchi identities and field equations}

This expression for the near-horizon fields is similar to that for the IIA case in the previous chapter and \cite{iiaindex} though their dependence on the gauge potentials is different.  The massive theory contains an additional parameter $m$, the mass term, and the fields and both the gravitino and dilatino KSEs depend on it. The dependence on the coordinates $u,r$ is given explicitly and all the fields depend on the coordinates $y^I$
of the spatial horizon section ${\cal S}$ defined by $u=r=0$.

Adapting  Gaussian null coordinates \cite{isen, gnull} near massive IIA Killing horizons, one finds
\be
G&=& \bbe^+ \wedge \bbe^- \wedge X +r \bbe^+ \wedge Y + \tilde G~,~~
\cr
H &=&\bbe^+ \wedge \bbe^- \wedge L+ r \bbe^+ \wedge M + \tilde H~,~~~
\cr
F&=& \bbe^+ \wedge \bbe^- S + r \bbe^+ \wedge T+ \tilde F~,
\la{miiahormetrx}
\ee
where  $\Delta$ is a function, $h$, $L$ and $T$ are 1-forms, $X$, $M$ and $\tilde F$ are 2-forms,
$Y, \tilde H$ are 3-forms and $\tilde G$ is a 4-form on  the spatial horizon section ${\cal S}$. 
The basis introduced in ({\ref{basis1}}) is used. The dilaton $\Phi$ is also taken as a function
on ${\cal S}$.
Substituting the fields (\ref{miiahormetrx}) into the Bianchi identities  of massive IIA supergravity, one finds that
\bea
\label{miiabian2}
M&=&d_h L~,~~~T=d_h S - mL~,~~~Y=d_h X-L\wedge \tilde F-S \tilde H~,~~~
\cr
d\tilde G&=&\tilde H\wedge \tilde F~,~~~d\tilde H=0, \, d\tilde F=m\tilde{H}~,
\ee
where  $d_h \theta \equiv d \theta-
h \wedge \theta$ for any form $\theta$. The Bianchi identities relate some of the components of the near-horizon fields; in particular, $M$, $T$ and $Y$ are not independent.
Similarly, the independent field equations of the near-horizon fields are as follows. The 2-form field equation ({\ref{miiageq1}) gives
\bea
\label{miiafeq1}
\tn^i {\tilde{F}}_{ik}-h^i {\tilde{F}}_{ik}
+T_k -L^i X_{ik} +{1 \over 6} {\tilde{H}}^{\ell_1 \ell_2
\ell_3} {\tilde{G}}_{\ell_1 \ell_2 \ell_3 k}=0~,
\ee
the 3-form field equation  ({\ref{miiaeineq}}) gives
\bea
\label{miiafeq2}
\tn^i (e^{-2 \Phi}L_i)- mS -{1 \over 2} {\tilde{F}}^{ij}
X_{ij} +{1 \over 1152}
\epsilon^{\ell_1 \ell_2 \ell_3 \ell_4 \ell_5 \ell_6 \ell_7
\ell_8} {\tilde{G}}_{\ell_1 \ell_2 \ell_3 \ell_4}
{\tilde{G}}_{\ell_5 \ell_6 \ell_7 \ell_8}=0~,
\ee
and
\bea
\label{miiafeq3}
&&\tn^i(e^{-2 \Phi} {\tilde{H}}_{imn})-m\tilde{F}_{m n}
-e^{-2 \Phi} h^i {\tilde{H}}_{imn}
+e^{-2 \Phi} M_{mn}+S X_{mn}-{1 \over 2}
{\tilde{F}}^{ij} {\tilde{G}}_{ijmn}
\nonumber \\
&&-{1 \over 48} \epsilon_{mn}{}^{\ell_1
\ell_2 \ell_3 \ell_4 \ell_5 \ell_6} X_{\ell_1 \ell_2}
{\tilde{G}}_{\ell_3 \ell_4 \ell_5 \ell_6} =0~,
\ee
and the 4-form field equation ({\ref{miiaceq1}}) gives
\bea
\label{miiafeq4}
\tn^i X_{ik} +{1 \over 144} \epsilon_k{}^{\ell_1 \ell_2
\ell_3 \ell_4 \ell_5 \ell_6 \ell_7}
{\tilde{G}}_{\ell_1 \ell_2 \ell_3 \ell_4} {\tilde{H}}_{\ell_5 \ell_6 \ell_7} =0~,
\ee
and
\bea
\label{miiafeq5}
\tn^i {\tilde{G}}_{ijkq}+Y_{jkq}-h^i {\tilde{G}}_{ijkq}
-{1 \over 12} \epsilon_{jkq}{}^{\ell_1 \ell_2 \ell_3
\ell_4 \ell_5} X_{\ell_1 \ell_2} {\tilde{H}}_{\ell_3 \ell_4 \ell_5}-{1 \over 24}\epsilon_{jkq}{}^{\ell_1 \ell_2 \ell_3
\ell_4 \ell_5} {\tilde{G}}_{\ell_1 \ell_2 \ell_3 \ell_4}
L_{\ell_5} =0~,
\ee
where $\tilde \nabla$ is the Levi-Civita connection of the metric on ${\cal S}$.
In addition, the dilaton field equation ({\ref{miiadileq}}) becomes
\bea
\label{miiafeq6}
\tn^i \tn_i \Phi - h^i \tn_i \Phi &=&
2 \tn_i \Phi \tn^i \Phi +{1 \over 2} L_i L^i
-{1 \over 12} {\tilde{H}}_{\ell_1 \ell_2 \ell_3}
{\tilde{H}}^{\ell_1 \ell_2 \ell_3}-{3 \over 4} e^{2 \Phi}S^2
\nonumber \\
&+&{3 \over 8} e^{2 \Phi} {\tilde{F}}_{ij}
{\tilde{F}}^{ij} -{1 \over 8} e^{2 \Phi} X_{ij}X^{ij}
+{1 \over 96} e^{2 \Phi} {\tilde{G}}_{\ell_1 \ell_2 \ell_3
\ell_4} {\tilde{G}}^{\ell_1 \ell_2 \ell_3 \ell_4}
\nonumber \\
 &+& \frac{5}{4}e^{2\Phi}m^2~.
\ee
It remains to evaluate the Einstein field equation. This gives
\bea
\label{miiafeq7}
{1 \over 2} \tn^i h_i -\Delta -{1 \over 2}h^2
&=& h^i \tn_i \Phi -{1 \over 2} L_i L^i -{1 \over 4} e^{2 \Phi} S^2 -{1 \over 8} e^{2 \Phi} X_{ij} X^{ij}
\nonumber \\
&-&{1 \over 8} e^{2 \Phi} {\tilde{F}}_{ij} {\tilde{F}}^{ij}
-{1 \over 96} e^{2 \Phi} {\tilde{G}}_{\ell_1 \ell_2
\ell_3 \ell_4} {\tilde{G}}^{\ell_1 \ell_2 \ell_3 \ell_4} - \frac{1}{4}e^{2\Phi}m^2~,
\ee
and
\bea
\label{miiafeq8}
{\tilde{R}}_{ij} &=& -\tn_{(i} h_{j)}
+{1 \over 2} h_i h_j -2 \tn_i \tn_j \Phi
-{1 \over 2} L_i L_j +{1 \over 4} {\tilde{H}}_{i
\ell_1 \ell_2} {\tilde{H}}_j{}^{\ell_1 \ell_2}
\nonumber \\
&+&{1 \over 2} e^{2 \Phi} {\tilde{F}}_{i \ell}
{\tilde{F}}_j{}^\ell -{1 \over 2} e^{2 \Phi} X_{i \ell}
X_j{}^\ell+{1 \over 12} e^{2 \Phi} {\tilde{G}}_{i \ell_1 \ell_2 \ell_3} {\tilde{G}}_j{}^{\ell_1 \ell_2 \ell_3}
\nonumber \\
&+&\delta_{ij} \bigg({1 \over 4} e^{2 \Phi} S^2 - \frac{1}{4}e^{2\Phi}m^2
-{1 \over 8} e^{2 \Phi} {\tilde{F}}_{\ell_1 \ell_2}
{\tilde{F}}^{\ell_1 \ell_2} +{1 \over 8} e^{2 \Phi}
X_{\ell_1 \ell_2}X^{\ell_1 \ell_2}-{1 \over 96} e^{2 \Phi} {\tilde{G}}_{\ell_1 \ell_2
\ell_3 \ell_4} {\tilde{G}}^{\ell_1 \ell_2 \ell_3 \ell_4}
\bigg)~,
\nonumber \\
\ee
where $\tilde R$ denotes the Ricci tensor of ${\cal S}$.
There are additional Bianchi identities and field equations which however are not independent of those we have stated above. We give these because they are useful
in many of the intermediate computations. In particular, we have the additional Bianchi identities
\bea
dT + S dh + dS \wedge h + m dL&=&0~,
\nonumber \\
dM+L \wedge dh -h \wedge dL &=&0~,
\nonumber \\
dY + dh \wedge X - h \wedge dX + h \wedge (S {\tilde{H}}
+{\tilde{F}} \wedge L)+T \wedge {\tilde{H}}+{\tilde{F}} \wedge M &=&0~.
\ee
There are also additional field equations given by
\bea
\label{miiaauxeq1}
-\tn^i T_i + h^i T_i -{1 \over 2} dh^{ij} {\tilde{F}}_{ij}
-{1 \over 2} X_{ij}M^{ij} -{1 \over 6} Y_{ijk}{\tilde{H}}^{ijk}=0~,
\ee
\bea
\label{miiaauxeq2}
- \tn^i (e^{-2 \Phi}M_{ik}) + e^{-2 \Phi} h^i M_{ik}
-{1 \over 2} e^{-2 \Phi} dh^{ij} {\tilde{H}}_{ijk}
-T^i X_{ik} -{1 \over 2} {\tilde{F}}^{ij} Y_{ijk}
\nonumber \\
-mT_k -{1 \over 144} \epsilon_k{}^{\ell_1
\ell_2 \ell_3 \ell_4 \ell_5 \ell_6 \ell_7}
Y_{\ell_1 \ell_2 \ell_3} {\tilde{G}}_{\ell_4 \ell_5 \ell_6 \ell_7} =0~,
\ee
\bea
\label{miiaauxeq3}
- \tn^i Y_{imn}+h^i Y_{imn}-{1 \over 2} dh^{ij}
{\tilde{G}}_{ijmn}
+{1 \over 36} \epsilon_{mn}{}^{\ell_1 \ell_2 \ell_3
\ell_4 \ell_5 \ell_6} Y_{\ell_1 \ell_2 \ell_3}
{\tilde{H}}_{\ell_4 \ell_5 \ell_6}
\nonumber \\
+{1 \over 48} \epsilon_{mn}{}^{\ell_1 \ell_2 \ell_3
\ell_4 \ell_5 \ell_6} {\tilde{G}}_{\ell_1 \ell_2 \ell_3
\ell_4} M_{\ell_5 \ell_6}=0~,
\ee
corresponding to equations obtained from
the $+$ component of ({\ref{miiageq1}}),
the $k$ component of ({\ref{miiaeineq}}) and
the $mn$ component of ({\ref{miiaceq1}}) respectively.
However, ({\ref{miiaauxeq1}}), ({\ref{miiaauxeq2}}) and
({\ref{miiaauxeq3}}) are implied by ({\ref{miiafeq1}})-
({\ref{miiafeq5}}) together with the Bianchi identities
({\ref{miiabian2}}).
Note also that the $++$ and $+i$ components of the
Einstein equation, which are
\bea
\label{miiaauxeq4}
{1 \over 2} \tn^i \tn_i \Delta -{3 \over 2} h^i \tn_i \Delta-{1 \over 2} \Delta \tn^i h_i
+ \Delta h^2 +{1 \over 4} dh_{ij} dh^{ij}
&=&(\tn^i \Delta - \Delta h^i)\tn_i \Phi 
\cr
&+&{1 \over 4} M_{ij}
M^{ij}
+{1 \over 2} e^{2 \Phi} T_i T^i
\nonumber \\
&+&{1 \over 12} e^{2 \Phi} Y_{ijk} Y^{ijk} \ ,
\ee
and
\bea
\label{miiaauxeq5}
{1 \over 2} \tn^j dh_{ij}-dh_{ij} h^j - \tn_i \Delta + \Delta h_i
&=& dh_i{}^j \tn_j \Phi -{1 \over 2} M_i{}^j L_j
+{1 \over 4} M_{\ell_1 \ell_2} {\tilde{H}}_i{}^{\ell_1 \ell_2}
\nonumber \\
&-&{1 \over 2} e^{2 \Phi} S T_i
+{1 \over 2} e^{2 \Phi} T^j {\tilde{F}}_{ij}
\nonumber \\
&-&{1 \over 4} e^{2 \Phi} Y_i{}^{\ell_1 \ell_2}
X_{\ell_1 \ell_2} +{1 \over 12} e^{2 \Phi}
Y_{\ell_1 \ell_2 \ell_3} {\tilde{G}}_i{}^{\ell_1 \ell_2
\ell_3}~,
\ee
are implied by ({\ref{miiafeq6}}), ({\ref{miiafeq7}}), ({\ref{miiafeq8}}),
together with ({\ref{miiafeq1}})-({\ref{miiafeq5}}),
and the Bianchi identities ({\ref{miiabian2}}).

\subsection{Integration of KSEs along lightcone}

The KSEs of massive IIA supergravity can be solved along the lightcone directions. The solution is
\bea\label{miialightconesol}
\e=\e_++\e_-~,~~~\e_+=\phi_+(u,y)~,~~~\e_-=\phi_-+r \Gamma_-\Theta_+ \phi_+~,
\ee
and
\bea
\phi_-=\eta_-~,~~~\phi_+=\eta_++ u \Gamma_+ \Theta_-\eta_-~,
\ee
where
\bea
\Theta_\pm &=& {1\over4} h_i\Gamma^i\mp{1\over4} \Gamma_{11} L_i \Gamma^i-{1\over16} e^{\Phi} \Gamma_{11} (\pm 2 S+\tilde F_{ij} \Gamma^{ij})
\cr
&-& {1\over8 \cdot 4!} e^{\Phi} (\pm 12 X_{ij} \Gamma^{ij}
+\tilde G_{ijkl} \Gamma^{ijkl}) - \frac{1}{8}e^{\Phi}m \ ,
\ee
 $\Gamma_\pm\epsilon_\pm=0$, and $\eta_\pm=\eta_\pm(y)$ depend only on the coordinates $y$ of the spatial horizon section ${\cal S}$. Both $\eta_\pm$ are sections of the $Spin(8)$ bundle over ${\cal S}$ associated
  with the Majorana representation.
Substituting the solution (\ref{miialightconesol}) of the KSEs along the light cone directions back into the gravitino KSE (\ref{miiaGKSE}), and appropriately expanding in the $r$ and $u$ coordinates, we find that
for  the $\mu = \pm$ components, one obtains  the additional conditions
\bea
\label{miiaint1}
&&\bigg({1\over2}\Delta - {1\over8}(dh)_{ij}\Gamma^{ij} + {1\over8}M_{ij}\Gamma_{11}\Gamma^{ij} + 2\big( {1\over 4} h_i \Gamma^{i} - {1\over 4} L_{i}\Gamma_{11}\Gamma^{i}
\cr
&&- \frac{1}{8\cdot4!}e^{\Phi}(12X_{i j}\Gamma^{i j} - \tilde{G}_{i j k l}\Gamma^{i j k l}) 
\cr
&&- {1\over 16} e^{\Phi}\Gamma_{11}(-2S + \tilde{F}_{i j}\Gamma^{i j}) + \frac{1}{8}e^{\Phi}m \big)\Theta_{+} \bigg)\phi_{+} = 0~,
\ee
\bea
\label{miiaint2}
&&\bigg(\frac{1}{4}\Delta h_i \Gamma^{i} - \frac{1}{4}\partial_{i}\Delta \Gamma^{i} + \big(-\frac{1}{8}(dh)_{ij}\Gamma^{ij} - \frac{1}{8}M_{ij}\Gamma^{ij}\Gamma_{11} 
\cr
&&- \frac{1}{4}e^{\Phi}T_{i}\Gamma^{i}\Gamma_{11} + \frac{1}{24}e^{\Phi}Y_{i j k}\Gamma^{i j k} \big) \Theta_{+} \bigg) \phi_{+} = 0~,
\ee
\bea
\label{miiaint3}
&&\bigg(-\frac{1}{2}\Delta - \frac{1}{8}(dh)_{ij}\Gamma^{ij} + \frac{1}{8}M_{ij}\Gamma^{ij}\Gamma_{11} - \frac{1}{4}e^{\Phi} T_{i}\Gamma^{i} \Gamma_{11} - \frac{1}{24}e^{\Phi}Y_{ijk}\Gamma^{ijk} 
\cr
&&+ 2\big( -{1\over4} h_i\Gamma^i -{1\over4} \Gamma_{11} L_i \Gamma^i+{1\over16} e^\phi \Gamma_{11} (2 S+\tilde F_{ij} \Gamma^{ij})
\cr
&&-{1\over8 \cdot 4!} e^\phi (12 X_{ij} \Gamma^{ij}
+\tilde G_{ijkl} \Gamma^{ijkl})-\frac{1}{8}e^{\Phi}m \big) \Theta_{-} \bigg)\phi_{-} = 0 \ .
\ee
Similarly the $\mu=i$ component of the gravitino KSEs gives
\bea
\label{miiaint4}
&&\nabla^{(\pm)}_i\phi_\pm=0~~~~
\ee
and
\bea
\label{miiaint5}
&&\tilde \nabla_i \tau_{+} + \bigg( -\frac{3}{4}h_i - \frac{1}{16}e^{\Phi}X_{l_1 l_2}\Gamma^{l_1 l_2}\Gamma_{i} - \frac{1}{8\cdot4!}e^{\Phi} \tilde G_{l_1\cdots l_4}\Gamma^{l_1 \cdots l_4}\Gamma_{i} - \frac{1}{8}e^{\Phi}m\Gamma_{i}
\cr
&&- \Gamma_{11}(\frac{1}{4}L_i + \frac{1}{8}\tilde{H}_{i j k}\Gamma^{j k} + \frac{1}{8}e^{\Phi} S \Gamma_{i} +    \frac{1}{16}e^{\Phi}\tilde{F}_{l_1 l_2}\Gamma^{l_1 l_2}\Gamma_{i})\bigg )\tau_{+}
\cr
&&+ \bigg(-\frac{1}{4}(dh)_{ij}\Gamma^{j} - \frac{1}{4}M_{ij}\Gamma^{j}\Gamma_{11} + \frac{1}{8}e^{\Phi}T_{j}\Gamma^{j}\Gamma_{i}\Gamma_{11} + \frac{1}{48}e^{\Phi}Y_{l_1 l_2 l_3}\Gamma^{l_1 l_2 l_3}\Gamma_{i} \bigg)\phi_{+} = 0~,
\nonumber \\
\ee
where we have set
\bea
\label{miiaint6}
\tau_{+} = \Theta_{+}\phi_{+} \ .
\ee
We shall demonstrate that all the above conditions are not independent and follow  upon using the field equations and the Bianchi identities from those in (\ref{miiacovr}).
Similarly,
substituting the solution of the KSEs (\ref{miialightconesol})  into the dilatino KSE (\ref{miiaAKSE}) and expanding appropriately in the $r$ and $u$ coordinates, we find
\bea
\label{miiaint7}
&&\partial_i \Phi \Gamma^i \phi_{\pm} -{1\over12} \Gamma_{11} (\mp 6 L_i \Gamma^i+\tilde H_{ijk} \Gamma^{ijk}) \phi_\pm+{3\over8} e^\Phi \Gamma_{11} (\mp2 S+\tilde F_{ij} \Gamma^{ij})\phi_\pm
\cr
&&
+{1\over 4\cdot 4!}e^{\Phi} (\mp 12 X_{ij} \Gamma^{ij}+\tilde G_{j_1j_2j_3j_4} \Gamma^{j_1j_2j_3j_4}) \phi_\pm + \frac{5}{4}e^{\Phi}m\phi_\pm=0 \ ,
\ee
\be
\label{miiaint8}
&&-\bigg( \partial_{i}\Phi\Gamma^{i} + \frac{1}{12}\Gamma_{11} (6L_i \Gamma^{i} + \tilde{H}_{ijk}\Gamma^{ijk}) + \frac{3}{8}e^{\Phi}\Gamma_{11}(2S + \tilde{F}_{ij}\Gamma^{ij})
\cr
&&- \frac{1}{4\cdot 4!}e^{\Phi}(12X_{ij}\Gamma^{ij} + \tilde{G}_{ijkl}\Gamma^{ijkl}) -  \frac{5}{4}e^{\Phi}m \bigg)\tau_{+}
\cr
&&+ \bigg(\frac{1}{4}M_{ij}\Gamma^{ij}\Gamma_{11} + \frac{3}{4}e^{\Phi}T_{i}\Gamma^{i}\Gamma_{11} + \frac{1}{24}e^{\Phi}Y_{ijk}\Gamma^{ijk}\bigg)\phi_{+}=0~.
\ee
Again, these are not  independent of those in (\ref{miiacovr}).

\subsection{The independent KSEs on $\cal{S}$}

To describe the remaining independent KSEs consider the operators
\bea
\nabla_{i}^{(\pm)}&=& \tilde{\nabla}_{i} + \Psi^{(\pm)}_{i}~,
\ee
with
\bea
\label{miiaalg1pm}
\Psi^{(\pm)}_{i} &=& \bigg( \mp \frac{1}{4}h_{i} \mp \frac{1}{16}e^{\Phi}X_{l_1 l_2}\Gamma^{l_1 l_2}\Gamma_{i} + \frac{1}{8 \cdot 4!}e^{\Phi}{\tilde{G}}_{l_1 l_2 l_3 l_4}\Gamma^{l_1 l_2 l_3 l_4}\Gamma_{i} + \frac{1}{8}e^{\Phi}m \Gamma_i \bigg)
\cr
&+& \Gamma_{11}\bigg(\mp \frac{1}{4}L_{i} + \frac{1}{8}{\tilde{H}}_{i l_1 l_2}\Gamma^{l_1 l_2}
\pm \frac{1}{8}e^{\Phi}S\Gamma_{i} - \frac{1}{16}e^{\Phi}{\tilde{F}}_{l_1 l_2}\Gamma^{l_1 l_2}\Gamma_{i}\bigg)~,
\ee
and
\bea
\label{miiaalg2pm}
\mathcal{A}^{(\pm)} &=& \partial_i \Phi \Gamma^i  + \bigg(\mp \frac{1}{8}e^{\Phi}X_{l_1 l_2}\Gamma^{l_1 l_2} + \frac{1}{4 \cdot 4!}e^{\Phi}{\tilde{G}}_{l_1 l_2 l_3 l_4}\Gamma^{l_1 l_2 l_3 l_4} + \frac{5}{4}e^{\Phi}m \bigg)
\cr
&+& \Gamma_{11}\bigg(\pm \frac{1}{2}L_{i}\Gamma^{i} - \frac{1}{12}{\tilde{H}}_{i j k}\Gamma^{i j k} \mp \frac{3}{4}e^{\Phi}S + \frac{3}{8}e^{\Phi}{\tilde{F}}_{i j}\Gamma^{i j}\bigg)~.
\ee
These are derived from the naive restriction of the supercovariant derivative and the dilatino KSE on ${\cal S}$.
\vskip 0.3cm
{\it Theorem:} The remaining independent KSEs are
\bea
\label{miiacovr}
\nabla_{i}^{(\pm)}\eta_{\pm}  = 0~,~~~\mathcal{A}^{(\pm)}\eta_{\pm} = 0~.
\ee
Moreover if $\eta_-$ solves the KSEs, then
\bea
\eta_+ = \Gamma_{+}\Theta_{-}\eta_{-}~,
\label{miiaepfem}
\ee
is also a solution.
\vskip 0.3cm

{\it Proof:} The proof is given in Appendix D.4.
\rightline{ $\square$}

\section{Supersymmetry enhancement}

\subsection{Horizon Dirac equations}
To proceed with the proof of the first part of the conjecture define the modified horizon Dirac operators   as
\bea
\label{miiaredef2}
{\mathscr D}^{(\pm)}={\cal D}^{(\pm)} -{\cal A}^{(\pm)}~,
\ee
 where
\bea
{\cal D}^{(\pm)} \equiv \Gamma^{i}\nabla_{i}^{(\pm)} = \Gamma^{i}\tilde{\nabla}_{i} + \Psi^{(\pm)}~,
\ee
with
\bea
\label{miiaalg3pm}
\Psi^{(\pm)} \equiv \Gamma^{i}\Psi^{(\pm)}_{i} &=& \mp\frac{1}{4}h_{i}\Gamma^{i}
\mp\frac{1}{4}e^{\Phi}X_{i j}\Gamma^{i j} + e^{\Phi}m
\cr
&+& \Gamma_{11}\bigg(\pm \frac{1}{4}L_{i}\Gamma^{i} - \frac{1}{8}{\tilde{H}}_{i j k}\Gamma^{i j k} \mp e^{\Phi}S + \frac{1}{4}e^{\Phi}{\tilde{F}}_{i j}\Gamma^{i j} \bigg)~,
\ee
are the horizon Dirac operators associated with the supercovariant derivatives $\nabla^{(\pm)}$.

\subsection{Lichnerowicz type theorems for ${\mathscr D}^{(\pm)}$ }
\label{lichthx}

\vskip 0.3cm
{\it Theorem:}
Let ${\cal S}$ and the fields satisfy the conditions for the maximum principle to apply, e.g.~the fields are smooth and ${\cal S}$ is
compact without boundary. Then there is a 1-1 correspondence between the zero modes of ${\mathscr D}^{(+)}$ and the $\eta_+$ Killing spinors, i.e.
\bea
\nabla_{i}^{(+)}\eta_+=0~,~~~{\cal A}^{(+)}\eta_+=0~\Longleftrightarrow~ {\mathscr D}^{(+)}\eta_+=0~.
\ee
Moreover $\parallel\eta_+\parallel^2$ is constant.
\vskip 0.3cm
{\it Proof:} It is evident that if $\eta_+$ is a Killing spinor, then it is a zero mode of ${\mathscr D}^{(+)}$. To prove the converse, assuming that
$\eta_+$ is a zero mode of ${\mathscr D}^{(+)}$ and after using the field equations and Bianchi identities, one can establish the identity, see Appendix D,
\bea
{\tilde{\nabla}}^{i}{\tilde{\nabla}}_{i}\parallel\eta_+\parallel^2 - (2\tilde{\nabla}^i \Phi +  h^i) {\tilde{\nabla}}_{i}\parallel\eta_+\parallel^2 = 2\parallel{\hat\nabla^{(+)}}\eta_{+}\parallel^2 + (-4\kappa - 16 \kappa^2)\parallel\mathcal{A}^{(+)}\eta_+\parallel^2~,
\label{miiamaxprin}
\ee
where
\bea
\label{miiaredef1}
\hat{\nabla}_{i}^{(\pm)}=\nabla_{i}^{(\pm)}+ \kappa \Gamma_i {\cal A}^{(\pm)}~,
\ee
for some $\kappa\in \bR$. Provided that $\kappa$ is chosen in the interval $(-{1\over4}, 0)$, the theorem follows as an application of the maximum
principle. 

Let us turn to investigate the relation between Killing spinors and the zero modes of the ${\mathscr D}^{(-)}$ operator.
\vskip 0.3cm
{\it Theorem:} Let ${\cal S}$ be compact without boundary and the horizon fields be smooth. There is a 1-1 correspondence between the
zero modes of ${\mathscr D}^{(-)}$ and the $\eta_-$ Killing spinors, i.e.
\bea
\nabla_{i}^{(-)}\eta_-=0~,~~~{\cal A}^{(-)}\eta_-=0~\Longleftrightarrow~ {\mathscr D}^{(-)}\eta_-=0~ \ .
\ee
\vskip 0.3cm
{\it Proof:} It is clear that if $\eta_-$ is a Killing spinor, then it is a zero mode of ${\mathscr D}^{(-)}$.  To prove the converse, if $\eta_-$
is a zero mode of ${\mathscr D}^{(-)}$, then upon using the field equations and Bianchi identities one can establish the formula, see Appendix D,
\bea
\label{miial2b}
{\tilde{\nabla}}^{i} \big( e^{-2 \Phi} V_i \big)
= -2 e^{-2 \Phi} \parallel{\hat\nabla^{(-)}}\eta_{-}\parallel^2 +   e^{-2 \Phi} (4 \kappa +16 \kappa^2) \parallel\mathcal{A}^{(-)}\eta_-\parallel^2~,
\ee
where $V=-d \parallel \eta_- \parallel^2 - \parallel \eta_- \parallel^2 h $. The theorem follows after integrating the above formula over ${\cal S}$ using Stokes' theorem
for  $\kappa\in (-{1\over4}, 0)$.
\subsection{Index theory and supersymmetry enhancement}
To prove the first part of the conjecture, we shall establish the theorem:

\vskip 0.3cm
{\it Theorem:} The number of supersymmetries preserved by massive IIA horizons is even.
\vskip 0.3cm

{\it Proof:} Let $N_\pm$ be the number of $\eta_\pm$ Killing spinors. As a consequence of the two theorems we have established in the previous section
$N_\pm=\mathrm{dim}\,\mathrm{Ker}\, {\mathscr D}^{(\pm)}$. The $Spin(9,1)$ bundle over the spacetime decomposes as  $S_+\oplus S_-$
upon restriction to ${\cal S}$. Furthermore $S_+$ and $S_-$ are isomorphic as $Spin(8)$ bundles as both are associated with the
Majorana representation.  The action of  ${\mathscr D}^{(+)}: \Gamma(S_+)\rightarrow \Gamma(S_+)$ on the section $ \Gamma(S_+)$  of $S_+$
is not  chirality preserving.  Since the principal symbol of ${\mathscr D}^{(+)}$ is the same as the principal symbol
of the standard Dirac operator acting on Majorana but not-Weyl spinors, the index vanishes \cite{atiyah1}.  Therefore
\bea
N_+=\mathrm{dim}\,\mathrm{Ker}\, {\mathscr D}^{(+)}= \mathrm{dim}\,\mathrm{Ker}\, ({\mathscr D}^{(+)})^\dagger~,
\ee
where $({\mathscr D}^{(+)})^\dagger$ is the adjoint of ${\mathscr D}^{(+)}$.  On the other hand, one can establish
\bea
\big(e^{2 \Phi} \Gamma_-\big) \big({\mathscr D}^{(+)}\big)^\dagger
= {\mathscr D}^{(-)} \big(e^{2 \Phi} \Gamma_-\big)~,
\ee
and so
\bea
N_-=\mathrm{dim}\,\mathrm{Ker}\, ({\mathscr D}^{(-)})=\mathrm{dim}\,\mathrm{Ker}\, ({\mathscr D}^{(+)})^\dagger~.
\ee
Therefore, we conclude that $N_+=N_-$ and so the number of supersymmetries of massive IIA horizons $N=N_++N_-=2 N_-$ is even.
\rightline{ $\square$}
\section{The $\mathfrak{sl}(2,\bR)$ symmetry of massive IIA horizons}
\subsection{ $\eta_+$ from $\eta_{-}$ Killing spinors}

We shall demonstrate the existence of the $\mathfrak{sl}(2,\bR)$ symmetry of massive IIA horizons by  directly constructing
the vector fields on the spacetime generated by the action of $\mathfrak{sl}(2,\bR)$. In turn the existence of such vector fields
is a consequence of the property that massive IIA horizons admit an even number of supersymmetries. We have seen that if $\eta_-$ is a Killing spinor, then  $\eta_+=\Gamma_+\Theta_-\eta_-$
is also a Killing spinor provided that $\eta_+\not=0$. It turns out that under certain conditions this is always possible.
\vskip 0.3cm
{\it Lemma:} Suppose that ${\cal S}$ and the fields satisfy the requirements for the maximum principle to apply. Then
\bea
\mathrm{Ker}\, \Theta_-=\{0\}~.
\label{miiakerz}
\ee
\vskip 0.3cm
{\it Proof:}
We shall prove this by contradiction. Assume that $\Theta_-$ has a non-trivial kernel, so there is $\eta_-\not=0$
such that $\Theta_- \eta_-=0$. In such a case,
  ({\ref{miiaint3}}) gives
$ \Delta \langle \eta_- , \eta_- \rangle =0$. Thus $\Delta =0$,
as $\eta_-$ is no-where vanishing.
Next the gravitino KSE $\nabla^{(-)}\eta_-=0$ together with $\langle \eta_{-}, \Gamma_{i}\Theta_{-}\eta_{-} \rangle=0$ imply that
\bea
\label{miianrm1a}
{\tilde{\nabla}}_i \parallel \eta_-\parallel^2 = - h_i  \parallel \eta_-\parallel^2~.
\ee
On taking the divergence of this expression, eliminating  ${\tilde{\nabla}}^i h_i$ upon using ({\ref{miiafeq7}}),
and after setting $\Delta=0$, one finds
\bea
\label{miianrm1ab}
{\tilde{\nabla}}^i {\tilde{\nabla}}_i  \parallel \eta_-\parallel^2 &=& 2\tilde{\nabla}^{i}\Phi {\tilde{\nabla}}_i \parallel \eta_-\parallel^2
+ \bigg(L^2+ \frac{1}{2}e^{2\Phi}S^2 + \frac{1}{4}e^{2\Phi}X^2 + \frac{1}{4}e^{2\Phi}\tilde{F}^2 
\cr
&+& \frac{1}{48}e^{2\Phi}\tilde{G}^2 + \frac{1}{2}e^{2\Phi}m^2\bigg)  \parallel \eta_-\parallel^2~.
\ee
The maximum principle implies that $\parallel \eta_- \parallel^2$ is constant. However, the remainder of
({\ref{miianrm1ab}}) can never vanish, due to the quadratic term in $m$. So there can be no solutions, with $m \neq 0$, such
that $\eta_- \neq 0$ is in the Kernel of $\Theta_-$, and so $\mathrm{Ker}\, \Theta_-=\{0\}$.
\subsection{Killing vectors}

\subsection{$\mathfrak{sl}(2,\bR)$ symmetry}
Using $\eta_-$ and $\eta_+=\Gamma_+\Theta_-\eta_-$ and the formula (\ref{miialightconesol}),
one can construct two linearly independent Killing spinors on the  spacetime as
\bea
\epsilon_1=\eta_-+u\eta_++ru \Gamma_-\Theta_+\eta_+~,~~~\epsilon_2=\eta_++r\Gamma_-\Theta_+\eta_+~.
\ee
It is known from the general theory of supersymmetric massive IIA backgrounds that for any Killing spinors $\zeta_1$ and $\zeta_2$ the dual vector field $K(\zeta_1, \zeta_2)$ of the 1-form
bilinear
\bea
\omega(\zeta_1, \zeta_2)=\langle(\Gamma_+-\Gamma_-) \zeta_1, \Gamma_a\zeta_2\rangle\, e^a~,
\label{miia1formbi}
\ee
is a Killing vector and leaves invariant all the other fields of the theory.
Evaluating, the vector field bilinears of the Killing spinors $\epsilon_1$ and $\epsilon_2$, we find that
\bea
K_1(\epsilon_1, \epsilon_2)&=&-2u \parallel\eta_+\parallel^2 \partial_u+ 2r \parallel\eta_+\parallel^2 \partial_r+ \tilde V~,
\cr
K_2(\epsilon_2, \epsilon_2)&=&-2 \parallel\eta_+\parallel^2 \partial_u~,
\cr
K_3(\epsilon_1, \epsilon_1)&=&-2u^2 \parallel\eta_+\parallel^2 \partial_u +(2 \parallel\eta_-\parallel^2+ 4ru \parallel\eta_+\parallel^2)\partial_r+ 2u \tilde V~,
\label{miiakkk}
\ee
where we have set
\bea
\label{miiavii}
\tilde V =  \langle \Gamma_+ \eta_- , \Gamma^i \eta_+ \rangle\, \tilde \partial_i~,
\ee
is a vector field on ${\cal S}$.
To derive the above expressions for the Killing vector fields, we have used the identities
\bea
- \Delta\, \parallel\eta_+\parallel^2 +4  \parallel\Theta_+ \eta_+\parallel^2 =0~,~~~\langle \eta_+ , \Gamma_i \Theta_+ \eta_+ \rangle  =0~,
\ee
which follow from the first integrability condition in ({\ref{miiaint1}}),  $\parallel\eta_+\parallel=\mathrm{const}$ and the KSEs of $\eta_+$.
\vskip 0.3cm
{\it Theorem:} The Lie bracket algebra of  $K_1$, $K_2$ and $K_3$  is $\mathfrak{sl}(2,\bR)$.
\vskip 0.3cm
{\it Proof:} Using the identities summarised in Appendix D, one can demonstrate after a direct computation that
\bea
[K_1,K_2]=2 \parallel\eta_+\parallel^2 K_2,~[K_2, K_3]=-4 \parallel\eta_+\parallel^2 K_1,~[K_3,K_1]=2 \parallel\eta_+\parallel^2 K_3~. 
\ee
This proves the theorem and the last part of the {\it horizon conjecture}.
\rightline{ $\square$}

\section{The geometry and isometries of ${\cal S}$}
It is known that the vector fields associated with the 1-form Killing spinor bilinears given in (\ref{miia1formbi}) leave invariant all the fields of
massive IIA supergravity.  In particular for massive IIA horizons we have that ${\cal L}_{K_a} g=0$ and ${\cal L}_{K_a} F=0$, $a=1,2,3$, where $F$ denotes collectively all the
fluxes of massive IIA supergravity, where $K_a$ are given in (\ref{miiakkk}).  Solving these conditions by expanding in $u,r$, one finds that
\bea
\tilde\nabla_{(i} \tilde V_{j)}=0~,~~~\tilde {\cal L}_{\tilde V} h=\tilde {\cal L}_{\tilde V}\Delta=0~,~~~ \tilde {\cal L}_{\tilde V} \Phi=0~,
\nonumber \\
\tilde {\cal L}_{\tilde V} X=\tilde {\cal L}_{\tilde V} \tilde G=\tilde {\cal L}_{\tilde V} L=\tilde {\cal L}_{\tilde V} \tilde H=
\tilde {\cal L}_{\tilde V} S=\tilde {\cal L}_{\tilde V} \tilde F=0~.
\ee
Therefore $V$ is an isometry of ${\cal S}$ and leaves all the fluxes on ${\cal S}$ invariant.
Furthermore, one can establish the identities
\bea
&&-2 \parallel\eta_+\parallel^2-h_i \tilde V^i+2 \langle\Gamma_+\eta_-, \Theta_+\eta_+\rangle=0~,~~~i_{\tilde V} (dh)+2 d \langle\Gamma_+\eta_-, \Theta_+\eta_+\rangle=0~,
\cr
&& 2 \langle\Gamma_+\eta_-, \Theta_+\eta_+\rangle-\Delta \parallel\eta_-\parallel^2=0~,~~~
{\tilde V}+ \parallel\eta_-\parallel^2 h+d \parallel\eta_-\parallel^2=0~,
\label{miiaconconx}
\ee
which imply that ${\cal L}_{\tilde V}\parallel\eta_-\parallel^2=0$. These conditions are similar to those established for M-theory and IIA theory horizons
in \cite{11index} and \cite{iiaindex}, respectively,  but of course the dependence of the various tensors on the fields is different. In the special case that $\tilde V=0$, the
horizons are warped products of $AdS_2$ with ${\cal S}$.

	\chapter{$D=5$ Supergravity Coupled to  Vector Multiplets}

In this section, we will establish the horizon conjecture for black
holes in five dimensional gauged, and ungauged, supergravity coupled
to an arbitrary number of vector multiplets. 

Five dimensional supergravity is interesting from several points of view.
It can be constructed by compactifying the eleven dimensional 
supergravity, on some six dimensional manifolds e.g. $CY_3$ or $T^6$.
 There are several interesting supersymmetric
 solutions for the ungauged five dimensional supergravity
 which preserve half of the supersymmetry. In particular, there are
various types of half-BPS black objects corresponding to
 black holes, black strings, black rings, black lens and solutions with a topologically non-trivial $S^2$-cycle outside the horizon \cite{BMPV1} \cite{gauntlett} \cite{BR1} \cite{blacklens} \cite{lucietti} \cite{Horowitz:2017fyg}. Most recently all the possible solutions of the minimal theory with a $U(1)^2$ isometry, generated by two commuting rotational Killing fields have been classified \cite{u1proof}, of which there are infinitely many solutions.

The black hole near-horizon geometry is the maximally
supersymmetric near-horizon BMPV solution, which in the
case of static black holes is simply AdS$_2\times S^3$.
The near-horizon geometry of both the black string and the black ring
is the maximally supersymmetric AdS$_3\times S^2$ solution.
Each of these solutions has a specific charge configuration.
A black hole has only electric charges, a black string has only
magnetic charges, while a black ring has both electric and magnetic
charges. The black ring solution can be uplifted to $D=11$ supergravity.
 From this point of view, the electric and
magnetic  charges correspond to $M2$ and
 $M5$-branes respectively wrapping nontrivial cycles of the internal space. The black ring is the first example of a
 black object with a compact horizon spatial section, which has non-spherical horizon topology. It is asymptotically flat and carries  angular momentum.
 Furthermore, the existence of
 this solution implies that the black hole uniqueness theorems can not
 be extended to five dimensions, except in the static case \cite{gibbons}.
 
 The analysis for ungauged $D=5$ supersymmetric horizons determines the near-horizon geometry to be either that of the BMPV solution, with either a squashed
or round $S^3$ horizon spatial cross-section \cite{BMPV1, BMPV2, BMPV3};
or $AdS_3 \times S^2$, which is near-horizon geometry of the black ring (also
the black string), with $S^1 \times S^2$ horizon spatial cross section \cite{Chamseddine:1999qs, BR1, BS1}; or ${\mathbb{R}}^{4,1}$ with
spatial horizon cross-section $T^3$.

 Previous work has also been done on the classification of near-horizon geometries for
five dimensional ungauged supergravity in \cite{gutbh, reall}. However there an additional assumption was made on assuming the vector bilinear matching condition i.e the black hole Killing horizon associated with a Killing vector field is identified as a Killing spinor bilinear. We do not make this assumption here, and we prove the results on (super)symmetry enhancement in full generality. The only assumptions we make are that all the fields are smooth (or at least $C^2$ differentiable) and the spatial horizon section ${\cal S}$ is compact, connected and without boundary. These assumptions are made in order that  various global techniques can be applied to the analysis.

\section{Near-horizon geometry of the BMPV black holes and black rings}

Before proceeding with the analysis
of the Killing spinor equations, we shall briefly summarize properties of the two key solutions
which arise in the minimal ungauged theory, which are the BMPV black hole \cite{BMPV1}, \cite{gauntlett} and the supersymmetric black ring \cite{BR1}. The general form for the black hole and black ring solutions is given by the following metric
\bea
\label{5dstring} 
ds^2 = -f^2 (dt + \omega)^2 + f^{-1}ds^2({\cal M}_4) \ .
\ee
Here $V=\partial_t$ is a timelike Killing vector which can be constructed as a bilinear of the preserved supersymmetry parameter, and ${\cal M}_4$
 is a hyper-K\"ahler space and $f$ and $\omega$ are a scalar and one-form on ${\cal M}_4$, which satisfy,
\bea
dG^{+} = 0,~~ \Delta f^{-1} = \frac{4}{9}(G^+)^2 \ ,
\ee
where $G^+ = \frac{1}{2} f (d\omega + \star d\omega)$ with $\star$ the Hodge dual on ${\cal M}_4$ and $\Delta$ is the Laplacian on ${\cal M}_4$. The two-form field strength is given by,
\bea
F = \frac{\sqrt{3}}{2}d[f(dt + \omega)] - \frac{1}{\sqrt{3}}G^{+} \ .
\ee
We will only be interested in the case when ${\cal M}_4$ is flat space, $\bb{R}^4$, and it will be useful to use the following coordinates.
\bea
ds^2(\mathbb{R}^4) &=& H[dx^i dx^i] + H^{-1}(d\psi + \chi_i dx^i )^2
\nonumber \\
&=& H[dr^2 + r^2(d\theta^2 + \sin^2 (\theta) d\phi^2)] + H^{-1}(d\psi + \cos(\theta) d\phi)^2 \ ,
\ee
with $H=1/r$. We will also demand that the tri-holomorphic vector field $\partial_{\psi}$ is a Killing vector of the five-dimensional metric ({}\ref{5dstring}}). This will allow us the express the most general solution in terms of harmonic functions $K, L$ and $M$ on $\mathbb{R}^3$ as,
\bea
f^{-1} = H^{-1}K^2 + L \ ,
\ee
and
\bea
\omega = \bigg(H^{-2}K^3 + \frac{3}{2}H^{-1}K L + M\bigg)(d\psi + \cos \theta d \phi) + \hat{\omega} \ ,
\ee
where $\hat{\omega}$ is a 1-form on $\mathbb{R}^3$ which satisfies,
\bea
\label{curlw}
\nabla \times \hat{\omega} = H\nabla M - M \nabla H + \frac{3}{2}(K\nabla L - L\nabla K) \ .
\ee
For a single black-ring solution, by writing $f$ and $\omega$ in Gibbons-Hawking coordinates $(r,\theta,\psi, \phi)$ we can determine the harmonic functions $K, L$ and $M$. They can be expressed in terms of a single harmonic function $h_1$ given by,
\bea
h_1 = \frac{1}{|\mathbf{x} - \mathbf{x_1}|} \ ,
\ee
with a single centre on the negative $z$-axis given by $\mathbf{x_1} = (0,0,-R^2/4)$. In particular we have,
\bea
K &=& -\frac{q}{2}h_1 \ ,
\cr
L &=& 1 + \frac{Q - q^2}{4}h_1 \ ,
\cr
M &=& \frac{3q}{4} - \frac{3qR^2}{16} h_1 \ ,
\ee
with $\mathbf{x} = \mathbf{x_1}$ a coordinate singularity corresponding to the event horizon of the black ring with topology $S^1 \times S^2$. The radius of the $S^2$ is $q/2$ and that of the $S^1$ is $l$ defined by,
\bea
l \equiv \sqrt{3\bigg[\frac{(Q-q^2)^2}{4q^2} - R^2\bigg]} \ ,
\ee
where we demand that $l > 0$ to ensure that the solution does not contain closed time-like curves. We can also write ({\ref{curlw}}) explicitly in terms of $h_1$ as,
\bea
\label{curlw2}
\nabla \times \hat{\omega} = \frac{3}{4} q \bigg(\nabla h_1 - \nabla\bigg(\frac{1}{r}\bigg) \bigg) - \frac{3q}{16}R^2 \bigg(\frac{1}{r}\nabla h_1 - h_1 \nabla \bigg(\frac{1}{r}\bigg) \bigg) \ .
\ee
 The ADM charges of this solution are given by
 \bea
 M_{ADM} &=&\frac{3\pi}{4G_5}Q \ ,
 \nonumber \\
 J_1 &=& \frac{\pi}{8G_5}q[6R^2+3Q-q^2] \ ,
 \nonumber \\
 J_2 &=& \frac{\pi}{8G_5}q(3Q-q^2) \ .
 \label{J}
 \ee
where we write the metric on ${\mathbb{R}}^4$ as two copies of the metric on ${\mathbb{R}}^2$;
\bea
{1 \over r} \big(dr^2+r^2 d \theta^2+r^2\sin^2 \theta d \phi^2 \big)
+r\big(d \psi + \cos \theta d \phi \big)^2&=&dr_1^2+r_1^2 d \phi_1^2
\nonumber \\
&+&dr_2^2+r_2^2 d \phi_2^2 \ ,
\ee
and the angular momenta $J_1$ and $J_2$ are the asymptotic charges associated with the isometries
${\partial \over \partial \phi_1}$ and ${\partial \over \partial \phi_2}$. 
\vspace{2mm}
\begin{eg}
For the BMPV black hole we have $R=0$ with $h_1 = 1/r$ and the solution can be written explicitly as,
\bea 
f^{-1} =  1 + \frac{Q}{4r} \ ,
\ee
and from ({\ref{curlw2}}) we also have $\hat{\omega} = 0$ and,
\bea
\omega = \frac{q}{16r}(q^2 - 3Q)(d\psi + \cos \theta d\phi) \ .
\ee
The angular momentum also coincide $J_1 = J_2 = J$ and the ADM charges are given by,
\bea
M_{ADM} &=&\frac{3\pi}{4G_5}Q,
\nonumber \\
J &=& \frac{\pi}{8G_5}q[3Q-q^2] \ .
\label{J2}
\ee
The metric is explicitly given as,
\bea
ds^2 &=& -r\bigg(r + \frac{Q}{4}\bigg)^{-2} \bigg(dt + \frac{q}{16r}(q^2 - 3Q)(d\psi + \cos \theta d\phi)\bigg)^2 
\cr
&+& \bigg(r + \frac{Q}{4}\bigg)\bigg({1 \over r^2}dr^2+ d \theta^2+\sin^2 \theta d \phi^2
+\big(d \psi + \cos \theta d \phi \big)^2\bigg) \ .
\ee
This be written in Gaussian null coordinates under the transformation $(t, r, \theta, \phi, \psi) \rightarrow (u, r, \theta, \phi, \tilde{\psi})$ with $t \rightarrow u + \lambda_1(r),~ \psi \rightarrow \tilde{\psi} + \lambda_2(r)$ and taking
\bea
\lambda_1(r) &=& \frac{1}{16}\int p(r) r^{-2} dr
\cr
\lambda_2(r) &=& -q(3Q -q^2)\int p(r)^{-1} r^{-1} dr \ ,
\ee
with
\bea
p(r) = (-q^6 + 6Q q^4 - 9Q^2 q^2 + 4Q^3 + 48Q^2 r + 192 Q r^2 + 256 r^3)^{\tfrac{1}{2}} \ .
\ee
The geometry of the spatial horizon cross section is that of a squashed $S^3$. Now we take the NHL with
\bea
r\rightarrow \epsilon r,~ u\rightarrow \epsilon^{-1} u,~ y^I \rightarrow y^I,~ \epsilon \rightarrow 0 \ ,
\ee
and making the further change with,
\bea
r \rightarrow -\bigg(\frac{(Q-q^2)\sqrt{(4Q-q^2)}}{4Q}\bigg)r \ .
\ee
The near-horizon metric can be written in GNC as,
\begin{eqnarray}
ds^2 &=& 2(dr + r h_{2} d\phi + r h_{3} d\psi - \frac{1}{2}r^2 \Delta du) du 
+ \gamma_{1 1}d\theta^2 + \gamma_{2 2}d\phi^2 + \gamma_{3 3}d\psi^2 + 2\gamma_{2 3}d\phi d\psi \ ,
\nonumber \\
\end{eqnarray}
with the near-horizon data is given by,
\bea
\Delta &=& \frac{(Q-q^2)^2(4Q-q^2)}{Q^4}
\cr
h_{2} &=& -\frac{q(3Q-q^2)(Q-q^2)\sqrt{(4Q-q^2)}}{4Q^3}
\cr
h_{3} &=& -\frac{q\cos \theta (3Q-q^2)(Q-q^2)\sqrt{(4Q-q^2)}}{4Q^3}
\cr
\gamma_{1 1} &=& \frac{Q}{4}
\cr
\gamma_{2 2} &=& \frac{(Q-q^2)^2(4Q-q^2)}{16Q^2}
\cr
\gamma_{2 3} &=& \frac{\cos \theta(Q-q^2)^2(4Q-q^2)}{16Q^2}
\cr
\gamma_{3 3} &=& \frac{4Q^3 - q^2 \cos^2 \theta  (3Q-q^2)^2}{16Q^2} \ ,
\ee
which is known as the BMPV near-horizon geometry. The Killing vectors $\{K_1, \dots, K_7\}$ of the near-horizon metric are given by,
\bea
K_1 &=& \partial_{u},~K_{2} = -u\partial_{u} + r\partial_{r}+\bigg(\frac{q(3Q-q^2)}{(Q-q^2)\sqrt{(4Q-q^2)}}\bigg) \partial_\phi ,~ 
\nonumber \\
K_3 &=& -\frac{u^2}{2}\partial_u +\bigg(\frac{Q}{4}+ur\bigg)\partial_r + \bigg(\frac{q(3Q-q^2)u}{(Q-q^2)\sqrt{(4Q-q^2)}}\bigg)\partial_\phi \ ,
\nonumber \\
K_4 &=& \partial_{\phi},~K_5 = \partial_{\psi} \ ,
\nonumber \\
K_6 &=& 
\sin \psi \partial_\theta - \cos \psi (\sin \theta)^{-1} \partial_\phi + \cos \psi \cot \theta \partial_\psi \ ,
\nonumber \\
K_7 &=& 
\cos \psi \partial_\theta + \sin \psi (\sin \theta)^{-1} \partial_\phi - \sin \psi \cot \theta \partial_\psi \ ,
\ee
with the Lie algebra ${\mathfrak{u}}(1) \times {\mathfrak{sl}}(2,\mathbb{R}) \times {\mathfrak{so}}(3)$,
\bea
[K_1,K_2] &=& -K_1,~[K_1,K_3]=K_2,~ [K_2,K_3] = -K_3,
\cr
[K_5,K_6]&=&K_7,~ [K_5,K_7]=-K_6,~[K_6,K_7]=K_5
\ee
The isometries $\{ K_1, K_2, K_3 \}$ generate a ${\mathfrak{sl}}(2,\mathbb{R})$
algebra, whereas the isometries $\{ K_5, K_6, K_7 \}$ generate a ${\mathfrak{so}}(3)$
algebra.
\end{eg}
\vspace{2mm}
\begin{eg}
Now we will consider the black ring solution with $R \neq 0$, here we will state the result given in \cite{BR1}. We first make the coordinate transformation,
\bea
dt = du - B(r)dr,~~~ d\phi = d\phi' - C(r)dr,~~~
d\psi = d\psi' - C(r)dr \ ,
\ee
where
\bea
B(r) &=& \frac{B_2}{r^2} + \frac{B_1}{r} + B_0 \ ,
\cr
C(r) &=& \frac{C_1}{r} + C_0 \ .
\ee
The constants $B_i$ and $C_i$ are chosen so that all metric components remain finite as $r \rightarrow 0$. We choose,
\bea
B_0 &=& \frac{q^2 l}{8 R^3} + \frac{2l}{3R} - \frac{R}{2 l} + \frac{3R^3}{2l^3} + 3\frac{(Q-q^2)^3}{16q^2 R l^3} \ ,
\cr
B_1 &=&\frac{(Q+2q^2)}{4l} + \frac{l(Q-q^2)}{3R^2} \ ,
\cr
B_2 &=& \frac{q^2 l}{4R} \ ,
\ee
and
\bea
C_0 &=& -\frac{(Q-q^2)^3}{8 q^3 r l^3} \ ,
\cr
C_1 &=& -\frac{q}{2l} \ .
\ee
The metric becomes
\bea
ds^2 &=& -\frac{16r^4}{q^4}du^2 + \frac{2R}{l}du dr + \frac{4r^3 \sin^2 \theta}{Rq} du d\phi'+ \frac{4R r}{q}du d\psi' + \frac{3q r \sin^2 \theta}{l}dr d\phi'
\cr
&+& 2 \bigg(\frac{q l}{2R}\cos \theta + \frac{3qR}{2l} + \frac{(Q-q^2)(3R^2 - 2l^2)}{3q R l}\bigg)dr d\psi' 
\nonumber \\
&+& l^2d\psi'^2 + \frac{q^2}{4}[d\theta^2 + \sin^2 \theta (d\phi' - d\psi')^2] + ... \ ,
\ee
where the dots are terms which involve subleading powers of $r$ in all the metric components given explicitly, as well as terms in $g_{r r}$ starting at $\mathcal{O}(r)$. The spatial horizon cross section clearly has the geometry $S^1 \times S^2$. Now we take the NHL with,
\bea
r \rightarrow \bigg(\frac{l}{R}\bigg)\epsilon r,~ u\rightarrow \epsilon^{-1} u,~ \epsilon \rightarrow 0 \ .
\ee
The near-horizon metric can be written as,
\begin{eqnarray}
ds^2 = 2(dr + r h_{3} d\psi) du + \gamma_{1 1}d\theta^2 + \gamma_{2 2}d\chi^2 + \gamma_{3 3}d\psi^2
\ ,
\end{eqnarray}
where $\chi\equiv \phi' - \psi' = \phi - \psi$ and the near-horizon data,
\bea
h_3 = \frac{2l}{q},~ \gamma_{1 1} = \frac{q^2}{4}, \gamma_{2 2} = \frac{q^2}{4}\sin^2 \theta,~ \gamma_{3 3} = l^2 \ .
\ee
This metric is that of $AdS_3 \times S^2$, where the co-ordinates
$\{ r, u,  \psi \}$ parametrize the $AdS_3$, and $\{ \theta, \chi \}$ are
spherical polar co-ordinates on $S^2$. The Killing vectors $\{K_1, \dots, K_9\}$ of the near-horizon metric are given by,
\bea
K_1 &=& \partial_{u},~K_{2} = -u\partial_{u} + r\partial_{r} -{q \over 2 l} \partial_\psi ,~ 
\nonumber \\
K_3 &=& -\frac{u^2}{2}\partial_u +\bigg(\frac{q^2}{4}+ur\bigg)\partial_r - \frac{qu}{2l}\partial_\psi \ ,
\nonumber \\
K_4 &=& e^{\frac{-2 l \psi}{q}}\partial_r \ ,
\nonumber \\ 
K_5 &=& e^{\frac{2 l \psi}{q}}\bigg(\frac{q^2}{4}\partial_u + \frac{1}{2}r^2 \partial_r - \frac{q r}{2l}\partial_\psi \bigg) \ ,
\nonumber \\
K_6 &=& \frac{q}{2l}\partial_{\psi},~ K_7 = \partial_{\chi} \ ,
\nonumber \\
K_8 &=&\sin{\chi}\partial_{\theta} + \cos{\chi}\cot{\theta}\partial_{\chi} \ ,
\nonumber \\
K_9 &=& \cos{\chi}\partial_{\theta} - \sin{\chi}\cot{\theta}\partial_{\chi} \ ,
\ee
Since $\psi$ is periodic and $K_4, K_5$ are not globally defined, the Lie algebra is ${\mathfrak{u}}(1) \times {\mathfrak{sl}}(2,\mathbb{R}) \times {\mathfrak{so}}(3)$,
\bea
[K_1, K_2] &=& -K_1,~ [K_1, K_3] = K_2,~ [K_2, K_3] = -K_3 \ ,
\cr
[K_7, K_8] &=& K_9,~ [K_7, K_9] = -K_8,~ [K_8, K_9] = K_7 \ ,
\ee
We obtain a ${\mathfrak{sl}}(2,\mathbb{R})$ algebra generated by
the isometries $\{ K_1, K_2, K_3 \}$ and the isometry $K_6$ generates the residual $\mathfrak{u}(1)$ broken from ${\mathfrak{sl}}(2,\mathbb{R})$ due to the periodicity of $\psi$. The isometries $\{ K_7, K_8, K_9 \}$ generate an ${\mathfrak{so}}(3)$ algebra.
\end{eg}
The BMPV black hole and the supersymmetric black ring preserve 4 out of 8 preserved real supersymmetries, but their near-horizon geometries are maximally supersymmetric. This is due to the mechanism of supersymmetry enhancement which occurs on the horizon and it is known that all $D = 5$ supergravity black holes undergo supersymmetry enhancement in the near-horizon limit from the classification of \cite{reall}.

\section{Horizon fields and KSEs}

We shall consider the horizon conjecture in the context of both gauged and ungauged
5-dimensional supergravities, coupled to arbitrary many vector multiplets. 
These theories are summarized in Chapter 3.  We next proceed to
consider how the bosonic fields and their associated Bianchi identities and field equations are written in the near-horizon limit,
prior to solving the Killing spinor equations.
\vspace{2mm}
\begin{eg}
In the case of the $STU$ model, which has $C_{123}=1$, and $X^1 X^2 X^3=1$, the non-vanishing components of
the gauge coupling are given by
\begin{eqnarray}
Q_{11}={1 \over 2(X^1)^2}, \quad Q_{22}={1 \over 2(X^2)^2}, 
\quad Q_{33}={1 \over 2(X^3)^2} \ ,
\end{eqnarray}
with scalar potential
\begin{eqnarray}
U= 18 \bigg({V_1 V_2 \over X^3}+{V_1 V_3 \over X^2}+{V_2 V_3 \over X^1}\bigg) \ .
\end{eqnarray}
\end{eg}
When considering near-horizon solutions for the gauged theory, conditions which are sufficient
to ensure that $U \geq 0$ are
that $V_I \geq 0$ for $I=1,2,3$, and also that there exists a point on
the horizon section at which $X^I >0$ for $I=1,2,3$.\footnote{As we shall assume
that the scalars are smooth functions on (and outside of) the horizon, this implies that $X^I>0$ everywhere on the horizon.}

\subsection{Near-horizon fields}

For $N=2$, $D=5$ supergravity, in addition to the metric, there are also gauge field strengths and
scalars. We will assume that these are also analytic in $r$ and regular at the horizon, 
and that there is also a consistent near-horizon limit for these matter fields:
\bea
A^{I} &=& -r \alpha^I \bbe^+ + {\tilde{A}}^{I}
\nonumber \\
F^I &=& \bbe^+ \wedge \bbe^- \alpha^I + r \bbe^+ \wedge \beta^I + {\tilde F}^I~,
\la{5dhormetr}
\ee
where $F^{I} = dA^{I}$ and we use the frame introduced in ({\ref{basis1}}).

We can also express the near-horizon fields $F$ and $G^I$ (\ref{decom}) in this frame as
\bea
\label{nhfs}
F &=& \bbe^+ \wedge \bbe^- \alpha + r \bbe^+ \wedge \beta + {\tilde F}
\nonumber \\
G^I &=& \bbe^+ \wedge \bbe^- L^I + r \bbe^+ \wedge M^I + {\tilde G}^I \ ,
\ee
where $X_I L^I = X_I M^I = X_I \tilde{G}^I = 0$ and we set $\alpha = X_I \alpha^I$, $\tilde{F} = X_I \tilde{F}^I$ and $\beta = X_I \beta^I$.

\subsection{Horizon Bianchi indentities and field equations}

Substituting the fields (\ref{5dhormetr}) into the the Bianchi identity $dF^I = 0$ implies
\bea
\label{beqo}
\beta^I = (d_h \alpha^I), ~~d\tilde{F}^I = 0 \ ,
\ee
and
\bea
\label{beqo2}
d\beta^{I} + \alpha^{I}dh + d\alpha^{I} \wedge h = 0 \ .
\ee
Note that (\ref{beqo2}) is implied (\ref{beqo}). Similarly, the independent field equations of the near-horizon fields are as follows. The Maxwell gauge equations (\ref{maxwell}) are given by,
\bea
d_h(Q_{I J}\star_3 \tilde{F}^J) - Q_{I J}\star_3\beta^J =  \frac{1}{2}C_{I J K}\alpha^J \tilde{F}^K \ .
\ee
In components this can be expressed as,
\bea
\label{5dfeq1o}
\tilde{\nabla}^j(Q_{I J}\tilde{F}^J{}_{j i}) - Q_{I J}h^j \tilde{F}^J{}_{j i} + Q_{I J}\beta^J{}_i + \frac{1}{4}C_{I J K}\epsilon_i{}^{\ell_1 \ell_2}\alpha^J \tilde{F}^K{}_{\ell_1 \ell_2} = 0 \ ,
\ee
which corresponds to the $i$-component of (\ref{maxeq}). There is another equation given by the $+$-component of (\ref{maxeq}) but this is implied by (\ref{5dfeq1o}) and is not used in the analysis at any stage.
The $+-$ and $ij$-component of the Einstein equation (\ref{eins}) gives
\bea
\label{5dfeq2o}
- \Delta - \frac{1}{2}h^2 + \frac{1}{2}\tilde{\nabla}^i(h_i) = -Q_{I J}\bigg(\frac{2}{3}\alpha^I \alpha^J + \frac{1}{6}\tilde{F}^I{}_{\ell_1 \ell_2}\tilde{F}^{J \ell_1 \ell_2}\bigg) - \frac{2}{3}\chi^2 U \ ,
\ee
and
\bea
\label{5dfeq3o}
\tilde{R}_{i j} &=& -\tilde{\nabla}_{(i}h_{j)} + \frac{1}{2}h_i h_j -\frac{2}{3}\chi^2 U \delta_{i j} 
\nonumber \\
&+& Q_{I J}\bigg[\tilde{F}^I{}_{i \ell}\tilde{F}^{J}{}_{j}{}^{\ell} + \tilde{\nabla}_i X^I \tilde{\nabla}_j X^J 
+ \delta_{i j}\bigg(\frac{1}{3}\alpha^I\alpha^J - \frac{1}{6}\tilde{F}^I{}_{\ell_1 \ell_2}\tilde{F}^{J \ell_1 \ell_2}\bigg) \bigg] \ .
\ee
The scalar field equation (\ref{scalareq1}) gives
\bea
\label{5dfeq4o}
&&\tilde{\nabla}^{i}\tilde{\nabla}_{i}{X_{I}} - h^i\tilde{\nabla}_i X_I + \tilde{\nabla}_{i}{X^{M}} \tilde{\nabla}^{i}{X^{N}} \left(\frac{1}{2}C_{M N K}X_{I} X^{K} - \frac{1}{6}C_{I M N}\right) 
\nonumber \\
&&+ \bigg[\frac{1}{2}\tilde{F}^{M}{}_{\ell_1 \ell_2} \tilde{F}^{N \ell_1 \ell_2} -\alpha^M \alpha^N \bigg]\bigg(C_{I N P} X_{M} X^{P} - \frac{1}{6}C_{I M N} - 6X_{I} X_{M} X_{N}+\frac{1}{6}C_{M N J} X_{I} X^{J}\bigg)  
\nonumber \\
&&+ 3 \chi^2 V_{M} V_{N}\bigg(\frac{1}{2}C_{I J K}Q^{M J}Q^{N K} + X_{I}(Q^{M N} - 2 X^{M}X^{N})\bigg) = 0 \ .
\ee
We remark that the $++$ and $+i$ components of the
Einstein equations, which are
\bea
\label{5dauxeq1o}
{1 \over 2} \tn^i \tn_i \Delta -{3 \over 2} h^i \tn_i \Delta-{1 \over 2} \Delta \tn^i h_i
+ \Delta h^2 +{1 \over 4} dh_{ij} dh^{ij} - Q_{I J}\beta^{I}{}_{\ell}\beta^{J \ell} = 0 \ ,
\ee
and
\bea
\label{5dauxeq2o}
{1 \over 2} \tn^j dh_{ij}-dh_{ij} h^j - \tn_i \Delta + \Delta h_i
+ Q_{I J}\alpha^{I}\beta^{J}{}_{i} - Q_{I J}\beta^{I}{}_{\ell}\tilde{F}^{J}{}_{i}{}^{\ell}  = 0 \ ,
\ee
are implied by ({\ref{5dfeq2o}}), ({\ref{5dfeq3o}}), ({\ref{5dfeq4o}}),
together with ({\ref{5dfeq1o}}).
and the Bianchi identities ({\ref{beqo}}).
\subsection{Gauge field decomposition}
Using the decomposition $F^I = FX^I + G^I$ with $F = X_I F^I$, $X_I G^I = 0$ and $dF^I = 0$ implies
\bea
dF &=& -X_I dG^I \ ,
\nonumber \\
(\delta^I{}_J - X^I X_J) dG^J &=& -dX^I \wedge F \ .
\ee
We write the near-horizon fields as
\bea
\label{decomp}
\alpha^I &=& \alpha X^I + L^I \ ,
\nonumber \\
\beta^I &=& \beta X^I + M^I \ ,
\nonumber \\
{\tilde F}^I &=& {\tilde F} X^I + {\tilde G}^I \ .
\ee
where $X_I L^I = X_I M^I = X_I \tilde{G}^I = 0$ and $\alpha = X_I \alpha^I$, $\tilde{F} = X_I \tilde{F}^I$, $\beta = X_I \beta^I$. By using (\ref{decomp}) we can express the Bianchi identities (\ref{beqo}) as
\bea
\label{beq}
\beta &=& d_h \alpha - L^I dX_I \ ,
\nonumber \\
d\tilde{F} &=& -X_I d\tilde{G}^I \ ,
\nonumber \\
(\delta^I{}_J - X^I X_J)(d_h L^{J} -M^{J}) &=& -dX^{I}\alpha \ ,
\nonumber \\
(\delta^I{}_J - X^I X_J) d\tilde{G}^J &=& -dX^I \wedge \tilde{F} \ ,
\ee
and corresponding to (\ref{beqo2})
\bea
\label{5dbeqo2}
d M^{I} - h \wedge M^{I} + L^{I} dh  + dX^{I} \wedge \beta &=& 0 \ ,
\nonumber \\
d \beta - h \wedge \beta + \alpha  dh  + dX_{I} \wedge M^{I} &=& 0 \ .
\ee
However, (\ref{5dbeqo2}) is implied by (\ref{beq}). The field equations can also be decomposed using (\ref{decomp}) as follows. The Maxwell gauge equation (\ref{5dfeq1o}) gives
\bea
\label{5dfeq1}
&&\frac{3}{2}X_I \tilde{\nabla}^j(\tilde{F}_{j i}) + \tilde{\nabla}^j(Q_{I J}\tilde{G}^J{}_{j i}) 
+ \frac{3}{2}\tilde{\nabla}^jX_I \tilde{F}_{j i} -\frac{3}{2}X_I h^j \tilde{F}_{j i} + \frac{3}{2}X_I \beta_i + Q_{I J}M^J{}_i  - Q_{I J}h^j \tilde{G}^J{}_{j i}
\cr
&&+ \frac{1}{4}\epsilon_i{}^{\ell_1 \ell_2}\bigg(6 X_I \alpha \tilde{F}_{\ell_1 \ell_2} - 2 Q_{I J}\alpha \tilde{G}^J{}_{\ell_1 \ell_2} - 2 Q_{I J}\tilde{F}_{\ell_1 \ell_2}L^J + C_{I J K}L^J \tilde{G}^K{}_{\ell_1 \ell_2}\bigg) = 0 \ ,
\ee
where we have used the identity $\tilde{\nabla}_i(Q_{I J})X^J = 3\tilde{\nabla}_i X_I$. By contracting with $X^I$ this gives,
\bea
\label{5dfeq2}
\tilde{\nabla}^j(\tilde{F}_{j i}) + \tilde{\nabla}^j(X_J) \tilde{G}^J{}_{j i} - h^j \tilde{F}_{j i} + \beta_i 
+ \epsilon_i{}^{\ell_1 \ell_2} \alpha \tilde{F}_{\ell_1 \ell_2} - \frac{1}{3}Q_{I J} \epsilon_i{}^{\ell_1 \ell_2}L^I\tilde{G}^J{}_{\ell_1 \ell_2} = 0 \ .
\ee
The Einstein equation (\ref{5dfeq2o}) gives
\bea
\label{5dfeq3}
- \Delta - \frac{1}{2}h^2 + \frac{1}{2}\tilde{\nabla}^i(h_i) = -\bigg[\alpha^2 + \frac{1}{4}\tilde{F}_{\ell_1 \ell_2}\tilde{F}^{\ell_1 \ell_2} + \frac{2}{3}\chi^2 U + Q_{I J}\bigg(\frac{2}{3}L^I L^J + \frac{1}{6}\tilde{G}^I{}_{\ell_1 \ell_2}\tilde{G}^{J \ell_1 \ell_2}\bigg)\bigg] \ ,
\nonumber \\
\ee
and (\ref{5dfeq3o})
\bea
\label{5dfeq4}
\tilde{R}_{i j} &=& -\tilde{\nabla}_{(i}h_{j)} + \frac{1}{2}h_i h_j + \frac{3}{2}\tilde{F}_{i k}\tilde{F}_j{}^k + \delta_{i j}\bigg(\frac{1}{2}\alpha^2 - \frac{1}{4}\tilde{F}_{\ell_1 \ell_2}\tilde{F}^{\ell_1 \ell_2} -\frac{2}{3}\chi^2 U \bigg)
\nonumber \\
&+& Q_{I J}\bigg[\tilde{G}^{I}{}_{i \ell}\tilde{G}^{J}{}_{j}{}^{\ell} + \tilde{\nabla}_{i}{X^I} \tilde{\nabla}_{j}{X^J} + \delta_{i j}\bigg(\frac{1}{3}L^I L^J - \frac{1}{6}\tilde{G}^I{}_{\ell_1 \ell_2}\tilde{G}^{J \ell_1 \ell_2}\bigg)\bigg] \ .
\ee
The scalar field equations (\ref{5dfeq4o}) give,
\bea
\label{5dfeq5}
&&\tilde{\nabla}^{i}{\tilde{\nabla}_{i}{X_{I}}} - h^i\tilde{\nabla}_i X_I + \tilde{\nabla}_{i}{X^{M}}\tilde{\nabla}^{i}{X^{N}} \bigg(\frac{1}{2}C_{M N K}X_{I}X^{K} - \frac{1}{6}C_{M N I}\bigg) 
\nonumber \\
&&+ \frac{2}{3}Q_{I J}\bigg(2\alpha L^{J} - \tilde{F}_{\ell_1 \ell_2}\tilde{G}^{J \ell_1 \ell_2}\bigg)
\cr 
&&-\frac{1}{12}\bigg[\tilde{G}^{M}{}_{\ell_1 \ell_2}\tilde{G}^{N \ell_1 \ell_2} - 2 L^{M}L^{N}\bigg]\bigg(C_{M N I} - X_{I}C_{M N J}X^{J}\bigg) 
\nonumber \\
&&+ 3 \chi^2 V_{M} V_{N}\bigg(\frac{1}{2}C_{I J K}Q^{M J}Q^{N K} + X_{I}(Q^{M N} - 2 X^{M}X^{N})\bigg) = 0 \ .
\ee
Furthermore (\ref{5dauxeq1o}) gives
\bea
\label{5dauxeq1}
{1 \over 2} \tn^i \tn_i \Delta -{3 \over 2} h^i \tn_i \Delta-{1 \over 2} \Delta \tn^i h_i
+ \Delta h^2 +{1 \over 4} dh_{ij} dh^{ij}  = \frac{3}{2}\beta^2 + Q_{I J}M^{I}{}_{\ell}M^{J \ell} \ ,
\ee
and ({\ref{5dauxeq2o}}) gives
\bea
\label{5dauxeq2}
{1 \over 2} \tn^j dh_{ij}-dh_{ij} h^j - \tn_i \Delta + \Delta h_i
= \frac{3}{2}\bigg(\beta_{\ell}\tilde{F}_{i}{}^{\ell}-\alpha \beta_{i} \bigg) + Q_{I J}\bigg(M^{I}{}_{\ell} \tilde{G}^{J}{}_{i}{}^{\ell} - L^{I}M^{J}{}_{i}\bigg) \ .
\ee
The conditions ({\ref{5dauxeq1}}) and ({\ref{5dauxeq2}}) correspond to the $++$ and $+i$-component of the Einstein equation and we remark that these are both implied by ({\ref{5dfeq3}}), ({\ref{5dfeq4}}), ({\ref{5dfeq5}}), together with ({\ref{5dfeq1}}) and ({\ref{5dfeq2}}) and the Bianchi identities ({\ref{beq}}).

\subsection{Integration of the KSEs along the lightcone}
For supersymmetric near-horizon horizons we assume there exists an $\epsilon \neq 0$ which is a solution to the KSEs. In this section, we will determine the neccessary conditions on the Killing spinor. To do this we first integrate along the two lightcone directions i.e.~we integrate the KSEs along the $u$ and $r$ coordinates. To do this, we decompose $\epsilon$ as
\bea
\e=\e_++\e_-~,
\label{5dksp1}
\ee
where $\Gamma_\pm\epsilon_\pm=0$. To begin, we consider the $\mu = -$ component of the gravitino KSE (\ref{5dgGkse}) which can be integrated to obtain, 
\bea\label{5dlightconesol}
\e_+=\phi_+(u,y)~,~~~\e_-=\phi_-+r \Gamma_-\Theta_+ \phi_+~,
\ee
where $\partial_r \phi_{\pm} = 0$. Now we consider the $\mu=+$ component; on evaluating this component at $r=0$ we get,
\bea
\phi_-=\eta_-~,~~~\phi_+=\eta_++ u \Gamma_+ \Theta_-\eta_-~,
\ee
where $\partial_{u}\eta_{\pm} = \partial_r \eta_{\pm} = 0$ and,
\bea
\label{thetapm}
\Theta_\pm &=& {1\over4} h_i\Gamma^i - \frac{i}{8}(\tilde{F}_{j k}\Gamma^{j k}  \pm 4\alpha) - \frac{1}{2}\chi V_I X^I \ ,
\ee
and $\eta_\pm$ depend only on the coordinates of the spatial horizon section ${\cal S}$.
Substituting the solution (\ref{5dlightconesol}) of the KSEs along the light cone directions back into the gravitino KSE (\ref{5dgGkse}), and appropriately expanding in the $r$ and $u$ coordinates, we find that
for  the $\mu = \pm$ components, one obtains  the additional conditions
\bea
\label{5dint1}
&& \bigg({1\over2}\Delta - {1\over8}(dh)_{ij}\Gamma^{ij} -\frac{i}{4}\beta_i \Gamma^i + \frac{3i}{2}\chi V_{I}\alpha^I \bigg)\phi_+ 
\nonumber \\
&&+ 2\bigg({1\over4} h_i\Gamma^i - \frac{i}{8}(-\tilde{F}_{j k}\Gamma^{j k} + 4\alpha) + \frac{1}{2} \chi V_{I}X^{I}\bigg)\tau_+ = 0~,
\ee
\bea
\label{5dint2}
&& \bigg(\frac{1}{4}\Delta h_i \Gamma^{i} - \frac{1}{4}\partial_{i}\Delta \Gamma^{i}\bigg)\phi_+ + \bigg(-\frac{1}{8}(dh)_{ij}\Gamma^{ij} +\frac{3i}{4}\beta{}_i\Gamma^i  + \frac{3i}{2}\chi V_{I}\alpha^I\bigg) \tau_+ = 0~,
\ee
\bea
\label{5dint3}
&&\bigg(-\frac{1}{2}\Delta - \frac{1}{8}(dh)_{ij}\Gamma^{ij} -\frac{3i}{4}\beta{}_i \Gamma^i + \frac{3i}{2}\chi V_{I}\alpha^I 
\nonumber \\
&&+ 2\big(-{1\over4} h_i\Gamma^i - \frac{i}{8}(\tilde{F}_{j k}\Gamma^{j k} + 4\alpha) - \frac{1}{2}\chi V_{I}X^{I}\big) \Theta_{-} \bigg)\phi_{-} = 0 \ .
\ee
Similarly the $\mu=i$ component of the gravitino KSEs gives
\bea
\label{5dint4}
\tilde{\nabla}_i \phi_\pm + \bigg( \mp \frac{1}{4}h_i  \mp \frac{i}{4}\alpha \Gamma_i  + \frac{i}{8}\tilde{F}_{j k}\Gamma_i{}^{j k} - \frac{i}{2}\tilde{F}_{i j}\Gamma^j - \frac{3i}{2}\chi V_{I}\tilde{A}^{I}{}_{i} + \frac{1}{2}\chi V_I X^{I}\Gamma_i \bigg)\phi_\pm=0 \ ,
\ee
and
\bea
\label{5dint5}
&&\tilde \nabla_i \tau_{+} + \bigg( -\frac{3}{4}h_i - \frac{i}{4}\alpha\Gamma_i - \frac{i}{8}\tilde{F}_{j k}\Gamma_i{}^{j k} + \frac{i}{2}\tilde{F}_{i j}\Gamma^j - \frac{3i}{2}\chi V_{I}\tilde{A}^{I}{}_{i} - \frac{1}{2}\chi V_I X^{I}\Gamma_i\bigg )\tau_{+}
\nonumber \\
&&+ \bigg(-\frac{1}{4}(dh)_{ij}\Gamma^{j} - \frac{i}{4}\beta_j \Gamma_i{}^j + \frac{i}{2}\beta_i   \bigg)\phi_{+} = 0~,
\ee
where we have set
\bea
\label{5dint6}
\tau_{+} = \Theta_{+}\phi_{+} \ .
\ee
Similarly, substituting the solution of the KSEs (\ref{5dlightconesol})  into the algebraic KSE (\ref{5d2Akse}) and expanding appropriately in the $u$ and $r$ coordinates, we find
\bea
\label{5dint7}
\bigg[\tilde{G}^I{}_{i j}\Gamma^{i j} \mp 2 L^I + 2i\tilde{\nabla}_i X^I \Gamma^i - 6i\chi \bigg(Q^{I J} - \frac{2}{3}X^{I}X^{J}\bigg)V_J\bigg]\phi_\pm = 0  \ ,
\ee
\be
\label{5dint8}
&&\bigg[\tilde{G}^I{}_{i j}\Gamma^{i j} + 2 L^I - 2i\tilde{\nabla}_i X^I \Gamma^i - 6i\chi \bigg(Q^{I J} - \frac{2}{3}X^{I}X^{J}\bigg)V_J \bigg]\tau_{+} + 2 M^I{}_i\Gamma^i \phi_{+}=0~.
\ee
In the next section, we will demonstrate that many of the above conditions are redundant as they are implied by the independent KSEs\footnote{These are given by the naive restriction of the KSEs on ${\cal S}.$} (\ref{5dcovr}), upon using the field equations and Bianchi identities.
\subsection{The independent KSEs on $\cal{S}$}
The integrability conditions of the KSEs in any supergravity theory are known to imply some of the Bianchi identities and field equations. Also, the KSEs are first order differential equations which are usually easier to solve than the field equations which are second order. As a result, the standard approach to find solutions is to first solve all the KSEs and then impose the remaining independent components of the field equations and Bianchi identities as required.
We will take a different approach here because of the difficulty of solving the KSEs and the algebraic conditions which include the $\tau_+$ spinor given in (\ref{5dint6}). Furthermore,  we are particularly interested
in the minimal set of conditions required for supersymmetry, in order to systematically analyse the necessary and
sufficient conditions for supersymmetry enhancement. 
In particular, the conditions  (\ref{5dint1}), (\ref{5dint2}), (\ref{5dint5}), and (\ref{5dint8}) which contain $\tau_+$ are implied from those containing $\phi_+$, along with some of the field equations and Bianchi identities. Furthermore, (\ref{5dint3}) and the terms linear in $u$ in (\ref{5dint4}) and (\ref{5dint7}) from the $+$ component are implied by the field equations, Bianchi identities and the $-$ component of (\ref{5dint4}) and (\ref{5dint7}).
Details of the calculations used to show this are presented in Appendix E. On taking this into account, it follows that, on making use of the field equations and Bianchi identities, the independent KSEs are
\bea
\label{5dcovr}
\nabla^{(\pm)}_{i} \eta_{\pm} = 0, \qquad {\cal A}^{I, (\pm)}\eta_{\pm} = 0 \ ,
\ee
where
\bea
\nabla^{(\pm)}_{i} = \tilde{\nabla}_{i} + \Psi^{(\pm)}_{i} \ ,
\ee
with
\bea
\label{5dalg1pm}
\Psi^{(\pm)}_{i} &=& \mp \frac{1}{4}h_i  \mp \frac{i}{4}\alpha \Gamma_i  + \frac{i}{8}\tilde{F}_{j k}\Gamma_i{}^{j k} - \frac{i}{2}\tilde{F}_{i j}\Gamma^j - \frac{3i}{2}\chi V_{I}\tilde{A}^{I}{}_{i} + \frac{1}{2}\chi V_I X^{I}\Gamma_i \ ,
\ee
and
\bea
\label{5dalg2pm}
\mathcal{A}^{I, (\pm)} &=& \tilde{G}^I{}_{i j}\Gamma^{i j} \mp 2 L^I + 2i\tilde{\nabla}_i X^I \Gamma^i - 6i\chi \bigg(Q^{I J} - \frac{2}{3}X^{I}X^{J}\bigg)V_J \ .
\ee
These are derived from the naive restriction of the supercovariant derivative and the algebraic KSE on ${\cal S}$.
Furthermore, if $\eta_{-}$ solves $(\ref{5dcovr})$ then
\bea
\eta_+ = \Gamma_{+}\Theta_{-}\eta_{-}~,
\label{5depfem}
\ee
also solves $(\ref{5dcovr})$. However, further analysis using global techniques, is required in order to determine if $\Theta_-$
has a non-trivial kernel.

\section{Supersymmetry enhancement}
\subsection{Horizon Dirac equation}
To proceed further we shall now define the horizon Dirac operators associated with the supercovariant derivatives following from the gravitino KSE as
\bea
{\cal D}^{(\pm)} \equiv \Gamma^{i}\nabla_{i}^{(\pm)} = \Gamma^{i}\tilde{\nabla}_{i} + \Psi^{(\pm)}~,
\ee
where
\bea
\label{5ddalg3pm}
\Psi^{(\pm)} \equiv \Gamma^{i}\Psi^{(\pm)}_{i} = \mp \frac{1}{4}h_i\Gamma^{i}  \mp \frac{3i}{4}\alpha  - \frac{3i}{8}\tilde{F}_{\ell_1 \ell_2}\Gamma^{\ell_1 \ell_2}
- \frac{3i}{2}\chi V_{I}\tilde{A}^{I}{}_{i}\Gamma^{i} + \frac{3}{2}\chi V_I X^{I} \ .
\ee
We can generalise the gravitino KSE and define an equivalent set of KSEs to (\ref{5dcovr}) as,
\bea
\label{5ddcovr2}
\hat{\nabla}^{(\pm)}_{i} \eta_{\pm} = 0, ~ {\cal A}^{I, (\pm)}\eta_{\pm} = 0 \ ,
\ee
where
\bea
\label{5ddgdirac}
\hat{\nabla}^{(\pm)}_{i} = \nabla^{(\pm)}_{i} + \kappa_{I}\Gamma_{i}{\cal A}^{I, (\pm)}
\ee
and a modified horizon Dirac operator as
\bea
{\mathscr D}^{(\pm)}= {\cal D}^{(\pm)}+ q_I \mathcal{A}^{I, (\pm)}
\ee
for some $\kappa_{I}, q_{I} \in \mathbb{R}$. Clearly if $q_{I} = 3\kappa_{I}$ then ${\mathscr D}^{(\pm)} = \Gamma^{i}\hat{\nabla}^{(\pm)}_{i}$. However, we shall not assume this in general and $q_I$, $\kappa_I$ will be generic. As we shall see in the following section, $q_I$ will be fixed by the requirement of the analysis and there will be appropriate choices of $\kappa_{I}$ such that the zero modes of ${\mathscr D}^{(\pm)}$ satisfy (\ref{5ddcovr2}) and are Killing spinors on $\cal{S}$.

\subsection{Lichnerowicz type theorems for ${\mathscr D}^{(\pm)}$}

{\it Theorem:}
Let ${\cal S}$ and the fields satisfy the conditions for the maximum principle to apply, e.g.~the fields are smooth and ${\cal S}$ is
compact without boundary. Then there is a 1-1 correspondence between the zero modes of ${\mathscr D}^{(+)}$ and the $\eta_+$ Killing spinors, i.e.
\bea
\nabla_{i}^{(+)}\eta_+=0~,~~~{\cal A}^{I,(+)}\eta_+=0~\Longleftrightarrow~ {\mathscr D}^{(+)}\eta_+=0~.
\ee
Moreover $\parallel\eta_+\parallel^2$ is constant.
\vskip 0.3cm
{\it Proof:} It is evident that if $\eta_+$ is a Killing spinor, then it is a zero mode of ${\mathscr D}^{(+)}$. To prove the converse, assuming that
$\eta_+$ is a zero mode of ${\mathscr D}^{(+)}$ and after using the field equations and Bianchi identities, one can establish the identity, see Appendix E,
\bea
\label{5ddl1}
{\tilde{\nabla}}^{i}{\tilde{\nabla}}_{i}\parallel\eta_+\parallel^2 - \, h^i {\tilde{\nabla}}_{i}\parallel\eta_+\parallel^2 &=& 2\parallel{\hat{\nabla}^{(+)}}\eta_{+}\parallel^2
\nonumber \\
&+&~  \bigg(\frac{1}{16}Q_{I J} - 3\kappa_{I}\kappa_{J}\bigg){\rm Re } \langle {\cal A}^{I, (+)} \eta_+, {\cal A}^{J, (+)} \eta_+ \rangle  \ ,
\ee
provided that $q_I = 0$. For the details of the symmetric and positive definite $Spin(3)$-invariant inner product see Appendix B and E. The maximum principle can only be applied for sufficient choices of $\kappa_I$ such that the second term in (\ref{5ddl1}) is non-negative. This is clearly true for $\kappa_I = 0$, which is the case with ${\mathscr D}^{(\pm)} = \Gamma^{i}\hat{\nabla}^{(\pm)}_{i}$. 

For such choices of $\kappa_I$, the maximum principle thus implies that $\eta_+$ are Killing spinors i.e ${\hat{\nabla}^{(+)}}\eta_{+}=0,~ {\cal A}^{I, (+)}\eta_{+} = 0$ and $\parallel\eta_+\parallel=\mathrm{const}$. 
Let us turn to investigate the relation between Killing spinors and the zero modes of the ${\mathscr D}^{(-)}$ operator.

{\it Theorem:} Let ${\cal S}$ be compact without boundary and the horizon fields be smooth. There is a 1-1 correspondence between the
zero modes of ${\mathscr D}^{(-)}$ and the $\eta_-$ Killing spinors, i.e.
\bea
\nabla_{i}^{(-)}\eta_-=0~,~~~{\cal A}^{(-)}\eta_-=0~\Longleftrightarrow~ {\mathscr D}^{(-)}\eta_-=0~ \ .
\ee
\vskip 0.3cm
{\it Proof:} It is clear that if $\eta_-$ is a Killing spinor, then it is a zero mode of ${\mathscr D}^{(-)}$.  To prove the converse, if $\eta_-$
is a zero mode of ${\mathscr D}^{(-)}$, then upon using the field equations and Bianchi identities one can establish the formula, see Appendix E.4,
\bea
\label{l2}
{\tilde{\nabla}}^{i} \bigg(\tilde{\nabla}_{i} \parallel \eta_- \parallel^2 + \parallel \eta_- \parallel^2  h_{i}\bigg)
&=& 2 \parallel{\hat{\nabla}^{(-)}}\eta_{-}\parallel^2 
\nonumber \\
&+&~  \bigg(\frac{1}{16}Q_{I J} - 3\kappa_{I}\kappa_{J}\bigg){\rm Re } \langle {\cal A}^{I,(-)} \eta_-, {\cal A}^{J,(-)}\eta_- \rangle ~,
\ee
provided that $q_I = 0$. On integrating this over ${\cal{S}}$ and assuming that ${\cal{S}}$ is compact and without boundary, the LHS vanishes since it is a total derivative and one finds that $\eta_{-}$ are Killing spinors i.e ${\hat{\nabla}^{(-)}}\eta_{-}=0,~ {\cal A}^{I, (-)}\eta_{-} = 0$.

\subsection{Index theory and supersymmetry enhancement}
In this section we will consider the counting of the number of supersymmetries, which will differ slightly in the ungauged and gauged case.  We will denote by $N_\pm$ the number
of linearly independent (over $\bC$) $\eta_\pm$ Killing spinors i.e,
\bea
N_\pm={{\rm dim}}_{\mathbb{C}}  \  {\rm{Ker}} \{ {{\nabla}^{(\pm)}}, {\cal A}^{I, (\pm)} \} ~.
\ee
Consider a spinor $\eta_+$ satisfying the corresponding KSEs in ({\ref{5dcovr}}).
In the ungauged theory, the spinor $C*\eta_+$ also satisfies the same KSEs, and $C* \eta_+$ is linearly independent from $\eta_+$, where $C*$ denotes charge conjugation\footnote{This corresponds to a complex conjugation and a matrix multiplication by $C$} for which the details are given in Appendix E. So in the ungauged theory, $N_+$ must be even. However, in the gauged theory $C*\eta_+$ is not parallel and so $N_+$ need not be even.
It is straightforward to see this from the KSEs, in particular\footnote{The charge conjugation matrix satisfies $\Gamma_{\mu}C* + C*\Gamma_{\mu} = 0$}
\bea
\nabla^{(+)}_{i} (C*\eta_{+}) &=& C*\big(\tilde{\nabla}_{i}{\eta_{+}} +\Sigma_i \eta_+ \big)
\nonumber \\
{\cal A}^{I, (+)}(C*\eta_{+}) &=& C*P^I \eta_+
\ee
where,
\bea
\label{extr1}
\Sigma_i = - \frac{1}{4}h_i  - \frac{i}{4}\alpha \Gamma_i  + \frac{i}{8}\tilde{F}_{j k}\Gamma_i{}^{j k} - \frac{i}{2}\tilde{F}_{i j}\Gamma^j + \frac{3i}{2}\chi V_{I}\tilde{A}^{I}{}_{i} - \frac{1}{2}\chi V_I X^{I}\Gamma_i
\ee
and
\bea
\label{exrtr2}
P^I = \tilde{G}^I{}_{i j}\Gamma^{i j} - 2 L^I + 2i\tilde{\nabla}_i X^I \Gamma^i + 6i\chi \bigg(Q^{I J} - \frac{2}{3}X^{I}X^{J}\bigg)V_J
\ee
For the ungauged theory with $\chi = 0$ we have $\Sigma_i = \Psi^{(+)}_i$ and $P^I = {\cal A}^{I, (+)}$ and therefore,
\bea
\nabla^{(+)}_{i} (C*\eta_{+}) = C*\nabla^{(+)}_{i}\eta_+ = 0,~~ {\cal A}^{I, (+)}(C*\eta_+) = C*{\cal A}^{I, (+)}\eta_+ = 0
\ee 
Thus $C* \eta_+$ satisfies the KSEs for the ungauged theory. This is clearly not the case of the gauged theory since the extra terms which depend on $\chi$ have the wrong sign in the last two terms of (\ref{extr1}) and the last term of (\ref{exrtr2}).

The spinors in the KSEs of $N=2, D=5$ (un)gauged supergravity horizons with an arbitrary number of vector multiplets are Dirac spinors. In terms of the spinors $\eta_\pm$ restricted to ${\cal{S}}$, for the ungauged theory the spin bundle $\bS$ decomposes
as $\bS = \bS^+ \oplus \bS^-$ where the signs refer to the projections with respect to $\Gamma_\pm$,
and $\bS^\pm$ are $Spin(3)$ bundles.
For the gauged theory, the spin bundle $\bS \otimes {\cal L}$, where ${\cal L}$ is a $U(1)$ bundle on ${\cal{S}}$, decomposes as $\bS \otimes {\cal L} = \bS^+ \otimes {\cal L} \oplus \bS^- \otimes {\cal L}$ where $\bS^\pm \otimes {\cal L}$ are $Spin_{c}(3) = Spin(3) \times U(1)$.
To proceed further, we will show that the analysis which we have developed implies that the
number of real supersymmetries of near-horizon geometries is $4 N_+$. This is because the number
of real supersymmetries is $N=2 (N_+ + N_-)$ and we shall establish that $N_+ = N_-$ via the following global analysis. In particular, utilizing the Lichnerowicz type theorems which we have established previously, we have
\bea
N_\pm=\mathrm{dim}\,\mathrm{Ker}\, {\mathscr D}^{(\pm)}~.
\ee
Next let us focus on the index of the ${\mathscr D}^{(+)}$ operator. Since ${\mathscr D}^{(+)}$  is defined on the odd dimensional manifold ${\cal S}$, the index vanishes \cite{atiyah1}.  As a result, we conclude that
\bea
\mathrm{dim}\,\mathrm{Ker}\, {\mathscr D}^{(+)}= \mathrm{dim}\,\mathrm{Ker}\, ({\mathscr D}^{(+)})^\dagger \ ,
\ee
where $({\mathscr D}^{(+)})^\dagger$ is the adjoint of ${\mathscr D}^{(+)}$.  Furthermore observe that
\bea
\Gamma_- ({\mathscr D}^{(+)})^\dagger= {\mathscr D}^{(-)} \Gamma_-~,
\ee
and so
\bea
N_-=\mathrm{dim}\,\mathrm{Ker}\, ({\mathscr D}^{(-)})=\mathrm{dim}\,\mathrm{Ker}\, ({\mathscr D}^{(+)})^\dagger~.
\ee
Therefore, we conclude that $N_+=N_-$ and so the number of (real) supersymmetries of such horizons is $N=2 (N_++N_-) = 4 N_+$.

\section{The $\mathfrak{sl}(2,\bR)$ symmetry of $D=5$ horizons}
\subsection{Algebraic relationship between $\eta_+$ and $\eta_{-}$ spinors}
\label{kernal}
We shall exhibit the existence of the $\mathfrak{sl}(2,\bR)$ symmetry of gauged $D=5$ vector multiplet horizons by directly constructing
the vector fields on the spacetime which generate the action of $\mathfrak{sl}(2,\bR)$. The existence of these vector fields
is a direct consequence of the doubling of the supersymmetries. We have seen that if $\eta_-$ is a Killing spinor, then  $\eta_+=\Gamma_+\Theta_-\eta_-$
is also a Killing spinor provided that $\eta_+\not=0$. It turns out that under certain conditions this is always possible. To consider this we must investigate the kernel
of $\Theta_-$.
\vskip 0.3cm
{\it Lemma:} Suppose that ${\cal S}$ and the fields satisfy the requirements for the maximum principle to apply,
and that 
\bea
\mathrm{Ker}\, \Theta_- \neq \{0\}~.
\label{5dkerz}
\ee
\vskip 0.3cm
Then the near-horizon data is trivial, i.e. all fluxes vanish and the scalars are constant.
\vskip2mm
{\it Proof:}
Suppose that there is $\eta_-\not=0$
such that $\Theta_- \eta_-=0$. In such a case,
  ({\ref{5dint3}}) gives
$ \Delta {\rm Re } \langle \eta_- , \eta_- \rangle =0$. Thus $\Delta =0$,
as $\eta_-$ is no-where vanishing. Next, the gravitino KSE $\nabla^{(-)}\eta_-=0$, together with ${\rm Re } \langle \eta_{-}, \Gamma_{i}\Theta_{-}\eta_{-} \rangle=0$, imply that
\begin{eqnarray}
\label{5dnrm1a}
{\tilde{\nabla}}_i \parallel \eta_-\parallel^2 = - h_i  \parallel \eta_-\parallel^2~.
\end{eqnarray}
On taking the divergence of this expression, eliminating  ${\tilde{\nabla}}^i h_i$ upon using ({\ref{5dfeq3}}),
and after setting $\Delta=0$, one finds
\begin{eqnarray}
\label{5dnrm1ab}
{\tilde{\nabla}}^i {\tilde{\nabla}}_i  \parallel \eta_-\parallel^2 &=&
 \bigg(2\alpha^2 + \frac{1}{2}\tilde{F}^2 + \frac{4}{3}Q_{I J}L^{I}L^{J} 
 + \frac{1}{3}Q_{I J}\tilde{G}^{I \ell_1 \ell_2}\tilde{G}^{J}{}_{\ell_1 \ell_2}
 + \frac{4}{3}\chi^2 U \bigg)  \parallel \eta_-\parallel^2~.
 \nonumber \\
\end{eqnarray}
As we have assumed that $Q_{IJ}$ is positive definite, and that $U \geq 0$, the maximum principle implies that $\parallel \eta_- \parallel^2$ is constant. We conclude
that $\alpha = \tilde{F} = L^{I} = \tilde{G}^{I} = U = 0$ and from ({\ref{5dint7}}) that $X^I$ is constant. Also $U = 0$ implies $V_{I} = 0$. Furthermore, ({\ref{5dnrm1a}}) implies that $dh=0$, and then (\ref{5dauxeq1}) implies that $\beta = M^{I} = 0$.
Finally, integrating ({\ref{5dfeq3}}) over the horizon section implies that $h=0$. Thus, all the fluxes vanish, and the scalars are constant. 

We remark that in the ungauged theory, if $\mathrm{Ker}\, \Theta_- \neq \{0\}$, triviality
of the near-horizon data implies that the spacetime geometry is
${\mathbb{R}}^{1,1} \times T^3$. In the case of the gauged theory, imposing $\mathrm{Ker}\, \Theta_- \neq \{0\}$ leads directly to a contradiction. To see this, note that the condition $U=0$ implies
that
\bea
V_I V_J (X^I X^J-{1 \over 2} Q^{IJ})=0 \ .
\ee
However the algebraic KSE imply that
\bea
V_I V_J (Q^{IJ}-{2 \over 3} X^I X^J)=0 \ .
\ee
These conditions cannot hold simultaneously, so there is a contradiction.
Hence, to exclude both the trivial ${\mathbb{R}}^{1,1} \times T^3$
solution in the ungauged theory, and the contradiction in the gauged theory, we shall henceforth take $\mathrm{Ker}\, \Theta_-=\{0\}$.

\subsection{Killing vectors}
Having established how to obtain $\eta_+$ type spinors from $\eta_-$ spinors, we next proceed
to determine the $\mathfrak{sl}(2,\bR)$ spacetime symmetry.
First note that the spacetime Killing spinor $\epsilon$ can be expressed in terms of $\eta_\pm$ as
\begin{eqnarray}
\epsilon= \eta_++ u \Gamma_+\Theta_-\eta_-+ \eta_-+r \Gamma_-\Theta_+\eta_++ru \Gamma_-\Theta_+\Gamma_+\Theta_-\eta_-~.
\label{5dgensolkse}
\end{eqnarray}
Since the $\eta_-$ and $\eta_+$ Killing spinors appear in pairs for supersymmetric horizons, let us choose a $\eta_-$ Killing spinor.  Then from the previous results, horizons with non-trivial fluxes also admit $\eta_+=\Gamma_+\Theta_-\eta_-$ as a Killing spinor. Taking $\eta_-$ and $\eta_+=\Gamma_+\Theta_-\eta_-$,
one can construct two linearly independent Killing spinors on the  spacetime as
\bea
\epsilon_1=\eta_-+u\eta_++ru \Gamma_-\Theta_+\eta_+~,~~~\epsilon_2=\eta_++r\Gamma_-\Theta_+\eta_+~.
\ee
It is known from the general theory of supersymmetric $D=5$ backgrounds that for any Killing spinors $\zeta_1$ and $\zeta_2$ the dual vector field $K(\zeta_1, \zeta_2)$ of the 1-form
bilinear
\bea
\omega(\zeta_1, \zeta_2) &=& {\rm Re } \langle(\Gamma_+-\Gamma_-) \zeta_1, \Gamma_a\zeta_2\rangle\, e^a 
\label{5d1formbi}
\ee
is a Killing vector which leaves invariant all the other bosonic fields of the theory, i.e.
\bea
{\cal L}_{K} g = {\cal L}_{K} X^{I} = {\cal L}_{K} F^{I} = 0 \ .
\ee
Evaluating the 1-form bilinears of the Killing spinor $\epsilon_1$ and $\epsilon_2$, we find that
\begin{eqnarray}
\label{bforms}
 \omega_1(\epsilon_1, \epsilon_2)&=& (2r {\rm Re} \langle\Gamma_+\eta_-, \Theta_+\eta_+\rangle+  4 u r^2  \parallel \Theta_{+}\eta_+\parallel^2) \,{\bf{e}}^+-2u \parallel\eta_+\parallel^2\, {\bf{e}}^-
\cr
&+& ({\rm Re} \langle \Gamma_+\eta_{-}, \Gamma_{i}\eta_{+}\rangle + 4 u r {\rm Re} \langle \eta_+, \Gamma_{i} \Theta_+ \eta_+ \rangle) {\bf{e}}^i \ ,
 \cr
 \omega_2(\epsilon_2, \epsilon_2)&=& 4 r^2 \parallel \Theta_+ \eta_+\parallel^2 \,{\bf{e}}^+-2 \parallel\eta_+\parallel^2 {\bf{e}}^- + 4 r{\rm Re}  \langle \eta_{+}, \Gamma_{i} \Theta_+ \eta_+ \rangle {\bf{e}}^i~,
 \cr
 \omega_3(\epsilon_1, \epsilon_1)&=&(2\parallel\eta_-\parallel^2+4r u {\rm Re} \langle\Gamma_+\eta_-, \Theta_+\eta_+\rangle+ 4 r^2 u^2 \parallel \Theta_+ \eta_+\parallel^2 ) {\bf{e}}^+
 \cr
 &-& 2u^2 \parallel\eta_+\parallel^2 {\bf{e}}^-+(2u {\rm Re } \langle \Gamma_+ \eta_- , \Gamma_i \eta_+ \rangle
+ 4 u^2 r{\rm Re} \langle \eta_+, \Gamma_i \Theta_+ \eta_+ \rangle) {\bf{e}}^i~.
\end{eqnarray}
Moreover, we can establish the following identities
\begin{eqnarray}
\label{ident1}
- \Delta\, \parallel\eta_+\parallel^2 +4  \parallel\Theta_+ \eta_+\parallel^2 =0~,~~~{\rm Re } \langle \eta_+ , \Gamma_i \Theta_+ \eta_+ \rangle  =0~,
\end{eqnarray}
which follow from the first integrability condition in ({\ref{5dint1}}),  $\parallel\eta_+\parallel=\mathrm{const}$ and the KSEs of $\eta_+$.
Further simplification to the bilinears can be obtained by making use of (\ref{ident1}).
We then obtain
\begin{eqnarray}
 \omega_1(\epsilon_1, \epsilon_2)&=& (2r {\rm Re} \langle\Gamma_+\eta_-, \Theta_+\eta_+\rangle+  u r^2 \Delta \parallel \eta_+\parallel^2) \,{\bf{e}}^+-2u \parallel\eta_+\parallel^2\, {\bf{e}}^-+ \tilde V_i {\bf{e}}^i~,
 \cr
 \omega_2(\epsilon_2, \epsilon_2)&=& r^2 \Delta\parallel\eta_+\parallel^2 \,{\bf{e}}^+-2 \parallel\eta_+\parallel^2 {\bf{e}}^-~,
 \cr
 \omega_3(\epsilon_1, \epsilon_1)&=&(2\parallel\eta_-\parallel^2+4r u {\rm Re} \langle\Gamma_+\eta_-, \Theta_+\eta_+\rangle+ r^2 u^2 \Delta \parallel\eta_+\parallel^2) {\bf{e}}^+
 -2u^2 \parallel\eta_+\parallel^2 {\bf{e}}^-+2u \tilde V_i {\bf{e}}^i~,
 \label{5db1forms}
\nonumber \\
\end{eqnarray}
where we have set,
\begin{eqnarray}
\label{5dvii}
\tilde V_i =  {\rm Re } \langle \Gamma_+ \eta_- , \Gamma_i \eta_+ \rangle\, ~.
\end{eqnarray}
\subsection{$\mathfrak{sl}(2,\bR)$ symmetry}
To uncover explicitly the $\mathfrak{sl}(2,\mathbb{R})$ symmetry of such horizons it remains to compute the Lie bracket algebra of the vector fields $K_1$, $K_2$ and $K_3$ which are dual to  the 1-form spinor bilinears $\omega_1, \omega_2$ and $\omega_3$. In simplifying the resulting expressions, we shall make use of the following identities
\begin{eqnarray}
&&-2 \parallel\eta_+\parallel^2-h_i \tilde V^i+2 {\rm Re } \langle\Gamma_+\eta_-, \Theta_+\eta_+\rangle=0~,~~~i_{\tilde V} (dh)+2 d {\rm Re } \langle\Gamma_+\eta_-, \Theta_+\eta_+\rangle=0~,
\cr
&& 2 {\rm Re } \langle\Gamma_+\eta_-, \Theta_+\eta_+\rangle-\Delta \parallel\eta_-\parallel^2=0~,~~~
{\tilde V}+ \parallel\eta_-\parallel^2 h+d \parallel\eta_-\parallel^2=0~.
\label{5dconconx}
\end{eqnarray}
We then obtain the following dual Killing vector fields:
\begin{eqnarray}
K_1 &=&-2u \parallel\eta_+\parallel^2 \partial_u+ 2r \parallel\eta_+\parallel^2 \partial_r+ \tilde V~,
\cr
K_2 &=&-2 \parallel\eta_+\parallel^2 \partial_u~,
\cr
K_3 &=&-2u^2 \parallel\eta_+\parallel^2 \partial_u +(2 \parallel\eta_-\parallel^2+ 4ru \parallel\eta_+\parallel^2)\partial_r+ 2u \tilde V~.
\label{5dkkk}
\end{eqnarray}
As we have previously mentioned, each of these Killing vectors also leaves invariant all the other bosonic fields in the theory. It is then straightforward to determine the
algebra satisfied by these isometries:

{\it Theorem:} The Lie bracket algebra of  $K_1$, $K_2$ and $K_3$  is $\mathfrak{sl}(2,\bR)$.
\vskip 0.3cm
{\it Proof:} Using the identities summarised above, one can demonstrate after a direct computation that
\begin{eqnarray}
[K_1,K_2]=2 \parallel\eta_+\parallel^2 K_2,~[K_2, K_3]=-4 \parallel\eta_+\parallel^2 K_1,~[K_3,K_1]=2 \parallel\eta_+\parallel^2 K_3~. 
\end{eqnarray}
\section{The geometry and isometries of ${\cal S}$}
It is known that the vector fields associated with the 1-form Killing spinor bilinears given in (\ref{5d1formbi}) leave invariant all the fields of
gauged $D=5$ supergravity with vector multiplets.  In particular suppose that $\tilde V \neq 0$. The isometries $K_a$ ($a=1,2,3$) leave all the bosonic fields invariant:
\bea
{\cal L}_{K_a} g=0, \qquad {\cal L}_{K_a} F^I=0, \qquad {\cal L}_{K_a} X^I=0 \ .
\ee
Imposing these conditions and expanding in $u,r$, and also making use of the identities
({\ref{5dconconx}}), one finds that
\begin{eqnarray}
\tilde\nabla_{(i} \tilde V_{j)}=0~,~ {\cal L}_{\tilde V} h= {\cal L}_{\tilde V}\Delta=0~,~ {\cal L}_{\tilde V} X^{I} = 0~,
\cr
 {\cal L}_{\tilde V} \tilde{F}= {\cal L}_{\tilde V} \alpha= {\cal L}_{\tilde V} L^{I}= {\cal L}_{\tilde V} \tilde{G}^{I}=0~.
\end{eqnarray}
Therefore $\tilde V$ is an isometry of ${\cal S}$ and leaves all the fluxes on ${\cal S}$ invariant. In fact,${\tilde{V}}$ is a spacetime
isometry as well. Furthermore, the conditions ({\ref{5dconconx}}) imply that ${\cal L}_{\tilde V}\parallel\eta_-\parallel^2=0$. 

\subsection{Classification of the geometry of ${\cal S}$ in the ungauged theory}
Here we will consider further restrictions on the geometry of ${\cal S}$ for the ungauged theory with $\chi = 0$,
which will turn out to be sufficient to determine the geometry completely.
In fact, as the isometry $K_2$ generated by the spinor $\epsilon_2$ is proportional
to ${\partial \over \partial u}$, it follows from the analysis of \cite{gutbh}
that the near-horizon geometries are either those of the BMPV solution (rotating, or static),
or $AdS_3 \times S^2$, which corresponds to the black ring/string near-horizon geometry,
or Minkowski space $\bR^{1,4}$. However, it is also possible to derive this classification by making direct use of the conditions
we have obtained so far, and we will do this here. We begin by explicitly expanding out the identities established in (\ref{ident1}) in terms of bosonic fields and using (\ref{5dconconx}) along with the field equations (\ref{5dfeq1})-(\ref{5dauxeq2}) and Bianchi identities (\ref{beq}). On expanding ({\ref{ident1}}) we obtain,
\bea
\label{bos1}
\Delta\, \parallel\eta_+\parallel^2 = {\rm Re } \langle \eta_+ , \bigg(\frac{1}{4}h^2 + \frac{1}{8}\tilde{F}^2 + \alpha^2 - \frac{i}{4}\tilde{F}_{\ell_1 \ell_2}h_{\ell_3}\Gamma^{\ell_1 \ell_2 \ell_3}\bigg) \eta_+ \rangle \ ,
\ee
and
\bea
\label{bos2}
{\rm Re } \langle \eta_+ , \Gamma_i \Theta_+ \eta_+ \rangle = {\rm Re } \langle \eta_+ , \bigg(\frac{1}{4}h_{i}  - \frac{i}{8}\tilde{F}_{\ell_1 \ell_2}\Gamma_{i}{}^{\ell_1 \ell_2}\bigg) \eta_+ \rangle = 0 \ .
\ee
Contracting (\ref{bos2}) with $h^i$ and substituting in (\ref{bos1}) we get,
\bea
\Delta \parallel\eta_+\parallel^2 = \bigg(-\frac{1}{4}h^2 + \frac{1}{8}\tilde{F}^2 + \alpha^2\bigg)\parallel\eta_+\parallel^2 \ .
\ee
Since $\eta_+ \neq 0$ the norm is non-vanishing, thus
\bea
\label{Delta1} 
\Delta  = -\frac{1}{4}h^2 + \frac{1}{8}\tilde{F}^2 + \alpha^2 \ .
\ee
Using (\ref{conv}) with (\ref{bos2}) we obtain,
\bea
\label{htoF}
h_{i} = \frac{1}{2}\epsilon_{i}{}^{\ell_1 \ell_2}\tilde{F}_{\ell_1 \ell_2} \iff \tilde{F}_{\ell_1 \ell_2} = h_{i}\epsilon^{i}{}_{\ell_1 \ell_2} \implies h^2 = \frac{1}{2}\tilde{F}^2 \ .
\ee
Substituting (\ref{htoF}) into ({\ref{Delta1}) we obtain,
	\bea
	\label{Delta2}
	\Delta = \alpha^2 \ .
	\ee
	Using (\ref{5dfeq3}) we have,
	\bea
	\frac{1}{2}\tilde{\nabla}^{i}{h_{i}} = -Q_{I J}\bigg(\frac{2}{3}L^{I}L^{J} + \frac{1}{6}\tilde{G}^{I}{}_{\ell_1 \ell_2}\tilde{G}^{J \ell_1 \ell_2}\bigg) \ .
	\ee
	Integrating both sides over ${\cal S}$ the left hand side vanishes and we obtain,
	\bea
	\int_{{\cal S}} Q_{I J}\bigg(\frac{2}{3}L^{I}L^{J} + \frac{1}{6}\tilde{G}^{I}{}_{\ell_1 \ell_2}\tilde{G}^{J \ell_1 \ell_2}\bigg) = 0 \ .
	\ee
	This implies that 
	\bea
	\label{LGh}
	L^{I} = 0 \ , \ \tilde{G}^{I}{}_{\ell_1 \ell_2} = 0 \ , \ \ \  {\rm and} \ \  \tilde{\nabla}^{i}{h_i} = 0 \ .
	\ee
	Substituting this into the algebraic KSE (\ref{5dint7}), we obtain $\tilde{\nabla}_{i}{X^{I}} = 0$ i.e $X^{I}$ is constant.
	The third line of the Bianchi identity (\ref{beq}) or the condition (\ref{5dint8}) also gives that $M^{I} = 0$. This implies that we have $G^{I}{}_{\mu \nu} = 0$ and $F^{I} = X^{I}F$ thus we are reduced to the minimal theory and the algebraic KSE is satisfied trivially.
	Now let us consider the third condition in (\ref{5dconconx}). In components we have
	\bea
	{\tilde V}_{i} + \parallel\eta_-\parallel^2 h_i + \tilde{\nabla}_{i}{\parallel\eta_-\parallel^2}=0 \ .
	\ee
	On taking the divergence, and using the fact that $\tilde{V}$ is an isometry as established in the previous section and (\ref{LGh}) we get
	\bea
	\tilde{\nabla}^{i}{\tilde{\nabla}_i}{\parallel\eta_-\parallel^2} + h^{i}\tilde{\nabla}_{i}{\parallel\eta_-\parallel^2} = 0 \ .
	\ee
	The maximum principle implies that $\parallel\eta_-\parallel^2$ is a (nonzero) constant. Thus $h$ is an isometry and a symmetry of the full solution where,
	\bea
	\label{Vtoh}
	\tilde{V}_{i} = -\parallel\eta_-\parallel^2 h_i \ ,
	\ee
	on using (\ref{htoF}) and (\ref{Delta2}) with the Einstein equation (\ref{5dfeq4}) the Ricci tensor of
	${\cal{S}}$ simplifies to,
	\bea
	\tilde{R}_{i j} = - h_{i}h_{j} + \delta_{i j}\bigg(\frac{1}{2}\Delta + h^2 \bigg) \ ,
	\ee
	and hence
	\bea
	\tilde{R} = 2h^2 + \frac{3}{2}\Delta \geq 0 \ .
	\ee
	
	Since ${\cal S}$ is a spatial 3-manifold, this completely determines the curvature of ${\cal S}$. The second condition in (\ref{5dconconx}), using (\ref{Vtoh}), gives
	\bea
	\label{hD}
	\tilde{\nabla}_{i}{(h^2 + 2\Delta)} = 0 \ .
	\ee
	Now let us consider the gauge field equation (\ref{5dfeq2}). On substituting (\ref{htoF}) and (\ref{LGh}) we obtain
	\bea
	\label{nnaux2b}
	-\frac{1}{2}\epsilon_{i}{}^{\ell_1 \ell_2} (dh)_{\ell_1 \ell_2} + \tilde{\nabla}_{i}{\alpha} + \alpha h_{i} = 0 \ .
	\ee
	On taking the divergence and using (\ref{LGh}) we obtain,
	\bea
	\tilde{\nabla}^{i}{\tilde{\nabla}_i}{\alpha} + h^{i}\tilde{\nabla}_{i}{\alpha} = 0
	\ee
	On multiplying by $\alpha$, and integrating both sides over ${\cal S}$ using integration by parts, we conclude that $\alpha$ is constant. From (\ref{Delta2}) this implies that $\Delta$ is constant. Finally on using (\ref{hD}), we also conclude that $h^2$ is constant. Hence all the scalars are constant on $\cal{S}$ which confirms the attractor behaviour.
	
	Having obtained these results, it is now possible to fully classify the near-horizon geometries. There are a number of possible cases:
	
	\begin{itemize}
		\item[(I)] $\Delta \neq 0, h \neq 0.$ Then ${\cal{S}}$ is a squashed $S^3$. This corresponds to the
		near-horizon geometry of the rotating BMPV solution.

		\item[(II)] $\Delta \neq 0, h =0.$ In this case, ${\cal{S}}$ is a (round) $S^3$, which corresponds to the
		near-horizon geometry $AdS_2 \times S^3$ of the static BMPV solution. 
		
		\item[(III)] $\Delta = 0, h \neq 0$.
		The condition ({\ref{nnaux2b}}) implies that $dh = 0$ and thus $h$ is covariantly constant, i.e $\tilde{\nabla}{h} = 0$. The Ricci tensor of ${\cal{S}}$ is then given by
		\bea
		\tilde{R}_{i j} = - h_{i}h_{j} + \delta_{i j}h^2  \ .
		\ee
		In this case ${\cal{S}}=S^1 \times S^2$. The near-horizon geometry is locally $AdS_3 \times S^2$ which corresponds to the geometry of both the black string, and also the supersymmetric black ring solution.
		\item[(IV)] $\Delta = 0, h = 0.$ In this case, all the fluxes vanish, with ${\cal{S}}=T^3$. The near-horizon geometry is locally isometric to Minkowski space $\bR^{1,4}$.
	\end{itemize}

\subsection{Analysis of the geometry of ${\cal S}$ in the gauged theory}

It is also possible to investigate properties of the geometry of
${\cal S}$ in the gauged theory, with $\chi \neq 0$. In particular, it is
useful to define the 1-form $Z$ as
\bea
\label{Zdef}
Z_{i} = \parallel\eta_+\parallel^{-2} {\rm Re } \langle \eta_+ , \Gamma_i  \eta_+ \rangle \ .
\ee
The covariant derivative of $Z$ is then given by
	\bea
	\label{covZ}
	\tilde{\nabla}_{i} Z_{j} = -Z_{i}h_{j} - \frac{1}{2}\alpha \epsilon_{i j k}Z^{k} + \delta_{i j}h_{k}Z^{k} + 3 \chi V_{I}X^{I} \bigg(Z_{i}Z_{j} - \delta_{i j}\bigg) \ ,
	\ee
	and hence the divergence is
	\bea
	\label{divZ}
	\tilde{\nabla}^{i}Z_{i} = 2 Z_{i}h^{i} - 6\chi V_{I}X^{I} \ .
	\ee
Using these identities it is straightforward to show that there are no
near-horizon geometries for which $h=0$, in contrast to the case of
the ungauged theory. To see this, if $h=0$,
then on integrating ({\ref{divZ}}) over ${\cal{S}}$ one obtains the condition
	\bea
	\int_{{\cal{S}}} -6\chi V_I X^I =0 \ .
	\ee
So there must exist a point on ${\cal{S}}$ at which $V_I X^I=0$. However, at such a point $U=-{9 \over 2} Q^{IJ} V_I V_J <0$, in contradiction to our assumption that $U \geq 0$ on ${\cal{S}}$. Hence, it follows that there are no near-horizon geometries with $h=0$ in the gauged theory.

Similar reasoning can be used to prove that there are also no solutions with 
${\tilde{V}}=0$ in the gauged theory. In this case,  the group action generated by $K_1, K_2$ and $K_3$ has only 2-dimensional orbits. A direct substitution of this condition in (\ref{5dconconx}) reveals that
	\begin{eqnarray}
	\label{zva}
	\Delta \parallel\eta_-\parallel^2=2 \parallel\eta_+\parallel^2~,~~~h=\Delta^{-1} d\Delta~.
	\end{eqnarray}
	Since $h$ is exact, such horizons are static. A coordinate transformation $r\rightarrow \Delta r$ reveals that the  geometry is a warped product of $AdS_2$ with ${\cal S}$, $AdS_2\times_w {\cal S}$. 
	We note that ({\ref{zva}}) implies that $\Delta$ is positive everywhere on ${\cal{S}}$. On making use of ({\ref{divZ}}) and ({\ref{zva}}) to eliminate $h$ in terms of $d \Delta$, one obtains the condition
	\begin{eqnarray}
	{\tilde{\nabla}}^i \big( \Delta^{-2} Z_i \big) = -6 \Delta^{-2} \chi V_I X^I \ ,
	\end{eqnarray}
	on setting $Z^2=1$, which follows on making use of a Fierz identity. Integrating this expression over ${\cal{S}}$ gives
	\begin{eqnarray}
	\int_{{\cal{S}}} \Delta^{-2} \chi V_I X^I =0 \ .
	\end{eqnarray}
	As before this is in contradiction to our assumption that $U \geq 0$ on ${\cal{S}}$. Hence, it follows that there are no near-horizon geometries in the gauged theory
for which ${\tilde{V}}=0$. For the gauged theory we also find,
\bea
\Delta = \alpha^2~,~~~L^I = 0~,~~~M^I = \alpha dX^I~,~~~\beta = d_h \alpha \ ,
\ee
and
\bea
{\tilde{V}}_i = \bigg[\chi V_{I}X^{I}\bigg(Z_{i} + W_{i} \bigg) - h_{i} \bigg] \parallel\eta_-\parallel^2~,
\ee
with
\bea
W_{i} = \parallel\eta_-\parallel^{-2}  {\rm Re } \langle \eta_- ,  \Gamma_{i}\eta_{-} \rangle~,
\ee
and we can also write the gauge field ${\tilde{F}}^I = X^I{\tilde{F}} + {\tilde{G}}^I$ as,
\bea
\tilde{F}^{I} = *_3 \bigg(dX^{I} + h X^{I} - 3 \chi Q^{I J}V_{J}Z\bigg) \ .
\ee
It would be interesting to determine if the supersymmetry enhancement automatically produces {\it rotational} isometries. A full classification of the possible geometries of ${\cal{S}}$ will be given elsewhere. We remark that in the case of the minimal gauged theory, such a classification has been constructed in \cite{grover},
and supersymmetric black rings in this theory have been excluded. However,
as noted in \cite{Kunduri:2007qy}, by considering solutions with two commuting
rotational isometries, it may be the case that there exist solutions
in the gauged theory coupled to vector multiplets which cannot be reduced
to solutions of the minimal theory. In particular, it is known that there
exists a case for which ${\cal{S}}=S^1 \times S^2$ which could correspond
to the near-horizon geometry of an inherently {\it non-minimal} supersymmetric
black ring. It would be interesting to investigate this further.

	% body of thesis comes here
	
	\chapter{Conclusion}

\label{ch:conclusions}

In this thesis we have investigated the conditions on the geometry 
of regular supersymmetric Killing horizons in various supergravity theories
which arise as a consequence of supersymmetry. This analysis
was performed in the near-horizon limit. A given near-horizon
geometry may not necessarily extend into to bulk uniquely to give a
well-defined full black hole solution. In some cases, such as for various
theories in five dimensions, it is also possible to fully classify the
near-horizon solutions. For more complicated higher dimensional
supergravities, such a complete classification has yet to be found.
However, the conditions obtained 
on the near-horizon geometry produce restrictions on the types of
possible black hole solutions.

We have demonstrated that smooth IIA horizons with compact spatial sections without boundary always admit an even number of supersymmetries. This result also applies to the massive IIA case. In addition, the near-horizon resolutions
with non-trivial fluxes admit an $\mathfrak{sl}(2,\mathbb{R})$ symmetry subalgebra.
The above result together with those obtained in \cite{5dindex, 11index} and \cite{iibindex} provide further evidence in support the conjecture
of \cite{iibindex} regarding the (super)symmetries of supergravity horizons. It also emphases that the (super)symmetry enhancement
that is observed near the horizons of supersymmetric black holes is a consequence of the smoothness of the fields.
Apart from exhibiting an $\mathfrak{sl}(2,\mathbb{R})$ symmetry, IIA horizons are further geometrically restricted.  This is because we have not explored
all the restrictions imposed by the KSEs and the field equations of the theory -- in this thesis we only explored enough to establish the $\mathfrak{sl}(2,\mathbb{R})$ symmetry.
However, the understanding of the horizons admitting two supersymmetries is within the capability of the technology developed so far for the classification of supersymmetric IIA backgrounds \cite{iiacla} and it will be explored
elsewhere.

The understanding of all IIA horizons is a more involved problem. As such spaces preserve an even number of supersymmetries and there are no IIA horizons
with non-trivial fluxes preserving 32 supersymmetries, which follows from the classification of maximally supersymmetric backgrounds in \cite{maxsusy}, there are potentially 15 different
cases to examine. Of course, all IIA horizons preserving more than 16 supersymmetries are homogenous spaces as a consequence of the results of \cite{josefof}. 
It is also now known that there exist no supersymmetric near-horizon geometries
preserving exactly $N$ supersymmetries for $16<N<32$ in all $D=10$ and $D=11$ supergravities, as a consequence of the analysis of the superalgebras
associated with such solutions in \cite{Gran:2017}. Hence, it remains to classify
the solutions with $N \leq 16$ supersymmetries, for even $N$, which is an avenue for future research.

We have also investigated the supersymmetry preserved by horizons in $N=2, D=5$ gauged, and ungauged, supergravity with an arbitrary number of vector multiplets. Making use of global techniques, we have demonstrated that such horizons always admit $N=4N_+$ (real) supersymmetries. Furthermore, in the ungauged theory,
we have shown that $N_+$ must be even. Therefore, all supersymmetric near-horizon geometries in the ungauged theory must be maximally supersymmetric. We have also shown that the near-horizon geometries possess a $\mathfrak{sl}{(2, \mathbb{R})}$ symmetry group. The analysis that we have conducted is further evidence that this type of symmetry enhancement is a generic property of supersymmetric black holes. 
In fact, the complete classification of the geometries in the ungauged theory is quite straightforward, because the identity
\bea
K_2=-2 \parallel\eta_+\parallel^2 \partial_u \ ,
\ee
implies that the timelike isometry $\partial_u$ can be written as a spinor bilinear. All supersymmetric near-horizon geometries in the ungauged theory for which $\partial_u$
can be written as a spinor bilinear in this fashion have been fully classified in
\cite{gutbh}. In particular, the solutions
reduce to those of the minimal ungauged theory and the scalars are constant. 
The supersymmetry enhancement in this case therefore automatically imposes  an attractor-type mechanism, whereby the scalars take constant values on the horizon.
The possible near-horizon geometries
in the ungauged theory are therefore $\bR^{1,1} \times T^3$; and $AdS_3 \times S^2$, 
corresponding to the near-horizon black string/ring geometry \cite{Chamseddine:1999qs, BS1, BR1}; and the near-horizon BMPV solution \cite{BMPV1, BMPV2}.

For near-horizon solutions in the gauged theory, the total number of supersymmetries is either 4 or 8. In the case of maximal supersymmetry,
the geometry is locally isometric to $AdS_5$, with $F^I=0$ and constant scalars.{\footnote{As observed in \cite{preons}, there also exist discrete quotients of $AdS_5$ preserving 6 out of 8 supersymmetries. In this case, the spinors which are excluded are not smooth due to the periodic identification.}} 
It remains to 
classify the geometries of $N=4$ solutions in the gauged theory; details of this will be given elsewhere. By analysing the conditions on the geometry, we demonstrated that there are no static solution in the gauged theory for $h=0$ and there exists at least one Killing vector on the horizon section as we have also shown there are no solutions with $\tilde{V} = 0$. The analysis in \cite{Kunduri:2007qy} provides a complete classification of near-horizon geometries of supersymmetric black holes of $U(1)^3$-gauged supergravity with vector multiplets, assuming the existence of two rotational isometries on the horizon section. It would be interesting to determine if, in the case
of supersymmetric near-horizon geometries, the supersymmetry enhancement
automatically produces such rotational isometries.
The classification for the geometry of the horizon, in the cases for which such isometries are assumed, shows that it is either spherical $S^3$, $S^1 \times S^2$ or a $T^3$ \cite{Kunduri:2007qy} - the last two have no analogue in the minimal gauged theory, corresponding to the near-horizon geometry $AdS_3 \times S^2$ and $AdS_3 \times T^2$. The difference between
the minimal theory and the $STU$ theory in this context is encoded in the parameter
\bea
\lambda = Q^{IJ} V_I V_J - (V_I X^I)^2 \ .
\ee
The near-horizon geometries constructed in \cite{Kunduri:2007qy} for which
$S^1 \times S^2$ arises as a solution are required to have $\lambda>0$ as
a consequence of the analysis of the geometry. This condition can be satisfied
in the $STU$ theory, but not in minimal gauged supergravity. In fact,
supersymmetric $AdS_5$ black rings have been excluded from minimal gauged supergravity in \cite{5dindex}. This analysis did not assume the existence of two commuting
rotational isometries which had been done earlier \cite{ads5}, rather it derived the existence of such isometries via
the supersymmetry enhancement mechanism. The possibility of 
an $AdS_5$ black ring remains for the gauged $STU$ theory. 
As we have noted,
a regular supersymmetric near-horizon geometry with $S^1 \times S^2$ event horizon
topology is known to exist in the gauged $STU$ theory. There are no known 
obstructions, analogous to the stability analysis considered in \cite{tdef},
to extending the near-horizon solution into the bulk, and it is unknown if
a supersymmetric $AdS_5$ black ring exists. Although supersymmetric black rings have been excluded from minimal gauged supergravity in \cite{5dindex}, it is still not
known if there exists a supersymmetric black ring in a non-minimal gauged supergravity. 

Another avenue for further research is higher derivative supergravity. In general, higher derivative supergravity theories have extremely complicated field equations, which makes
a systematic analysis of the near-horizon geometries challenging e.g  $\alpha'$ corrections of $D=11$, type IIA and IIB supergravity are not easy to deal with. One theory for which the
field equations are relatively simple is heterotic supergravity with $\alpha'$ corrections, 
the near-horizon analysis in this theory has already been considered in \cite{Fontanella}. 
In the context of $D=5$ theories, higher derivative theories have been constructed in \cite{Hanaki},
and the near-horizon analysis has been considered in \cite{Gutowskihd}, however the analysis 
in this case assumes that the black hole timelike isometry ${\partial \over \partial u}$
arises as a Killing spinor bilinear. The analysis of the KSEs is relatively straightforward, because the gravitino equation has the same form as in the 2-derivative theory. However,
the 2-form which appears in the gravitino equation is an auxiliary field which is related to the
Maxwell field strengths via highly nonlinear auxiliary field equations. This makes the
analysis of the geometric conditions particularly involved. Despite these difficulties,
it would nevertheless be interesting to investigate supersymmetry enhancement of near-horizon geometries
in higher derivative supergravity.

	\appendix
	\appendixpage
	\addappheadtotoc
	
	\chapter{Regular Coordinate Systems and Curvature}

In this Appendix, we derive the form for the GNC and consider other regular coordinate systems around the horizon. For a full derivation for this coordinate system, see \cite{gnull}. We also summarize the Riemann curvature of generic near-horizon geometries.

\section{Gaussian null coordinates}

Let $({\cal{M}},g)$ be a $D$ dimensional spacetime and $\cal{H}$ be a smooth co-dimension 1 null hypersurface in $\cal{M}$. Let $V$ be a vector field normal to $\cal{H}$ such that the integral curves of $V$ are future directed null geodesic generators of $\cal{H}$. Let $\cal{S}$ be a smooth spacelike co-dimension 2 cross-section of $\cal{H}$ such that each integral curve of $V$ crosses exactly once, we can assign local coordinates ($y^I$) with $I=1, .. , D-2$ to $\cal{S}$. 

Starting from point $p \in \cal{S}$ we assign a point $q \in \cal{H}$ lying a parameter (need not be affine) value $u$ away along the integral curve of $V$ the coordinates $(u, y^I)$, keeping the functions $y^I$ constant along the curve. Thus $(u, y^I)$ describe the coordinate system on $\cal{H}$ in the neighbourhood of the integral curves of $N$ through $\cal{S}$ with $V = \frac{\partial}{\partial u}$.

Recall that $V$ is normal $\cal{H}$ and is null on $\cal{H}$, so we have the metric functions $g_{u u} = V.V = 0$ and $g_{u I} = V.X_{I} = 0$ with $X_{I} = \partial_{I} = \frac{\partial}{\partial y^I}$ on $\cal{H}$. Now at every point $q \in \cal{H}$, let $W$ be a unique past directed null vector satisfying the normalisation $V.W=1$ and orthogonality $W.X_{I} = 0$. Starting from the point $q$, the point $s \in \cal{M}$ lying affine parameter value $r$ along the null geodesic with tangent $W$ is assigned coordinates $(u, r, y^I)$ with functions $u$ and $y^I$ kept constant along the geodesic as they are extended into $\cal{M}$. 

Therefore $(u, r, y^I)$ describe a coordinate system into the neighbourhood of $\cal{H}$ in $\cal{M}$, where the null hypersurface $\cal{H}$ is located at ${r=0}$ and $W = \frac{\partial}{\partial r}$. Using these coordinates, we can also extend the definitions of the vector fields $V, W, X_{I}$ into $\cal{M}$, and since these are coordinate vector fields, they all commute. By construction $W$ is null and $\nabla_{W} W = 0$ as the integral curves of $W$ are null geodesics, so we have $g_{r r} = W.W =0$ everywhere. Consider the directional derivatives,
\bea
\nabla_{W}(W.V) &=& W.\nabla_{W}{V} = W.([W,V] + \nabla_{V}{W}) = \frac{1}{2}\nabla_{V}{(W.W)} = 0 \ ,
\nonumber \\
\nabla_{W}{(W.X_{I})} &=& W.\nabla_{W}X_{I} = W.([W,X_{I}] + \nabla_{I}W) = \frac{1}{2}\nabla_{I}{(W.W)} = 0 \ .
\ee
Therefore we also have $g_{u r}=V.W = 1$ and $g_{r I} = W.X_{I}$ for all $r$ in the open set where the coordinates are defined, not only on $\cal{H}$. Nevertheless, $g_{u u} = V.V$ and $g_{u I} = V.X_{I}$ need not vanish outside $\cal{H}$. Hence the metric in Gaussian null coordinates takes the form \cite{gnull},
\bea
ds^2 &=&  -r f(y,r) du^2 + 2 du dr + 2 r h_{I}(y,r) dy^{I} du + \gamma_{I J}(y,r)dy^{I}dy^{J} \ .
\ee

\section{Other regular co-ordinate systems}

Consider an arbitrary metric which is regular at the horizon $r=0$\footnote{All the metric components are smooth functions in $r$} written in the coordinates $(u,r,y^I)$ with a Killing vector $V = \partial_u$ generating the Killing horizon,
\bea
\label{ametric}
ds^2 &=& g_{u u}(y,r)du^2 + g_{r r}(y,r)dr^2 + 2 g_{u r}(y,r) du dr 
\cr
&+& 2 g_{u I}(y,r)du dy^I + 2g_{r I}(y,r)dr dy^I + g_{I J}(y,r)dy^Idy^J \ ,
\ee
where GNC corresponds to making the choice,
\bea
g_{u u}(y,r) &=& -rf(y,r),~ g_{r r}(y,r) = 0,~ g_{u r}(y,r) = 1, 
\cr
g_{u I}(y,r)&=&r h_{I}(y,r),~ g_{I J} = \gamma_{I J} \ .
\ee
Here we will not consider this choice, but we will see what the necessary conditions are on the metric components for the near-horizon limit to be well defined. If we take the near-horizon limit with ({\ref{ametric}), we get,
\bea
ds^2 &=& g_{u u}(y,\epsilon r)\epsilon^{-2} du^2 + g_{r r}(y,\epsilon r)\epsilon^2 dr^2 + 2 g_{u r}(y,\epsilon r) du dr 
\cr
&+& 2 g_{u I}(y,\epsilon r)\epsilon^{-1}du dy^I + 2g_{r I}(y,\epsilon r)\epsilon dr dy^I + g_{I J}(y,\epsilon r)dy^Idy^J \ .
\ee
The near-horizon limit only exists when,\footnote{The first and third are always true in GNC, while the second is satisfied for an extremal black hole}
\bea
g_{u u}(y,0) = 0,~ \partial_r g_{u u}(y,0) = 0,~ g_{u I}(y,0) = 0 \ ,
\ee
and in this case the metric becomes,
\bea
ds^2 = \frac{r^2}{2}\partial^2_r g_{u u}(y,0) du^2  + 2 g_{u r}(y,0) du dr +2r \partial_r g_{u I}(y,0)du dy^I  + g_{I J}(y,0)dy^Idy^J \ .
\ee
If we have $g_{ur}(y,0) = 1$ then this is the near-horizon metric of the black hole solution ({\ref{nhmetricf}}) expressed in GNC if we identify with the near-horizon data,
\bea
\Delta = -\frac{1}{2}\partial^2_r g_{u u}(y,0),~ h_{I} = \partial_r g_{u I}(y,0),~  \gamma_{I J} = g_{I J}(y,0) \ .
\ee

\section{Curvature of near-horizon geometries}

We now summarize the Riemann curvature of generic near-horizon geometries, with components with respect to the basis ({\ref{basis1}}).
The spin connection written as,
\bea
\Omega_{A B}=\Omega_{C,AB} \bbe^C \ ,
\ee
satisfies
\bea
d\bbe^+ + \Omega^{+}{}_{B} \wedge \bbe^B &=& \Omega_{-, + -}\bbe^+ \wedge \bbe^- + (\Omega_{+,-i} + \Omega_{i, + -})\bbe^+ \wedge \bbe^i + \Omega_{-,-i}\bbe^- \wedge \bbe^i  \ ,
\nonumber \\
&+& \Omega_{i, - j}\bbe^i \wedge \bbe^j = 0
\nonumber \\ 
d\bbe^- + \Omega^{+}{}_{B}\wedge \bbe^B &=& (r\Delta  + \Omega_{+,+-})\bbe^+ \wedge \bbe^- + \bigg(\frac{1}{2}r^2 (\tilde{\nabla}_i {\Delta} - \Delta h_i ) + \Omega_{+, + i} \bigg)\bbe^+ \wedge \bbe^i \ ,
\nonumber \\
&+& (h_i + \Omega_{i, + -} - \Omega_{-,+i})\bbe^- \wedge \bbe^i + \bigg(\frac{1}{2} r(dh)_{i j} + \Omega_{i, + j}\bigg)\bbe^i \wedge \bbe^j = 0
\nonumber \\
d\bbe^i + \Omega^{i}{}_{B}\wedge \bbe^B &=& \delta^{\ell i}\bigg[ (\Omega_{+,\ell -} - \Omega_{j, \ell +} )\bbe^+ \wedge \bbe^- + (\Omega_{+, \ell j}-\Omega_{j, \ell +})\bbe^+ \wedge \bbe^j 
\nonumber \\
&+& (\Omega_{-, \ell j} - \Omega_{j,\ell -})\bbe^- \wedge \bbe^j + (\Omega_{k, \ell j} - \tilde{\Omega}_{k, \ell j})\bbe^k \wedge \bbe^j\bigg] = 0 \ .
\ee
With respect to the basis introduced in ({\ref{basis1}}), the curvature 2-forms
of an extremal near-horizon geometry are given by
\bea
\rho_{AB}= d\Omega_{AB}+ \Omega_{AC} \wedge \Omega^{C}_{\phantom{C}B} = \frac{1}{2} R_{ABCD} \bbe^C \wedge \bbe^D \ ,
\ee
with components
\bea
\rho_{ij} &=& \tilde{\rho}_{ij} + \tilde{\nabla}_{[i} h_{j]}\bbe^+ \wedge \bbe^-  
\cr
&+& r\bigg( - h_l \tilde{\nabla}_{[i} h_{j]}+ \tilde{\nabla}_l \tilde{\nabla}_{[i} h_{j]} 
+ \frac{1}{2}h_i \tilde{\nabla}_{[j} h_{l]} - \frac{1}{2}h_j \tilde{\nabla}_{[i} h_{l]} \bigg) \bbe^+ \wedge \bbe^l , 
\nonumber \\
\rho_{+-} &=& \bigg( \frac{1}{4}h^2 + \Delta \bigg)\bbe^+\wedge \bbe^- + r\bigg( \partial_j \Delta -  \Delta h_j-\frac{1}{2}h_i \tilde{\nabla}_{[i} h_{j]} \bigg) \bbe^+ \wedge \bbe^j  + \frac{1}{2} \tilde{\nabla}_{[i} h_{j]} \bbe^i \wedge \bbe^j,
\nonumber \\
\rho_{+i} &=& r^2  \bigg[ \bigg(\frac{1}{2} \tilde{\nabla}_l - h_l \bigg)( \partial_i \Delta -\Delta h_i) - \frac{1}{2}\Delta \tilde{\nabla}_{[i} h_{l]} + \tilde{\nabla}_{[k} h_{i]} \tilde{\nabla}_{[k} h_{l]} - \frac{1}{2} h_{[i} \tilde{\nabla}_{l]} \Delta \bigg] \bbe^+ \wedge \bbe^l
\nonumber \\ 
&+& r  \bigg( \partial_i \Delta - h_i \Delta - \frac{1}{2} h_j \tilde{\nabla}_{[j} h_{i]} \bigg)\bbe^+ \wedge \bbe^- + \frac{1}{2} \bigg( \tilde{\nabla}_i h_j- \frac{1}{2}h_ih_j \bigg)\bbe^- \wedge \bbe^j 
\nonumber \\ 
&+& r  \bigg( - \tilde{\nabla}_l \tilde{\nabla}_{[i} h_{j]} + \frac{1}{2} h_i \tilde{\nabla}_{[l} h_{j]} - \frac{1}{2} h_j \tilde{\nabla}_{[i} h_{l]} \bigg)\bbe^j \wedge \bbe^l, 
\nonumber \\ 
\rho_{-i} &=& \frac{1}{2} \bigg( \tilde{\nabla}_j h_i - \frac{1}{2}h_ih_j \bigg)\bbe^+ \wedge \bbe^j \, ,
\ee
where $\tilde{\rho}_{ab}$ is the curvature of $\tilde{\Omega}_{ab}$ on ${\cal S}$.

	\chapter{Clifford Algebras and Gamma Matrices}
In this Appendix, we will consider Clifford algebras and gamma matrices of arbitrary spacetime dimensions $d=t+s$ or as a signature $(t,s)$, with $t$ timelike
directions and $s$ spacelike directions using the conventions \cite{cliffconv1, cliffconv2}. We will focus on the cases where we have $(t,s) = (0,d)$ known as the Euclidean signature and  $(t,s) = (1,d-1)$ for the Lorentz signature for spinor representations of $Spin(d)$ and $Spin(1,d-1)$ which are double coverings of $SO(d)$ and $SO(1,d-1)$ respectively. The Clifford algebra is defined as,\footnote{We use the mostly positive signature $(-+\cdots +)$ for Lorentzian.}
\bea
\{ \Gamma_{\mu}, \Gamma_{\nu} \} = \Gamma_\mu \Gamma_\nu + \Gamma_\nu \Gamma_\mu = 2 g_{\mu \nu} \ ,
\label{defClifford}
\ee
In order to define these in arbitrary spacetime dimensions, we first introduce the Hermitian Pauli matrices given by,
\bea
 \sigma_1=\left(\begin{array}{cc} 0&1\\1&0\end{array}\right )\hspace{0.5cm}
\sigma_2=\left(\begin{array}{cc} 0&-i\\ i&0\end{array}\right
)\hspace{0.5cm} \sigma_3=\left(\begin{array}{cc}
1&0\\0&-1\end{array}\right )\,.
\ee
The only relevant properties are that they square to $\mathds{1}$ and $\sigma_1\sigma_2=i\sigma_3$ with cyclic\footnote{$\sigma^i \sigma^j = \delta^{ij} \bI_2 + i \epsilon^{ijk} \sigma^k,~ i=1,2,3$}. Let us define the $2k +1$ matrices of size $2k \times 2k$ by the tensor products of $k$ Pauli matrices and the identity $\mathds{1}$ as,
\bea
\Sigma_1&=&\sigma_1\otimes \underbrace{\mathds{1} \otimes \mathds{1} \otimes \ldots \otimes \mathds{1} \otimes \mathds{1}}_{k-1} \ ,
\nonumber\\ \Sigma_2&=&\sigma_2\otimes \mathds{1} \otimes \mathds{1} \otimes
\ldots \otimes \mathds{1} \otimes \mathds{1} \ , \nonumber\\ \Sigma_3&=&\sigma_3\otimes \sigma_1 \otimes \mathds{1}
\otimes \ldots \otimes \mathds{1} \otimes \mathds{1} \ , \nonumber\\ \Sigma_4&=&\sigma_3\otimes \sigma_2
\otimes \mathds{1} \otimes \ldots \otimes \mathds{1} \otimes \mathds{1} \ , \nonumber\\ \Sigma_5&=&\sigma_3\otimes
\sigma_3 \otimes \sigma_1 \otimes \underbrace{\ldots \otimes \mathds{1} \otimes \mathds{1}}_{k-3} \ ,
\nonumber\\ 
&\cdots&
\nonumber \\
\Sigma_{2k-3} &=& \underbrace{\sigma_3 \otimes \sigma_3 \otimes \sigma_3 \otimes \ldots}_{k-2} \otimes ~ \sigma_1 \otimes \mathds{1} \ ,
\nonumber \\
\Sigma_{2k-2} &=& \sigma_3 \otimes \sigma_3 \otimes \sigma_3 \otimes \ldots \otimes \sigma_2 \otimes \mathds{1} \ ,
\nonumber \\
\Sigma_{2k-1} &=& \sigma_3 \otimes \sigma_3 \otimes \sigma_3 \otimes \ldots \otimes \sigma_3 \otimes \sigma_1 \ ,
\nonumber \\
\Sigma_{2k} &=& \sigma_3 \otimes \sigma_3 \otimes \sigma_3 \otimes \ldots \otimes \sigma_3 \otimes \sigma_2 \ ,
\nonumber \\
\Sigma_{2k+1} &=& \sigma_3 \otimes \sigma_3 \otimes \sigma_3 \otimes \ldots \otimes \sigma_3 \otimes \sigma_3 \ ,
\label{representationClifford}
\ee
where $\otimes$ is the Kronecker product with the properties,
\bea
(A \otimes B)^{\dagger} &=& A^{\dagger}  \otimes B^{\dagger} \ ,
\nonumber \\
(A \otimes B)(C \otimes D) &=& AC \otimes BD \ .
\ee
In a Euclidean signature $(0,d)$ the representation of the Clifford algebra can be given by the gamma matrices defined as,
\bea
\Gamma_{i} = \Sigma_{i},~~ i=1,\cdots,d \ .
\ee
These gamma matrices are Hermitian, with respect to the (Euclidean) Dirac $Spin$-invariant inner product, and the generators of rotations in $Spin(d)$ are given by $\Gamma^{i j}$ which satisfy the $\mathfrak{so}(d)$ Lie algebra. For even and odd dimensions we let $d=2k$ and $d=2k+1$ respectively. It is easy to see that any of two gamma matrices anti-commute, while the square of any one is the identity matrix $\mathds{1}$. Therefore for even dimensions, this gives a representation of Clifford algebra for $Spin(2k)$. In fact, for odd dimensions this is a representation of Clifford algebra for $Spin(2k + 1)$ as well, by including $\Gamma^{2k+1} = \Sigma^{2k+1} \equiv \Gamma_{*}$.

In the Lorentzian signature $(1,d-1)$ the representation is given by,
\bea
\Gamma_0 &=& i\Sigma_1,
\nonumber \\
\Gamma_{k} &=& \Sigma_{k},~~ k=1,\cdots, d-1 \ ,
\ee
$\Gamma^0$ are anti-Hermitian and the other gamma matrices are Hermitian. This can also be expressed collectively as,
\bea
(\Gamma^{\mu})^{\dagger} = \Gamma^{0}\Gamma^{\mu}\Gamma^{0},~~ \mu = 0, \cdots, d-1 \ .
\ee
This preserves the $Spin(1,D-1)$-invariant Dirac inner product $\langle \phantom{i},\phantom{i} \rangle$ with,
\bea
\langle \psi, \phi \rangle = \bar{\psi} \phi,~~ \bar{\psi} = \psi^{\dagger}\Gamma^{0} \ .
\ee
We note that the $Spin(1,d-1)$-invariant inner product restricted to the particular representation is positive definite and real, and so symmetric. The generators of rotations in $Spin(1,d-1)$ are given by $\Gamma^{\mu \nu}$ which satisfy the $\mathfrak{so}(1,d-1)$ Lie algebra. Similar to the Euclidean signature, for even spacetime dimensions $d$ with signature $(1,d-1)$, this gives a representation of Clifford algebra for $Spin(1, 2k-1)$ and odd dimensions for $Spin(1, 2k)$ by including $\Gamma^{2k+1} = \Sigma^{2k+1} \equiv \Gamma_{*}$.

We can define Weyl spinors when $d$ is even in both signatures, but Majorana and Majorana-Weyl spinors occur in different dimensions to the Euclidean signature case. Now we state some useful identities for calculations in arbitrary dimensions \cite{gamiden}; for either signature ($t=0,1$) we can write $\Gamma_{*}$ as,
\bea
\Gamma_{*} =(-i)^{\lfloor d/2\rfloor +t}\Gamma_1 \cdots \Gamma_d,~~ (\Gamma_{*})^2 = 1 \ .
\ee
The anti-symmetrised product of gamma matrices in terms of the Levi-Civita tensor $\epsilon$ as
\bea
\Gamma_{a_1 \cdots a_n} = \frac{1}{(d-n)!}\epsilon_{a_1 \cdots a_d}i^{\lfloor d/2\rfloor + t}(\Gamma_{*})^{d-1} \Gamma^{a_d \cdots a_{n+1}} \ .
\ee
For a product of two anti-symmetrized gamma matrices we have,
\bea
\Gamma_{i_1 \cdots i_n}\Gamma^{j_1 \cdots j_m} = \sum_{k=0}^{min(n,m)}\frac{m! n!}{(m-k)! (n-k)! k!}\Gamma_{[i_1 \cdots i_{n-k}}^{\phantom{i} [j_{k+1}\cdots j_{m}}\delta^{j_1\cdots}_{i_n}\delta^{j_k]}_{i_{n-k+1}]} \ .
\ee

	\chapter{IIA Supergravity Calculations}

In this Appendix, we present  technical details of the analysis of the KSE
for the near-horizon solutions in IIA supergravity.

\section{Integrability}

First we will state the supercovariant connection $\cal{R}_{\mu \nu}$ given by,
\bea
[{\cal D}_{\mu}, {\cal D}_{\nu}]\epsilon \equiv \cal{R}_{\mu \nu}\epsilon \ ,
\ee 
where,
\bea
{\cal{R}}_{\mu \nu} &=& \frac{1}{4}R_{\mu \nu, \rho \sigma}\Gamma^{\rho \sigma} + \frac{1}{192}e^{\Phi} \Gamma_{\nu}\,^{\rho \sigma \kappa \lambda} \nabla_{\mu}{G_{\rho \sigma \kappa \lambda}}+\frac{1}{192}e^{\Phi}G^{\rho \sigma \kappa \lambda} \Gamma_{\nu \rho \sigma \kappa \lambda}  \nabla_{\mu}{\Phi} 
\nonumber \\
&-& \frac{1}{48}e^{\Phi}\Gamma^{\rho \sigma \kappa}  \nabla_{\mu}{G_{\nu \rho \sigma \kappa}} - \frac{1}{48}e^{\Phi} G_{\nu}\,^{\rho \sigma \kappa} \Gamma_{\rho \sigma \kappa}  \nabla_{\mu}{\Phi} - \frac{1}{192}e^{\Phi} \Gamma_{\mu}\,^{\rho \sigma \kappa \lambda}  \nabla_{\nu}{G_{\rho \sigma \kappa \lambda}} 
\nonumber \\
&-& \frac{1}{192}e^{\Phi}G^{\rho \sigma \kappa \lambda} \Gamma_{\mu \rho \sigma \kappa \lambda}  \nabla_{\nu}{\Phi}+\frac{1}{48}e^{\Phi}\Gamma^{\rho \sigma \kappa}  \nabla_{\nu}{G_{\mu \rho \sigma \kappa}}+\frac{1}{48}e^{\Phi} G_{\mu}\,^{\rho \sigma \kappa} \Gamma_{\rho \sigma \kappa}  \nabla_{\nu}{\Phi}
\nonumber \\
&-& \frac{1}{8}H_{\mu}\,^{\rho \sigma} H_{\nu \rho}\,^{\kappa} \Gamma_{\sigma \kappa} - \frac{1}{64}e^{\Phi}F^{\rho \sigma} H_{\mu}\,^{\kappa \lambda} \Gamma_{\nu \rho \sigma \kappa \lambda} +\frac{1}{8}e^{\Phi} F^{\rho \sigma} H_{\mu \nu \rho} \Gamma_{\sigma} 
\nonumber \\
&+&\frac{1}{32}e^{\Phi} F^{\rho \sigma} H_{\mu \rho \sigma} \Gamma_{\nu} +\frac{1}{32}e^{\Phi} F_{\nu}\,^{\rho} H_{\mu}\,^{\sigma \kappa} \Gamma_{\rho \sigma \kappa} +\frac{1}{18432} e^{2\Phi}  G^{\rho \sigma \kappa \lambda} G^{\tau h i j} \Gamma_{\mu \nu \rho \sigma \kappa \lambda \tau h i j}
\nonumber \\
&-& \frac{1}{4608}e^{2\Phi} G_{\nu}\,^{\rho \sigma \kappa} G^{\lambda \tau h i} \Gamma_{\mu \rho \sigma \kappa \lambda \tau h i}  - \frac{1}{384} e^{2\Phi} G_{\nu}\,^{\rho \sigma \kappa} G_{\rho}\,^{\lambda \tau h} \Gamma_{\mu \sigma \kappa \lambda \tau h} 
\nonumber \\
&-& \frac{1}{256}e^{2\Phi}G^{\rho \sigma \kappa \lambda} G_{\rho \sigma}\,^{\tau h} \Gamma_{\mu \nu \kappa \lambda \tau h} +\frac{1}{384}e^{2\Phi} G_{\mu}\,^{\rho \sigma \kappa} G_{\rho}\,^{\lambda \tau h} \Gamma_{\nu \sigma \kappa \lambda \tau h} 
\nonumber \\
&+&\frac{1}{128}e^{2\Phi} G_{\nu}\,^{\rho \sigma \kappa} G_{\rho \sigma}\,^{\lambda \tau} \Gamma_{\mu \kappa \lambda \tau}  - \frac{1}{128}e^{2\Phi} G_{\mu}\,^{\rho \sigma \kappa} G_{\rho \sigma}\,^{\lambda \tau} \Gamma_{\nu \kappa \lambda \tau} 
\nonumber \\
&+&\frac{1}{192}e^{2\Phi} G_{\nu}\,^{\rho \sigma \kappa} G_{\rho \sigma \kappa}\,^{\lambda} \Gamma_{\mu \lambda} +\frac{1}{768} e^{2\Phi} \Gamma_{\mu \nu}  {G}^{2} - \frac{1}{192}e^{2\Phi} G_{\mu}\,^{\rho \sigma \kappa} G_{\rho \sigma \kappa}\,^{\lambda} \Gamma_{\nu \lambda} 
\nonumber \\
&+&\frac{1}{96}e^{2\Phi}G_{\mu \nu}\,^{\rho \sigma} G_{\rho}\,^{\kappa \lambda \tau} \Gamma_{\sigma \kappa \lambda \tau} +\frac{1}{4608}e^{2\Phi} G_{\mu}\,^{\rho \sigma \kappa} G^{\lambda \tau h i} \Gamma_{\nu \rho \sigma \kappa \lambda \tau h i} 
\nonumber \\
&+&\frac{1}{64}e^{\Phi}F^{\rho \sigma} H_{\nu}\,^{\kappa \lambda} \Gamma_{\mu \rho \sigma \kappa \lambda}  - \frac{1}{32}e^{\Phi} F^{\rho \sigma} H_{\nu \rho \sigma} \Gamma_{\mu}  - \frac{1}{128}e^{2\Phi} F^{\rho \sigma} F^{\kappa \lambda} \Gamma_{\mu \nu \rho \sigma \kappa \lambda} 
\nonumber \\
&+&\frac{1}{64}e^{2\Phi} F_{\nu}\,^{\rho} F^{\sigma \kappa} \Gamma_{\mu \rho \sigma \kappa} +\frac{1}{32}e^{2\Phi} F_{\nu}\,^{\rho} F_{\rho}\,^{\sigma} \Gamma_{\mu \sigma} +\frac{1}{64}e^{2\Phi} \Gamma_{\mu \nu}  {F}^{2} - \frac{1}{32}e^{2\Phi} F_{\mu}\,^{\rho} F_{\rho}\,^{\sigma} \Gamma_{\nu \sigma} 
\nonumber \\
&-& \frac{1}{32}e^{\Phi}  F_{\mu}\,^{\rho} H_{\nu}\,^{\sigma \kappa} \Gamma_{\rho \sigma \kappa}  - \frac{1}{64}e^{2\Phi} F_{\mu}\,^{\rho} F^{\sigma \kappa} \Gamma_{\nu \rho \sigma \kappa} 
\nonumber \\
&+&\Gamma_{11}\bigg(\frac{1}{8}\Gamma^{\rho \sigma} \nabla_{\mu}{H_{\nu \rho \sigma}} - \frac{1}{16}e^{\Phi}\Gamma_{\nu}\,^{\rho \sigma}  \nabla_{\mu}{F_{\rho \sigma}} - \frac{1}{16} e^{\Phi}F^{\rho \sigma} \Gamma_{\nu \rho \sigma} \nabla_{\mu}{\Phi}+\frac{1}{8}e^{\Phi}\Gamma^{\rho}  \nabla_{\mu}{F_{\nu \rho}}
\nonumber \\
&+&\frac{1}{8}e^{\Phi} F_{\nu}\,^{\rho} \Gamma_{\rho}  \nabla_{\mu}{\Phi} - \frac{1}{8}\Gamma^{\rho \sigma} \nabla_{\nu}{H_{\mu \rho \sigma}}+\frac{1}{16}e^{\Phi} \Gamma_{\mu}\,^{\rho \sigma}  \nabla_{\nu}{F_{\rho \sigma}}+\frac{1}{16}e^{\Phi} F^{\rho \sigma} \Gamma_{\mu \rho \sigma}  \nabla_{\nu}{\Phi} 
\nonumber \\
&-& \frac{1}{8}e^{\Phi} \Gamma^{\rho}  \nabla_{\nu}{F_{\mu \rho}} - \frac{1}{8}e^{\Phi} F_{\mu}\,^{\rho} \Gamma_{\rho}  \nabla_{\nu}{\Phi}+\frac{1}{768}e^{\Phi} G^{\rho \sigma \kappa \lambda} H_{\mu}\,^{\tau h} \Gamma_{\nu \rho \sigma \kappa \lambda \tau h} 
\nonumber \\
&-& \frac{1}{48}e^{\Phi} G^{\rho \sigma \kappa \lambda} H_{\mu \nu \rho} \Gamma_{\sigma \kappa \lambda}  - \frac{1}{64}e^{\Phi}G^{\rho \sigma \kappa \lambda} H_{\mu \rho \sigma} \Gamma_{\nu \kappa \lambda}  - \frac{1}{192}e^{\Phi} G_{\nu}\,^{\rho \sigma \kappa} H_{\mu}\,^{\lambda \tau} \Gamma_{\rho \sigma \kappa \lambda \tau} 
\nonumber \\
&+&\frac{1}{32}e^{\Phi}G_{\nu}\,^{\rho \sigma \kappa} H_{\mu \rho \sigma} \Gamma_{\kappa}  - \frac{1}{768}e^{\Phi} G^{\rho \sigma \kappa \lambda} H_{\nu}\,^{\tau h} \Gamma_{\mu \rho \sigma \kappa \lambda \tau h} +\frac{1}{64}e^{\Phi} G^{\rho \sigma \kappa \lambda} H_{\nu \rho \sigma} \Gamma_{\mu \kappa \lambda} 
\nonumber \\
&-& \frac{1}{384}e^{2\Phi} F^{\rho \sigma} G_{\nu}\,^{\kappa \lambda \tau} \Gamma_{\mu \rho \sigma \kappa \lambda \tau} +\frac{1}{96}e^{2\Phi}  F^{\rho \sigma} G_{\rho}\,^{\kappa \lambda \tau} \Gamma_{\mu \nu \sigma \kappa \lambda \tau} 
\nonumber \\
&-& \frac{1}{768}e^{2\Phi} F_{\mu}\,^{\rho} G^{\sigma \kappa \lambda \tau} \Gamma_{\nu \rho \sigma \kappa \lambda \tau} - \frac{1}{64}e^{2\Phi} F^{\rho \sigma} G_{\nu \rho}\,^{\kappa \lambda} \Gamma_{\mu \sigma \kappa \lambda}  - \frac{1}{192}e^{2\Phi} F_{\mu}\,^{\rho} G_{\rho}\,^{\sigma \kappa \lambda} \Gamma_{\nu \sigma \kappa \lambda} 
\nonumber \\
&+&\frac{1}{64}e^{2\Phi}F^{\rho \sigma} G_{\nu \rho \sigma}\,^{\kappa} \Gamma_{\mu \kappa} 
+\frac{1}{768}e^{2\Phi}F_{\nu}\,^{\rho} G^{\sigma \kappa \lambda \tau} \Gamma_{\mu \rho \sigma \kappa \lambda \tau} +\frac{1}{192}e^{2\Phi} F_{\nu}\,^{\rho} G_{\rho}\,^{\sigma \kappa \lambda} \Gamma_{\mu \sigma \kappa \lambda} 
\nonumber \\
&+&\frac{1}{384}e^{2\Phi} F_{\mu \nu} G^{\rho \sigma \kappa \lambda} \Gamma_{\rho \sigma \kappa \lambda} +\frac{1}{192}e^{\Phi} G_{\mu}\,^{\rho \sigma \kappa} H_{\nu}\,^{\lambda \tau} \Gamma_{\rho \sigma \kappa \lambda \tau}  - \frac{1}{32}e^{\Phi} G_{\mu}\,^{\rho \sigma \kappa} H_{\nu \rho \sigma} \Gamma_{\kappa} 
\nonumber \\
&+&\frac{1}{384}e^{2\Phi} F^{\rho \sigma} G_{\mu}\,^{\kappa \lambda \tau} \Gamma_{\nu \rho \sigma \kappa \lambda \tau}  - \frac{1}{64}e^{2\Phi} F^{\rho \sigma} G_{\mu \nu}\,^{\kappa \lambda} \Gamma_{\rho \sigma \kappa \lambda} +\frac{1}{64}e^{2\Phi} F^{\rho \sigma} G_{\mu \rho}\,^{\kappa \lambda} \Gamma_{\nu \sigma \kappa \lambda}  
\nonumber \\
&-& \frac{1}{64}e^{2\Phi} F^{\rho \sigma} G_{\mu \rho \sigma}\,^{\kappa} \Gamma_{\nu \kappa}  +\frac{1}{32}e^{2\Phi} F^{\rho \sigma} G_{\mu \nu \rho \sigma} \bigg) \ ,
\ee
we can also relate the field equations to the supersymmetry variations. Consider,
\bea
\Gamma^{\nu}[{\cal D}_{\mu}, {\cal D}_{\nu}]\epsilon &-&[{\cal D}_{\mu}, {\cal A}]\epsilon + \Phi_{\mu}{\cal A}\epsilon 
\nonumber \\
&=& \bigg[\bigg(-\frac{1}{2}E_{\mu \nu}\Gamma^{\nu}
-\frac{1}{48}e^{\Phi}FG_{\lambda_1 \lambda_2 \lambda_3}\Gamma_{\mu}{}^{\lambda_1 \lambda_2 \lambda_3} + \frac{1}{16}e^{\Phi}FG_{\mu \lambda_1 \lambda_2}\Gamma^{\lambda_1 \lambda_2} 
\nonumber \\
&-&\frac{5}{192}e^{\Phi}BG_{\mu \lambda_1 \lambda_2 \lambda_3 \lambda_4}\Gamma^{\lambda_1 \lambda_2 \lambda_3 \lambda_4}
+ \frac{1}{192}e^{\Phi}BG_{\lambda_1 \lambda_2 \lambda_3 \lambda_4 \lambda_5}\Gamma_{\mu}{}^{\lambda_1 \lambda_2 \lambda_3 \lambda_4 \lambda_5}
 \bigg)
\nonumber \\
&+& \Gamma_{11}\bigg(\frac{1}{16}e^{\Phi}BF_{\lambda_1 \lambda_2 \lambda_3}\Gamma_{\mu}{}^{\lambda_1 \lambda_2 \lambda_3} - \frac{3}{16}e^{\Phi}BF_{\mu \lambda_1 \lambda_2}\Gamma^{\lambda_1 \lambda_2} - \frac{1}{8}e^{\Phi}FF_{\lambda}\Gamma_{\mu}{}^{\lambda} 
\nonumber \\
&+& \frac{1}{8}e^{\Phi}FF_{\mu} -\frac{1}{6}BH_{\mu \lambda_1 \lambda_2 \lambda_3}\Gamma^{\lambda_1 \lambda_2 \lambda_3} - \frac{1}{4}FH_{\mu \lambda}\Gamma^{\lambda} \bigg)\bigg]\epsilon \ ,
\ee
where,
\bea
\Phi_{\mu} &=& \bigg( \frac{1}{192}e^{\Phi}G_{\lambda_1 \lambda_2 \lambda_3 \lambda_4}\Gamma^{\lambda_1 \lambda_2 \lambda_3 \lambda_4}\Gamma_{\mu}\bigg) + \Gamma_{11}\bigg(\frac{1}{4}H_{\mu \lambda_1 \lambda_2}\Gamma^{\lambda_1 \lambda_2} 
- \frac{1}{16}e^{\Phi}F_{\lambda_1 \lambda_2}\Gamma^{\lambda_1 \lambda_2}\Gamma_{\mu}\bigg) \ ,
\nonumber \\
\ee
and
\bea
\Gamma^{\mu}[{\cal D}_{\mu}, {\cal A}]\epsilon &+& \theta {\cal A}\epsilon 
\nonumber \\
&=& \bigg(F\Phi  - \frac{1}{24}e^{\Phi}FG_{\lambda_1 \lambda_2 \lambda_3}\Gamma^{\lambda_1 \lambda_2 \lambda_3} + \frac{1}{96}e^{\Phi}BG_{\lambda_1 \lambda_2 \lambda_3 \lambda_4 \lambda_5}\Gamma^{\lambda_1 \lambda_2 \lambda_3 \lambda_4 \lambda_5}\bigg)\epsilon
\nonumber \\
&+& \Gamma_{11}\bigg( \frac{3}{4}e^{\Phi}FF_{\lambda}\Gamma^{\lambda} - \frac{3}{8}e^{\Phi}BF_{\lambda_1 \lambda_2 \lambda_3}\Gamma^{\lambda_1 \lambda_2 \lambda_3} + \frac{1}{12}BH_{\lambda_1 \lambda_2 \lambda_3 \lambda_4}\Gamma^{\lambda_1 \lambda_2 \lambda_3 \lambda_4} 
\nonumber \\
&+& \frac{1}{4}FH_{\lambda_1 \lambda_2}\Gamma^{\lambda_1 \lambda_2} \bigg)\epsilon \ ,
\ee
and
\bea
\theta = \bigg(-2\nabla_{\mu}{\Phi}\Gamma^{\mu} \bigg) + \Gamma_{11}\bigg(\frac{1}{12}H_{\lambda_1 \lambda_2 \lambda_3}\Gamma^{\lambda_1 \lambda_2 \lambda_3} - \frac{1}{2}e^{\Phi}F_{\lambda_1 \lambda_2}\Gamma^{\lambda_1 \lambda_2} \bigg) \ .
\ee
The field equations and Bianchi identities are 
\bea
\label{iiaeineq}
E_{\mu \nu} &=& R_{\mu \nu} + 2 \nabla_\mu \nabla_\nu \Phi
-{1 \over 4} H_{\mu \nu}^2
-{1 \over 2} e^{2 \Phi} F_{\mu \nu}^2
-{1 \over 12} e^{2 \Phi} G_{\mu \nu}^2 
\nonumber \\
&+& g_{\mu \nu} \bigg({1 \over 8}e^{2 \Phi}
F^2
+{1 \over 96}e^{2 \Phi} G^2  \bigg) = 0 ~,
\ee
\bea
F\Phi = \nabla^2 \Phi
- 2 (d \Phi)^2
+{1 \over 12} H^2-{3 \over 8} e^{2 \Phi}
F^2 -{1 \over 96} e^{2 \Phi} G^2  = 0 ~,
\ee
\bea
FF_{\mu} = \nabla^\lambda F_{\mu \lambda} -{1 \over 6} H^{\lambda_1 \lambda_2 \lambda_3} G_{\lambda_1 \lambda_2 \lambda_3 \mu} = 0 ~,
\ee
\bea
FH_{\mu \nu} &=& e^{2\Phi}\nabla^\lambda \bigg( e^{-2 \Phi} H_{\mu \nu \lambda}\bigg) - {1 \over 2} e^{2\Phi}G_{\mu \nu \lambda_1 \lambda_2} F^{\lambda_1 \lambda_2} 
\nonumber \\
&+&{1 \over 1152}e^{2\Phi} \epsilon_{\mu \nu}{}^{\lambda_1 \lambda_2
\lambda_3 \lambda_4 \lambda_5 \lambda_6 \lambda_7 \lambda_8}
G_{\lambda_1 \lambda_2 \lambda_3 \lambda_4}
G_{\lambda_5 \lambda_6 \lambda_7 \lambda_8} = 0 ~,
\ee
\bea
FG_{\mu \nu \rho} = \nabla^\lambda G_{\mu \nu \rho \lambda}-{1 \over 144} \epsilon_{\mu \nu \rho}{}^{\lambda_1 \lambda_2 \lambda_3 \lambda_4 \lambda_5 \lambda_6 \lambda_7}G_{\lambda_1 \lambda_2 \lambda_3 \lambda_4}H_{\lambda_5 \lambda_6 \lambda_7} = 0 ~.
\ee

\bea
\label{iiabian1}
BF_{\mu \nu \rho} = \nabla_{[\mu}F_{\nu \rho]}  = 0 \ ,
\ee

\bea
BH_{\mu \nu \rho \lambda} = \nabla_{[\mu}H_{\nu \rho \lambda]} = 0 \ ,
\ee

\bea
\label{iiabian3}
BG_{\mu \nu \rho \lambda \kappa} = \nabla_{[\mu}G_{\nu \rho \lambda \kappa]} - 2F_{[\mu \nu}H_{\rho \lambda \kappa]} = 0 \ .
\ee

\section{Alternative derivation of dilaton field equation}
The dilaton field equation is implied by the Einstein equation and all other field equations and Bianchi identities, up to a constant.
\bea
R = -2\nabla^2 \Phi + \frac{1}{4}H^2 - \frac{3}{4}e^{2\Phi}F^2 - \frac{1}{48}e^{2\Phi}G^2 \ .
\ee
On taking the Divergence of (\ref{iiaeineq})\footnote{For a $p$-form $\omega$ we write $\omega^2 = \omega_{\lambda_1 \cdots \lambda_p}\omega^{\lambda_1 \cdots \lambda_p}$ and $\omega^2_{\mu \nu} = \omega_{\mu \lambda_1 \cdots \lambda_{p-1}} \omega_{\nu}{}^{\lambda_1 \cdots \lambda_{p-1}}$},
\bea
\nabla^{\mu}R_{\mu \nu} &=& -2\nabla^2 \nabla_{\nu} \Phi + \frac{1}{4}\nabla^{\mu}H_{\mu \nu}^2
+ \frac{1}{2}\nabla^{\mu}(e^{2\Phi}F_{\mu \nu}^2) + \frac{1}{12}\nabla^{\mu}(e^{2\Phi}G_{\mu \nu}^2) 
\nonumber \\
&-& \frac{1}{8}\nabla_{\nu}(e^{2\Phi}F^2) - \frac{1}{96}\nabla_{\nu}(e^{2\Phi}G^2) \ ,
\ee
We can rewrite the first term as
\bea
-2\nabla^2 \nabla_{\nu} \Phi &=& -2\nabla_{\nu} \nabla^2 \Phi - 2R_{\mu \nu}\nabla^{\mu}\Phi
\nonumber \\
&=& -2\nabla_{\nu} \nabla^2 \Phi + 2\nabla_{\nu}(d\Phi)^2 - \frac{1}{2}H_{\mu \nu}^2 \nabla^{\mu}\Phi - e^{2\Phi}F_{\mu \nu}^2 \nabla^{\mu}\Phi 
\nonumber \\
&-& \frac{1}{6}e^{2\Phi}G_{\mu \nu}^2\nabla^{\mu}\Phi + \frac{1}{4}e^{2\Phi}F^2 \nabla_{\nu}\Phi + \frac{1}{48}e^{2\Phi} G^2 \nabla_{\nu}\Phi \ .
\ee
Where we have used (\ref{iiaeineq}) again. This gives,
\bea
\nabla^{\mu}R_{\mu \nu} &=& -2\nabla_{\nu} \nabla^2 \Phi + 2\nabla_{\nu}(d\Phi)^2 + \frac{1}{4}e^{2\Phi}\nabla^{\mu}(e^{-2\Phi}H_{\mu \nu}^2)
+ \frac{1}{2}e^{2\Phi}\nabla^{\mu}(F_{\mu \nu}^2) 
\nonumber \\
&+& \frac{1}{12}e^{2\Phi}\nabla^{\mu}(G_{\mu \nu}^2) - \frac{1}{8}\nabla_{\nu}(e^{2\Phi}F^2) + \frac{1}{4}e^{2\Phi}\nabla_{\nu}\Phi F^2 
\nonumber \\
&-& \frac{1}{96}\nabla_{\nu}(e^{2\Phi}G^2) + \frac{1}{48}e^{2\Phi}\nabla_{\nu}\Phi G^2 \ .
\ee
On the other hand,
\bea
\nabla^{\mu} R_{\mu \nu} &=& \frac{1}{2}\nabla_{\nu}R
\nonumber \\
&=& -\nabla_{\nu} \nabla^2 \Phi + \frac{1}{8}\nabla_{\nu}H^2 - \frac{3}{8}\nabla_{\nu}(e^{2\Phi}F^2) - \frac{1}{96}\nabla_{\nu}(e^{2\Phi}G^2) \ .
\ee
Rearranging the Einstein equation we obtain,
\bea
\label{iiaeins1}
\nabla_{\nu}\nabla^2\Phi &=& 2\nabla_{\nu}(d\Phi)^2  - \frac{1}{8}\nabla_{\nu}H^2 + \frac{1}{4}e^{2\Phi}\nabla^{\mu}(e^{-2\Phi}H_{\mu \nu}^2)
+ \frac{1}{2}e^{2\Phi}\nabla^{\mu}(F_{\mu \nu}^2)  
\nonumber \\
&+& \frac{1}{4}\nabla_{\nu}(e^{2\Phi}F^2) + \frac{1}{4}e^{2\Phi}\nabla_{\nu}\Phi F^2   + \frac{1}{12}e^{2\Phi}\nabla^{\mu}(G_{\mu \nu}^2) + \frac{1}{48}e^{2\Phi}\nabla_{\nu}\Phi G^2) \ .
\ee
We can compute certain terms by using the Field equations (\ref{iiageq1})-(\ref{iiaceq1}) and Bianchi identities (\ref{iiabian1})
\bea
\frac{1}{2}e^{2\Phi}\nabla^{\mu}(F_{\mu \nu}^2) = \frac{1}{2}e^{2\Phi}\nabla^{\mu}(F_{\mu \lambda})F_\nu{}^{\lambda} + \frac{1}{2}e^{2\Phi}F_{\mu \lambda}\nabla^{\mu}F_\nu{}^{\lambda} \ ,
\ee
\bea
\frac{1}{4}e^{2\Phi}\nabla^{\mu}(e^{-2\Phi}H_{\mu \nu}^2) = \frac{1}{4}e^{2\Phi}\nabla^{\mu}(e^{-2\Phi}H_{\mu \lambda_1 \lambda_2})H_\nu{}^{\lambda_1 \lambda_2} + \frac{1}{4}H_{\mu \lambda_1 \lambda_2}\nabla^{\mu}H_\nu{}^{\lambda_1  \lambda_2} \ ,
\ee
\bea
\frac{1}{12}e^{2\Phi}\nabla^{\mu}(G_{\mu \nu}^2) = \frac{1}{12}e^{2\Phi}\nabla^{\mu}(G_{\mu \lambda_1 \lambda_2 \lambda_3})G_\nu{}^{\lambda_1 \lambda_2 \lambda_3} + \frac{1}{12}e^{2\Phi}G_{\mu \lambda_1 \lambda_2 \lambda_3}\nabla^{\mu} G_\nu{}^{\lambda_1 \lambda_2 \lambda_3} \ .
\ee
The Bianchi identities (\ref{iiabian1}) imply,
\bea
F_{\mu \lambda}\nabla^{\mu}F_\nu{}^{\lambda} = \frac{1}{4}\nabla_{\nu}F^2 \ ,
\ee
\bea
H_{\mu \lambda_1 \lambda_2}\nabla^{\mu}H_\nu{}^{\lambda_1 \lambda_2} = \frac{1}{6}\nabla_{\nu}H^2 \ ,
\ee
\bea
G_{\mu \lambda_1 \lambda_2 \lambda_3}\nabla^{\mu} G_\nu{}^{\lambda_1 \lambda_2 \lambda_3} = \frac{1}{8}\nabla_{\nu}G^2 + \frac{5}{2}g_{\nu \kappa}G_{\mu \lambda_1 \lambda_2 \lambda_3}F^{[\mu \kappa}H^{\lambda_1 \lambda_2 \lambda_3]} \ .
\ee
Substituting this back into (\ref{iiaeins1}) we obtain,
\bea
\nabla_{\nu}\nabla^2\Phi &=& \nabla_{\nu}\bigg(2(d\Phi)^2 - \frac{1}{12}H^2 + \frac{3}{8}e^{2\Phi}F^2 + \frac{1}{96}e^{2\Phi}G^2  \bigg)
\nonumber \\
&+& \frac{1}{2}e^{2\Phi}\nabla^{\mu}(F_{\mu \lambda})F_\nu{}^{\lambda} +  \frac{1}{4}e^{2\Phi}\nabla^{\mu}(e^{-2\Phi}H_{\mu \lambda_1 \lambda_2})H_\nu{}^{\lambda_1 \lambda_2}
\nonumber \\
&+& \frac{1}{12}e^{2\Phi}\nabla^{\mu}(G_{\mu \lambda_1 \lambda_2 \lambda_3})G_\nu{}^{\lambda_1 \lambda_2 \lambda_3} + \frac{5}{24}g_{\nu \kappa}G_{\mu \lambda_1 \lambda_2 \lambda_3}F^{[\mu \kappa}H^{\lambda_1 \lambda_2 \lambda_3]} \ .
\ee
On applying the field equations (\ref{iiageq1})-(\ref{iiaceq1}),
\bea
\nabla_{\nu}\nabla^2\Phi &=& \nabla_{\nu}\bigg(2(d\Phi)^2 - \frac{1}{12}H^2 + \frac{3}{8}e^{2\Phi}F^2 + \frac{1}{96}e^{2\Phi}G^2 \bigg) \ .
\ee
%Since,
%\bea
%&&\frac{1}{4608}e^{2\Phi}\epsilon_{\lambda_1 \lambda_2}{}^{\tau_1 \tau_2 \tau_3 \tau_4 \tau_5 \tau_6 \tau_7 \tau_8}G_{\tau_1 \tau_2 \tau_3 \tau_4}G_{\tau_5 \tau_6 \tau_7 \tau_8}H_\nu{}^{\lambda_1 \lambda_2}
%\nonumber \\
%&+& \frac{1}{1728}e^{2\Phi}\epsilon_{\lambda_1 \lambda_2 \lambda_3}{}^{\tau_1 \tau_2 \tau_3 \tau_4 \tau_5 \tau_6 \tau_7}G_{\tau_1 \tau_2 \tau_3 \tau_4}H_{\tau_5 \tau_6 \tau_7}G_\nu{}^{\lambda_1 \lambda_2 \lambda_3} = 0
%\ee
%\bea
%- \frac{1}{12}e^{2\Phi}H^{\lambda_1 \lambda_2 \lambda_3} G_{\lambda_1 \lambda_2 \lambda_3 \lambda}F_\nu{}^{\lambda} +  \frac{1}{8}e^{2\Phi}G_{\lambda_1 \lambda_2}{}^{\tau_1 \tau_2}F_{\tau_1 \tau_2}H_\nu{}^{\lambda_1 \lambda_2} + \frac{5}{24}g_{\nu \kappa}G_{\mu \lambda_1 \lambda_2 \lambda_3}F^{[\mu \kappa}H^{\lambda_1 \lambda_2 \lambda_3]} = 0
%\nonumber \\
%\ee
This implies the dilaton field equation (\ref{iiadileq}) up to a constant.
In terms of the field equations and Bianchi identities, one gets
\bea
\nabla^{\mu}{R_{\mu \nu}} - \frac{1}{2}\nabla_{\nu}{R} &=& -\nabla_{\nu}{(F\Phi)} - 2 E_{\nu \lambda}\nabla^{\lambda}{\Phi} 
+ \nabla^{\mu}{(E_{\mu \nu})} - \frac{1}{2}\nabla_{\nu}{(E^{\mu}{}_{\mu})}
\nonumber \\
&-& \frac{1}{3}BH_{\nu}{}^{\lambda_1 \lambda_2 \lambda_3} H_{\lambda_1 \lambda_2 \lambda_3} 
+ \frac{1}{4}FH_{\lambda_1 \lambda_2}H_{\nu}{}^{\lambda_1 \lambda_2}
- \frac{3}{4}e^{2\Phi}BF_{\nu}{}^{\lambda_1 \lambda_2}F_{\lambda_1 \lambda_2} 
\nonumber \\
&-& \frac{1}{2}e^{2\Phi}FF_{\lambda} F_{\nu}{}^{\lambda}
- \frac{5}{48}e^{2\Phi}BG_{\nu}{}^{\lambda_1 \lambda_2 \lambda_3 \lambda_4}G_{\lambda_1 \lambda_2 \lambda_3 \lambda_4} 
\nonumber \\
&-& \frac{1}{12}e^{2\Phi}FG_{\lambda_1 \lambda_2 \lambda_3}G_{\nu}{}^{\lambda_1 \lambda_2 \lambda_3} = 0 \ .
\ee

\section{Invariance of IIA fluxes}

In this Appendix we will give a proof to show that the bilinears constructed from Killing spinors are Killing vectors and preserve all the fluxes. We will make use of the Killing spinor equations, field equations and Bianchi identities. We will use the following notation for the bilinears,
\bea
&& \alpha^{IJ}_{\mu_1 \cdots \mu_k} \equiv B(\epsilon^I, \Gamma_{\mu_1  \cdots \mu_k}\epsilon^J) ~, \notag \\
&& \tau^{IJ}_{\mu_1 \cdots \mu_k} \equiv B(\epsilon^I, \Gamma_{\mu_1  \cdots \mu_k} \Gamma_{11}\epsilon^J) ~,
\ee
with the inner product $B(\epsilon^I, \epsilon^J) \equiv \langle  \Gamma_0 C* \epsilon^I, \epsilon^J \rangle$, where $C=\Gamma_{6789}$, is antisymmetric, i.e.~$B(\epsilon^I, \epsilon^J)=-B(\epsilon^J, \epsilon^I)$ and all $\Gamma$-matrices are anti-Hermitian with respect to this inner product, i.e.~$B(\Gamma_\mu \epsilon^I,\epsilon^J)=-B(\epsilon^I,\Gamma_\mu \epsilon^J)$. The bilinears have the symmetry properties
\bea
\alpha^{IJ}_{\mu_1 \cdots \mu_k} &=& \alpha^{JI}_{\mu_1 \cdots \mu_k}  \qquad  k=1,2,5
\cr
\alpha^{IJ}_{\mu_1 \cdots \mu_k} &=& -\alpha^{JI}_{\mu_1 \cdots \mu_k} \quad ~ k=0,3,4
\ee
and
\bea
\tau^{IJ}_{\mu_1 \cdots \mu_k} &=& \tau^{JI}_{\mu_1 \cdots \mu_k}  \qquad  k=0,1,4,5
\cr
\tau^{IJ}_{\mu_1 \cdots \mu_k} &=& -\tau^{JI}_{\mu_1 \cdots \mu_k}\quad  ~k=2,3 ~.
\ee
First we show that the 1-form bi-linears whose associated vectors are Killing. We use the gravitino KSE to replace covariant derivatives with terms which are linear in the fluxes. The 1-form bilinears associated with the Killing vectors are $\alpha^{IJ}_\mu e^\mu$, we let $\nabla_{\mu} \epsilon = -\Psi_{\mu}\epsilon$ from the KSEs, and compute;
\bea
\nabla_\mu \alpha^{IJ}_\nu &=& \nabla_\mu B(\epsilon^{I}, \Gamma_\nu \epsilon^{J})  
\nonumber \\
&=& B(\nabla_\mu \epsilon^{I}, \Gamma_\nu  \epsilon^{J}) +B( \epsilon^{I}, \Gamma_\nu \nabla_\mu\epsilon^{J}) 
\nonumber \\
&=& -B(\Psi_\mu \epsilon^{I}, \Gamma_\nu \epsilon^{J}) -B( \epsilon^{I}, \Gamma_\nu \Psi_\mu\epsilon^{J}) 
\nonumber \\
&=& B(\Gamma_\nu \epsilon^{J},\Psi_\mu \epsilon^{I}) -B( \epsilon^{I}, \Gamma_\nu  \Psi_\mu\epsilon^{J})  
\nonumber \\
&=&  -B(\epsilon^{J},  \Gamma_\nu \Psi_\mu \epsilon^{I}) -B( \epsilon^{I}, \Gamma_\nu  \Psi_\mu\epsilon^{J}) 
\nonumber \\
&=& -2  B(\epsilon^{(I}, \Gamma_{B}\Psi_\mu \epsilon^{J)})
\nonumber \\
&=&\Big( \frac{1}{8}e^{\Phi}G_{\mu \nu}{}^{ \lambda_1 \lambda_2}\alpha^{IJ}_{\lambda_1 \lambda_2} + \frac{1}{96}e^{\Phi}G^{\lambda_1 \lambda_2 \lambda_3 \lambda_4} \alpha^{IJ}_{\mu \nu \lambda_1 \lambda_2 \lambda_3 \lambda_4} \nonumber \\
&& +\frac{1}{4} e^{\Phi}F_{\mu \nu} \tau^{IJ} - \frac{1}{2} H_{\mu \nu}{}^\lambda \tau^{IJ}_\lambda +\frac{1}{8} e^{\Phi}F^{\lambda_1 \lambda_2} \tau^{IJ}_{\mu \nu \lambda_1 \lambda_2}\Big) \ .
\ee
Since the resulting expression is antisymmetric in its free indices we find that $\nabla_{(\mu} \alpha^{IJ}_{\nu)}=0 $ and hence the vectors associated with $\alpha^{IJ}_\mu e^\mu$ are Killing.
Note that the dilatino KSE (\ref{iiaAKSE}) imply that
\bea
0  = B(\epsilon^{(I},{\cal A} \epsilon^{J)}) = \alpha^{IJ}_\mu \partial^\mu \Phi~,
\ee
and hence $i_K d\Phi=0$, where $K=\alpha^{IJ}_\mu e^\mu$ denotes the 1-forms associated with the Killing vectors with the $IJ$ indices suppressed. With this relation it follows that the Killing vectors preserve the dilaton:
\bea
{\cal L}_K \Phi := i_K d\Phi + d (i_K \Phi) =  0 ~,
\ee
since $i_K \Phi \equiv 0$.
To see that the 3-form flux $H$ is preserved we need to analyse the 1-form bi-linears which are not related to the Killing vectors, i.e. $\tau^{IJ}_\mu e^\mu$. As above, we find that
\bea
\nabla_{[\mu} \tau^{IJ}_{\nu]} = -2 B(\epsilon^{(I},\Gamma_{11} \Gamma_{[\mu} \Psi_{\nu]} \epsilon^{J)}) = - \frac{1}{2} H_{\mu \nu}{}^\lambda \alpha^{IJ}_\lambda ~,
\label{iia3formeq}
\ee
where we have indicated the degree of the form $\tau$ and suppressed the indices labelling the Killing spinors.
By taking the exterior derivative of (\ref{iia3formeq}), and using the Bianchi identity for $H$ with $dH=0$,
it follows that
\bea
{\cal L}_K H = 0 \ ,
\ee
and hence the Killing vectors preserve also the $H$ flux.
We now turn to the 2-form flux $F$. Computing the covariant derivative of the scalar $\tau^{IJ}$, and making use of the gravitino KSE as above, we find
\bea
\nabla_{\mu} \tau^{I J} = (i_K F)_{\mu} \ .
\label{iiaFexpr}
\ee
Acting with another derivative on (\ref{iiaFexpr}), and re-substituting (\ref{iiaFexpr}) into the resulting expression, we obtain
\bea
{\cal L}_K F =  i_K (dF)  = 0  ~,
\ee
where in the second step we have used  (\ref{iia3formeq}) and the Bianchi identity for $F$, i.e. $dF = 0$. For the field strength $G$ computing the covariant derivative of $\alpha_{\mu \nu}^{IJ}$ leads to
\bea
\nabla_{[\mu} \alpha^{I J}_{\nu \rho]} =
 \frac{1}{3}(i_K G)_{\mu \nu \rho}  + \frac{1}{3}\tau^{IJ}H_{\mu \nu \rho} - F_{[\mu \nu} \tau^{IJ}_{\rho]} ~.
\label{iiaGexpr}
\ee
Acting with an exterior derivative on (\ref{iiaGexpr}) and re-substituting (\ref{iiaGexpr}) into the resulting expression, and using (\ref{iia3formeq}), (\ref{iiaFexpr}) and the Bianchi identity for $F$ we obtain,
\bea
{\cal L}_K G =   i_K (dG)  + i_K F \wedge  H + F \wedge i_K H = 0 \ ,
\ee
where in the second step we have used the Bianchi identity for $G$, i.e. $dG = F \wedge H$.

\section{Independent KSEs}
It is well known that the KSEs imply some of the Bianchi identities and field equations
of a theory. Because of this, to find solutions it is customary to solve the KSEs and then
impose the remaining field equations and Bianchi identities. However, we shall not do
this here because of the complexity of solving the KSEs (\ref{iiaint1}), (\ref{iiaint2}), (\ref{iiaint5}), and (\ref{iiaint8})
which contain the $\tau_+$ spinor as expressed in (\ref{iiaint6}). Instead, we shall first show that all
the KSEs which contain $\tau_+$ are actually implied from those containing $\phi_+$, i.e. (\ref{iiaint4}) and
(\ref{iiaint7}), and some of the field equations and Bianchi identities.
Then we also show that (\ref{iiaint3}) and the terms linear in u from the $+$ components of (\ref{iiaint4}) and (\ref{iiaint7}) are implied
by the field equations, Bianchi identities and the $-$ components of (\ref{iiaint4}) and (\ref{iiaint7}).
\subsection{The (\ref{iiaint5}) condition}
The (\ref{iiaint5}) component of the KSEs is implied by (\ref{iiaint4}), (\ref{iiaint6}) and (\ref{iiaint7}) together with a number of field equations and Bianchi identities. First evaluate the LHS of (\ref{iiaint5}) by substituting in (\ref{iiaint6}) to eliminate $\tau_+$, and use (\ref{iiaint4}) to evaluate the supercovariant derivative of $\phi_+$. Also, using (\ref{iiaint4}) one can compute
\bea
(\tilde{\nabla}_{j}\tilde{\nabla}_{i} - \tilde{\nabla}_{i}\tilde{\nabla}_{j})\phi_{+} &=& {1\over 4} \tilde \nabla_j (h_i) \phi_+ + {1\over4} \Gamma_{11} \tilde \nabla_j(L_i) \phi_+ -{1\over8} \Gamma_{11} \tilde \nabla_j (\tilde H_{i l_1 l_2}) \Gamma^{l_1 l_2} \phi_+
\cr
&+&{1\over16} e^\Phi \Gamma_{11} (- 2\tilde \nabla_j (S)+ \tilde \nabla_j (\tilde F_{kl}) \Gamma^{kl}) \Gamma_i \phi_+ 
\nonumber \\
&-& {1\over 8\cdot 4!} e^\Phi (- 12 \tilde \nabla_j (X_{kl}) \Gamma^{kl}+ \tilde \nabla_j (\tilde G_{j_1j_2j_3j_4}) \Gamma^{j_1j_2j_3j_4}) \Gamma_i \phi_+
\cr
&+&{1\over16} \tilde \nabla_j \Phi e^\Phi \Gamma_{11} (- 2 S+  \tilde F_{kl} \Gamma^{kl}) \Gamma_i \phi_+ 
\cr
&-& {1\over 8\cdot 4!} \tilde \nabla_j \Phi e^\Phi (- 12  X_{kl} \Gamma^{kl}+  \tilde G_{j_1j_2j_3j_4} \Gamma^{j_1j_2j_3j_4}) \Gamma_i \phi_+
\cr
&+& \big( {1\over 4} h_i  + {1\over4} \Gamma_{11} L_i  -{1\over8} \Gamma_{11} \tilde H_{ijk} \Gamma^{jk} +{1\over16} e^\Phi \Gamma_{11} (- 2 S+ \tilde F_{kl} \Gamma^{kl}) \Gamma_i
\cr
&-&{1\over 8\cdot 4!} e^\Phi (- 12 X_{kl} \Gamma^{kl}+ \tilde G_{j_1j_2j_3j_4} \Gamma^{j_1j_2j_3j_4}) \Gamma_i\big)\tilde \nabla_{j} \phi_+ - (i \leftrightarrow j)  \ .
\label{iiaDDphicond}
\nonumber \\
\ee
Then consider the following, where the first terms cancels from the definition of curvature,
\bea
\bigg(\frac{1}{4}\tilde{R}_{ij}\Gamma^{j} - \frac{1}{2}\Gamma^{j}(\tilde{\nabla}_{j}\tilde{\nabla}_{i} - \tilde{\nabla}_{i}\tilde{\nabla}_{j}) \bigg)\phi_+ + \frac{1}{2}\tilde{\nabla}_{i}(\mathcal{A}_1) + \frac{1}{2}\Psi_{i} \mathcal{A}_1 = 0~,
\label{iiaB5intcond}
\ee
where
\bea
\mathcal{A}_1 &=& \partial_i \Phi \Gamma^i \phi_{+} -{1\over12} \Gamma_{11} (- 6 L_i \Gamma^i+\tilde H_{ijk} \Gamma^{ijk}) \phi_++{3\over8} e^\Phi \Gamma_{11} (-2 S+\tilde F_{ij} \Gamma^{ij})\phi_+
\cr
&
+&{1\over 4\cdot 4!}e^{\Phi} (- 12 X_{ij} \Gamma^{ij}+\tilde G_{j_1j_2j_3j_4} \Gamma^{j_1j_2j_3j_4}) \phi_+
\label{iiaA1cond}
\ee
and
\bea
\Psi_{i} &=& - \frac{1}{4}h_{i} + \Gamma_{11}(\frac{1}{4}L_{i} - \frac{1}{8}\tilde{H}_{i j k}\Gamma^{j k})~.
\label{iiaPsiicond}
\ee
The expression in (\ref{iiaA1cond}) vanishes on making use of (\ref{iiaint7}), as $\mathcal{A}_1 = 0$ is equivalent to the $+$ component of (\ref{iiaint7}). However a non-trivial identity is obtained by using (\ref{iiaDDphicond}) in (\ref{iiaB5intcond}), and expanding out the $\mathcal{A}_1$ terms. Then, on adding (\ref{iiaB5intcond}) to the LHS of (\ref{iiaint5}), with $\tau_+$ eliminated in favour of $\eta_+$ as described above, one obtains the following
 \bea
&&\frac{1}{4}\bigg(\tilde{R}_{ij} + \tn_{(i} h_{j)}
-{1 \over 2} h_i h_j + 2 \tn_i \tn_j \Phi
+{1 \over 2} L_i L_j -{1 \over 4} {\tilde{H}}_{i
l_1 l_2} {\tilde{H}}_j{}^{l_1 l_2}
\cr
&-&{1 \over 2} e^{2 \Phi} {\tilde{F}}_{i l}
{\tilde{F}}_j{}^l + {1 \over 8} e^{2 \Phi} {\tilde{F}}_{l_1 l_2}
{\tilde{F}}^{l_1 l_2}\delta_{ij} + {1 \over 2} e^{2 \Phi} X_{i l}
X_j{}^l - {1 \over 8} e^{2 \Phi}
X_{l_1 l_2}X^{l_1 l_2}\delta_{ij}
\cr
&-&{1 \over 12} e^{2 \Phi} {\tilde{G}}_{i \ell_1 \ell_2 \ell_3} {\tilde{G}}_j{}^{\ell_1 \ell_2 \ell_3}
+ {1 \over 96} e^{2 \Phi} {\tilde{G}}_{\ell_1 \ell_2
\ell_3 \ell_4} {\tilde{G}}^{\ell_1 \ell_2 \ell_3 \ell_4}\delta_{ij} - {1 \over 4} e^{2 \Phi} S^2\delta_{ij}\bigg)\Gamma^{j}=0~.
\ee
This vanishes identically on making use of the Einstein equation (\ref{iiafeq8}). Therefore it follows that (\ref{iiaint5}) is implied by the $+$ component of (\ref{iiaint4}), (\ref{iiaint6}) and (\ref{iiaint7}), the Bianchi identities (\ref{iiabian2}) and the gauge field equations (\ref{iiafeq1})-(\ref{iiafeq5}).
\subsection{The (\ref{iiaint8}) condition}
Let us define
\bea
\mathcal{A}_2 =
&&-\bigg( \partial_{i}\Phi\Gamma^{i} + \frac{1}{12}\Gamma_{11} (6L_i \Gamma^{i} + \tilde{H}_{ijk}\Gamma^{ijk}) + \frac{3}{8}e^{\Phi}\Gamma_{11}(2S + \tilde{F}_{ij}\Gamma^{ij})
\cr
&&- \frac{1}{4\cdot 4!}e^{\Phi}(12X_{ij}\Gamma^{ij} + \tilde{G}_{ijkl}\Gamma^{ijkl})\bigg)\tau_{+}
\cr
&&+ \bigg(\frac{1}{4}M_{ij}\Gamma^{ij}\Gamma_{11} + \frac{3}{4}e^{\Phi}T_{i}\Gamma^{i}\Gamma_{11} + \frac{1}{24}e^{\Phi}Y_{ijk}\Gamma^{ijk}\bigg)\phi_{+}~,
\ee
where $\mathcal{A}_2$ equals the expression in (\ref{iiaint8}).
One obtains the following identity
\bea
\mathcal{A}_2 = -\frac{1}{2}\Gamma^{i}\tilde{\nabla}_i \mathcal{A}_1 + \Psi_1\mathcal{A}_1 ~,
\ee
where
\bea
\Psi_1 &&= \tilde{\nabla}_{i}\Phi\Gamma^{i} + \frac{3}{8}h_{i}\Gamma^{i} + \frac{1}{16}e^{\Phi}X_{l_1 l_2}\Gamma^{l_1 l_2} - \frac{1}{192}e^{\Phi}\tilde{G}_{l_1 l_2 l_3 l_4}\Gamma^{l_1 l_2 l_3 l_4}
\cr
&&+ \Gamma_{11}\bigg(\frac{1}{48}\tilde{H}_{l_1 l_2 l_3}\Gamma^{l_1 l_2 l_3} - \frac{1}{8}L_{i}\Gamma^{i} + \frac{1}{16}e^{\Phi}\tilde{F}_{l_1 l_2}\Gamma^{l_1 l_2} - \frac{1}{8}e^{\Phi}S \bigg)~.
\ee
We have made use of the $+$ component of (\ref{iiaint4}) in order to evaluate the covariant derivative in the above expression. In addition we have made use of the Bianchi identities (\ref{iiabian2}) and the field equations (\ref{iiafeq1})-(\ref{iiafeq6}).
\subsection{The (\ref{iiaint1}) condition}
\label{iiaB1sec}
In order to show that (\ref{iiaint1}) is implied from the independent KSEs we can compute the following,
\bea
&&\bigg(-\frac{1}{4}\tilde{R} - \Gamma^{i j}\tilde{\nabla}_{i}\tilde{\nabla}_{j}\bigg)\phi_{+} - \Gamma^{i}\tilde{\nabla}_i(\mathcal{A}_1)
\cr
&+& \bigg(\tilde{\nabla}_i\Phi \Gamma^{i} + \frac{1}{4}h_{i}\Gamma^{i} + \frac{1}{16}e^{\Phi}X_{l_1 l_2}\Gamma^{l_1 l_2} - \frac{1}{192}e^{\Phi}\tilde{G}_{l_1 l_2 l_3 l_4}\Gamma^{l_1 l_2 l_3 l_4}
\cr
&+& \Gamma_{11}(-\frac{1}{4}L_{l}\Gamma^{l} - \frac{1}{24}\tilde{H}_{l_1 l_2 l_3}\Gamma^{l_1 l_2 l_3} - \frac{1}{8}e^{\Phi}S
+ \frac{1}{16}e^{\Phi}\tilde{F}_{l_1 l_2}\Gamma^{l_1 l_2}) \bigg)\mathcal{A}_1 = 0~,
\ee
where
\bea
\tilde{R} &&= -2\Delta - 2h^{i}\tilde{\nabla}_{i}\Phi - 2\tilde{\nabla}^2\Phi - \frac{1}{2}h^2 + \frac{1}{2}L^2 + \frac{1}{4}\tilde{H}^{2} + \frac{5}{2}e^{2\Phi}S^2
\cr
&&- \frac{1}{4}e^{2\Phi}\tilde{F}^2 + \frac{3}{4}e^{2\Phi}X^2 + \frac{1}{48}e^{2\Phi}\tilde{G}^2 \ ,
\ee
and where we use the $+$ component of (\ref{iiaint4}) to evaluate the covariant derivative terms. In order to obtain (\ref{iiaint1}) from these expressions we make use of the Bianchi identities (\ref{iiabian2}), the field equations (\ref{iiafeq1})-(\ref{iiafeq6}), in particular in order to eliminate the $(\tilde{\nabla} \Phi)^2$ term. We have also made use of the $+-$ component of the Einstein equation (\ref{iiafeq7}) in order to rewrite the scalar curvature $\tilde{R}$ in terms of $\Delta$. Therefore (\ref{iiaint1}) follows from (\ref{iiaint4}) and (\ref{iiaint7}) together with the field equations and Bianchi identities mentioned above.
\subsection{The + (\ref{iiaint7}) condition linear in $u$}
Since $\phi_+ = \eta_+ + u\Gamma_{+}\Theta_{-}\eta_-$, we must consider the part of the $+$ component of (\ref{iiaint7}) which is linear in $u$. On defining
\bea
\mathcal{B}_1 &=& \partial_i \Phi \Gamma^i \eta_{-} -{1\over12} \Gamma_{11} ( 6 L_i \Gamma^i+\tilde H_{ijk} \Gamma^{ijk}) \eta_-+{3\over8} e^\Phi \Gamma_{11} (2 S+\tilde F_{ij} \Gamma^{ij})\eta_-
\cr
&&
+{1\over 4\cdot 4!}e^{\Phi} ( 12 X_{ij} \Gamma^{ij}+\tilde G_{j_1j_2j_3j_4} \Gamma^{j_1j_2j_3j_4}) \eta_- \ ,
\ee
one finds that the $u$-dependent part of (\ref{iiaint7}) is proportional to
\bea
-\frac{1}{2}\Gamma^{i}\tilde{\nabla}_{i}(\mathcal{B}_1) + \Psi_2 \mathcal{B}_1~,
\ee
where
\bea
\Psi_2 &&= \tilde{\nabla}_{i}\Phi\Gamma^{i} + \frac{1}{8}h_{i}\Gamma^{i} - \frac{1}{16}e^{\Phi}X_{l_1 l_2}\Gamma^{l_1 l_2} - \frac{1}{192}e^{\Phi}\tilde{G}_{l_1 l_2 l_3 l_4}\Gamma^{l_1 l_2 l_3 l_4}
\cr
&&+ \Gamma_{11}\bigg(\frac{1}{48}\tilde{H}_{l_1 l_2 l_3}\Gamma^{l_1 l_2 l_3} + \frac{1}{8}L_{i}\Gamma^{i} + \frac{1}{16}e^{\Phi}\tilde{F}_{l_1 l_2}\Gamma^{l_1 l_2} + \frac{1}{8}e^{\Phi}S \bigg)~.
\ee
We have made use of the $-$ component of (\ref{iiaint4}) in order to evaluate the covariant derivative in the above expression. In addition we have made use of the Bianchi identities (\ref{iiabian2}) and the field equations (\ref{iiafeq1})-(\ref{iiafeq6}).
\subsection{The (\ref{iiaint2}) condition }
In order to show that (\ref{iiaint2}) is implied from the independent KSEs we will show that it follows from (\ref{iiaint1}). First act on (\ref{iiaint1}) with the Dirac operator $\Gamma^{i}\tilde{\nabla}_{i}$ and use the field equations (\ref{iiafeq1}) - (\ref{iiafeq6}) and the Bianchi identities to eliminate the terms which contain derivatives of the fluxes and then use (\ref{iiaint1}) to rewrite the $dh$-terms in terms of $\Delta$. Then use the conditions (\ref{iiaint4}) and (\ref{iiaint5}) to eliminate the $\partial_i \phi$-terms from the resulting expression, some of the remaining terms will vanish as a consequence of (\ref{iiaint1}). After performing these calculations, the condition (\ref{iiaint2}) is obtained, therefore it follows from section \ref{iiaB1sec} above that (\ref{iiaint2}) is implied by (\ref{iiaint4}) and (\ref{iiaint7}) together with the field equations and Bianchi identities mentioned above.
\subsection{The (\ref{iiaint3}) condition }
In order to show that (\ref{iiaint3}) is implied by the independent KSEs we can compute the following,
\bea
&&\bigg(\frac{1}{4}\tilde{R} + \Gamma^{i j}\tilde{\nabla}_{i}\tilde{\nabla}_{j}\bigg)\eta_- + \Gamma^{i}\tilde{\nabla}_i(\mathcal{B}_1)
\cr
&+& \bigg(-\tilde{\nabla}_i\Phi \Gamma^{i} + \frac{1}{4}h_{i}\Gamma^{i} + \frac{1}{16}e^{\Phi}X_{l_1 l_2}\Gamma^{l_1 l_2} + \frac{1}{192}e^{\Phi}\tilde{G}_{l_1 l_2 l_3 l_4}\Gamma^{l_1 l_2 l_3 l_4}
\cr
&+& \Gamma_{11}(-\frac{1}{4}L_{l}\Gamma^{l} + \frac{1}{24}\tilde{H}_{l_1 l_2 l_3}\Gamma^{l_1 l_2 l_3} - \frac{1}{8}e^{\Phi}S
- \frac{1}{16}e^{\Phi}\tilde{F}_{l_1 l_2}\Gamma^{l_1 l_2}) \bigg)\mathcal{B}_1 = 0~,
\ee
where we use the $-$ component of (\ref{iiaint4}) to evaluate the covariant derivative terms. The expression above vanishes identically since the $-$ component of (\ref{iiaint7}) is equivalent to $\mathcal{B}_1 = 0$. In order to obtain (\ref{iiaint3}) from these expressions we make use of the Bianchi identities (\ref{iiabian2}) and the field equations (\ref{iiafeq1})-(\ref{iiafeq6}). Therefore (\ref{iiaint3}) follows from (\ref{iiaint4}) and (\ref{iiaint7}) together with the field equations and Bianchi identities mentioned above.
\subsection{The + (\ref{iiaint4}) condition linear in $u$}
Next consider the part of the $+$ component of (\ref{iiaint4}) which is linear in $u$. First compute
\bea
\bigg(\Gamma^{j}(\tilde{\nabla}_{j}\tilde{\nabla}_{i} - \tilde{\nabla}_{i}\tilde{\nabla}_{j})  -\frac{1}{2}\tilde{R}_{ij}\Gamma^{j}\bigg)\eta_- - \tilde{\nabla}_{i}(\mathcal{B}_1) - \Psi_{i} \mathcal{B}_1 = 0~,
\ee
where
\bea
\Psi_{i} &=& \frac{1}{4}h_{i} - \Gamma_{11}(\frac{1}{4}L_{i} + \frac{1}{8}\tilde{H}_{i j k}\Gamma^{j k}) \ ,
\ee
and where we have made use of the $-$ component of (\ref{iiaint4}) to evaluate the covariant derivative terms. The resulting expression corresponds to the expression obtained by expanding out the $u$-dependent part of the $+$ component of (\ref{iiaint4}) by using the $-$ component of (\ref{iiaint4}) to evaluate the covariant derivative. We have made use of the Bianchi identities (\ref{iiabian2}) and the field equations (\ref{iiafeq1})-(\ref{iiafeq5}).

\section{Calculation of Laplacian of $\parallel \eta_\pm \parallel^2$}
\label{iiamaxpex}
In this Appendix, we calculate the Laplacian of $\parallel \eta_\pm \parallel^2$, which will be particularly useful in the analysis
of the global properties of IIA horizons in Section 3.
We shall consider the modified gravitino KSE ({\ref{iiaredef1}})
defined in section 3.1, and we
shall assume throughout that the modified Dirac equation ${\mathscr D}^{(\pm)}\eta_\pm=0$ holds, where ${\mathscr D}^{(\pm)}$ is defined in
({\ref{iiaredef2}}). Also, $\Psi^{(\pm)}_i$ and $\mathcal{A}^{(\pm)}$
are defined by ({\ref{iiaalg1pm}}) and ({\ref{iiaalg2pm}}), and
$\Psi^{(\pm)}$ is defined by ({\ref{iiaalg3pm}}).
To proceed, we compute the Laplacian
\bea
\tilde{\nabla}^i \tilde{\nabla}_i ||\eta_{\pm}||^2 = 2\langle\eta_\pm,\tilde{\nabla}^i \tilde{\nabla}_i\eta_\pm\rangle + 2 \langle\tilde{\nabla}^i \eta_\pm, \tilde{\nabla}_i \eta_\pm\rangle \ .
\ee
To evaluate this expression note that
\bea
\tilde{\nabla}^i \tilde{\nabla}_i \eta_\pm &=& \Gamma^{i}\tilde{\nabla}_{i}(\Gamma^{j}\tilde{\nabla}_j \eta_\pm) -\Gamma^{i j}\tilde{\nabla}_i \tilde{\nabla}_j \eta_\pm
\nonumber \\
&=& \Gamma^{i}\tilde{\nabla}_{i}(\Gamma^{j}\tilde{\nabla}_j \eta_\pm) + \frac{1}{4}\tilde{R}\eta_\pm
\nonumber \\
&=& \Gamma^{i}\tilde{\nabla}_{i}(-\Psi^{(\pm)}\eta_\pm -q\mathcal{A}^{(\pm)}\eta_{\pm}) + \frac{1}{4}\tilde{R} \eta_\pm \ .
\ee
It follows that
\bea
\langle\eta_\pm,\tilde{\nabla}^i \tilde{\nabla}_i\eta_\pm \rangle &=& \frac{1}{4}\tilde{R}\parallel \eta_\pm \parallel^2
+ \langle\eta_\pm, \Gamma^{i}\tilde{\nabla}_i(-\Psi^{(\pm)} - q\mathcal{A}^{(\pm)})\eta_\pm\rangle
\nonumber \\
&+& \langle\eta_\pm, \Gamma^{i}(-\Psi^{(\pm)} - q\mathcal{A}^{(\pm)})\tilde{\nabla}_i \eta_\pm \rangle~,
\ee
and also
\bea
\langle\tilde{\nabla}^i \eta_\pm, \tilde{\nabla}_i \eta_\pm\rangle &=& \langle{\hat\nabla^{(\pm)i}} \eta_\pm, {\hat\nabla^{(\pm)}_{i}} \eta_\pm\rangle - 2\langle\eta_\pm, (\Psi^{(\pm)i} + \kappa\Gamma^{i}\mathcal{A}^{(\pm)})^{\dagger} \tilde{\nabla}_i \eta_\pm \rangle
\nonumber \\
&-& \langle\eta_\pm, (\Psi^{(\pm)i} + \kappa\Gamma^{i}\mathcal{A}^{(\pm)})^{\dagger} (\Psi^{(\pm)}_i + \kappa \Gamma_{i} \, \mathcal{A}^{(\pm)}) \eta_\pm \rangle
\nonumber \\
&=& \parallel {\hat\nabla^{(\pm)}}\eta_{\pm} \parallel^2 - 2\langle \eta_{\pm}, \Psi^{(\pm)i\dagger}\tilde{\nabla}_{i}\eta_{\pm}\rangle
- 2\kappa \langle \eta_{\pm}, \mathcal{A}^{(\pm)\dagger}\Gamma^{i}\tilde{\nabla}_{i}\eta_{\pm}\rangle
\nonumber \\
&-&  \langle \eta_\pm, (\Psi^{(\pm)i\dagger}\Psi^{(\pm)}_i + 2\kappa \mathcal{A}^{(\pm)\dagger}\Psi^{(\pm)} + 8\kappa^2\mathcal{A}^{(\pm)\dagger}\mathcal{A}^{(\pm)})\eta_\pm \rangle
\nonumber \\
&=& \parallel {\hat\nabla^{(\pm)}}\eta_{\pm} \parallel^2 - 2\langle \eta_{\pm}, \Psi^{(\pm)i\dagger}\tilde{\nabla}_{i}\eta_{\pm}\rangle - \langle \eta_\pm , \Psi^{(\pm)i\dagger}\Psi^{(\pm)}_i \eta_\pm \rangle
\nonumber \\
&+& (2\kappa q - 8\kappa^2)\parallel \mathcal{A}^{(\pm)}\eta_\pm \parallel^2 \ .
\ee
Therefore,
\bea
\label{iiaextralap1b}
\frac{1}{2}\tilde{\nabla}^i \tilde{\nabla}_i ||\eta_{\pm}||^2 &=& \parallel {\hat\nabla^{(\pm)}}\eta_{\pm} \parallel^2 + \, (2\kappa q - 8\kappa^2)\parallel \mathcal{A}^{(\pm)}\eta_\pm \parallel^2
\nonumber \\
&+& \langle \eta_\pm, \bigg(\frac{1}{4}\tilde{R} + \Gamma^{i}\tilde{\nabla}_i(-\Psi^{(\pm)} - q\mathcal{A}^{(\pm)}) - \Psi^{(\pm)i\dagger}\Psi^{(\pm)}_i \bigg) \eta_\pm \rangle
\nonumber \\
&+& \langle \eta_\pm, \bigg( \Gamma^{i}(-\Psi^{(\pm)} - q\mathcal{A}^{(\pm)}) - 2\Psi^{(\pm)i\dagger}\bigg)\tilde{\nabla}_i \eta_\pm \rangle \ .
\ee
In order to simplify the expression for the Laplacian, we shall attempt to rewrite the third line in ({\ref{iiaextralap1b}}) as
\bea
\label{iiabilin}
\langle \eta_\pm, \bigg( \Gamma^{i}(-\Psi^{(\pm)} - q\mathcal{A}^{(\pm)}) - 2\Psi^{(\pm)i\dagger}\bigg)\tilde{\nabla}_i \eta_\pm \rangle &=& \langle \eta_\pm, \mathcal{F}^{(\pm)}\Gamma^{i}\tilde{\nabla}_i \eta_\pm \rangle + W^{(\pm)i}\tilde{\nabla}_i \parallel \eta_\pm \parallel^2~,
\nonumber \\
\ee
where $\mathcal{F}^{(\pm)}$ is linear in the fields and $W^{(\pm)i}$ is a vector. This expression is particularly advantageous, because the
first term on the RHS can be rewritten using the horizon
Dirac equation, and the second term is consistent with the application
of the maximum principle/integration by parts arguments which
are required for the generalized Lichnerowicz theorems. In order to rewrite ({\ref{iiabilin}}) in this fashion, note that
\bea
\Gamma^{i}(\Psi^{(\pm)} + q\mathcal{A}^{(\pm)}) &+& 2\Psi^{(\pm)i\dagger} 
\nonumber \\
&=& \big(\mp h^i \mp (q+1)\Gamma_{11}L^{i} + {1 \over 2}(q+1)\Gamma_{11}\tilde{H}^{i}{}_{\ell_1 \ell_2}\Gamma^{\ell_1 \ell_2}
+ 2q\tilde{\nabla}^i \Phi \big)
\nonumber \\
&+& \big(\pm \frac{1}{4}h_{j}\Gamma^{j} \pm (\frac{q}{2} + \frac{1}{4})\Gamma_{11} L_{j}\Gamma^{j}
\nonumber \\
&-& (\frac{q}{12} + \frac{1}{8})\Gamma_{11}\tilde{H}_{\ell_1 \ell_2 \ell_3}\Gamma^{\ell_1 \ell_2 \ell_3}- q\tilde{\nabla}_j \Phi \Gamma^{j}\big)\Gamma^{i}
\nonumber \\
&\mp &{1 \over 8}(q+1)e^{\Phi}X_{\ell_1 \ell_2}\Gamma^{i}\Gamma^{\ell_1 \ell_2} +{1 \over 96} (q+1)e^{\Phi}\tilde{G}_{\ell_1 \ell_2 \ell_3 \ell_4}\Gamma^{i}\Gamma^{\ell_1 \ell_2 \ell_3 \ell_4}
\nonumber \\
&+& (q+1)\Gamma_{11}\bigg(\pm {3 \over 4} e^{\Phi}S\Gamma^{i}
-{3 \over 8}e^{\Phi}\tilde{F}_{\ell_1 \ell_2}\Gamma^{i}\Gamma^{\ell_1 \ell_2}\bigg) \ .
\ee
One finds that (\ref{iiabilin}) is only
possible for $q=-1$ and thus we have
\bea
W^{(\pm)i} = \frac{1}{2}(2\tilde{\nabla}^i \Phi \pm h^i) \ ,
\ee
\bea
\mathcal{F}^{(\pm)} = \mp \frac{1}{4}h_{j}\Gamma^{j} - \tilde{\nabla}_{j}\Phi \Gamma^{j} + \Gamma_{11}\bigg(\pm \frac{1}{4}L_{j}\Gamma^{j} +  \frac{1}{24}\tilde{H}_{\ell_1 \ell_2 \ell_3}\Gamma^{\ell_1 \ell_2 \ell_3}\bigg) \ .
\ee
We remark that  $\dagger$ is the adjoint with respect to the $Spin(8)$-invariant inner product $\langle \phantom{i},\phantom{i} \rangle$. In order to compute the adjoints above we note that the $Spin(8)$-invariant inner product restricted to the Majorana representation is positive definite and real, and so symmetric. With respect to this the gamma matrices are Hermitian and thus the skew symmetric products $\Gamma^{[k]}$ of $k$ $Spin(8)$ gamma matrices are Hermitian for $k=0 \, (\text{mod }4)$ and $k = 1 \, (\text{mod }4)$ while they are anti-Hermitian for $k = 2 \, (\text{mod }4)$ and $k = 3 \, (\text{mod }4)$. The $\Gamma_{11}$ matrix is also Hermitian since it is a product of the first 10 gamma matrices and we take $\Gamma_0$ to be anti-Hermitian. It also follows that $\Gamma_{11}\Gamma^{[k]}$ is Hermitian for $k=0\, (\text{mod }4)$ and $k=3\, (\text{mod }4)$ and anti-Hermitian for $k= 1 \, (\text{mod }4)$ and $k=2\, (\text{mod }4)$. This also implies the following identities
\bea
\label{iiahermiden1}
\langle \eta_+, \Gamma^{[k]} \eta_+ \rangle = 0, \qquad
k = 2\, (\text{mod }4) \ {\rm and} \  k=3\, (\text{mod }4) \ ,
\ee
and
\bea
\label{iiahermiden2}
\langle \eta_+, \Gamma_{11}\Gamma^{[k]} \eta_+ \rangle = 0,
\qquad k = 1\, (\text{mod }4) \ {\rm and} \  k= 2\, (\text{mod }4) \ .
\ee
It follows that
\bea
\label{iialaplacian}
\frac{1}{2}\tilde{\nabla}^i \tilde{\nabla}_i ||\eta_{\pm}||^2 &=& \parallel {\hat\nabla^{(\pm)}}\eta_{\pm} \parallel^2 + \, (-2\kappa  - 8\kappa^2)\parallel \mathcal{A}^{(\pm)}\eta_\pm \parallel^2
+ W^{(\pm)i}\tilde{\nabla}_{i}\parallel \eta_\pm \parallel^2
\nonumber \\
&+& \langle \eta_\pm, \bigg(\frac{1}{4}\tilde{R} + \Gamma^{i}\tilde{\nabla}_i(-\Psi^{(\pm)} + \mathcal{A}^{(\pm)}) - \Psi^{(\pm)i\dagger}\Psi^{(\pm)}_i  + \mathcal{F}^{(\pm)}(-\Psi^{(\pm)} + \mathcal{A}^{(\pm)})\bigg) \eta_\pm \rangle \ .
\nonumber \\
\ee
It is also useful to evaluate ${\tilde{R}}$ using (\ref{iiafeq8}) and the dilaton field equation (\ref{iiafeq6}); we obtain
\bea
\tilde{R} &=& -\tilde{\nabla}^{i}(h_i) + \frac{1}{2}h^2 - 4(\tilde{\nabla}\Phi)^2 - 2h^{i}\tilde{\nabla}_{i}\Phi - \frac{3}{2}L^2 + \frac{5}{12}\tilde{H}^2
\nonumber \\
&+& \frac{7}{2}e^{2\Phi}S^2 - \frac{5}{4}e^{2\Phi}\tilde{F}^2 + \frac{3}{4}e^{2\Phi}X^2 - \frac{1}{48}e^{2\Phi}\tilde{G}^2 \ .
\ee
One obtains, upon using the field equations and Bianchi identities,
\bea
\label{iiaquad}
\bigg(\frac{1}{4}\tilde{R} &+& \Gamma^{i}\tilde{\nabla}_i(-\Psi^{(\pm)} + \mathcal{A}^{(\pm)}) - \Psi^{(\pm)i\dagger}\Psi^{(\pm)}_i  + \mathcal{F}^{(\pm)}(-\Psi^{(\pm)} + \mathcal{A}^{(\pm)})\bigg)\eta_\pm
\nonumber \\
&=& \bigg[ \big(\pm \frac{1}{4}\tilde{\nabla}_{\ell_1}(h_{\ell_2}) \mp \frac{1}{16}\tilde{H}^{i}{}_{\ell_1 \ell_2}L_{i}\big)\Gamma^{\ell_1 \ell_2}+  \big( \pm \frac{1}{8}\tilde{\nabla}_{\ell_1}(e^{\Phi}X_{\ell_2 \ell_3}) 
\nonumber \\
&+& \frac{1}{24}\tilde{\nabla}^{i}(e^{\Phi}\tilde{G}_{i \ell_1 \ell_2 \ell_3}) \mp \frac{1}{96}e^{\Phi}h^{i}\tilde{G}_{i \ell_1 \ell_2 \ell_3} -\frac{1}{32}e^{\Phi}X_{\ell_1 \ell_2}h_{\ell_3}  \mp \frac{1}{8}e^{\Phi}\tilde{\nabla}_{\ell_1}\Phi X_{\ell_2 \ell_3}
\nonumber \\
&-& \frac{1}{24}e^{\Phi}\tilde{\nabla}^{i}\Phi \tilde{G}_{i \ell_1 \ell_2 \ell_3}
\mp \frac{1}{32}e^{\Phi}\tilde{F}_{\ell_1 \ell_2}L_{\ell_3}
\mp \frac{1}{96}e^{\Phi}S\tilde{H}_{\ell_1 \ell_2 \ell_3} - \frac{1}{32}e^{\Phi}\tilde{F}^{i}{}_{\ell_1}\tilde{H}_{i \ell_2 \ell_3}\big)\Gamma^{\ell_1 \ell_2 \ell_3}
\nonumber \\
&+& \Gamma_{11}\bigg(\big(\mp \frac{1}{4}\tilde{\nabla}_{\ell}(e^{\Phi}S) - \frac{1}{4}\tilde{\nabla}^{i}(e^{\Phi}\tilde{F}_{i \ell}) +\frac{1}{16}e^{\Phi}S h_{\ell} \pm \frac{1}{16}e^{\Phi}h^{i}\tilde{F}_{i \ell} \pm \frac{1}{4}e^{\Phi}\tilde{\nabla}_{\ell}\Phi S
\nonumber \\
&+& \frac{1}{4}e^{\Phi}\tilde{\nabla}^{i}\Phi\tilde{F}_{i \ell} + \frac{1}{16}e^{\Phi}L^i X_{i \ell}
\mp \frac{1}{32}e^{\Phi}\tilde{H}^{i j}{}_{\ell}X_{i j}
- \frac{1}{96}e^{\Phi}\tilde{G}^{i j k}{}_{\ell}\tilde{H}_{i j k}\big)\Gamma^{\ell}
\nonumber \\
&+& \big(\mp \frac{1}{4}\tilde{\nabla}_{\ell_1}(L_{\ell_2}) - \frac{1}{8}\tilde{\nabla}^{i}(\tilde{H}_{i \ell_1 \ell_2}) + \frac{1}{4}\tilde{\nabla}^{i}\Phi \tilde{H}_{i \ell_1 \ell_2} \pm  \frac{1}{16}h^{i}\tilde{H}_{i \ell_1 \ell_2}\big)\Gamma^{\ell_1 \ell_2}
\nonumber \\
&+& \big(\pm \frac{1}{384}e^{\Phi}\tilde{G}_{\ell_1 \ell_2 \ell_3 \ell_4}L_{\ell_5} \pm  \frac{1}{192}e^{\Phi}\tilde{H}_{\ell_1 \ell_2 \ell_3}X_{\ell_4 \ell_5}
\nonumber \\
&+& \frac{1}{192}e^{\Phi}\tilde{G}^{i}{}_{\ell_1 \ell_2 \ell_3}\tilde{H}_{i \ell_4 \ell_5}\big)\Gamma^{\ell_1 \ell_2 \ell_3 \ell_4 \ell_5}\bigg)
\bigg] \eta_\pm
\nonumber \\
&+& {1 \over 2} \big(1 \mp 1\big) \bigg(h^i {\tilde{\nabla}}_i \Phi
-{1 \over 2} {\tilde{\nabla}}^i h_i \bigg) \eta_\pm \ .
\ee
Note that with the exception of the final line of the RHS of ({\ref{iiaquad}}), all terms on the RHS of the above expression
give no contribution to the second line of (\ref{iialaplacian}),
using (\ref{iiahermiden1}) and (\ref{iiahermiden2}), since all these terms in (\ref{iiaquad}) are anti-Hermitian and thus the bilinears vanish.
Furthermore, the contribution to the Laplacian of $\parallel \eta_+ \parallel^2$ from the final line of ({\ref{iiaquad}}) also vanishes;
however the final line of ({\ref{iiaquad}}) {\it does} give a contribution
to the second line of ({\ref{iialaplacian}}) in the case of the
Laplacian of $\parallel \eta_- \parallel^2$.
We  proceed
to consider the Laplacians
of $\parallel \eta_\pm \parallel^2$ separately, as the analysis
of the conditions imposed by the global properties of ${\cal{S}}$
differs slightly in the two cases.
For the Laplacian
of $\parallel \eta_+ \parallel^2$, we obtain from ({\ref{iialaplacian}}):
\bea
\label{iial1}
{\tilde{\nabla}}^{i}{\tilde{\nabla}}_{i}\parallel\eta_+\parallel^2 - (2\tilde{\nabla}^i \Phi +  h^i) {\tilde{\nabla}}_{i}\parallel\eta_+\parallel^2 = 2\parallel{\hat\nabla^{(+)}}\eta_{+}\parallel^2 - (4\kappa + 16 \kappa^2)\parallel\mathcal{A}^{(+)}\eta_+\parallel^2 \ .
\ee
This proves (\ref{iiamaxprin}).
The Laplacian of $\parallel \eta_- \parallel^2$
is calculated from ({\ref{iialaplacian}}), on taking account of the contribution to the second line of
({\ref{iialaplacian}}) from the final line of ({\ref{iiaquad}}). One
obtains
\bea
\label{iial2}
{\tilde{\nabla}}^{i} \big( e^{-2 \Phi} V_i \big)
= -2 e^{-2 \Phi} \parallel{\hat\nabla^{(-)}}\eta_{-}\parallel^2 +   e^{-2 \Phi} (4 \kappa +16 \kappa^2) \parallel\mathcal{A}^{(-)}\eta_-\parallel^2~,
\ee
where
\bea
V=-d \parallel \eta_- \parallel^2 - \parallel \eta_- \parallel^2 h \ .
\ee
This proves (\ref{iial2b}) and completes the proof. 
It should be noted that in the $\eta_-$ case, one does not have to set $q=-1$. In fact, a formula similar to (\ref{iial1}) can be established 
for arbitrary $q$. However some terms get modified and the end result does not have the simplicity of (\ref{iial1}).  For example, 
 the numerical coefficient in front of the $\parallel\mathcal{A}^{(-)}\eta_-\parallel^2$ is modified to
$-2-4\kappa q+ 16 \kappa^2+2q^2$ and of course reduces to that of (\ref{iial1}) upon setting  $q=-1$.

\section{Lichnerowicz theorem for $\mathcal{D}^{(-)}$}
In this section we shall give the proof of the Lichnerowicz type theorem decribed in section 4.9.
\bea
\nabla^{(-)}_{i}\eta_{-} = \tilde{\nabla}_{i}\eta_{-} + \Psi^{(-)}_{i} \eta_{-}  = 0 \ ,
\ee
where
\bea
\Psi^{(-)}_{i} &=& \bigg(\frac{1}{4}h_{i} + \frac{1}{16}e^{\Phi}X_{\ell_1 \ell_2}\Gamma^{\ell_1 \ell_2}\Gamma_{i} + \frac{1}{8.4!}e^{\Phi}{\tilde{G}}_{\ell_1 \ell_2 \ell_3 \ell_4}\Gamma^{\ell_1 \ell_2 \ell_3 \ell_4}\Gamma_{i}\bigg)
\cr
&+& \Gamma_{11}\bigg(\frac{1}{4}L_{i} + \frac{1}{8}{\tilde{H}}_{i \ell_1 \ell_2}\Gamma^{\ell_1 \ell_2}
- \frac{1}{8}e^{\Phi}S\Gamma_{i} - \frac{1}{16}e^{\Phi}{\tilde{F}}_{\ell_1 \ell_2}\Gamma^{\ell_1 \ell_2}\Gamma_{i}\bigg) \ ,
\ee 
and
\bea
\mathcal{A}^{(-)}\eta_{-} = \tilde{\nabla}_{i}\Phi \Gamma^{i} \eta_{-} + \rho^{(-)}\eta_{-} = 0 \ ,
\ee
where
\bea
\rho^{(-)} &=& \bigg(\frac{1}{8}e^{\Phi}X_{\ell_1 \ell_2}\Gamma^{\ell_1 \ell_2} + \frac{1}{4.4!}e^{\Phi}{\tilde{G}}_{\ell_1 \ell_2 \ell_3 \ell_4}\Gamma^{\ell_1 \ell_2 \ell_3 \ell_4}\bigg)
\cr
&+& \Gamma_{11}\bigg(- \frac{1}{2}L_{i}\Gamma^{i} - \frac{1}{12}{\tilde{H}}_{i j k}\Gamma^{i j k} + \frac{3}{4}e^{\Phi}S + \frac{3}{8}e^{\Phi}{\tilde{F}}_{i j}\Gamma^{i j}\bigg) \ .
\ee
We also rewrite the associated horizon Dirac equation (4.3) as
\bea
\mathcal{D}^{(-)}\eta_{-} = \Gamma^{i}\tilde{\nabla}_{i}\eta_{-} + \Psi^{(-)}\eta_{-} = 0 \ ,
\ee
with
\bea
\Psi^{(-)} \equiv \Gamma^{i}\Psi^{(-)}_{i} &=& \frac{1}{4}h_{i}\Gamma^{i}
+\frac{1}{4}e^{\Phi}X_{i j}\Gamma^{i j}
\cr
&+& \Gamma_{11}\bigg(- \frac{1}{4}L_{i}\Gamma^{i} - \frac{1}{8}{\tilde{H}}_{i j k}\Gamma^{i j k} + e^{\Phi}S + \frac{1}{4}e^{\Phi}{\tilde{F}}_{i j}\Gamma^{i j} \bigg) \ .
\nonumber \\
\ee
We define
\bea
\label{lichn1}
{\cal{I}} &=& \int_{{\cal{S}}} \big ( \parallel \nabla^{(-)}\eta_-\parallel^2-\parallel{\cal D}^{(-)}\eta_-\parallel^2\big)~,
 \ee
and decompose
\bea
\label{lichner1}
{\cal{I}}= {\cal{I}}_1 + {\cal{I}}_2 + {\cal{I}}_3~,
\ee
where
\bea
\label{termx1}
{\cal{I}}_1 = \int_{{\cal{S}}} \langle {\tilde{\nabla}}_i \eta_- ,
{\tilde{\nabla}}^i \eta_- \rangle
- \langle \Gamma^i {\tilde{\nabla}}_i \eta_- ,
 \Gamma^j {\tilde{\nabla}}_j \eta_- \rangle~.
 \ee
 and
 \bea
 \label{termx2}
 {\cal{I}}_2  = 2 \bigg( \int_{{\cal{S}}} \langle {\tilde{\nabla}}_i \eta_-, \Psi{}^{(-)}{}^i \eta_-  \rangle
 - \langle \Gamma^i {\tilde{\nabla}}_i \eta_- , \Psi{}^{(-)} \eta_-  \rangle \bigg)~,
\ee
and
\bea
\label{termx3}
{\cal{I}}_3 = \int_{{\cal{S}}} \langle  \Psi{}^{(-)}_i \eta_-  ,
 \Psi{}^{(-)}{}^i \eta_-  \rangle
- \langle  \Psi{}^{(-)} \eta_-  , \Psi{}^{(-)} \eta_- \rangle~.
\ee
where $ \langle\cdot, \cdot \rangle$ is the Dirac inner product of $Spin(8)$ which can be identified
with the standard Hermitian inner product on $\Lambda^*(\mathbb{C}^4)$ restricted on the real subspace of Majorana spinors and $\parallel \cdot \parallel$ is the associated norm. Therefore, $ \langle\cdot, \cdot \rangle$ is a real and positive definite. The $Spin(8)$ gamma matrices are Hermitian with respect to
$ \langle\cdot, \cdot \rangle$.
Then, on integrating by parts, one can rewrite
\bea
{\cal{I}}_1 = \int_{{\cal{S}}} -\tilde{\nabla}_{i}\langle \eta_{-}, \Gamma^{i j}\tilde{\nabla}_{j}\eta_{-} \rangle + \int_{{\cal{S}}} \langle \eta_{-}, \Gamma^{i j}\tilde{\nabla}_i \tilde{\nabla}_j \eta_{-}\rangle \ ,
\ee
and
\bea
\label{ttr1}
{\cal{I}}_2 &=& \int_{{\cal{S}}} \tilde{\nabla}_{i}\langle \eta_{-}, (\Psi^{(-)i} - \Gamma^{i}\Psi^{(-)}) \eta_{-} \rangle
\nonumber \\
&+& \int_{\cal{S}} \langle \eta_-,(\Gamma^i {\tilde{\nabla}}_i \Psi^{(-)} - ({\tilde{\nabla}}^i \Psi^{(-)}_i))  \eta_- \rangle
\nonumber \\
&+& \int_{\cal{S}} \langle \eta_- , \big( (\Psi{}^{(-)}{}^i)^\dagger - \Psi{}^{(-)}{}^i - (\Psi{}^{(-)}{}^\dagger -\Psi{}^{(-)}) \Gamma^i \big) \tn_i \eta_- \rangle
\nonumber \\
&+& \int_{\cal{S}} \langle \eta_- , \big( \Gamma^i \Psi{}^{(-)} - \Psi{}^{(-)} \Gamma^i \big) \tn_i \eta_- \rangle \ ,
\ee
and
\bea
\label{termx3b}
{\cal{I}}_3 &=& \int_{\cal{S}} \langle \eta_- , \bigg( \Psi{}^{(-)\dagger}_i \Psi{}^{(-)}{}^i - \Psi{}^{(-)}{}^\dagger \Psi{}^{(-)}\bigg)\eta_- \rangle \ .
\ee
Let us now define
\bea
{\hat\nabla^{(-)}_{i}}\eta_{-} &=& \nabla^{(-)}_{i}\eta_{-} + \kappa\Gamma_{i} \,\mathcal{A}^{(-)}\eta_{-}
\cr
&=& \tilde{\nabla}_{i}\eta_{-} + (\Psi^{(-)}_{i} + \kappa\Gamma_{i} \,\mathcal{A}^{(-)})\eta_{-} \ ,
\ee
and
\bea
\hat{\mathcal{D}}^{(-)}\eta_{-} &=& \mathcal{D}^{(-)}\eta_{-} + q\mathcal{A}^{(-)}\eta_{-}
\cr
&=& \Gamma^{i}\nabla^{(-)}_{i}\eta_{-} + (\Psi^{(-)} + q\mathcal{A}^{(-)})\eta_{-} \ ,
\ee
\bea
\label{lichn2}
\hat{\cal{I}} &=& \int_{{\cal{S}}} \big ( \parallel \hat{\nabla}^{(-)}\eta_-\parallel^2-\parallel\hat{\cal D}^{(-)}\eta_-\parallel^2\big)~,
 \ee
Replacing $\Psi^{(-)}$ and $\Psi^{(-)}_i$ with  ${\hat \Psi^{(-)}} =  \Psi^{(-)} + q\mathcal{A}^{(-)}$ and ${\hat \Psi^{(-)}_i} = \Psi^{(-)}_{i} + \kappa\Gamma_{i} \,\mathcal{A}^{(-)}$ in (5.60),(5.61) and (5.62) one obtains
\bea
\hat{\cal{I}} &=& {\cal{I}} + 2( \kappa-q)\int_{\cal{S}} \langle \eta_-,\mathcal{A}^{(-)\dagger}\hat{D}^{(-)}\eta_{-} \rangle + (8\kappa^2 + q^2 - 2q\kappa)\int_{\cal{S}} \parallel \mathcal{A}^{(-)}\eta_{-} \parallel^2 \ ,
\nonumber \\
&=& {\cal{I}}_1 + {\cal{I}}_2 + {\cal{I}}_3 \ ,
\nonumber \\
&+& 2( \kappa-q)\int_{\cal{S}} \langle \eta_-,\mathcal{A}^{(-)\dagger}\hat{D}^{(-)}\eta_{-} \rangle + (8\kappa^2 + q^2 - 2q\kappa)\int_{\cal{S}} \parallel \mathcal{A}^{(-)}\eta_{-} \parallel^2 \ .
\ee
It is straightforward to evaluate ${\cal{I}}_1$, to obtain
\bea
\label{termx2c}
{\cal{I}}_1 &=& \int_{\cal{S}} - {\tilde{\nabla}}_i \langle \eta_-, \Gamma^{ij} {\tilde{\nabla}}_j \eta_- \rangle
-{1 \over 4} \int_{\cal{S}} h^i {\tilde{\nabla}}_i \langle \eta_- , \eta_- \rangle
\nonumber \\
&+& \int_{\cal{S}} \langle \eta_- , \bigg( -{1 \over 8} h^2 + \frac{1}{2}\tilde{\nabla}^2 \Phi + \frac{1}{8}L^2 - \frac{1}{16}\tilde{H}^2
+ \frac{1}{8}e^{2\Phi}\tilde{F}^2 - \frac{1}{8}e^{2\Phi}X^2 - \frac{1}{2}e^{2\Phi}S^2
\bigg) \eta_- \rangle \ ,
\nonumber \\
&=& \int_{\cal{S}} - {\tilde{\nabla}}_i \langle \eta_-, \Gamma^{ij} {\tilde{\nabla}}_j \eta_- \rangle
-{1 \over 4} \int_{\cal{S}} h^i {\tilde{\nabla}}_i \langle \eta_- , \eta_- \rangle
\nonumber \\
&+& \int_{\cal{S}} \langle \eta_- , \bigg(-\frac{1}{8}h^2 + (\tilde{\nabla}\Phi)^2 + \frac{1}{2}h^{i}\tilde{\nabla}_{i}\Phi
+ \frac{3}{8}L^2 - \frac{5}{48}\tilde{H}^2- \frac{7}{8}e^{2\Phi}S^2 + \frac{5}{16}e^{2\Phi}\tilde{F}^2
\nonumber \\
&-& \frac{3}{16}e^{2\Phi}X^2 + \frac{1}{192}e^{2\Phi}\tilde{G}^2 \bigg)\eta_{-} \rangle \ ,
\ee
where we have used the Einstein equations ({\ref{iiafeq8}}) to compute
\bea
\tilde{R} &=& -\tilde{\nabla}^{i}(h_i) + \frac{1}{2}h_i h^i - 2\tilde{\nabla}^{i}\tilde{\nabla}_{i}\Phi - \frac{1}{2}L^i L_i + \frac{1}{4}\tilde{H}_{\ell_1 \ell_2 \ell_3}\tilde{H}^{\ell_1 \ell_2 \ell_3} 
\cr
&-& \frac{1}{2}e^{2\Phi}\tilde{F}_{\ell_1 \ell_2}\tilde{F}^{\ell_1 \ell_2} + \frac{1}{2}e^{2\Phi}X_{\ell_1 \ell_2}X^{\ell_1 \ell_2} + 2e^{2\Phi}S^2 \ ,
\ee
and the dilaton field equation to eliminate the $\tilde{\nabla}^2 \Phi$ term, and we recall. Now we evaluate,
\bea
\Gamma^{ij} {\tilde{\nabla}}_i {\tilde{\nabla}}_j \eta_- = -{1 \over 4} {\tilde{R}} \eta_- \ .
\ee
\bea
(\Psi^{(-)}{}_i^\dagger \Psi^{(-)i} - \Psi^{(-) \dagger} \Psi^{(-)}) &=& -\frac{5}{16}e^{\Phi}h^i X_{i l}\Gamma^{l} + \frac{1}{384}e^{\Phi}\tilde{G}_{\ell_1 \ell_2 \ell_3 \ell_4}h_{\ell_5}\Gamma^{\ell_1 \ell_2 \ell_3 \ell_4 \ell_5} 
\nonumber \\
&+& \frac{1}{16}e^{2\Phi}X_{\ell_1 \ell_2}X_{\ell_3 \ell_4}\Gamma^{\ell_1 \ell_2 \ell_3 \ell_4} - \frac{1}{16}e^{2\Phi}X^2 
\nonumber \\
&-& \frac{1}{4608}e^{2\Phi}\tilde{G}_{\ell_1 \ell_2 \ell_3 \ell_4}\tilde{G}_{\ell_5 \ell_6 \ell_7 \ell_8}\Gamma^{\ell_1 \ell_2 \ell_3 \ell_4 \ell_5 \ell_6 \ell_6 \ell_8} + \frac{1}{192}e^{2\Phi}\tilde{G}^2
\nonumber \\
&+& \frac{1}{8}L^2 + \frac{3}{16}L^{i}\tilde{H}_{i \ell_1 \ell_2}\Gamma^{\ell_1 \ell_2} - \frac{1}{16}e^{\Phi}S L_{i}\Gamma^{i} 
\nonumber \\
&-& \frac{1}{8}e^{\Phi}\tilde{F}_{\ell_1 \ell_2}L_{\ell_3}\Gamma^{\ell_1 \ell_2 \ell_3} + \frac{1}{16}e^{\Phi}L^{i}\tilde{F}_{i l}\Gamma^{l} 
\nonumber \\
&+& \frac{1}{8}\tilde{H}^i{}_{\ell_1 \ell_2}\tilde{H}_{i \ell_3 \ell_4}\Gamma^{\ell_1 \ell_2 \ell_3 \ell_4} - \frac{1}{16}\tilde{H}^2
\nonumber \\
&-& \frac{3}{64}e^{\Phi}\tilde{F}_{\ell_1 \ell_2}\tilde{H}_{\ell_3 \ell_4 \ell_5} \Gamma^{\ell_1 \ell_2 \ell_3 \ell_4 \ell_5} + \frac{13}{32}e^{\Phi}\tilde{F}^{i j}\tilde{H}_{i j l}\Gamma^{l} - \frac{7}{8}e^{2\Phi}S^2 
\nonumber \\
&+& \frac{1}{16}e^{2\Phi}\tilde{F}_{\ell_1 \ell_2}\tilde{F}_{\ell_3 \ell_4}\Gamma^{\ell_1 \ell_2 \ell_3 \ell_4}
- \frac{1}{16}e^{2\Phi}\tilde{F}^2
\nonumber \\
&+& \Gamma_{11}\bigg( \frac{1}{8}L^i h_{i} + \frac{3}{32}e^{\Phi}\tilde{F}_{\ell_1 \ell_2}h_{\ell_3}\Gamma^{\ell_1 \ell_2 \ell_3} + \frac{1}{32}e^{\Phi}L_{\ell_1}X_{\ell_2 \ell_3}\Gamma^{\ell_1 \ell_2 \ell_3}
\nonumber \\
&+& \frac{11}{32}e^{\Phi}\tilde{H}^i{}_{\ell_1 \ell_2}X_{i \ell_3}\Gamma^{\ell_1 \ell_2 \ell_3} + \frac{1}{8}e^{2\Phi}\tilde{F}_{\ell_1 \ell_2}X_{\ell_3 \ell_4}\Gamma^{\ell_1 \ell_2 \ell_3 \ell_4} 
\nonumber \\
&-& \frac{1}{8}e^{2\Phi}\tilde{F}^{i j}X_{i j} - \frac{1}{768}e^{\Phi}\tilde{G}_{\ell_1 \ell_2 \ell_3 \ell_4}\tilde{H}_{\ell_5 \ell_6 \ell_7}\Gamma^{\ell_1 \ell_2 \ell_3 \ell_4 \ell_5 \ell_6 \ell_7} 
\nonumber \\
&+& \frac{1}{4}e^{\Phi}L^i X_{i l}\Gamma^{l} - \frac{1}{64}e^{\Phi}\tilde{G}_{i j \ell_1 \ell_2}\tilde{H}^{i j}{}_{\ell_3}\Gamma^{\ell_1 \ell_2 \ell_3} 
\nonumber \\
&-& \frac{1}{96}e^{\Phi}L^i \tilde{G}_{i \ell_1 \ell_2 \ell_3}\Gamma^{\ell_1 \ell_2 \ell_3} + \frac{1}{8}L_i h_j \Gamma^{i j} + \frac{1}{16}\tilde{H}_{\ell_1 \ell_2 \ell_3}h_{\ell_4}\Gamma^{\ell_1 \ell_2 \ell_3 \ell_4}\bigg) \ ,
\nonumber \\
\ee
\bea
{\cal{I}}_3 &=& \int_{\cal{S}} \langle \eta_{-}, \bigg( -\frac{1}{16}e^{2\Phi}X^2 + \frac{1}{192}e^{2\Phi}\tilde{G}^2 + \frac{1}{8}L^2
- \frac{1}{16}\tilde{H}^2 - \frac{7}{8}e^{2\Phi}S^2 - \frac{1}{16}e^{2\Phi}\tilde{F}^2
\nonumber \\
&+& (-\frac{5}{16}e^{\Phi}h^{i}X_{i l} - \frac{1}{16}e^{\Phi}S L_{l}
+ \frac{1}{16}e^{\Phi}L^i \tilde{F}_{i l} + \frac{13}{32}e^{\Phi}\tilde{F}^{i j}\tilde{H}_{i j l})\Gamma^{l}
\nonumber \\
&+& (\frac{1}{16}e^{2\Phi}X_{\ell_1 \ell_2}X_{\ell_3 \ell_4} + \frac{1}{8}\tilde{H}^i{}_{\ell_1 \ell_2}\tilde{H}_{i \ell_3 \ell_4} + \frac{1}{16}e^{2\Phi}\tilde{F}_{\ell_1 \ell_2}\tilde{F}_{\ell_3 \ell_4})\Gamma^{\ell_1 \ell_2 \ell_3 \ell_4}
\nonumber \\
&+& (\frac{1}{384}e^{\Phi}\tilde{G}_{\ell_1 \ell_2 \ell_3 \ell_4}h_{\ell_5} - \frac{3}{64}\tilde{F}_{\ell_1 \ell_2}\tilde{H}_{\ell_3 \ell_4 \ell_5})\Gamma^{\ell_1 \ell_2 \ell_3 \ell_4 \ell_5} 
\nonumber \\
&-& \frac{1}{4608}e^{2\Phi}\tilde{G}_{\ell_1 \ell_2 \ell_3 \ell_4}\tilde{G}_{\ell_5 \ell_6 \ell_7 \ell_8}\Gamma^{\ell_1 \ell_2 \ell_3 \ell_4 \ell_5 \ell_6 \ell_7 \ell_8} \bigg) \eta_{-} \rangle
\nonumber \\
&+& \int_{\cal{S}} \langle \eta_{-}, \Gamma_{11}\bigg( \frac{1}{8}L_{i}h^{i} - \frac{1}{8}e^{2\Phi}\tilde{F}^{i j}X_{i j} + (\frac{3}{32}e^{\Phi}\tilde{F}_{\ell_1 \ell_2}h_{\ell_3} + \frac{1}{32}e^{\Phi}L_{\ell_1}X_{\ell_2 \ell_3} 
\nonumber \\
&+& \frac{11}{32}e^{\Phi}\tilde{H}^i{}_{\ell_1 \ell_2}X_{i \ell_3} 
- \frac{1}{64}e^{\Phi}\tilde{G}^{i j \ell_1 \ell_2}\tilde{H}^{i j}{}_{\ell_3} - \frac{1}{96}e^{\Phi}L^i \tilde{G}_{i \ell_1 \ell_2 \ell_3})\Gamma^{\ell_1 \ell_2 \ell_3} 
\nonumber \\
&+& (\frac{1}{8}e^{2\Phi}\tilde{F}_{\ell_1 \ell_2}X_{\ell_3 \ell_4} + \frac{1}{16}\tilde{H}_{\ell_1 \ell_2 \ell_3}h_{\ell_4})\Gamma^{\ell_1 \ell_2 \ell_3 \ell_4}- \frac{1}{768}\tilde{G}_{\ell_1 \ell_2 \ell_3 \ell_4}\tilde{H}_{\ell_5 \ell_6 \ell_7}\Gamma^{\ell_1 \ell_2 \ell_3 \ell_4 \ell_5 \ell_6 \ell_7}\bigg)\eta_{-} \rangle \ ,
\nonumber \\
\ee
where we have made use of the identities
\bea
\langle \eta_{-}, \Gamma^{\ell_1 \ell_2} \eta_{-} \rangle = \langle \eta_{-}, \Gamma^{\ell_1 \ell_2 \ell_3} \eta_{-} \rangle = 0 \ ,
\ee
and
\bea
\langle \eta_{-}, \Gamma_{11}\Gamma^{l} \eta_{-} \rangle = \langle \eta_{-}, \Gamma_{11}\Gamma^{\ell_1 \ell_2}\eta_{-} \rangle = 0 \ ,
\ee
\bea
\Gamma^i {\tilde{\nabla}}_i \Psi^{(-)} - ({\tilde{\nabla}}^i \Psi^{(-)}_i)
&=& \frac{1}{4}{\tilde{\nabla}}_{i}(h_j)\Gamma^{i j} + \frac{3}{16}{\tilde{\nabla}}_{i}(e^{\Phi}X_{\ell_1 \ell_2})\Gamma^{i \ell_1 \ell_2} + \frac{5}{8}{\tilde{\nabla}}^{i}(e^{\Phi}X_{i l})\Gamma^{l} 
\cr
&-& \frac{1}{8.4!}{\tilde{\nabla}}_{i}(e^{\Phi}\tilde{G}_{\ell_1 \ell_2 \ell_3 \ell_4})\Gamma^{i \ell_1 \ell_2 \ell_3 \ell_4} + \frac{1}{2.4!}{\tilde{\nabla}}^{i}(e^{\Phi}\tilde{G}_{i \ell_1 \ell_2 \ell_3})\Gamma^{\ell_1 \ell_2 \ell_3}
\cr
&+& \Gamma_{11}\bigg(\frac{1}{4}{\tilde{\nabla}}_{i}(L_j)\Gamma^{i j} + \frac{1}{8}{\tilde{\nabla}}_{i}(\tilde{H}_{\ell_1 \ell_2 \ell_3})\Gamma^{i \ell_1 \ell_2 \ell_3} 
\nonumber \\
&+& \frac{1}{4}{\tilde{\nabla}}^{i}(\tilde{H}_{i \ell_1 \ell_2})\Gamma^{\ell_1 \ell_2}- \frac{7}{8}{\tilde{\nabla}}_{i}(e^{\Phi}S)\Gamma^{i} 
\cr
&-&\frac{3}{16}{\tilde{\nabla}}_{i}(e^{\Phi}\tilde{F}_{\ell_1 \ell_2})\Gamma^{i \ell_1 \ell_2} - \frac{5}{8}{\tilde{\nabla}}^{i}(e^{\Phi}\tilde{F}_{i l})\Gamma^{l} \bigg) \ .
\ee
In order to compute ${\cal{I}}_2$ we note that
\bea
\langle \eta_{-}, \bigg(\Gamma^i {\tilde{\nabla}}_i \Psi^{(-)} - ({\tilde{\nabla}}^i \Psi^{(-)}_i)\bigg) \eta_{-}\rangle &=& \langle \eta_{-}, \bigg(\frac{5}{8}{\tilde{\nabla}}^{i}(e^{\Phi}X_{i l})\Gamma^{l} -\frac{1}{8.4!}{\tilde{\nabla}}_{i}(e^{\Phi}\tilde{G}_{\ell_1 \ell_2 \ell_3 \ell_4})\Gamma^{i \ell_1 \ell_2 \ell_3 \ell_4}\bigg)\eta_{-}\rangle
\cr
&+& \langle \eta_{-}, \Gamma_{11}\bigg(\frac{1}{8}{\tilde{\nabla}}_{i}(\tilde{H}_{\ell_1 \ell_2 \ell_3})\Gamma^{i \ell_1 \ell_2 \ell_3} 
\nonumber \\
&-& \frac{3}{16}{\tilde{\nabla}}_{i}(e^{\Phi}\tilde{F}_{\ell_1 \ell_2})\Gamma^{i \ell_1 \ell_2}\bigg)\eta_{-} \rangle \ .
\ee
On imposing the Bianchi identities and the field equations, 
\bea
\tilde{\nabla}^{i}(e^{\Phi}X_{i l})\Gamma^{l} &=& e^{\Phi}\tilde{\nabla}^{i}\Phi X_{i l}\Gamma^{l} + e^{\Phi}\tilde{\nabla}^{i}(X_{i l})\Gamma^{l} \ ,
\nonumber \\ 
&=& e^{\Phi}\tilde{\nabla}^{i}\Phi X_{i l}\Gamma^{l} - \frac{1}{144}e^{\Phi}\tilde{G}_{\ell_1 \ell_2 \ell_3 \ell_4}\tilde{H}_{\ell_5 \ell_6 \ell_7}\Gamma_{11}\Gamma^{\ell_1 \ell_2 \ell_3 \ell_4 \ell_5 \ell_6 \ell_7} \ ,
\ee
\bea
\tilde{\nabla}_{i}(e^{\Phi}\tilde{G}_{\ell_1 \ell_2 \ell_3 \ell_4})\Gamma^{i \ell_1 \ell_2 \ell_3 \ell_4} = e^{\Phi}\tilde{\nabla}_{i}\Phi \, \tilde{G}_{\ell_1 \ell_2 \ell_3 \ell_4}\Gamma^{i \ell_1 \ell_2 \ell_3 \ell_4} + 2e^{\Phi}\tilde{H}_{\ell_1 \ell_2 \ell_3}\tilde{F}_{\ell_4 \ell_5}\Gamma^{\ell_1 \ell_2 \ell_3 \ell_4 \ell_5} \ ,
\ee
\bea
\tilde{\nabla}_{i}(e^{\Phi}\tilde{F}_{\ell_1 \ell_2})\Gamma^{\ell_1 \ell_2} = e^{\Phi}\tilde{\nabla}_{i}\Phi \tilde{F}_{\ell_1 \ell_2}\Gamma^{i \ell_1 \ell_2} \ ,
\ee
One obtains the following expression
\bea
\langle \eta_{-}, (\Gamma^i {\tilde{\nabla}}_i \Psi^{(-)} - ({\tilde{\nabla}}^i \Psi^{(-)}_i)) \eta_{-}\rangle &=& \langle \eta_{-}, \bigg(\frac{5}{8}e^{\Phi}\tilde{\nabla}^{i}\Phi X_{i l} 
- \frac{1}{8.4!}e^{\Phi}\tilde{\nabla}_{i}\tilde{G}_{\ell_1 \ell_2 \ell_3 \ell_4}\Gamma^{i \ell_1 \ell_2 \ell_3 \ell_4}
\nonumber \\
&-& \frac{1}{4.4!}e^{\Phi}\tilde{H}_{\ell_1 \ell_2 \ell_3}\tilde{F}_{\ell_4 \ell_5}\Gamma^{\ell_1 \ell_2 \ell_3 \ell_4 \ell_5}\bigg) \eta_{-} \rangle 
\nonumber \\
&+& \langle \eta_{-}, \Gamma_{11}\bigg(-\frac{5}{1152}e^{\Phi}\tilde{G}_{\ell_1 \ell_2 \ell_3 \ell_4}\tilde{H}_{\ell_5 \ell_6 \ell_7}\Gamma^{\ell_1 \ell_2 \ell_3 \ell_4 \ell_5 \ell_6 \ell_7} 
\nonumber \\
&-& \frac{3}{16}e^{\Phi}\tilde{\nabla}_{i}\tilde{F}_{\ell_1 \ell_2}\Gamma^{i \ell_1 \ell_2}\bigg)\eta_{-} \rangle \ ,
\ee
\bea
(\Psi^{(-)i \dagger} - \Psi^{(-)i} - (\Psi^{(-)\dagger} - \Psi^{(-)})\Gamma^{i}) &=& -\frac{1}{16}e^{\Phi}X_{\ell_1 \ell_2}(\Gamma^{i}\Gamma^{\ell_1 \ell_2} + \Gamma^{\ell_1 \ell_2}\Gamma^{i}) 
\cr
&+& \frac{1}{2}e^{\Phi}X_{\ell_1 \ell_2}\Gamma^{\ell_1 \ell_2}\Gamma^{i}
\cr
&+& \frac{1}{8.4!}e^{\Phi}\tilde{G}_{\ell_1 \ell_2 \ell_3 \ell_4}(\Gamma^{i}\Gamma^{\ell_1 \ell_2 \ell_3 \ell_4} - \Gamma^{\ell_1 \ell_2 \ell_3 \ell_4}\Gamma^{i})
\cr
&+& \Gamma_{11}\bigg(-\frac{1}{2}L_{j}\Gamma^{j}\Gamma^{i} - \frac{1}{4}\tilde{H}^i{}_{\ell_1 \ell_2}\Gamma^{\ell_1 \ell_2} 
\cr
&+& \frac{1}{4}e^{\Phi}S\Gamma^{i}
- \frac{1}{16}e^{\Phi}\tilde{F}_{\ell_1 \ell_2}(\Gamma^{i}\Gamma^{\ell_1 \ell_2} - \Gamma^{\ell_1 \ell_2}\Gamma^{i}) 
\nonumber \\
&+& \frac{1}{2}e^{\Phi}\tilde{F}_{\ell_1 \ell_2}\Gamma^{\ell_1 \ell_2}\Gamma^{i} \bigg) \ ,
\ee
\bea
(\Gamma^{i}\Psi^{(-)} - \Psi^{(-)}\Gamma^{i}) &=& \frac{1}{4}h_{j}(\Gamma^{i}\Gamma^{j} - \Gamma^{j}\Gamma^{i}) + \frac{1}{4}e^{\Phi}X_{\ell_1 \ell_2}(\Gamma^{i}\Gamma^{\ell_1 \ell_2} - \Gamma^{\ell_1 \ell_2}\Gamma^{i})
\cr
&+& \Gamma_{11}\bigg(\frac{1}{4}L_{j}(\Gamma^{i}\Gamma^{j} + \Gamma^{j}\Gamma^{i})
+ \frac{1}{8}\tilde{H}_{\ell_1 \ell_2 \ell_3}(\Gamma^
{i}\Gamma^{\ell_1 \ell_2 \ell_3} + \Gamma^{\ell_1 \ell_2 \ell_3}\Gamma^{i}) 
\cr
&-& 2e^{\Phi} S \Gamma^{i} - \frac{1}{4}e^{\Phi}\tilde{F}_{\ell_1 \ell_2}(\Gamma^{i}\Gamma^{\ell_1 \ell_2} + \Gamma^{\ell_1 \ell_2}\Gamma^{i}) \bigg) \ .
\ee
Also one has
\bea
(\Psi^{(-)i \dagger} - \Psi^{(-)i} &-& (\Psi^{(-)\dagger} - \Psi^{(-)})\Gamma^{i})\tilde{\nabla}_{i}\eta_{-} + (\Gamma^{i}\Psi^{(-)} - \Psi^{(-)}\Gamma^{i})\tilde{\nabla}_{i}\eta_{-}  
\nonumber \\
&=& \bigg( \big( -\frac{1}{2}h_{j}\Gamma^{j} + \frac{3}{8}e^{\Phi}X_{\ell_1 \ell_2}\Gamma^{\ell_1 \ell_2} - \frac{1}{4.4!}e^{\Phi}\tilde{G}_{\ell_1 \ell_2 \ell_3 \ell_4}\Gamma^{\ell_1 \ell_2 \ell_3 \ell_4} 
\nonumber \\
&+& \Gamma_{11}(\frac{1}{4}\tilde{H}_{\ell_1 \ell_2 \ell_3}\Gamma^{\ell_1 \ell_2 \ell_3} - \frac{7}{4}e^{\Phi}S + \frac{5}{8}e^{\Phi}\tilde{F}_{\ell_1 \ell_2}\Gamma^{\ell_1 \ell_2}) \big)\Gamma^{i}
\nonumber \\
&+& \frac{1}{2}h^{i} + \frac{3}{4}e^{\Phi}X^{i}{}_{l}\Gamma^{l} + \Gamma_{11}(\frac{1}{2}L_{j}\Gamma^{i j} + \frac{1}{4}\tilde{H}_{\ell_1 \ell_2 \ell_3}\Gamma^{i \ell_1 \ell_2 \ell_3} 
\nonumber \\
&-& \frac{1}{4}\tilde{H}^{i}{}_{\ell_1 \ell_2}\Gamma^{\ell_1 \ell_2} - \frac{5}{8}e^{\Phi}\tilde{F}_{\ell_1 \ell_2}\Gamma^{i \ell_1 \ell_2}) \bigg) \tilde{\nabla}_{i}\eta_{-} \ .
\ee
Note that
\bea
\int_{\cal{S}} \langle \eta_{-}, h^{i}\tilde{\nabla}_{i}\eta_{-} \rangle = \frac{1}{2}\int_{\cal{S}}h^{i}\tilde{\nabla}_{i}\langle \eta_{-}, \eta_{-} \rangle \ .
\ee
On integrating by parts
\bea
\int_{\cal{S}} \langle \eta_{-}, e^{\Phi}X^{i}{}_{l}\Gamma^{l}\tilde{\nabla}_{i}\eta_{-} \rangle = -\frac{1}{2}\int_{\cal{S}} \langle \eta_{-},\tilde{\nabla}_{i}(e^{\Phi}X^{i}{}_{l})\Gamma^{l}\eta_{-} \rangle \ .
\ee
The Algebraic KSE gives
\bea
\mathcal{A}^{(-)\dagger}\mathcal{A}^{(-)} &=& (\tilde{\nabla}\Phi)^2 + \frac{1}{2}e^{\Phi}\tilde{\nabla}^{i}X_{i \ell}\Gamma^{\ell} + \frac{1}{48}e^{\Phi}\tilde{\nabla}_{i}\Phi\tilde{G}_{\ell_1 \ell_2 \ell_3 \ell_4}\Gamma^{i \ell_1 \ell_2 \ell_3 \ell_4} 
\nonumber \\
&-& \frac{1}{64}e^{2\Phi}X_{\ell_1 \ell_2}X_{\ell_3 \ell_4}\Gamma^{\ell_1 \ell_2 \ell_3 \ell_4} + \frac{1}{32}e^{2\Phi}X^2 - \frac{1}{48}e^{2\Phi}\tilde{G}^{i}{}_{\ell_1 \ell_2 \ell_3}X_{i \ell_4}\Gamma^{\ell_1 \ell_2 \ell_3 \ell_4} 
\nonumber \\
&+& \frac{1}{9216}e^{2\Phi}\tilde{G}_{\ell_1 \ell_2 \ell_3 \ell_4}\tilde{G}_{\ell_5 \ell_6 \ell_7 \ell_8}\Gamma^{\ell_1 \ell_2 \ell_3 \ell_4 \ell_5 \ell_6 \ell_7 \ell_8}- \frac{1}{128}e^{2\Phi}\tilde{G}^{i j}{}_{\ell_1 \ell_2}\tilde{G}_{i j \ell_3 \ell_4}\Gamma^{\ell_1 \ell_2 \ell_3 \ell_4} 
\nonumber \\
&+& \frac{1}{384}e^{2\Phi}\tilde{G}^2 + \frac{1}{4}L^2 - \frac{1}{12}\tilde{H}_{\ell_1 \ell_2 \ell_3}L_{\ell_4}\Gamma^{\ell_1 \ell_2 \ell_3 \ell_4}
\nonumber \\
&-& \frac{3}{4}e^{\Phi}S L_{\ell}\Gamma^{\ell} - \frac{3}{4}e^{\Phi}L^{i}\tilde{F}_{i \ell}\Gamma^{\ell} - \frac{1}{16}\tilde{H}^{i}{}_{\ell_1 \ell_2}\tilde{H}_{i \ell_3 \ell_4}\Gamma^{\ell_1 \ell_2 \ell_3 \ell_4} + \frac{1}{24}\tilde{H}^2 
\nonumber \\
&+& \frac{1}{16}e^{\Phi}\tilde{F}_{\ell_1 \ell_2}\tilde{H}_{\ell_3 \ell_4 \ell_5}\Gamma^{\ell_1 \ell_2 \ell_3 \ell_4 \ell_5}
- \frac{3}{8}e^{\Phi}\tilde{F}^{i j}\tilde{H}_{i j \ell}\Gamma^{\ell} + \frac{9}{16}e^{2\Phi}S^2 
\nonumber \\
&-& \frac{9}{64}e^{2\Phi}\tilde{F}_{\ell_1 \ell_2}\tilde{F}_{\ell_3 \ell_4}\Gamma^{\ell_1 \ell_2 \ell_3 \ell_4} + \frac{9}{32}e^{2\Phi}\tilde{F}^2 
\nonumber \\
&+& \Gamma_{11}\bigg(\tilde{\nabla}^{i}\Phi L_{i} + \frac{1}{6}\tilde{\nabla}_{i}\Phi \tilde{H}_{\ell_1 \ell_2 \ell_3}\Gamma^{i \ell_1 \ell_2 \ell_3} - \frac{3}{4}e^{\Phi}\tilde{\nabla}_{i}\Phi \tilde{F}_{\ell_1 \ell_2}\Gamma^{i \ell_1 \ell_2} + \frac{1}{8}e^{\Phi}L_{\ell_1}X_{\ell_2 \ell_3}\Gamma^{\ell_1 \ell_2 \ell_3}
\nonumber \\
&-& \frac{1}{8}e^{\Phi}\tilde{H}^{i}{}_{\ell_1 \ell_2}X_{i \ell_3}\Gamma^{\ell_1 \ell_2 \ell_3} - \frac{3}{32}\tilde{F}_{\ell_1 \ell_2}X_{\ell_3 \ell_4}\Gamma^{\ell_1 \ell_2 \ell_3 \ell_4} + \frac{3}{16}e^{2\Phi}\tilde{F}^{i j}X_{i j} 
\nonumber \\
&+& \frac{1}{24}e^{\Phi}L^{i}\tilde{G}_{i \ell_1 \ell_2 \ell_3}\Gamma^{\ell_1 \ell_2 \ell_3}
- \frac{1}{576}e^{\Phi}\tilde{G}_{\ell_1 \ell_2 \ell_3 \ell_4}\tilde{H}_{\ell_5 \ell_6 \ell_7}\Gamma^{\ell_1 \ell_2 \ell_3 \ell_4 \ell_5 \ell_6 \ell_7} 
\nonumber \\
&+& \frac{1}{16}e^{\Phi}\tilde{G}^{i j}{}_{\ell_1 \ell_2}\tilde{H}_{i j \ell_3}\Gamma^{\ell_1 \ell_2 \ell_3} + \frac{1}{64}e^{2\Phi}S \tilde{G}_{\ell_1 \ell_2 \ell_3 \ell_4}\Gamma^{\ell_1 \ell_2 \ell_3 \ell_4} 
\nonumber \\
&+& \frac{1}{16}e^{2\Phi}\tilde{F}^{i}{}_{\ell_1}\tilde{G}_{i \ell_2 \ell_3 \ell_4}\Gamma^{\ell_1 \ell_2 \ell_3 \ell_4}\bigg) \ .
\ee
Upon comparing terms with $\parallel \mathcal{A}^{(-)}\eta_{-} \parallel^2$ as above, one obtains the following expression
\bea
\hat{\cal{I}} = (-1 - 2\kappa q + 8 \kappa^2 + q^2)\int_{\cal{S}} \parallel \mathcal{A}^{(-)}\eta_{-} \parallel^2 +  (\kappa - q)\int_{\cal{S}} \langle \eta_{-}, \Psi \hat{\cal D}^{(-)} \eta_{-} \rangle \ .
\ee
where
\bea
\Psi &=& -\frac{1}{2}h_{j}\Gamma^{j} + 2q \tilde{\nabla}_{j}\Phi \Gamma^{j} + (-\frac{q}{8}-\frac{1}{8})e^{\Phi}X_{\ell_1 \ell_2}\Gamma^{\ell_1 \ell_2} + (\frac{q}{4.4!}+\frac{1}{4.4!})e^{\Phi}\tilde{G}_{\ell_1 \ell_2 \ell_3 \ell_4}\Gamma^{\ell_1 \ell_2 \ell_3 \ell_4}
\nonumber \\
&+& \Gamma_{11}\bigg( (\frac{3q}{4} + \frac{3}{4})e^{\Phi}S
+ (-\frac{3q}{8} - \frac{3}{8})e^{\Phi}\tilde{F}_{\ell_1 \ell_2}\Gamma^{\ell_1 \ell_2} + (q+\frac{1}{2})L_{j}\Gamma^{j}
+ (\frac{q}{6} + \frac{1}{4})\tilde{H}_{\ell_1 \ell_2 \ell_3}\Gamma^{\ell_1 \ell_2 \ell_3} \bigg) \ .
\nonumber \\
\ee
$\kappa$ and $q$ are arbitrary but we require $-1 - 2\kappa q + 8 \kappa^2 + q^2 < 0$ so that the coefficient of the first term is negative and thus we are able to prove the required result. The value of $q$ was fixed in section 4.11 by requiring that certain terms can be written as field bilinears. If we now take $q=-1$ we obtain
\bea
\hat{\cal{I}} = (2\kappa + 8 \kappa^2)\int_{\cal{S}} \parallel \mathcal{A}^{(-)}\eta_{-} \parallel^2 + (\kappa + 1)\int_{\cal{S}} \langle \eta_{-}, \Psi \hat{\cal D}^{(-)} \eta_{-} \rangle \ ,
\ee
where
\bea
\Psi = -\frac{1}{2}h_{j}\Gamma^{j} - 2 \tilde{\nabla}_{j}\Phi \Gamma^{j} + \Gamma_{11}\bigg(-\frac{1}{2}L_{j}\Gamma^{j}
+ \frac{1}{12}\tilde{H}_{\ell_1 \ell_2 \ell_3}\Gamma^{\ell_1 \ell_2 \ell_3} \bigg) \ .
\ee
Now we require $2\kappa + 8\kappa^2 < 0$ and thus we have $- \frac{1}{4} < \kappa < 0$. If we now choose $\kappa = - \frac{1}{8}$ so that $q=8\kappa$ and the Dirac operator is associated with the covariant derivative.
\bea
\hat{\cal{I}} = -\frac{1}{8}\int_{\cal{S}} \parallel \mathcal{A}^{(-)}\eta_{-} \parallel^2 + \frac{7}{8}\int_{\cal{S}} \langle \eta_{-}, \Psi \hat{\cal D}^{(-)} \eta_{-} \rangle \ .
\ee
Now supppose that we impose the improved horizon Dirac equation $(5.61)$, $\hat{\cal D}^{(-)}\eta_{-} = 0$. Then (5.102) implies that
\bea
\int_{{\cal{S}}} \parallel \hat{\nabla}^{(-)}\eta_-\parallel^2 = -\frac{1}{8}\int_{\cal{S}} \parallel \mathcal{A}^{(-)}\eta_{-} \parallel^2 \ .
\ee
As the LHS is non-negative and the RHS is non-positive, both sides must vanish. Therefore $\eta_-$ is a Killing spinor, ${\hat\nabla^{(-)}}\eta_{-} = \mathcal{A}^{(-)}\eta_- = 0$ which is equivalent to $\nabla^{(-)}\eta_{-} = \mathcal{A}^{(-)}\eta_- = 0$.
	\chapter{Massive IIA Supergravity Calculations}

In this Appendix, we present  technical details of the analysis of the KSE
for the near-horizon solutions in massive IIA supergravity.

\section{Integrability}
First we will state the supercovariant connection $\cal{R}_{\mu \nu}$ given by,
\bea
[{\cal D}_{\mu}, {\cal D}_{\nu}]\epsilon \equiv \cal{R}_{\mu \nu}\epsilon \ ,
\ee 
where,
\bea
{\cal{R}}_{\mu \nu} &=& \frac{1}{4}R_{\mu \nu, \rho \sigma}\Gamma^{\rho \sigma} + \frac{1}{8}e^{\Phi} m\Gamma_{\nu}  \nabla_{\mu}{\Phi}+\frac{1}{192}e^{\Phi}\Gamma_{\nu}\,^{\rho \sigma \kappa \lambda}  \nabla_{\mu}{G_{\rho \sigma \kappa \lambda}} 
\nonumber \\
&+& \frac{1}{192}e^{\Phi}G^{\rho \sigma \kappa \lambda} \Gamma_{\nu \rho \sigma \kappa \lambda}  \nabla_{\mu}{\Phi} - \frac{1}{48}e^{\Phi}\Gamma^{\rho \sigma \kappa}  \nabla_{\mu}{G_{\nu \rho \sigma \kappa}} - \frac{1}{48}e^{\Phi}G_{\nu}\,^{\rho \sigma \kappa} \Gamma_{\rho \sigma \kappa}  \nabla_{\mu}{\Phi} 
\nonumber \\
&-& \frac{1}{8}e^{\Phi}  m \Gamma_{\mu}  \nabla_{\nu}{\Phi} - \frac{1}{192}e^{\Phi}\Gamma_{\mu}\,^{\rho \sigma \kappa \lambda}  \nabla_{\nu}{G_{\rho \sigma \kappa \lambda}} - \frac{1}{192}e^{\Phi}G^{\rho \sigma \kappa \lambda} \Gamma_{\mu \rho \sigma \kappa \lambda}  \nabla_{\nu}{\Phi}
\nonumber \\
&+&\frac{1}{48}e^{\Phi}\Gamma^{\rho \sigma \kappa}  \nabla_{\nu}{G_{\mu \rho \sigma \kappa}}+\frac{1}{48}e^{\Phi} G_{\mu}\,^{\rho \sigma \kappa} \Gamma_{\rho \sigma \kappa}  \nabla_{\nu}{\Phi} - \frac{1}{8}H_{\mu}\,^{\rho \sigma} H_{\nu \rho}\,^{\kappa} \Gamma_{\sigma \kappa} 
\nonumber \\
&-& \frac{1}{64}e^{\Phi}F^{\rho \sigma} H_{\mu}\,^{\kappa \lambda} \Gamma_{\nu \rho \sigma \kappa \lambda} +\frac{1}{8}e^{\Phi} F^{\rho \sigma} H_{\mu \nu \rho} \Gamma_{\sigma} +\frac{1}{32}e^{\Phi}F^{\rho \sigma} H_{\mu \rho \sigma} \Gamma_{\nu} 
\nonumber \\
&+&\frac{1}{32}e^{\Phi}F_{\nu}\,^{\rho} H_{\mu}\,^{\sigma \kappa} \Gamma_{\rho \sigma \kappa} +\frac{1}{32}e^{2\Phi}\Gamma_{\mu \nu}  {m}^{2}+\frac{1}{384}e^{2\Phi} m G^{\rho \sigma \kappa \lambda} \Gamma_{\mu \nu \rho \sigma \kappa \lambda} 
\nonumber \\
&-& \frac{1}{192}e^{2\Phi} m G_{\nu}\,^{\rho \sigma \kappa} \Gamma_{\mu \rho \sigma \kappa}  +\frac{1}{18432} e^{2\Phi} G^{\rho \sigma \kappa \lambda} G^{\tau h i j} \Gamma_{\mu \nu \rho \sigma \kappa \lambda \tau h i j} 
\nonumber \\
&-& \frac{1}{4608}e^{2\Phi} G_{\nu}\,^{\rho \sigma \kappa} G^{\lambda \tau h i} \Gamma_{\mu \rho \sigma \kappa \lambda \tau h i}  - \frac{1}{384}e^{2\Phi} G_{\nu}\,^{\rho \sigma \kappa} G_{\rho}\,^{\lambda \tau h} \Gamma_{\mu \sigma \kappa \lambda \tau h}  
\nonumber \\
&-& \frac{1}{256}e^{2\Phi}G^{\rho \sigma \kappa \lambda} G_{\rho \sigma}\,^{\tau h} \Gamma_{\mu \nu \kappa \lambda \tau h} +\frac{1}{384}e^{2\Phi}G_{\mu}\,^{\rho \sigma \kappa} G_{\rho}\,^{\lambda \tau h} \Gamma_{\nu \sigma \kappa \lambda \tau h} 
\nonumber \\
&+&\frac{1}{128}e^{2\Phi}G_{\nu}\,^{\rho \sigma \kappa} G_{\rho \sigma}\,^{\lambda \tau} \Gamma_{\mu \kappa \lambda \tau}  - \frac{1}{128}e^{2\Phi}G_{\mu}\,^{\rho \sigma \kappa} G_{\rho \sigma}\,^{\lambda \tau} \Gamma_{\nu \kappa \lambda \tau} 
\nonumber \\
&+&\frac{1}{192}e^{2\Phi} G_{\nu}\,^{\rho \sigma \kappa} G_{\rho \sigma \kappa}\,^{\lambda} \Gamma_{\mu \lambda} +\frac{1}{768}e^{2\Phi}\Gamma_{\mu \nu}  {G}^{2} - \frac{1}{192}e^{2\Phi}G_{\mu}\,^{\rho \sigma \kappa} G_{\rho \sigma \kappa}\,^{\lambda} \Gamma_{\nu \lambda} 
\nonumber \\
&+&\frac{1}{96}e^{2\Phi} G_{\mu \nu}\,^{\rho \sigma} G_{\rho}\,^{\kappa \lambda \tau} \Gamma_{\sigma \kappa \lambda \tau} +\frac{1}{192}e^{2\Phi} m G_{\mu}\,^{\rho \sigma \kappa} \Gamma_{\nu \rho \sigma \kappa} 
\nonumber \\
&+&\frac{1}{4608}e^{2\Phi}G_{\mu}\,^{\rho \sigma \kappa} G^{\lambda \tau h i} \Gamma_{\nu \rho \sigma \kappa \lambda \tau h i} +\frac{1}{64}e^{\Phi}F^{\rho \sigma} H_{\nu}\,^{\kappa \lambda} \Gamma_{\mu \rho \sigma \kappa \lambda}  
\nonumber \\
&-& \frac{1}{32}e^{\Phi}F^{\rho \sigma} H_{\nu \rho \sigma} \Gamma_{\mu}  - \frac{1}{128}e^{2\Phi}F^{\rho \sigma} F^{\kappa \lambda} \Gamma_{\mu \nu \rho \sigma \kappa \lambda} +\frac{1}{64}e^{2\Phi}F_{\nu}\,^{\rho} F^{\sigma \kappa} \Gamma_{\mu \rho \sigma \kappa} 
\nonumber \\
&+&\frac{1}{32}e^{2\Phi}F_{\nu}\,^{\rho} F_{\rho}\,^{\sigma} \Gamma_{\mu \sigma} +\frac{1}{64}e^{2\Phi} \Gamma_{\mu \nu}  {F}^{2} - \frac{1}{32}e^{2\Phi}  F_{\mu}\,^{\rho} F_{\rho}\,^{\sigma} \Gamma_{\nu \sigma} 
\nonumber \\
&-& \frac{1}{32}e^{\Phi}F_{\mu}\,^{\rho} H_{\nu}\,^{\sigma \kappa} \Gamma_{\rho \sigma \kappa} - \frac{1}{64}e^{2\Phi} F_{\mu}\,^{\rho} F^{\sigma \kappa} \Gamma_{\nu \rho \sigma \kappa} 
\nonumber \\
&+&\Gamma_{11}\bigg(\frac{1}{8}\Gamma^{\rho \sigma} \nabla_{\mu}{H_{\nu \rho \sigma}} - \frac{1}{16}e^{\Phi}\Gamma_{\nu}\,^{\rho \sigma}  \nabla_{\mu}{F_{\rho \sigma}} - \frac{1}{16}e^{\Phi}F^{\rho \sigma} \Gamma_{\nu \rho \sigma}  \nabla_{\mu}{\Phi}
\nonumber \\
&+&\frac{1}{8}e^{\Phi}\Gamma^{\rho}  \nabla_{\mu}{F_{\nu \rho}}+\frac{1}{8}e^{\Phi}F_{\nu}\,^{\rho} \Gamma_{\rho}  \nabla_{\mu}{\Phi} - \frac{1}{8}\Gamma^{\rho \sigma} \nabla_{\nu}{H_{\mu \rho \sigma}}+\frac{1}{16}e^{\Phi} \Gamma_{\mu}\,^{\rho \sigma}  \nabla_{\nu}{F_{\rho \sigma}}
\nonumber \\
&+&\frac{1}{16}e^{\Phi}F^{\rho \sigma} \Gamma_{\mu \rho \sigma}  \nabla_{\nu}{\Phi} - \frac{1}{8}e^{\Phi} \Gamma^{\rho}  \nabla_{\nu}{F_{\mu \rho}} - \frac{1}{8}e^{\Phi}F_{\mu}\,^{\rho} \Gamma_{\rho}  \nabla_{\nu}{\Phi}+\frac{1}{32} e^{\Phi} m H_{\mu}\,^{\rho \sigma} \Gamma_{\nu \rho \sigma} 
\nonumber \\
&+&\frac{1}{768}e^{\Phi}G^{\rho \sigma \kappa \lambda} H_{\mu}\,^{\tau h} \Gamma_{\nu \rho \sigma \kappa \lambda \tau h}  - \frac{1}{48}e^{\Phi}G^{\rho \sigma \kappa \lambda} H_{\mu \nu \rho} \Gamma_{\sigma \kappa \lambda}  - \frac{1}{64}e^{\Phi}  G^{\rho \sigma \kappa \lambda} H_{\mu \rho \sigma} \Gamma_{\nu \kappa \lambda} 
\nonumber \\
&-& \frac{1}{192}e^{\Phi} G_{\nu}\,^{\rho \sigma \kappa} H_{\mu}\,^{\lambda \tau} \Gamma_{\rho \sigma \kappa \lambda \tau} +\frac{1}{32}e^{\Phi} G_{\nu}\,^{\rho \sigma \kappa} H_{\mu \rho \sigma} \Gamma_{\kappa}  - \frac{1}{32}e^{\Phi} m H_{\nu}\,^{\rho \sigma} \Gamma_{\mu \rho \sigma} 
\nonumber \\
&+&\frac{1}{16}e^{2\Phi} m F_{\mu \nu}  - \frac{1}{768} e^{\Phi} G^{\rho \sigma \kappa \lambda} H_{\nu}\,^{\tau h} \Gamma_{\mu \rho \sigma \kappa \lambda \tau h}  +\frac{1}{64} e^{\Phi}  G^{\rho \sigma \kappa \lambda} H_{\nu \rho \sigma} \Gamma_{\mu \kappa \lambda} 
\nonumber \\
&-& \frac{1}{384}e^{2\Phi} F^{\rho \sigma} G_{\nu}\,^{\kappa \lambda \tau} \Gamma_{\mu \rho \sigma \kappa \lambda \tau} +\frac{1}{96}e^{2\Phi} F^{\rho \sigma} G_{\rho}\,^{\kappa \lambda \tau} \Gamma_{\mu \nu \sigma \kappa \lambda \tau}  
\nonumber \\
&-& \frac{1}{768}e^{2\Phi}F_{\mu}\,^{\rho} G^{\sigma \kappa \lambda \tau} \Gamma_{\nu \rho \sigma \kappa \lambda \tau}  - \frac{1}{64}e^{2\Phi} F^{\rho \sigma} G_{\nu \rho}\,^{\kappa \lambda} \Gamma_{\mu \sigma \kappa \lambda} - \frac{1}{192}e^{2\Phi} F_{\mu}\,^{\rho} G_{\rho}\,^{\sigma \kappa \lambda} \Gamma_{\nu \sigma \kappa \lambda} 
\nonumber \\
&+&\frac{1}{64}e^{2\Phi}F^{\rho \sigma} G_{\nu \rho \sigma}\,^{\kappa} \Gamma_{\mu \kappa} +\frac{1}{768}e^{2\Phi}F_{\nu}\,^{\rho} G^{\sigma \kappa \lambda \tau} \Gamma_{\mu \rho \sigma \kappa \lambda \tau} +\frac{1}{192}e^{2\Phi} F_{\nu}\,^{\rho} G_{\rho}\,^{\sigma \kappa \lambda} \Gamma_{\mu \sigma \kappa \lambda} 
\nonumber \\
&+&\frac{1}{384}e^{2\Phi} F_{\mu \nu} G^{\rho \sigma \kappa \lambda} \Gamma_{\rho \sigma \kappa \lambda} +\frac{1}{192}e^{\Phi} G_{\mu}\,^{\rho \sigma \kappa} H_{\nu}\,^{\lambda \tau} \Gamma_{\rho \sigma \kappa \lambda \tau} \nonumber
\nonumber \\
&-& \frac{1}{32}e^{\Phi}G_{\mu}\,^{\rho \sigma \kappa} H_{\nu \rho \sigma} \Gamma_{\kappa}  +\frac{1}{384}e^{2\Phi} F^{\rho \sigma} G_{\mu}\,^{\kappa \lambda \tau} \Gamma_{\nu \rho \sigma \kappa \lambda \tau}  - \frac{1}{64}e^{2\Phi} F^{\rho \sigma} G_{\mu \nu}\,^{\kappa \lambda} \Gamma_{\rho \sigma \kappa \lambda} 
\nonumber \\
&+&\frac{1}{64}e^{2\Phi} F^{\rho \sigma} G_{\mu \rho}\,^{\kappa \lambda} \Gamma_{\nu \sigma \kappa \lambda}  - \frac{1}{64}e^{2\Phi} F^{\rho \sigma} G_{\mu \rho \sigma}\,^{\kappa} \Gamma_{\nu \kappa} +\frac{1}{32}e^{2\Phi} F^{\rho \sigma} G_{\mu \nu \rho \sigma} 
\nonumber \\
&+&\frac{1}{32}e^{2\Phi} m F_{\nu}\,^{\rho} \Gamma_{\mu \rho}  - \frac{1}{32}e^{2\Phi} m F_{\mu}\,^{\rho} \Gamma_{\nu \rho}  \bigg) \ ,
\ee
We can relate the field equations to the supersymmetry variations. Consider,
\bea
\Gamma^{\nu}[{\cal D}_{\mu}, {\cal D}_{\nu}]\epsilon &-&[{\cal D}_{\mu}, {\cal A}]\epsilon + \Phi_{\mu}{\cal A}\epsilon 
\nonumber \\
&=& \bigg[\bigg(-\frac{1}{2}E_{\mu \nu}\Gamma^{\nu}
-\frac{1}{48}e^{\Phi}FG_{\lambda_1 \lambda_2 \lambda_3}\Gamma_{\mu}{}^{\lambda_1 \lambda_2 \lambda_3} + \frac{1}{16}e^{\Phi}FG_{\mu \lambda_1 \lambda_2}\Gamma^{\lambda_1 \lambda_2} 
\nonumber \\
&-&\frac{5}{192}e^{\Phi}BG_{\mu \lambda_1 \lambda_2 \lambda_3 \lambda_4}\Gamma^{\lambda_1 \lambda_2 \lambda_3 \lambda_4}
+ \frac{1}{192}e^{\Phi}BG_{\lambda_1 \lambda_2 \lambda_3 \lambda_4 \lambda_5}\Gamma_{\mu}{}^{\lambda_1 \lambda_2 \lambda_3 \lambda_4 \lambda_5}
\bigg)
\nonumber \\
&+& \Gamma_{11}\bigg(\frac{1}{16}e^{\Phi}BF_{\lambda_1 \lambda_2 \lambda_3}\Gamma_{\mu}{}^{\lambda_1 \lambda_2 \lambda_3} - \frac{3}{16}e^{\Phi}BF_{\mu \lambda_1 \lambda_2}\Gamma^{\lambda_1 \lambda_2} - \frac{1}{8}e^{\Phi}FF_{\lambda}\Gamma_{\mu}{}^{\lambda} 
\nonumber \\
&+& \frac{1}{8}e^{\Phi}FF_{\mu} -\frac{1}{6}BH_{\mu \lambda_1 \lambda_2 \lambda_3}\Gamma^{\lambda_1 \lambda_2 \lambda_3} - \frac{1}{4}FH_{\mu \lambda}\Gamma^{\lambda} \bigg)\bigg]\epsilon \ ,
\ee
where,
\bea
\Phi_{\mu} &=& \bigg(\frac{1}{8}e^{\Phi}m\Gamma_{\mu} + \frac{1}{192}e^{\Phi}G_{\lambda_1 \lambda_2 \lambda_3 \lambda_4}\Gamma^{\lambda_1 \lambda_2 \lambda_3 \lambda_4}\Gamma_{\mu}\bigg) 
\nonumber \\
&+& \Gamma_{11}\bigg(\frac{1}{4}H_{\mu \lambda_1 \lambda_2}\Gamma^{\lambda_1 \lambda_2} 
- \frac{1}{16}e^{\Phi}F_{\lambda_1 \lambda_2}\Gamma^{\lambda_1 \lambda_2}\Gamma_{\mu}\bigg) \ ,
\ee
and
\bea
\Gamma^{\mu}[{\cal D}_{\mu}, {\cal A}]\epsilon &+& \theta {\cal A}\epsilon 
\nonumber \\
&=& \bigg(F\Phi  - \frac{1}{24}e^{\Phi}FG_{\lambda_1 \lambda_2 \lambda_3}\Gamma^{\lambda_1 \lambda_2 \lambda_3} + \frac{1}{96}e^{\Phi}BG_{\lambda_1 \lambda_2 \lambda_3 \lambda_4 \lambda_5}\Gamma^{\lambda_1 \lambda_2 \lambda_3 \lambda_4 \lambda_5}\bigg)\epsilon
\nonumber \\
&+& \Gamma_{11}\bigg( \frac{3}{4}e^{\Phi}FF_{\lambda}\Gamma^{\lambda} - \frac{3}{8}e^{\Phi}BF_{\lambda_1 \lambda_2 \lambda_3}\Gamma^{\lambda_1 \lambda_2 \lambda_3} + \frac{1}{12}BH_{\lambda_1 \lambda_2 \lambda_3 \lambda_4}\Gamma^{\lambda_1 \lambda_2 \lambda_3 \lambda_4} 
\nonumber \\
&+& \frac{1}{4}FH_{\lambda_1 \lambda_2}\Gamma^{\lambda_1 \lambda_2} \bigg)\epsilon \ ,
\ee
and
\bea
\theta = \bigg(-2\nabla_{\mu}{\Phi}\Gamma^{\mu} - e^{\Phi}m \bigg) + \Gamma_{11}\bigg(\frac{1}{12}H_{\lambda_1 \lambda_2 \lambda_3}\Gamma^{\lambda_1 \lambda_2 \lambda_3} - \frac{1}{2}e^{\Phi}F_{\lambda_1 \lambda_2}\Gamma^{\lambda_1 \lambda_2} \bigg) \ .
\ee
The field equations and Bianchi identities are 
\bea
\label{miiaeineq}
E_{\mu \nu} &=& R_{\mu \nu} + 2 \nabla_\mu \nabla_\nu \Phi
-{1 \over 4} H_{\mu \nu}^2
-{1 \over 2} e^{2 \Phi} F_{\mu \nu}^2
-{1 \over 12} e^{2 \Phi} G_{\mu \nu}^2 
\nonumber \\
&+& g_{\mu \nu} \bigg({1 \over 8}e^{2 \Phi}
F^2
+{1 \over 96}e^{2 \Phi} G^2  + \frac{1}{4}e^{2 \Phi} m^2 \bigg) = 0 ~,
\ee
\bea
F\Phi = \nabla^2 \Phi
- 2 (d \Phi)^2
+{1 \over 12} H^2-{3 \over 8} e^{2 \Phi}
F^2 -{1 \over 96} e^{2 \Phi} G^2 - \frac{5}{4}e^{2\Phi} m^2 = 0 ~,
\ee
\bea
FF_{\mu} = \nabla^\lambda F_{\mu \lambda} -{1 \over 6} H^{\lambda_1 \lambda_2 \lambda_3} G_{\lambda_1 \lambda_2 \lambda_3 \mu} = 0 ~,
\ee
\bea
FH_{\mu \nu} &=& e^{2\Phi}\nabla^\lambda \bigg( e^{-2 \Phi} H_{\mu \nu \lambda}\bigg) - {1 \over 2} e^{2\Phi}G_{\mu \nu \lambda_1 \lambda_2} F^{\lambda_1 \lambda_2} - e^{2\Phi} m F_{\mu \nu}
\nonumber \\
&+&{1 \over 1152}e^{2\Phi} \epsilon_{\mu \nu}{}^{\lambda_1 \lambda_2
\lambda_3 \lambda_4 \lambda_5 \lambda_6 \lambda_7 \lambda_8}
G_{\lambda_1 \lambda_2 \lambda_3 \lambda_4}
G_{\lambda_5 \lambda_6 \lambda_7 \lambda_8} = 0 ~,
\ee
\bea
FG_{\mu \nu \rho} = \nabla^\lambda G_{\mu \nu \rho \lambda}-{1 \over 144} \epsilon_{\mu \nu \rho}{}^{\lambda_1 \lambda_2 \lambda_3 \lambda_4 \lambda_5 \lambda_6 \lambda_7}G_{\lambda_1 \lambda_2 \lambda_3 \lambda_4}H_{\lambda_5 \lambda_6 \lambda_7} = 0 ~,
\ee

\bea
\label{miiabian1}
BF_{\mu \nu \rho} = \nabla_{[\mu}F_{\nu \rho]} - \frac{1}{3}m H_{\mu \nu \rho} = 0 \ ,
\ee

\bea
BH_{\mu \nu \rho \lambda} = \nabla_{[\mu}H_{\nu \rho \lambda]} = 0 \ ,
\ee

\bea
\label{miiabian3}
BG_{\mu \nu \rho \lambda \kappa} = \nabla_{[\mu}G_{\nu \rho \lambda \kappa]} - 2F_{[\mu \nu}H_{\rho \lambda \kappa]} = 0 \ .
\ee

\section{Alternative derivation of dilaton field equation}
The dilaton field equation is implied by the Einstein equation and all other field equations and Bianchi identities, up to a constant.
\bea
R = -2\nabla^2 \Phi + \frac{1}{4}H^2 - \frac{3}{4}e^{2\Phi}F^2 - \frac{1}{48}e^{2\Phi}G^2  - \frac{5}{2}e^{2\Phi}m^2 \ .
\ee
On taking the Divergence of (\ref{miiaeineq}),\footnote{For a $p$-form $\omega$ we write $\omega^2 = \omega_{\lambda_1 \cdots \lambda_p}\omega^{\lambda_1 \cdots \lambda_p}$ and $\omega^2_{\mu \nu} = \omega_{\mu \lambda_1 \cdots \lambda_{p-1}} \omega_{\nu}{}^{\lambda_1 \cdots \lambda_{p-1}}$} 
\bea
\nabla^{\mu}R_{\mu \nu} &=& -2\nabla^2 \nabla_{\nu} \Phi + \frac{1}{4}\nabla^{\mu}H_{\mu \nu}^2
+ \frac{1}{2}\nabla^{\mu}(e^{2\Phi}F_{\mu \nu}^2) + \frac{1}{12}\nabla^{\mu}(e^{2\Phi}G_{\mu \nu}^2) 
\nonumber \\
&-& \frac{1}{8}\nabla_{\nu}(e^{2\Phi}F^2) - \frac{1}{96}\nabla_{\nu}(e^{2\Phi}G^2) - \frac{1}{4}\nabla_{\nu}(e^{2\Phi}m^2) \ .
\ee
We can rewrite the first term as
\bea
-2\nabla^2 \nabla_{\nu} \Phi &=& -2\nabla_{\nu} \nabla^2 \Phi - 2R_{\mu \nu}\nabla^{\mu}\Phi
\nonumber \\
&=& -2\nabla_{\nu} \nabla^2 \Phi + 2\nabla_{\nu}(d\Phi)^2 - \frac{1}{2}H_{\mu \nu}^2 \nabla^{\mu}\Phi - e^{2\Phi}F_{\mu \nu}^2 \nabla^{\mu}\Phi 
\nonumber \\
&-& \frac{1}{6}e^{2\Phi}G_{\mu \nu}^2\nabla^{\mu}\Phi + \frac{1}{4}e^{2\Phi}F^2 \nabla_{\nu}\Phi + \frac{1}{48}e^{2\Phi} G^2 \nabla_{\nu}\Phi + \frac{1}{2}e^{2\Phi} m^2 \nabla_{\nu}\Phi \ .
\ee
Where we have used (\ref{miiaeineq}) again. This gives,
\bea
\nabla^{\mu}R_{\mu \nu} &=& -2\nabla_{\nu} \nabla^2 \Phi + 2\nabla_{\nu}(d\Phi)^2 + \frac{1}{4}e^{2\Phi}\nabla^{\mu}(e^{-2\Phi}H_{\mu \nu}^2)
+ \frac{1}{2}e^{2\Phi}\nabla^{\mu}(F_{\mu \nu}^2) 
\nonumber \\
&+& \frac{1}{12}e^{2\Phi}\nabla^{\mu}(G_{\mu \nu}^2) - \frac{1}{8}\nabla_{\nu}(e^{2\Phi}F^2) + \frac{1}{4}e^{2\Phi}\nabla_{\nu}\Phi F^2 - \frac{1}{96}\nabla_{\nu}(e^{2\Phi}G^2) + \frac{1}{48}e^{2\Phi}\nabla_{\nu}\Phi G^2 \ .
\nonumber \\
\ee
On the other hand,
\bea
\nabla^{\mu} R_{\mu \nu} &=& \frac{1}{2}\nabla_{\nu}R
\nonumber \\
&=& -\nabla_{\nu} \nabla^2 \Phi + \frac{1}{8}\nabla_{\nu}H^2 - \frac{3}{8}\nabla_{\nu}(e^{2\Phi}F^2) - \frac{1}{96}\nabla_{\nu}(e^{2\Phi}G^2) -  \frac{5}{4}\nabla_{\nu}(e^{2\Phi}m^2) \ .
\ee
Rearranging the Einstein equation we obtain,
\bea
\label{miiaeins1}
\nabla_{\nu}\nabla^2\Phi &=& 2\nabla_{\nu}(d\Phi)^2  - \frac{1}{8}\nabla_{\nu}H^2 + \frac{1}{4}e^{2\Phi}\nabla^{\mu}(e^{-2\Phi}H_{\mu \nu}^2)
+ \frac{1}{2}e^{2\Phi}\nabla^{\mu}(F_{\mu \nu}^2)  
\nonumber \\
&+& \frac{1}{4}\nabla_{\nu}(e^{2\Phi}F^2) + \frac{1}{4}e^{2\Phi}\nabla_{\nu}\Phi F^2   + \frac{1}{12}e^{2\Phi}\nabla^{\mu}(G_{\mu \nu}^2) + \frac{1}{48}e^{2\Phi}\nabla_{\nu}\Phi G^2 +  \frac{5}{4}\nabla_{\nu}(e^{2\Phi}m^2) \ .
\nonumber \\
\ee
We can compute certain terms by using the Field equations (\ref{miiageq1})-(\ref{miiaceq1}) and Bianchi identities (\ref{miiabian1})
\bea
\frac{1}{2}e^{2\Phi}\nabla^{\mu}(F_{\mu \nu}^2) = \frac{1}{2}e^{2\Phi}\nabla^{\mu}(F_{\mu \lambda})F_\nu{}^{\lambda} + \frac{1}{2}e^{2\Phi}F_{\mu \lambda}\nabla^{\mu}F_\nu{}^{\lambda} \ ,
\ee
\bea
\frac{1}{4}e^{2\Phi}\nabla^{\mu}(e^{-2\Phi}H_{\mu \nu}^2) = \frac{1}{4}e^{2\Phi}\nabla^{\mu}(e^{-2\Phi}H_{\mu \lambda_1 \lambda_2})H_\nu{}^{\lambda_1 \lambda_2} + \frac{1}{4}H_{\mu \lambda_1 \lambda_2}\nabla^{\mu}H_\nu{}^{\lambda_1 \lambda_2} \ ,
\ee
\bea
\frac{1}{12}e^{2\Phi}\nabla^{\mu}(G_{\mu \nu}^2) = \frac{1}{12}e^{2\Phi}\nabla^{\mu}(G_{\mu \lambda_1 \lambda_2 \lambda_3})G_\nu{}^{\lambda_1 \lambda_2 \lambda_3} + \frac{1}{12}e^{2\Phi}G_{\mu \lambda_1 \lambda_2 \lambda_3}\nabla^{\mu} G_\nu{}^{\lambda_1 \lambda_2 \lambda_3} \ .
\ee
The Bianchi identities (\ref{miiabian1}) imply,
\bea
F_{\mu \lambda}\nabla^{\mu}F_\nu{}^{\lambda} = \frac{1}{4}\nabla_{\nu}F^2 - \frac{1}{2} m H_{\nu \lambda_1 \lambda_2}F^{\lambda_1 \lambda_2} \ ,
\ee
and
\bea
H_{\mu \lambda_1 \lambda_2}\nabla^{\mu}H_\nu{}^{\lambda_1 \lambda_2} = \frac{1}{6}\nabla_{\nu}H^2 \ ,
\ee
and
\bea
G_{\mu \lambda_1 \lambda_2 \lambda_3}\nabla^{\mu} G_\nu{}^{\lambda_1 \lambda_2 \lambda_3} = \frac{1}{8}\nabla_{\nu}G^2 + \frac{5}{2}g_{\nu \kappa}G_{\mu \lambda_1 \lambda_2 \lambda_3}F^{[\mu \kappa}H^{\lambda_1 \lambda_2 \lambda_3]} \ .
\ee
Substituting this back into (\ref{miiaeins1}) we obtain,
\bea
\nabla_{\nu}\nabla^2\Phi &=& \nabla_{\nu}\bigg(2(d\Phi)^2 - \frac{1}{12}H^2 + \frac{3}{8}e^{2\Phi}F^2 + \frac{1}{96}e^{2\Phi}G^2 + \frac{5}{4}e^{2\Phi} m^2 \bigg)
\nonumber \\
&+& \frac{1}{2}e^{2\Phi}\nabla^{\mu}(F_{\mu \lambda})F_\nu{}^{\lambda} +  \frac{1}{4}e^{2\Phi}\nabla^{\mu}(e^{-2\Phi}H_{\mu \lambda_1 \lambda_2})H_\nu{}^{\lambda_1 \lambda_2} - \frac{1}{4} m H_{\nu \lambda_1 \lambda_2}F^{\lambda_1 \lambda_2}
\nonumber \\
&+& \frac{1}{12}e^{2\Phi}\nabla^{\mu}(G_{\mu \lambda_1 \lambda_2 \lambda_3})G_\nu{}^{\lambda_1 \lambda_2 \lambda_3} + \frac{5}{24}g_{\nu \kappa}G_{\mu \lambda_1 \lambda_2 \lambda_3}F^{[\mu \kappa}H^{\lambda_1 \lambda_2 \lambda_3]} \ .
\ee
On applying the field equations (\ref{miiageq1})-(\ref{miiaceq1}),
\bea
\nabla_{\nu}\nabla^2\Phi &=& \nabla_{\nu}\bigg(2(d\Phi)^2 - \frac{1}{12}H^2 + \frac{3}{8}e^{2\Phi}F^2 + \frac{1}{96}e^{2\Phi}G^2 + \frac{5}{4}e^{2\Phi} m^2\bigg) \ .
\ee
%Since,
%\bea
%&&\frac{1}{4608}e^{2\Phi}\epsilon_{\lambda_1 \lambda_2}{}^{\tau_1 \tau_2 \tau_3 \tau_4 \tau_5 \tau_6 \tau_7 \tau_8}G_{\tau_1 \tau_2 \tau_3 \tau_4}G_{\tau_5 \tau_6 \tau_7 \tau_8}H_\nu{}^{\lambda_1 \lambda_2}
%\nonumber \\
%&+& \frac{1}{1728}e^{2\Phi}\epsilon_{\lambda_1 \lambda_2 \lambda_3}{}^{\tau_1 \tau_2 \tau_3 \tau_4 \tau_5 \tau_6 \tau_7}G_{\tau_1 \tau_2 \tau_3 \tau_4}H_{\tau_5 \tau_6 \tau_7}G_\nu{}^{\lambda_1 \lambda_2 \lambda_3} = 0
%\ee
%\bea
%- \frac{1}{12}e^{2\Phi}H^{\lambda_1 \lambda_2 \lambda_3} G_{\lambda_1 \lambda_2 \lambda_3 \lambda}F_\nu{}^{\lambda} +  \frac{1}{8}e^{2\Phi}G_{\lambda_1 \lambda_2}{}^{\tau_1 \tau_2}F_{\tau_1 \tau_2}H_\nu{}^{\lambda_1 \lambda_2} + \frac{5}{24}g_{\nu \kappa}G_{\mu \lambda_1 \lambda_2 \lambda_3}F^{[\mu \kappa}H^{\lambda_1 \lambda_2 \lambda_3]} = 0
%\nonumber \\
%\ee
This implies the Dilaton field equation (\ref{miiadileq}) up to a constant.
In terms of the field equations and Bianchi identities, one gets
\bea
\nabla^{\mu}{R_{\mu \nu}} - \frac{1}{2}\nabla_{\nu}{R} &=& -\nabla_{\nu}{(F\Phi)} - 2 E_{\nu \lambda}\nabla^{\lambda}{\Phi} 
+ \nabla^{\mu}{(E_{\mu \nu})} - \frac{1}{2}\nabla_{\nu}{(E^{\mu}{}_{\mu})}
\nonumber \\
&-& \frac{1}{3}BH_{\nu}{}^{\lambda_1 \lambda_2 \lambda_3} H_{\lambda_1 \lambda_2 \lambda_3} + \frac{1}{4}FH_{\lambda_1 \lambda_2}H_{\nu}{}^{\lambda_1 \lambda_2}
- \frac{3}{4}e^{2\Phi}BF_{\nu}{}^{\lambda_1 \lambda_2}F_{\lambda_1 \lambda_2} 
\nonumber \\
&-& \frac{1}{2}e^{2\Phi}FF_{\lambda} F_{\nu}{}^{\lambda}
- \frac{5}{48}e^{2\Phi}BG_{\nu}{}^{\lambda_1 \lambda_2 \lambda_3 \lambda_4}G_{\lambda_1 \lambda_2 \lambda_3 \lambda_4} 
\nonumber \\
&-& \frac{1}{12}e^{2\Phi}FG_{\lambda_1 \lambda_2 \lambda_3}G_{\nu}{}^{\lambda_1 \lambda_2 \lambda_3} = 0 \ .
\ee
\section{Invariance of massive IIA fluxes}

In this Appendix we will give a proof to show that the bilinears constructed from Killing spinors are Killing vectors and preserve  all the fluxes. The proof will rely on the Killing spinor equations. In addition to the Killing spinor equations the proof will also rely on the field equations and Bianchi identities, and the result will thus hold in general for all supersymmetric supergravity solutions.  It is convenient to introduce the following notation
\bea
&& \alpha^{IJ}_{\mu_1 \cdots \mu_k} \equiv B(\epsilon^I, \Gamma_{\mu_1  \cdots \mu_k}\epsilon^J) ~, \notag \\
&& \tau^{IJ}_{\mu_1 \cdots \mu_k} \equiv B(\epsilon^I, \Gamma_{\mu_1  \cdots \mu_k} \Gamma_{11}\epsilon^J) ~,
\ee
with the inner product $B(\epsilon^I, \epsilon^J) \equiv \langle  \Gamma_0 C* \epsilon^I, \epsilon^J \rangle$, where $C=\Gamma_{6789}$, is antisymmetric, i.e.~$B(\epsilon^I, \epsilon^J)=-B(\epsilon^J, \epsilon^I)$ and all $\Gamma$-matrices are anti-Hermitian with respect to this inner product, i.e.~$B(\Gamma_\mu \epsilon^I,\epsilon^J)=-B(\epsilon^I,\Gamma_\mu \epsilon^J)$. The bilinears have the symmetry properties
\bea
\alpha^{IJ}_{\mu_1 \cdots \mu_k} &=& \alpha^{JI}_{\mu_1 \cdots \mu_k}  \qquad  k=1,2,5 
\cr
\alpha^{IJ}_{\mu_1 \cdots \mu_k} &=& -\alpha^{JI}_{\mu_1 \cdots \mu_k} \quad ~ k=0,3,4 \ .
\ee
and
\bea
\tau^{IJ}_{\mu_1 \cdots \mu_k} &=& \tau^{JI}_{\mu_1 \cdots \mu_k}  \qquad  k=0,1,4,5
\cr
\tau^{IJ}_{\mu_1 \cdots \mu_k} &=& -\tau^{JI}_{\mu_1 \cdots \mu_k}\quad  ~k=2,3 ~.
\ee
First we verify that there is a set of 1-form bi-linears whose associated vectors are Killing. We use the gravitino KSE to replace covariant derivatives with terms which are linear in the fluxes. The 1-form bilinears associated with the Killing vectors are $\alpha^{IJ}_\mu e^\mu$, and we let $\nabla_{\mu} \epsilon = -\Psi_{\mu} \epsilon$ from the KSEs
\bea
\nabla_\mu \alpha^{IJ}_\nu &=& \nabla_\mu B(\epsilon^{I}, \Gamma_\nu \epsilon^{J})  \notag \\
&=& B(\nabla_\mu \epsilon^{I}, \Gamma_\nu  \epsilon^{J}) +B( \epsilon^{I}, \Gamma_\nu \nabla_\mu\epsilon^{J}) \notag \\
&=& -B(\Psi_\mu \epsilon^{I}, \Gamma_\nu \epsilon^{J}) -B( \epsilon^{I}, \Gamma_\nu \Psi_\mu\epsilon^{J}) \notag \\
&=& B(\Gamma_\nu \epsilon^{J},\Psi_\mu \epsilon^{I}) -B( \epsilon^{I}, \Gamma_\nu  \Psi_\mu\epsilon^{J})  \notag \\
&=&  -B(\epsilon^{J},  \Gamma_\nu \Psi_\mu \epsilon^{I}) -B( \epsilon^{I}, \Gamma_\nu  \Psi_\mu\epsilon^{J}) \notag \\
&=& -2  B(\epsilon^{(I}, \Gamma_{B}\Psi_\mu \epsilon^{J)})\notag \\
&=&\Big(\frac{1}{4} m e^{\Phi} \alpha^{IJ}_{\mu \nu} + \frac{1}{8}e^{\Phi}G_{\mu \nu}{}^{ \lambda_1 \lambda_2}\alpha^{IJ}_{\lambda_1 \lambda_2} + \frac{1}{96}e^{\Phi}G^{\lambda_1 \lambda_2 \lambda_3 \lambda_4} \alpha^{IJ}_{\mu \nu \lambda_1 \lambda_2 \lambda_3 \lambda_4} \notag \\
&+& \frac{1}{4} e^{\Phi}F_{\mu \nu} \tau^{IJ} - \frac{1}{2} H_{\mu \nu}{}^C \tau^{IJ}_C +\frac{1}{8} e^{\Phi}F^{\lambda_1 \lambda_2} \tau^{IJ}_{\mu \nu \lambda_1 \lambda_2}\Big) \ .
\ee
Since the resulting expression is antisymmetric in its free indices we find that $\nabla_{(\mu} \alpha^{IJ}_{\nu)}=0 $ and hence the vectors associated with $\alpha^{IJ}_\mu e^\mu$ are Killing.
Note that the dilatino KSE (\ref{miiaAKSE}) imply that
\bea
0  = B(\epsilon^{(I},{\cal A} \epsilon^{J)}) = \alpha^{IJ}_\mu \partial^\mu \Phi~,
\ee
and hence $i_K d\Phi=0$, where $K=\alpha^{IJ}_\mu e^\mu$ denotes the 1-forms associated with the Killing vectors with the $IJ$ indices suppressed. With this relation it follows that the Killing vectors preserve the dilaton:
\bea
{\cal L}_K \Phi := i_K d\Phi + d (i_K \Phi) =  0 ~,
\ee
since $i_K \Phi \equiv 0$.
To see that the 3-form flux $H$ is preserved we need to analyse the 1-form bi-linears which are not related to the Killing vectors, i.e. $\tau^{IJ}_\mu e^\mu$. As above, we find that
\bea
\nabla_{[\mu} \tau^{IJ}_{\nu]} &=& -2 B(\epsilon^{(I},\Gamma_{11} \Gamma_{[\mu} \Psi_{\nu]} \epsilon^{J)}) \notag \\
&& = - \frac{1}{2} H_{\mu \nu}{}^\lambda \alpha^{IJ}_\lambda ~,
\label{3formeq}
\ee
where we have indicated the degree of the form $\tau$ and suppressed the indices labelling the Killing spinors.
By taking the exterior derivative of (\ref{3formeq}), and using the Bianchi identity for $H$ with $dH=0$,
it follows that
\bea
{\cal L}_K H = 0 \ ,
\ee
and hence the Killing vectors preserve also the $H$ flux.
We now turn to the 2-form flux $F$. Computing the covariant derivative of the scalar $\tau^{IJ}$, and making use of the gravitino KSE as above, we find
\bea
\nabla_{\mu} \tau^{I J} = (i_K F)_\mu  - m \tau_{\mu}^{IJ} ~.
\label{Fexpr}
\ee
Acting with another derivative on (\ref{Fexpr}), and re-substituting (\ref{Fexpr}) into the resulting expression, we obtain
\bea
{\cal L}_K F =  i_K (dF) + m d\tau_{\mu}^{IJ} \wedge dy^\mu = 0  ~,
\ee
where in the second step we have used  (\ref{3formeq}) and the Bianchi identity for $F$, i.e. $dF = mH$. For the field strength $G$ we compute the covariant derivative of $\alpha_{\mu \nu}^{IJ}$ leads to
\bea
\nabla_{[\mu} \alpha^{I J}_{\nu \rho]} =
\frac{1}{3}(i_K G)_{\mu \nu \rho}  + \frac{1}{3}\tau^{IJ}H_{\mu \nu \rho} - F_{[\mu \nu} \tau^{IJ}_{\rho]} ~.
\label{Gexpr}
\ee
Acting with an exterior derivative on (\ref{Gexpr}) and re-substituting (\ref{Gexpr}) into the resulting expression, and using (\ref{3formeq}), (\ref{Fexpr}) and the Bianchi identity for $F$ we obtain,
\bea
{\cal L}_K G =   i_K (dG)  + i_K F \wedge  H + F \wedge i_K H = 0 \ ,
\ee
where in the second step we have used the Bianchi identity for $G$, i.e. $dG = F \wedge H$.
Also since the mass parameter $m$ is constant, we have ${\cal L}_K m=0$.

\section{Independent KSEs}
\subsection{The (\ref{miiaint5}) condition}
The (\ref{miiaint5}) component of the KSEs is implied by (\ref{miiaint4}), (\ref{miiaint6}) and (\ref{miiaint7}) together with a number of field equations and Bianchi identities. First evaluate the LHS of (\ref{miiaint5}) by substituting in (\ref{miiaint6}) to eliminate $\tau_+$, and use (\ref{miiaint4}) to evaluate the supercovariant derivative of $\phi_+$. Also, using (\ref{miiaint4}) one can compute
\bea
(\tilde{\nabla}_{j}\tilde{\nabla}_{i} - \tilde{\nabla}_{i}\tilde{\nabla}_{j})\phi_{+}
 &=& {1\over 4} \tilde \nabla_j (h_i) \phi_+ + {1\over4} \Gamma_{11} \tilde \nabla_j(L_i) \phi_+ -{1\over8} \Gamma_{11} \tilde \nabla_j (\tilde H_{i l_1 l_2}) \Gamma^{l_1 l_2} \phi_+
\cr
&+&{1\over16} e^\Phi \Gamma_{11} (- 2\tilde \nabla_j (S)+ \tilde \nabla_j (\tilde F_{kl}) \Gamma^{kl}) \Gamma_i \phi_+ 
\nonumber \\
&-& {1\over 8\cdot 4!} e^\Phi (- 12 \tilde \nabla_j (X_{kl}) \Gamma^{kl}+ \tilde \nabla_j (\tilde G_{j_1j_2j_3j_4}) \Gamma^{j_1j_2j_3j_4}) \Gamma_i \phi_+
\cr
&+&{1\over16} \tilde \nabla_j \Phi e^\Phi \Gamma_{11} (- 2 S+  \tilde F_{kl} \Gamma^{kl}) \Gamma_i \phi_+ 
\nonumber \\
&-& {1\over 8\cdot 4!} \tilde \nabla_j \Phi e^\Phi (- 12  X_{kl} \Gamma^{kl}+  \tilde G_{j_1j_2j_3j_4} \Gamma^{j_1j_2j_3j_4}) \Gamma_i \phi_+
\cr
&-& \frac{1}{8}e^{\Phi}\tilde{\nabla}_j \Phi m\Gamma_i \phi_+ + \big( {1\over 4} h_i  + {1\over4} \Gamma_{11} L_i  -{1\over8} \Gamma_{11} \tilde H_{ijk} \Gamma^{jk} 
\nonumber \\
&+&{1\over16} e^\Phi \Gamma_{11} (- 2 S+ \tilde F_{kl} \Gamma^{kl}) \Gamma_i 
\nonumber \\
&-&{1\over 8\cdot 4!} e^\Phi (- 12 X_{kl} \Gamma^{kl}+ \tilde G_{j_1j_2j_3j_4} \Gamma^{j_1j_2j_3j_4}) \Gamma_i - \frac{1}{8}e^{\Phi} m\Gamma_i\big)\tilde \nabla_{j} \phi_+ - (i \leftrightarrow j) \ .
\label{miiaDDphicond}
\nonumber \\
\ee
Then consider the following, where the first terms cancel from the definition of curvature,
\bea
\bigg(\frac{1}{4}\tilde{R}_{ij}\Gamma^{j} - \frac{1}{2}\Gamma^{j}(\tilde{\nabla}_{j}\tilde{\nabla}_{i} - \tilde{\nabla}_{i}\tilde{\nabla}_{j}) \bigg)\phi_+ + \frac{1}{2}\tilde{\nabla}_{i}(\mathcal{A}_1) + \frac{1}{2}\Psi_{i} \mathcal{A}_1 = 0~,
\label{miiaB5intcond}
\ee
where
\bea
\mathcal{A}_1 &=& \partial_i \Phi \Gamma^i \phi_{+} -{1\over12} \Gamma_{11} (- 6 L_i \Gamma^i+\tilde H_{ijk} \Gamma^{ijk}) \phi_++{3\over8} e^\Phi \Gamma_{11} (-2 S+\tilde F_{ij} \Gamma^{ij})\phi_+
\cr
&
+&{1\over 4\cdot 4!}e^{\Phi} (- 12 X_{ij} \Gamma^{ij}+\tilde G_{j_1j_2j_3j_4} \Gamma^{j_1j_2j_3j_4}) \phi_+ +  \frac{5}{4}e^{\Phi} m\phi_+ \ ,
\label{miiaA1cond}
\ee
and
\bea
\Psi_{i} &=& - \frac{1}{4}h_{i} + \Gamma_{11}(\frac{1}{4}L_{i} - \frac{1}{8}\tilde{H}_{i j k}\Gamma^{j k})
\label{miiaPsiicond} \ .
\ee
The expression in (\ref{miiaA1cond}) vanishes on making use of (\ref{miiaint7}), as $\mathcal{A}_1 = 0$ is equivalent to the $+$ component of (\ref{miiaint7}). However a non-trivial identity is obtained by using (\ref{miiaDDphicond}) in (\ref{miiaB5intcond}), and expanding out the $\mathcal{A}_1$ terms. Then, on adding (\ref{miiaB5intcond}) to the LHS of (\ref{miiaint5}), with $\tau_+$ eliminated in favour of $\eta_+$ as described above, one obtains the following
 \bea
&&\frac{1}{4}\bigg(\tilde{R}_{ij} + \tn_{(i} h_{j)}
-{1 \over 2} h_i h_j + 2 \tn_i \tn_j \Phi
+{1 \over 2} L_i L_j -{1 \over 4} {\tilde{H}}_{i
l_1 l_2} {\tilde{H}}_j{}^{l_1 l_2}
\cr
&-&{1 \over 2} e^{2 \Phi} {\tilde{F}}_{i l}
{\tilde{F}}_j{}^l + {1 \over 8} e^{2 \Phi} {\tilde{F}}_{l_1 l_2}
{\tilde{F}}^{l_1 l_2}\delta_{ij} + {1 \over 2} e^{2 \Phi} X_{i l}
X_j{}^l - {1 \over 8} e^{2 \Phi}
X_{l_1 l_2}X^{l_1 l_2}\delta_{ij}
\cr
&-&{1 \over 12} e^{2 \Phi} {\tilde{G}}_{i \ell_1 \ell_2 \ell_3} {\tilde{G}}_j{}^{\ell_1 \ell_2 \ell_3}
+ {1 \over 96} e^{2 \Phi} {\tilde{G}}_{\ell_1 \ell_2
\ell_3 \ell_4} {\tilde{G}}^{\ell_1 \ell_2 \ell_3 \ell_4}\delta_{ij} - {1 \over 4} e^{2 \Phi} S^2\delta_{ij} + {1 \over 4} e^{2 \Phi} m^2\delta_{ij} \bigg)\Gamma^{j}=0~.
\nonumber \\
\ee
This vanishes identically on making use of the Einstein equation (\ref{miiafeq8}). Therefore it follows that (\ref{miiaint5}) is implied by the $+$ component of (\ref{miiaint4}), (\ref{miiaint6}) and (\ref{miiaint7}), the Bianchi identities (\ref{miiabian2}) and the gauge field equations (\ref{miiafeq1})-(\ref{miiafeq5}).

\subsection{The (\ref{miiaint8}) condition}
Let us define
\bea
\mathcal{A}_2 &=&
-\bigg( \partial_{i}\Phi\Gamma^{i} + \frac{1}{12}\Gamma_{11} (6L_i \Gamma^{i} + \tilde{H}_{ijk}\Gamma^{ijk}) + \frac{3}{8}e^{\Phi}\Gamma_{11}(2S + \tilde{F}_{ij}\Gamma^{ij})
\cr
&-& \frac{1}{4\cdot 4!}e^{\Phi}(12X_{ij}\Gamma^{ij} + \tilde{G}_{ijkl}\Gamma^{ijkl}) - \frac{5}{4}e^{\Phi}m\bigg)\tau_{+}
\cr
&+& \bigg(\frac{1}{4}M_{ij}\Gamma^{ij}\Gamma_{11} + \frac{3}{4}e^{\Phi}T_{i}\Gamma^{i}\Gamma_{11} + \frac{1}{24}e^{\Phi}Y_{ijk}\Gamma^{ijk}\bigg)\phi_{+}~,
\ee
where $\mathcal{A}_2$ equals the expression in (\ref{miiaint8}).
One obtains the following identity
\bea
\mathcal{A}_2 = -\frac{1}{2}\Gamma^{i}\tilde{\nabla}_i \mathcal{A}_1 + \Psi_1\mathcal{A}_1 ~,
\ee
where
\bea
\Psi_1 &=& \tilde{\nabla}_{i}\Phi\Gamma^{i} + \frac{3}{8}h_{i}\Gamma^{i} + \frac{1}{16}e^{\Phi}X_{l_1 l_2}\Gamma^{l_1 l_2} - \frac{1}{192}e^{\Phi}\tilde{G}_{l_1 l_2 l_3 l_4}\Gamma^{l_1 l_2 l_3 l_4} - \frac{1}{8}e^{\Phi}m
\cr
&+& \Gamma_{11}\bigg(\frac{1}{48}\tilde{H}_{l_1 l_2 l_3}\Gamma^{l_1 l_2 l_3} - \frac{1}{8}L_{i}\Gamma^{i} + \frac{1}{16}e^{\Phi}\tilde{F}_{l_1 l_2}\Gamma^{l_1 l_2} - \frac{1}{8}e^{\Phi}S \bigg)~.
\ee
We have made use of the $+$ component of (\ref{miiaint4}) in order to evaluate the covariant derivative in the above expression. In addition we have made use of the Bianchi identities (\ref{miiabian2}) and the field equations (\ref{miiafeq1})-(\ref{miiafeq6}).

\subsection{The (\ref{miiaint1}) condition}
\label{miiaB1sec}

In order to show that (\ref{miiaint1}) is implied by the independent KSEs we can compute the following,
\bea
&&\bigg(-\frac{1}{4}\tilde{R} - \Gamma^{i j}\tilde{\nabla}_{i}\tilde{\nabla}_{j}\bigg)\phi_{+} - \Gamma^{i}\tilde{\nabla}_i(\mathcal{A}_1)
\cr
&+& \bigg(\tilde{\nabla}_i\Phi \Gamma^{i} + \frac{1}{4}h_{i}\Gamma^{i} + \frac{1}{16}e^{\Phi}X_{l_1 l_2}\Gamma^{l_1 l_2} - \frac{1}{192}e^{\Phi}\tilde{G}_{l_1 l_2 l_3 l_4}\Gamma^{l_1 l_2 l_3 l_4} - \frac{1}{8}e^{\Phi}m
\cr
&+& \Gamma_{11}(-\frac{1}{4}L_{l}\Gamma^{l} - \frac{1}{24}\tilde{H}_{l_1 l_2 l_3}\Gamma^{l_1 l_2 l_3} - \frac{1}{8}e^{\Phi}S
+ \frac{1}{16}e^{\Phi}\tilde{F}_{l_1 l_2}\Gamma^{l_1 l_2}) \bigg)\mathcal{A}_1 = 0~,
\nonumber \\
\ee
where
\bea
\tilde{R} &=& -2\Delta - 2h^{i}\tilde{\nabla}_{i}\Phi - 2\tilde{\nabla}^2\Phi - \frac{1}{2}h^2 + \frac{1}{2}L^2 + \frac{1}{4}\tilde{H}^{2} + \frac{5}{2}e^{2\Phi}S^2
\cr
&-& \frac{1}{4}e^{2\Phi}\tilde{F}^2 + \frac{3}{4}e^{2\Phi}X^2 + \frac{1}{48}e^{2\Phi}\tilde{G}^2 - \frac{3}{2}e^{2\Phi}m^2 \ ,
\ee
and where we use the $+$ component of (\ref{miiaint4}) to evaluate the covariant derivative terms. In order to obtain (\ref{miiaint1}) from these expressions we make use of the Bianchi identities (\ref{miiabian2}), the field equations (\ref{miiafeq1})-(\ref{miiafeq6}), in particular in order to eliminate the $(\tilde{\nabla} \Phi)^2$ term. We have also made use of the $+-$ component of the Einstein equation (\ref{miiafeq7}) in order to rewrite the scalar curvature $\tilde{R}$ in terms of $\Delta$. Therefore (\ref{miiaint1}) follows from (\ref{miiaint4}) and (\ref{miiaint7}) together with the field equations and Bianchi identities mentioned above.
\subsection{The + (\ref{miiaint7}) condition linear in $u$}
Since $\phi_+ = \eta_+ + u\Gamma_{+}\Theta_{-}\eta_-$, we must consider the part of the $+$ component of (\ref{miiaint7}) which is linear in $u$. On defining
\bea
\mathcal{B}_1 &=& \partial_i \Phi \Gamma^i \eta_{-} -{1\over12} \Gamma_{11} ( 6 L_i \Gamma^i+\tilde H_{ijk} \Gamma^{ijk}) \eta_-+{3\over8} e^\Phi \Gamma_{11} (2 S+\tilde F_{ij} \Gamma^{ij})\eta_-
\cr
&+&{1\over 4\cdot 4!}e^{\Phi} ( 12 X_{ij} \Gamma^{ij}+\tilde G_{j_1j_2j_3j_4} \Gamma^{j_1j_2j_3j_4}) \eta_- + \frac{5}{4}e^{\Phi}m ~ \eta_- \ ,
\ee
one finds that the $u$-dependent part of (\ref{miiaint7}) is proportional to
\bea
-\frac{1}{2}\Gamma^{i}\tilde{\nabla}_{i}(\mathcal{B}_1) + \Psi_2 \mathcal{B}_1~,
\ee
where
\bea
\Psi_2 &=& \tilde{\nabla}_{i}\Phi\Gamma^{i} + \frac{1}{8}h_{i}\Gamma^{i} - \frac{1}{16}e^{\Phi}X_{l_1 l_2}\Gamma^{l_1 l_2} - \frac{1}{192}e^{\Phi}\tilde{G}_{l_1 l_2 l_3 l_4}\Gamma^{l_1 l_2 l_3 l_4} - \frac{1}{8}e^{\Phi}m
\cr
&+& \Gamma_{11}\bigg(\frac{1}{48}\tilde{H}_{l_1 l_2 l_3}\Gamma^{l_1 l_2 l_3} + \frac{1}{8}L_{i}\Gamma^{i} + \frac{1}{16}e^{\Phi}\tilde{F}_{l_1 l_2}\Gamma^{l_1 l_2} + \frac{1}{8}e^{\Phi}S \bigg)~.
\ee
We have made use of the $-$ component of (\ref{miiaint4}) in order to evaluate the covariant derivative in the above expression. In addition we have made use of the Bianchi identities (\ref{miiabian2}) and the field equations (\ref{miiafeq1})-(\ref{miiafeq6}).
\subsection{The (\ref{miiaint2}) condition }
In order to show that (\ref{miiaint2}) is implied by the independent KSEs we will show that it follows from (\ref{miiaint1}). First act on (\ref{miiaint1}) with the Dirac operator $\Gamma^{i}\tilde{\nabla}_{i}$ and use the field equations (\ref{miiafeq1}) - (\ref{miiafeq6}) and the Bianchi identities to eliminate the terms which contain derivatives of the fluxes and then use (\ref{miiaint1}) to rewrite the $dh$-terms in terms of $\Delta$. Then use the conditions (\ref{miiaint4}) and (\ref{miiaint5}) to eliminate the $\partial_i \Phi$-terms from the resulting expression, some of the remaining terms will vanish as a consequence of (\ref{miiaint1}). After performing these calculations, the condition (\ref{miiaint2}) is obtained, therefore it follows from section \ref{miiaB1sec} above that (\ref{miiaint2}) is implied by (\ref{miiaint4}) and (\ref{miiaint7}) together with the field equations and Bianchi identities mentioned above.
\subsection{The (\ref{miiaint3}) condition }
In order to show that (\ref{miiaint3}) is implied by the independent KSEs we can compute the following,
\bea
&&\bigg(\frac{1}{4}\tilde{R} + \Gamma^{i j}\tilde{\nabla}_{i}\tilde{\nabla}_{j}\bigg)\eta_- + \Gamma^{i}\tilde{\nabla}_i(\mathcal{B}_1)
\cr
&+& \bigg(-\tilde{\nabla}_i\Phi \Gamma^{i} + \frac{1}{4}h_{i}\Gamma^{i} + \frac{1}{16}e^{\Phi}X_{l_1 l_2}\Gamma^{l_1 l_2} + \frac{1}{192}e^{\Phi}\tilde{G}_{l_1 l_2 l_3 l_4}\Gamma^{l_1 l_2 l_3 l_4} + \frac{1}{8}e^{\Phi}m
\cr
&+& \Gamma_{11}(-\frac{1}{4}L_{l}\Gamma^{l} + \frac{1}{24}\tilde{H}_{l_1 l_2 l_3}\Gamma^{l_1 l_2 l_3} - \frac{1}{8}e^{\Phi}S
- \frac{1}{16}e^{\Phi}\tilde{F}_{l_1 l_2}\Gamma^{l_1 l_2}) \bigg)\mathcal{B}_1 = 0~,
\nonumber \\
\ee
where we use the $-$ component of (\ref{miiaint4}) to evaluate the covariant derivative terms. The expression above vanishes identically since the $-$ component of (\ref{miiaint7}) is equivalent to $\mathcal{B}_1 = 0$. In order to obtain (\ref{miiaint3}) from these expressions we make use of the Bianchi identities (\ref{miiabian2}) and the field equations (\ref{miiafeq1})-(\ref{miiafeq6}). Therefore (\ref{miiaint3}) follows from (\ref{miiaint4}) and (\ref{miiaint7}) together with the field equations and Bianchi identities mentioned above.

\subsection{The + (\ref{miiaint4}) condition linear in $u$}

Next consider the part of the $+$ component of (\ref{miiaint4}) which is linear in $u$. First compute
\bea
\bigg(\Gamma^{j}(\tilde{\nabla}_{j}\tilde{\nabla}_{i} - \tilde{\nabla}_{i}\tilde{\nabla}_{j})  -\frac{1}{2}\tilde{R}_{ij}\Gamma^{j}\bigg)\eta_- - \tilde{\nabla}_{i}(\mathcal{B}_1) - \Psi_{i} \mathcal{B}_1 = 0~,
\ee
where
\bea
\Psi_{i} = \frac{1}{4}h_{i} - \Gamma_{11}(\frac{1}{4}L_{i} + \frac{1}{8}\tilde{H}_{i j k}\Gamma^{j k}) \ ,
\ee
and where we have made use of the $-$ component of (\ref{miiaint4}) to evaluate the covariant derivative terms. The resulting expression corresponds to the expression obtained by expanding out the $u$-dependent part of the $+$ component of (\ref{miiaint4}) by using the $-$ component of (\ref{miiaint4}) to evaluate the covariant derivative. We have made use of the Bianchi identities (\ref{miiabian2}) and the field equations (\ref{miiafeq1})-(\ref{miiafeq5})

\section{Calculation of Laplacian of $\parallel \eta_\pm \parallel^2$}
\label{miiamaxpex}
To establish the Lichnerowicz type theorems in \ref{lichthx}, we calculate the Laplacian of $\parallel \eta_\pm \parallel^2$.
For this let us generalise the modified horizon Dirac operator as ${\mathscr D}^{(\pm)}= {\cal D}^{(\pm)}+ q {\cal A}^{(\pm)}$ and assume throughout that  ${\mathscr D}^{(\pm)}\eta_\pm=0$;
in section \ref{lichthx} we had set $q=-1$.

To proceed, we compute the Laplacian
\bea
\tilde{\nabla}^i \tilde{\nabla}_i ||\eta_{\pm}||^2 = 2\langle\eta_\pm,\tilde{\nabla}^i \tilde{\nabla}_i\eta_\pm\rangle + 2 \langle\tilde{\nabla}^i \eta_\pm, \tilde{\nabla}_i \eta_\pm\rangle \ .
\ee
To evaluate this expression note that
\bea
\tilde{\nabla}^i \tilde{\nabla}_i \eta_\pm &=& \Gamma^{i}\tilde{\nabla}_{i}(\Gamma^{j}\tilde{\nabla}_j \eta_\pm) -\Gamma^{i j}\tilde{\nabla}_i \tilde{\nabla}_j \eta_\pm
\nonumber \\
&=& \Gamma^{i}\tilde{\nabla}_{i}(\Gamma^{j}\tilde{\nabla}_j \eta_\pm) + \frac{1}{4}\tilde{R}\eta_\pm
\nonumber \\
&=& \Gamma^{i}\tilde{\nabla}_{i}(-\Psi^{(\pm)}\eta_\pm -q\mathcal{A}^{(\pm)}\eta_{\pm}) + \frac{1}{4}\tilde{R} \eta_\pm \ .
\ee
It follows that
\bea
\langle\eta_\pm,\tilde{\nabla}^i \tilde{\nabla}_i\eta_\pm \rangle &=& \frac{1}{4}\tilde{R}\parallel \eta_\pm \parallel^2
+ \langle\eta_\pm, \Gamma^{i}\tilde{\nabla}_i(-\Psi^{(\pm)} - q\mathcal{A}^{(\pm)})\eta_\pm\rangle
\nonumber \\
&+& \langle\eta_\pm, \Gamma^{i}(-\Psi^{(\pm)} - q\mathcal{A}^{(\pm)})\tilde{\nabla}_i \eta_\pm \rangle~,
\ee
and also
\bea
\langle\tilde{\nabla}^i \eta_\pm, \tilde{\nabla}_i \eta_\pm\rangle &=& \langle{\hat\nabla^{(\pm)i}} \eta_\pm, {\hat\nabla^{(\pm)}_{i}} \eta_\pm\rangle - 2\langle\eta_\pm, (\Psi^{(\pm)i} + \kappa\Gamma^{i}\mathcal{A}^{(\pm)})^{\dagger} \tilde{\nabla}_i \eta_\pm \rangle
\nonumber \\
&-& \langle\eta_\pm, (\Psi^{(\pm)i} + \kappa\Gamma^{i}\mathcal{A}^{(\pm)})^{\dagger} (\Psi^{(\pm)}_i + \kappa \Gamma_{i} \, \mathcal{A}^{(\pm)}) \eta_\pm \rangle
\nonumber \\
&=& \parallel {\hat\nabla^{(\pm)}}\eta_{\pm} \parallel^2 - 2\langle \eta_{\pm}, \Psi^{(\pm)i\dagger}\tilde{\nabla}_{i}\eta_{\pm}\rangle
- 2\kappa \langle \eta_{\pm}, \mathcal{A}^{(\pm)\dagger}\Gamma^{i}\tilde{\nabla}_{i}\eta_{\pm}\rangle
\nonumber \\
&-&  \langle \eta_\pm, (\Psi^{(\pm)i\dagger}\Psi^{(\pm)}_i + 2\kappa \mathcal{A}^{(\pm)\dagger}\Psi^{(\pm)} + 8\kappa^2\mathcal{A}^{(\pm)\dagger}\mathcal{A}^{(\pm)})\eta_\pm \rangle
\nonumber \\
&=& \parallel {\hat\nabla^{(\pm)}}\eta_{\pm} \parallel^2 - 2\langle \eta_{\pm}, \Psi^{(\pm)i\dagger}\tilde{\nabla}_{i}\eta_{\pm}\rangle - \langle \eta_\pm , \Psi^{(\pm)i\dagger}\Psi^{(\pm)}_i \eta_\pm \rangle
\nonumber \\
&+& (2\kappa q - 8\kappa^2)\parallel \mathcal{A}^{(\pm)}\eta_\pm \parallel^2 \ .
\ee
Therefore,
\bea
\label{miiaextralap1b}
\frac{1}{2}\tilde{\nabla}^i \tilde{\nabla}_i ||\eta_{\pm}||^2 &=& \parallel {\hat\nabla^{(\pm)}}\eta_{\pm} \parallel^2 + \, (2\kappa q - 8\kappa^2)\parallel \mathcal{A}^{(\pm)}\eta_\pm \parallel^2
\nonumber \\
&+& \langle \eta_\pm, \bigg(\frac{1}{4}\tilde{R} + \Gamma^{i}\tilde{\nabla}_i(-\Psi^{(\pm)} - q\mathcal{A}^{(\pm)}) - \Psi^{(\pm)i\dagger}\Psi^{(\pm)}_i \bigg) \eta_\pm \rangle
\nonumber \\
&+& \langle \eta_\pm, \bigg( \Gamma^{i}(-\Psi^{(\pm)} - q\mathcal{A}^{(\pm)}) - 2\Psi^{(\pm)i\dagger}\bigg)\tilde{\nabla}_i \eta_\pm \rangle \ .
\ee
In order to simplify the expression for the Laplacian, we shall attempt to rewrite the third line in ({\ref{miiaextralap1b}}) as
\bea
\label{miiabilin}
\langle \eta_\pm, \bigg( \Gamma^{i}(-\Psi^{(\pm)} - q\mathcal{A}^{(\pm)}) - 2\Psi^{(\pm)i\dagger}\bigg)\tilde{\nabla}_i \eta_\pm \rangle &=& \langle \eta_\pm, \mathcal{F}^{(\pm)}\Gamma^{i}\tilde{\nabla}_i \eta_\pm \rangle + W^{(\pm)i}\tilde{\nabla}_i \parallel \eta_\pm \parallel^2~,
\nonumber \\
\ee
where $\mathcal{F}^{(\pm)}$ is linear in the fields and $W^{(\pm)i}$ is a vector. This expression is particularly advantageous, because the
first term on the RHS can be rewritten using the horizon
Dirac equation, and the second term is consistent with the application
of the maximum principle/integration by parts arguments which
are required for the generalised Lichnerowicz theorems. In order to rewrite ({\ref{miiabilin}}) in this fashion, note that
\bea
\Gamma^{i}(\Psi^{(\pm)} + q\mathcal{A}^{(\pm)}) + 2\Psi^{(\pm)i\dagger} &=& \big(\mp h^i \mp (q+1)\Gamma_{11}L^{i} + {1 \over 2}(q+1)\Gamma_{11}\tilde{H}^{i}{}_{\ell_1 \ell_2}\Gamma^{\ell_1 \ell_2}
+ 2q\tilde{\nabla}^i \Phi \big)
\nonumber \\
&+& \big(\pm \frac{1}{4}h_{j}\Gamma^{j} \pm (\frac{q}{2} + \frac{1}{4})\Gamma_{11} L_{j}\Gamma^{j}
\nonumber \\
&-& (\frac{q}{12} + \frac{1}{8})\Gamma_{11}\tilde{H}_{\ell_1 \ell_2 \ell_3}\Gamma^{\ell_1 \ell_2 \ell_3}- q\tilde{\nabla}_j \Phi \Gamma^{j}\big)\Gamma^{i}
\nonumber \\
&+& (q+1) \bigg(\mp {1 \over 8}e^{\Phi}X_{\ell_1 \ell_2}\Gamma^{i}\Gamma^{\ell_1 \ell_2} + \frac{5}{4}e^{\Phi}m\Gamma^i 
\nonumber \\
&+&{1 \over 96} e^{\Phi}\tilde{G}_{\ell_1 \ell_2 \ell_3 \ell_4}\Gamma^{i}\Gamma^{\ell_1 \ell_2 \ell_3 \ell_4} \bigg) 
\nonumber \\
&+& (q+1)\Gamma_{11}\bigg(\pm {3 \over 4} e^{\Phi}S\Gamma^{i}
-{3 \over 8}e^{\Phi}\tilde{F}_{\ell_1 \ell_2}\Gamma^{i}\Gamma^{\ell_1 \ell_2}\bigg) \ .
\ee
One finds that (\ref{miiabilin}) is only
possible for $q=-1$ and thus we have
\bea
W^{(\pm)i} = \frac{1}{2}(2\tilde{\nabla}^i \Phi \pm h^i) \ ,
\ee
and
\bea
\mathcal{F}^{(\pm)} = \mp \frac{1}{4}h_{j}\Gamma^{j} - \tilde{\nabla}_{j}\Phi \Gamma^{j} + \Gamma_{11}\bigg(\pm \frac{1}{4}L_{j}\Gamma^{j} +  \frac{1}{24}\tilde{H}_{\ell_1 \ell_2 \ell_3}\Gamma^{\ell_1 \ell_2 \ell_3}\bigg) \ .
\ee
We remark that  $\dagger$ is the adjoint with respect to the $Spin(8)$-invariant inner product $\langle \phantom{i},\phantom{i} \rangle$. The choice of inner product is such that
\bea
\label{miiahermiden1}
\langle \eta_+, \Gamma^{[k]} \eta_+ \rangle &=& 0, \qquad
k = 2\, (\text{mod }4) \ {\rm and} \  k=3\, (\text{mod }4) \ ,
\cr
\langle \eta_+, \Gamma_{11}\Gamma^{[k]} \eta_+ \rangle &=& 0,
\qquad k = 1\, (\text{mod }4) \ {\rm and} \  k= 2\, (\text{mod }4) \ ,
\ee
where $\Gamma^{[k]}$ denote skew-symmetric products of k gamma matrices. For a more detailed explanation see \cite{iiaindex}.
It follows that
\bea
\label{miialaplacian}
\frac{1}{2}\tilde{\nabla}^i \tilde{\nabla}_i ||\eta_{\pm}||^2 &=& \parallel {\hat\nabla^{(\pm)}}\eta_{\pm} \parallel^2 + \, (-2\kappa  - 8\kappa^2)\parallel \mathcal{A}^{(\pm)}\eta_\pm \parallel^2
+ W^{(\pm)i}\tilde{\nabla}_{i}\parallel \eta_\pm \parallel^2
\nonumber \\
&+& \langle \eta_\pm, \bigg(\frac{1}{4}\tilde{R} + \Gamma^{i}\tilde{\nabla}_i(-\Psi^{(\pm)} + \mathcal{A}^{(\pm)}) - \Psi^{(\pm)i\dagger}\Psi^{(\pm)}_i  + \mathcal{F}^{(\pm)}(-\Psi^{(\pm)} + \mathcal{A}^{(\pm)})\bigg) \eta_\pm \rangle \ .
\nonumber \\
\ee
Using (\ref{miiafeq8}) and the dilaton field equation (\ref{miiafeq6}),  we get
\bea
\tilde{R} &=& -\tilde{\nabla}^{i}(h_i) + \frac{1}{2}h^2 - 4(\tilde{\nabla}\Phi)^2 - 2h^{i}\tilde{\nabla}_{i}\Phi - \frac{3}{2}L^2 + \frac{5}{12}\tilde{H}^2
\nonumber \\
&+& \frac{7}{2}e^{2\Phi}S^2 - \frac{5}{4}e^{2\Phi}\tilde{F}^2 + \frac{3}{4}e^{2\Phi}X^2 - \frac{1}{48}e^{2\Phi}\tilde{G}^2 - \frac{9}{2}e^{2\Phi}m^2 \ .
\ee
One obtains, upon using the field equations and Bianchi identities,
\bea
\label{miiaquad}
\bigg(\frac{1}{4}\tilde{R} &+& \Gamma^{i}\tilde{\nabla}_i(-\Psi^{(\pm)} + \mathcal{A}^{(\pm)}) - \Psi^{(\pm)i\dagger}\Psi^{(\pm)}_i  + \mathcal{F}^{(\pm)}(-\Psi^{(\pm)} + \mathcal{A}^{(\pm)})\bigg)\eta_\pm
\nonumber \\
&=& \bigg[ \big(\pm \frac{1}{4}\tilde{\nabla}_{\ell_1}(h_{\ell_2}) \mp \frac{1}{16}\tilde{H}^{i}{}_{\ell_1 \ell_2}L_{i}\big)\Gamma^{\ell_1 \ell_2}+  \big( \pm \frac{1}{8}\tilde{\nabla}_{\ell_1}(e^{\Phi}X_{\ell_2 \ell_3}) 
\nonumber \\
&+& \frac{1}{24}\tilde{\nabla}^{i}(e^{\Phi}\tilde{G}_{i \ell_1 \ell_2 \ell_3}) \mp \frac{1}{96}e^{\Phi}h^{i}\tilde{G}_{i \ell_1 \ell_2 \ell_3} -\frac{1}{32}e^{\Phi}X_{\ell_1 \ell_2}h_{\ell_3}  \mp \frac{1}{8}e^{\Phi}\tilde{\nabla}_{\ell_1}\Phi X_{\ell_2 \ell_3}
\nonumber \\
&-& \frac{1}{24}e^{\Phi}\tilde{\nabla}^{i}\Phi \tilde{G}_{i \ell_1 \ell_2 \ell_3}
\mp \frac{1}{32}e^{\Phi}\tilde{F}_{\ell_1 \ell_2}L_{\ell_3}
\mp \frac{1}{96}e^{\Phi}S\tilde{H}_{\ell_1 \ell_2 \ell_3} - \frac{1}{32}e^{\Phi}\tilde{F}^{i}{}_{\ell_1}\tilde{H}_{i \ell_2 \ell_3}\big)\Gamma^{\ell_1 \ell_2 \ell_3}
\nonumber \\
&+& \Gamma_{11}\bigg(\big(\mp \frac{1}{4}\tilde{\nabla}_{\ell}(e^{\Phi}S) - \frac{1}{4}\tilde{\nabla}^{i}(e^{\Phi}\tilde{F}_{i \ell}) +\frac{1}{16}e^{\Phi}S h_{\ell} \pm \frac{1}{16}e^{\Phi}h^{i}\tilde{F}_{i \ell} \pm \frac{1}{4}e^{\Phi}\tilde{\nabla}_{\ell}\Phi S
\nonumber \\
&+& \frac{1}{4}e^{\Phi}\tilde{\nabla}^{i}\Phi\tilde{F}_{i \ell} + \frac{1}{16}e^{\Phi}L^i X_{i \ell}
\mp \frac{1}{32}e^{\Phi}\tilde{H}^{i j}{}_{\ell}X_{i j}
- \frac{1}{96}e^{\Phi}\tilde{G}^{i j k}{}_{\ell}\tilde{H}_{i j k} \pm \frac{1}{16}e^{\Phi}m L_{\ell}\big)\Gamma^{\ell}
\nonumber \\
&+& \big(\mp \frac{1}{4}\tilde{\nabla}_{\ell_1}(L_{\ell_2}) - \frac{1}{8}\tilde{\nabla}^{i}(\tilde{H}_{i \ell_1 \ell_2}) + \frac{1}{4}\tilde{\nabla}^{i}\Phi \tilde{H}_{i \ell_1 \ell_2} \pm  \frac{1}{16}h^{i}\tilde{H}_{i \ell_1 \ell_2}\big)\Gamma^{\ell_1 \ell_2}
\nonumber \\
&+& \big(\pm \frac{1}{384}e^{\Phi}\tilde{G}_{\ell_1 \ell_2 \ell_3 \ell_4}L_{\ell_5} \pm  \frac{1}{192}e^{\Phi}\tilde{H}_{\ell_1 \ell_2 \ell_3}X_{\ell_4 \ell_5}
\nonumber \\
&+& \frac{1}{192}e^{\Phi}\tilde{G}^{i}{}_{\ell_1 \ell_2 \ell_3}\tilde{H}_{i \ell_4 \ell_5}\big)\Gamma^{\ell_1 \ell_2 \ell_3 \ell_4 \ell_5}\bigg)
\bigg] \eta_\pm
\nonumber \\
&+& {1 \over 2} \big(1 \mp 1\big) \bigg(h^i {\tilde{\nabla}}_i \Phi
-{1 \over 2} {\tilde{\nabla}}^i h_i \bigg) \eta_\pm \ .
\ee
Note that with the exception of the final line of the RHS of ({\ref{miiaquad}}), all terms on the RHS of the above expression
give no contribution to the second line of (\ref{miialaplacian}),
using (\ref{miiahermiden1}), since all these terms in (\ref{miiaquad}) are anti-Hermitian and thus the bilinears vanish.
Furthermore, the contribution to the Laplacian of $\parallel \eta_+ \parallel^2$ from the final line of ({\ref{miiaquad}}) also vanishes;
however the final line of ({\ref{miiaquad}}) {\it does} give a contribution
to the second line of ({\ref{miialaplacian}}) in the case of the
Laplacian of $\parallel \eta_- \parallel^2$.
We  proceed
to consider the Laplacians
of $\parallel \eta_\pm \parallel^2$ separately, as the analysis
of the conditions imposed by the global properties of ${\cal{S}}$
differs slightly in the two cases.
For the Laplacian
of $\parallel \eta_+ \parallel^2$, we obtain from ({\ref{miialaplacian}}):
\bea
\label{miial1}
{\tilde{\nabla}}^{i}{\tilde{\nabla}}_{i}\parallel\eta_+\parallel^2 - (2\tilde{\nabla}^i \Phi +  h^i) {\tilde{\nabla}}_{i}\parallel\eta_+\parallel^2 = 2\parallel{\hat\nabla^{(+)}}\eta_{+}\parallel^2 - (4\kappa + 16 \kappa^2)\parallel\mathcal{A}^{(+)}\eta_+\parallel^2 \ .
\ee
This proves (\ref{miiamaxprin}).
The Laplacian of $\parallel \eta_- \parallel^2$
is calculated from ({\ref{miialaplacian}}), on taking account of the contribution to the second line of
({\ref{miialaplacian}}) from the final line of ({\ref{miiaquad}}). One
obtains
\bea
\label{miial2}
{\tilde{\nabla}}^{i} \big( e^{-2 \Phi} V_i \big)
= -2 e^{-2 \Phi} \parallel{\hat\nabla^{(-)}}\eta_{-}\parallel^2 +   e^{-2 \Phi} (4 \kappa +16 \kappa^2) \parallel\mathcal{A}^{(-)}\eta_-\parallel^2~,
\ee
where
\bea
V=-d \parallel \eta_- \parallel^2 - \parallel \eta_- \parallel^2 h \ .
\ee
This proves (\ref{miial2b}) and completes the proof.

	\chapter{$D=5$ Supergravity Calculations}

In this Appendix, we present conventions \cite{5dindex} for $D=5$ supergravity coupled to vector multiplets, as well as technical details of the analysis of the KSE
for the near-horizon solutions.

\section{Supersymmetry conventions}
We first present a matrix representation of $\mathrm{Cliff}(4,1)$ adapted to the basis ({\ref{basis1}}).
The space of Dirac spinors is identified with $\bC^4$ and we set
\bea
\Gamma_i = \begin{pmatrix}  \sigma^i \ \ \ \ \  0  \\   \ \ 0 \ \ -\sigma^i \end{pmatrix}, \qquad
\Gamma_- = \begin{pmatrix} \ \ 0  \ \ \ \sqrt{2}\, \bI_2  \\  0 \ \ \ \ \ 0 \end{pmatrix}, \qquad
\Gamma_+ = \begin{pmatrix}  \ \ 0 \ \ \ \ \ 0  \\ \sqrt{2}\, \bI_2 \ \ \ \ 0 \end{pmatrix}
\ee
where $\sigma^i$, $i=1,2,3$ are the Hermitian Pauli matrices $\sigma^i \sigma^j = \delta^{ij} \bI_2 + i \epsilon^{ijk} \sigma^k$.
Note that
\bea
\Gamma_{+-} = \begin{pmatrix} -\bI_2 \ \ \ \ \ 0  \\ \ \ \ 0 \ \ \ \bI_2 \end{pmatrix}~,
\ee
and hence
\bea
\Gamma_{+-123} = -i \bI_4~.
\ee
It will be convenient to decompose the spinors into positive and negative chiralities
with respect to the lightcone directions as
\bea
\epsilon = \epsilon_+ + \epsilon_-~,
\ee
where
\bea
\Gamma_{+-} \epsilon_\pm = \pm \epsilon_\pm \ , \qquad {\rm or \ equivalently} \qquad \Gamma_\pm \epsilon_\pm =0~.
\ee
With these conventions, note that
\bea
\label{conv}
\Gamma_{ij} \epsilon_\pm = \mp i \epsilon_{ij}{}^k \Gamma_k \epsilon_\pm \ ,
\qquad \Gamma_{ijk} \epsilon_\pm = \mp i \epsilon_{ijk} \epsilon_\pm~.
\ee
The Dirac representation of $Spin(4,1)$ decomposes under $Spin(3)=SU(2)$ as $\bC^4=\bC^2\oplus \bC^2$ each subspace
specified by the lightcone projections $\Gamma_\pm$. On each $\bC^2$,
we have made use of the $Spin$-invariant inner product  ${\rm Re } \langle , \rangle$ which is identified with the real part of the standard Hermitian inner product. On $\bC^2\oplus \bC^2$, the Lie algebra of
$Spin(3)$ is spanned  by $\Gamma_{ij}$, $i,j=1,2,3$. In particular,  note that $(\Gamma_{ij})^\dagger = - \Gamma_{ij}$.
We can also introduce a non-degenerate $Spin(4,1) \times U(1)$ invariant inner product $\cal{B}$ by,
\bea
{\cal{B}}(\epsilon, \eta) = \langle (\Gamma_{+} -\Gamma_{-})\epsilon, \eta \rangle \ .
\ee
The charge conjugation operator $C$ 
can be chosen to be 
\bea
C=\begin{pmatrix}  i \sigma^2 \ \ \ \ \  0  \\   \ \ 0 \ \ -i \sigma^2 \end{pmatrix} = i \Gamma_{2}
\ee
and satisfies $C* \Gamma_\mu + \Gamma_\mu C* =0$.\footnote{$C*$ refers to taking the complex conjugate then a matrix multiplication by $C$} Furthermore, if $\epsilon$ is any Dirac spinor then
\bea
\langle \epsilon, C* \epsilon \rangle =0 \ .
\ee

\section{Integrability conditions of $D=5$ supergravity}
In this Appendix, we summarize the integrability conditions for the
five dimensional supergravity coupled to arbitrarily many vector multiplets.
We begin with the ungauged theory first, and then consider the effect of including
a $U(1)$ gauge term.
First we will state the supercovariant connection $\cal{R}_{\mu \nu}$ given by,
\bea
[{\cal D}_{\mu}, {\cal D}_{\nu}]\epsilon \equiv \cal{R}_{\mu \nu}\epsilon \ ,
\ee 
where,
\bea
{\cal R}_{\mu \nu} &=& \frac{1}{4}R_{\mu \nu, \rho \sigma}\Gamma^{\rho \sigma} + X_{I} \bigg(\frac{i}{2}\Gamma^{\rho} \nabla_{\nu}{F^{I}\,_{\mu \rho}}  - \frac{i}{2}\Gamma^{\rho} \nabla_{\mu}{F^{I}\,_{\nu \rho}}  - \frac{i}{8}\Gamma_{\mu}\,^{\rho \sigma} \nabla_{\nu}{F^{I}\,_{\rho \sigma}} 
\nonumber \\
&+&\frac{i}{8}\Gamma_{\nu}\,^{\rho \sigma} \nabla_{\mu}{F^{I}\,_{\rho \sigma}} +\frac{1}{8}F^{I}{}_{\nu \rho} F^{J}{}_{\sigma \kappa} X_{J} \Gamma_{\mu}{}^{ \rho \sigma \kappa} \bigg) +X_{I} X_{J} \bigg( - \frac{3}{8}F^{I}\,_{\mu}\,^{\rho} F^{J}\,_{\nu}\,^{\sigma} \Gamma_{\rho \sigma} 
\nonumber \\
&-& \frac{1}{8}F^{I}\,_{\mu}\,^{\rho} F^{J}\,_{\rho}\,^{\sigma} \Gamma_{\nu \sigma} - \frac{1}{8}F^{I}\,_{\mu}\,^{\rho} F^{J \sigma \kappa} \Gamma_{\nu \rho \sigma \kappa}+\frac{1}{8}F^{I}\,_{\nu}\,^{\rho} F^{J}\,_{\rho}\,^{\sigma} \Gamma_{\mu \sigma}
\nonumber \\
&+&\frac{1}{16}F^{I \rho \sigma} F^{J}\,_{\rho \sigma} \Gamma_{\mu \nu}\bigg)
+\frac{i}{2}F^{I}\,_{\mu}\,^{\rho} \Gamma_{\rho} \nabla_{\nu}{X_{I}}  - \frac{i}{2}F^{I}\,_{\nu}\,^{\rho} \Gamma_{\rho} \nabla_{\mu}{X_{I}} 
\nonumber \\
&-& \frac{i}{8}F^{I \rho \sigma} \Gamma_{\mu \rho \sigma} \nabla_{\nu}{X_{I}} +\frac{i}{8}F^{I \rho \sigma} \Gamma_{\nu \rho \sigma} \nabla_{\mu}{X_{I}} \ .
\ee
To proceed, we consider the identity
\bea
\Gamma^{\nu}[{\cal D}_{\mu}, {\cal D}_{\nu}]\e  &+& \Phi_{I \mu}{\cal A}^{I}\e 
\nonumber \\
&=& -\frac{1}{2}E_{\mu \nu}\Gamma^{\nu}\e
+ i X_{I}\bigg(-\frac{3}{4}BF^{I}{}_{\mu \nu \rho}\Gamma^{\nu \rho} +
\frac{1}{8}BF^{I}{}_{\nu \rho \lambda}\Gamma_{\mu}{}^{\nu \rho \lambda}
\bigg)\epsilon 
\nonumber \\
&+& iX^{I}\bigg(-\frac{1}{6}FF_{I \nu}\Gamma_{\mu}{}^{\nu}
+\frac{1}{3}FF_{I \mu}\bigg)\e
\nonumber\\
&=& -\frac{1}{2}E_{\mu \nu}\Gamma^{\nu}\e + i X^{J}\bigg(-\frac{1}{2}Q_{I J}BF^{J}{}_{\mu \nu \rho}\Gamma^{\nu \rho}
+ \frac{1}{12}Q_{I J}BF^{J}{}_{\nu \rho \lambda}\Gamma_{\mu}{}^{\nu \rho \lambda}
\nonumber \\
&-& \frac{1}{6}FF_{J \nu}\Gamma_{\mu}{}^{\nu} + \frac{1}{3}FF_{J \mu}\bigg)\e \ ,
\ee
where,
\bea
\Phi_{I \mu} = \frac{3i}{8}\nabla_{\mu}{X_{I}} + Q_{I J}\bigg(-\frac{1}{6}F^{J}{}_{\mu \nu}\Gamma^{\nu}
+ \frac{1}{24}F^{J}{}_{\nu \rho}\Gamma_{\mu}{}^{\nu \rho}\bigg) \ ,
\ee
and
\bea
\frac{i}{3}\Gamma^{\mu}[{\cal D}_{\mu}, {\cal A}_{I}]\e  &+& \theta_{I J} {\cal A}^{J}\e 
\nonumber \\
&=& 
FX_{I} \e + \frac{i}{3}\bigg[\bigg(Q_{I J}-\frac{3}{2}X_{I}X_{J}\bigg)BF^{J}{}_{\mu \nu \rho}\Gamma^{\mu \nu \rho}- 2\bigg(\delta^{J}{}_{I} - X^{J}X_{I}\bigg)FF_{J \mu}\Gamma^{\mu}\bigg]\e \ ,
\nonumber \\
\ee
where,
\bea
\theta_{I J} = X_{I}\bigg(-\frac{3i}{4}\nabla_{\mu}{X_{J}}\Gamma^{\mu} + \frac{1}{12}Q_{J K}F^{K}{}_{\mu \nu}\Gamma^{\mu \nu}\bigg)
+ \frac{1}{24}C_{I J K}{\cal A}^{K} \ ,
\ee
The field equations and Bianchi identities are 
\bea
\label{5dueineq}
E_{\mu \nu} = R_{\mu \nu}-  Q_{I J}\left(  \nabla_{\mu}X^{I}\nabla_{\nu}X^{J} +F^{I}{}_{\mu \rho}F^{J}{}_{\nu}{}^{\rho}
-\frac{1}{6}g_{\mu \nu}F^{I}{}_{\rho \lambda}F^{J \rho \lambda}\right) = 0 \ ,
\ee
\bea
\label{5duscalareq}
FX_{I} &=& \nabla^{\mu}\left(\nabla_{\mu}{X_{I}}\right) - \nabla_{\mu}{X^{M}} \nabla^{\mu}{X^{N}} \left(\frac{1}{6}C_{I M N} - \frac{1}{2}C_{M N K} X_{I} X^{K}\right) 
\nonumber \\
&+& \frac{1}{2}F^{M}{}_{\mu \nu} F^{N \mu \nu} \bigg(C_{I N P} X_{M} X^{P} - \frac{1}{6}C_{I M N}-6X_{I} X_{M} X_{N}+\frac{1}{6}C_{M N J} X_{I} X^{J}\bigg) = 0 \ ,
\nonumber \\
\ee
\bea
\label{5dufeq}
FF_{I \mu} = \nabla^{\nu}{(Q_{I J}F^{J}{}_{\mu \nu})} -  \frac{1}{16} \epsilon_{\mu}{}^{\nu \rho \lambda \kappa} C_{I J K}F^{J}{}_{\nu \rho}F^{K}{}_{\lambda \kappa} = 0 \ ,
\ee
\bea
\label{5dubian1}
BF^{I}{}_{\mu \nu \rho} = \nabla_{[\mu}F^{I}{}_{\nu \rho]} = 0 \ ,
\ee
We can decompose $F^I$ as 
\bea
F^I = FX^I + G^I \ ,
\ee
Where
\bea
X_I F^I &=& F \ ,
\cr
X_I G^I &=& 0 \ .
\ee
The KSEs (\ref{Gkseo}) and (\ref{Akseo}) become
\bea
{\cal D}_{\mu}\e&\equiv& \nabla_{\mu} \epsilon + \frac{i}{8}\bigg(\Gamma_{\mu}{}^{\nu \rho} - 4\delta_{\mu}{}^{\nu}\Gamma^{\rho}\bigg)F_{\nu \rho} \epsilon = 0 \ , \label{D2KSE}\\
{\cal A}^{I}\epsilon &=& G^{I}{}_{\mu \nu}\Gamma^{\mu \nu}\e + 2i\nabla_{\mu}{X^{I}}\Gamma^{\mu}\e = 0
\cr
{\cal A}_{I}\epsilon &=& Q_{I J}G^{J}{}_{\mu \nu}\Gamma^{\mu \nu}\e - 3i\nabla_{\mu}{X_{I}}\Gamma^{\mu}\e = 0 \ , \label{A2KSE}\\
\nonumber
\ee
The field equations (\ref{5dueineq}), (\ref{5duscalareq}) and (\ref{5dufeq}) can also be decomposed as:
\bea
\label{ein2eq}
E_{\mu \nu} &=& R_{\mu \nu} - \frac{3}{2}F_{\mu \rho}F_{\nu}{}^{\rho} + \frac{1}{4}\delta_{\mu \nu}F^2 
\nonumber \\
&-& Q_{I J}\bigg( \nabla_{\mu}{X^{I}}\nabla_{\nu}{X^{J}} + G^{I}{}_{\mu \rho}G^{J}{}_{\nu}{}^{\rho}- \frac{1}{6}\delta_{\mu \nu}G^{I}{}_{\rho d}G^{J c d}\bigg) = 0 \ ,
\ee
\bea
\label{scalar2eq}
FX_{I} &=& \nabla^{\mu}{\nabla_{\mu}{X_{I}}} - \bigg(\frac{1}{6}C_{M N I}- \frac{1}{2}C_{M N K}X_{I}X^{K}\bigg)\nabla_{\mu}{X^{M}}\nabla^{\mu}{X^{N}} 
\nonumber \\
&-& \frac{2}{3}Q_{I J} F_{\mu \nu}G^{J \mu \nu}- \frac{1}{12}\bigg(C_{M N I} - X_{I}C_{M N J}X^{J}\bigg)G^{M}{}_{\mu \nu}G^{N \mu \nu} = 0 \ ,
\ee
and
\bea
FF_{I \mu} = X_{I}FF_{\mu} + FG_{I \mu} = 0 \ ,
\ee
with
\bea
FF_{\mu} &=& \frac{3}{2}\nabla^{\nu}{F_{\mu \nu}} - \frac{3}{8}\epsilon_{\mu}{}^{\nu \rho \lambda \kappa} F_{\nu \rho} F_{\lambda \kappa}  + \frac{3}{2}G^{I}\,_{\mu \nu} \nabla^{\nu}{X_{I}} - \frac{1}{16}C_{I J K}X^{K} \epsilon_{\mu}{}^{\nu \rho \lambda \kappa} G^{I}{}_{\nu \rho} G^{J}{}_{\lambda \kappa} = 0 \ ,
\nonumber \\
\label{F2eq} \\
FG_{I \mu} &=& Q_{I J} \nabla^{\nu}{G^{J}\,_{\mu \nu}} + \frac{3}{2}F_{\mu}\,^{\nu} \nabla_{\nu}{X_{I}}- G^{J}\,_{\mu \nu} \nabla^{\nu}{X^{K}} \left(2Q_{J K} X_{I}+\frac{1}{2}C_{I J K}\right) 
\cr
&-&\epsilon_{\mu}\,^{\nu \rho \lambda \kappa} \bigg[\frac{1}{16}G^{J}\,_{\nu \rho} G^{K}\,_{\lambda \kappa} \bigg(C_{I J K}-C_{J K L} X_{I} X^{L}\bigg)+\frac{1}{8}C_{I J K} F_{\nu \rho} G^{J}\,_{\lambda \kappa} X^{K}\bigg] = 0 \ ,
\nonumber
\label{Geq} \\
\ee
and 
\bea
FF_{\mu} &=&  X^I FF_{I \mu} \ ,
\cr
X^I FG_{I \mu} &=& 0 \ .
\ee
Similarly the Bianchi identity (\ref{5dubian1}) can be decomposed as:
\bea
BF^{I}{}_{\mu \nu \rho} = X^{I}BF_{\mu \nu \rho} + BG^{I}{}_{\mu \nu \rho} = 0 \ ,
\ee
with
\bea
BF_{\mu \nu \rho} &=& \nabla_{[a}{F_{\nu \rho]}} - G^{I}{}_{[\mu \nu}\nabla_{\rho]}{X_{I}} = 0 \ , \label{5dbianF} \\
BG^{I}{}_{\mu \nu \rho} &=& \bigg(\delta^{I}{}_{J} - X^{I}X_{J}\bigg)\nabla_{[\mu}{G^{J}{}_{\nu \rho]}} + F_{[\mu \nu}\nabla_{\rho]}{X^{I}} = 0
\ee
and 
\bea
BF_{\mu \nu \rho} &=& X_I BF^{I}{}_{\mu \nu \rho}  \ ,
\cr
X_I BG^{I}{}_{\mu \nu \rho} &=& 0 \ .
\ee
The integrability conditions (\ref{5dint1}) and (\ref{5dint2}) become:
\bea
\label{intF}
\Gamma^{\nu}[{\cal D}_{\mu}, {\cal D}_{\nu}]\e  &+& \Phi_{I \mu}{\cal A}^{I}\e 
\nonumber \\
&=& -\frac{1}{2}E_{\mu \nu}\Gamma^{\nu}\e
+ i \bigg(-\frac{3}{4}BF_{\mu \nu \rho}\Gamma^{\nu \rho} +
\frac{1}{8}BF_{\nu \rho \lambda}\Gamma_{\mu}{}^{\nu \rho \lambda} -\frac{1}{4}FF_{\nu}\Gamma_{\mu}{}^{\nu}
+\frac{1}{2}FF_{\mu}\bigg)\e \ ,
\nonumber \\
\ee
where,
\bea
\Phi_{I \mu} = \frac{3i}{8}\nabla_{\mu}{X_{I}} + Q_{I J}\bigg(- \frac{1}{6}G^{J}{}_{\mu \nu}\Gamma^{\nu}
+ \frac{1}{24}G^{J}{}_{\nu \rho}\Gamma_{\mu}{}^{\nu \rho}\bigg) \ ,
\ee
and
\bea
\frac{i}{3}\Gamma^{\mu}[{\cal D}_{\mu}, {\cal A}_{I}]\e  &+& \theta_{I J} {\cal A}^{J}\e  \ ,
\nonumber \\
&=& 
FX_{I} \e + i \bigg(\frac{1}{3}Q_{I J}BG^{J}{}_{\mu \nu \rho}\Gamma^{\mu \nu \rho}
- \frac{2}{3}FG_{I \mu}\Gamma^{\mu}\bigg)\e \ ,
\ee
where,
\bea
\theta_{I J} = X_{I}\bigg(-\frac{3i}{4}\nabla_{\mu}{X_{J}}\Gamma^{\mu} + \frac{1}{12}Q_{J K}G^{K}{}_{\mu \nu}\Gamma^{\mu \nu}\bigg)
+ \frac{1}{24}C_{I J K}{\cal A}^{K} \ .
\ee

\subsection{Inclusion of a $U(1)$ gauge term}

Having considered the analysis of the integrability conditions for the ungauged theory, we next summarize the integrability conditions on including a $SU(1)$ gauge term.

Again, we proceed by first stating the supercovariant connection $\cal{R}_{\mu \nu}$ given by,
\bea
[{\cal D}_{\mu}, {\cal D}_{\nu}]\epsilon \equiv \cal{R}_{\mu \nu}\epsilon \ ,
\ee 
where,
\bea
{\cal R}_{\mu \nu} &=& \frac{1}{4}R_{\mu \nu, \rho \sigma}\Gamma^{\rho \sigma} + X_{I} \bigg(\frac{i}{2}\Gamma^{\rho} \nabla_{\nu}{F^{I}\,_{\mu \rho}}  - \frac{i}{2}\Gamma^{\rho} \nabla_{\mu}{F^{I}\,_{\nu \rho}}  - \frac{i}{8}\Gamma_{\mu}\,^{\rho \sigma} \nabla_{\nu}{F^{I}\,_{\rho \sigma}} 
\nonumber \\
&+&\frac{i}{8}\Gamma_{\nu}\,^{\rho \sigma} \nabla_{\mu}{F^{I}\,_{\rho \sigma}} +\frac{1}{8}F^{I}{}_{\nu \rho} F^{J}{}_{\sigma \kappa} X_{J} \Gamma_{\mu}{}^{ \rho \sigma \kappa} \bigg) +X_{I} X_{J} \bigg( - \frac{3}{8}F^{I}\,_{\mu}\,^{\rho} F^{J}\,_{\nu}\,^{\sigma} \Gamma_{\rho \sigma} 
\nonumber \\
&-& \frac{1}{8}F^{I}\,_{\mu}\,^{\rho} F^{J}\,_{\rho}\,^{\sigma} \Gamma_{\nu \sigma} - \frac{1}{8}F^{I}\,_{\mu}\,^{\rho} F^{J \sigma \kappa} \Gamma_{\nu \rho \sigma \kappa}+\frac{1}{8}F^{I}\,_{\nu}\,^{\rho} F^{J}\,_{\rho}\,^{\sigma} \Gamma_{\mu \sigma}
\nonumber \\
&+&\frac{1}{16}F^{I \rho \sigma} F^{J}\,_{\rho \sigma} \Gamma_{\mu \nu}\bigg)
+ i \chi X^{J} X_{I} \bigg(\frac{1}{2}F^{I}\,_{\mu}\,^{\rho} V_{J} \Gamma_{\nu \rho}   - \frac{1}{2}F^{I}\,_{\nu}\,^{\rho} V_{J} \Gamma_{\mu \rho}  
\nonumber \\
&+&\frac{1}{4}F^{I \rho \sigma} V_{J} \Gamma_{\mu \nu \rho \sigma}  \bigg)
+\frac{i}{2}F^{I}\,_{\mu}\,^{\rho} \Gamma_{\rho} \nabla_{\nu}{X_{I}}  - \frac{i}{2}F^{I}\,_{\nu}\,^{\rho} \Gamma_{\rho} \nabla_{\mu}{X_{I}}  
\nonumber \\
&-& \frac{i}{8}F^{I \rho \sigma} \Gamma_{\mu \rho \sigma} \nabla_{\nu}{X_{I}} +\frac{i}{8}F^{I \rho \sigma} \Gamma_{\nu \rho \sigma} \nabla_{\mu}{X_{I}}  - \frac{1}{2}\chi V_{I} \Gamma_{\mu} \nabla_{\nu}{X^{I}} +\frac{1}{2}\chi V_{I} \Gamma_{\nu} \nabla_{\mu}{X^{I}}
\nonumber \\
&+&\frac{3i}{2}\chi V_{I} \nabla_{\nu}{A^{I}\,_{\mu}}  - \frac{3i}{2}\chi V_{I} \nabla_{\mu}{A^{I}\,_{\nu}}  +\frac{1}{2}{\chi}^{2} V_{I} V_{J} X^{I} X^{J} \Gamma_{\mu \nu} \ .
\ee
Now consider,
\bea
\Gamma^{\nu}[{\cal D}_{\mu}, {\cal D}_{\nu}]\e  &+& \Phi_{I \mu}{\cal A}^{I}\e 
\nonumber \\
&=& -\frac{1}{2}E_{\mu \nu}\Gamma^{\nu}\e
+ i X_{I}\bigg(-\frac{3}{4}BF^{I}{}_{\mu \nu \rho}\Gamma^{\nu \rho} +
\frac{1}{8}BF^{I}{}_{\nu \rho \lambda}\Gamma_{\mu}{}^{\nu \rho \lambda}
\bigg)\epsilon 
\nonumber \\
&+& iX^{I}\bigg(-\frac{1}{6}FF_{I \nu}\Gamma_{\mu}{}^{\nu}
+\frac{1}{3}FF_{I \mu}\bigg)\e
\nonumber\\
&=& -\frac{1}{2}E_{\mu \nu}\Gamma^{\nu}\e + i X^{J}\bigg(-\frac{1}{2}Q_{I J}BF^{J}{}_{\mu \nu \rho}\Gamma^{\nu \rho}
+ \frac{1}{12}Q_{I J}BF^{J}{}_{\nu \rho \lambda}\Gamma_{\mu}{}^{\nu \rho \lambda}
\nonumber \\
&-& \frac{1}{6}FF_{J \nu}\Gamma_{\mu}{}^{\nu} + \frac{1}{3}FF_{J \mu}\bigg)\e \ ,
\ee
where
\bea
\Phi_{I \mu} = \frac{3i}{8}\nabla_{\mu}{X_{I}} + Q_{I J}\bigg(-\frac{1}{6}F^{J}{}_{\mu \nu}\Gamma^{\nu}
+ \frac{1}{24}F^{J}{}_{\nu \rho}\Gamma_{\mu}{}^{\nu \rho}\bigg) + \frac{1}{4} i \chi V_{I} \Gamma_{\mu} \ ,
\ee
and
\bea
\frac{i}{3}\Gamma^{\mu}[{\cal D}_{\mu}, {\cal A}_{I}]\e  &+& \theta_{I J} {\cal A}^{J}\e \ ,
\nonumber \\
&=&
FX_{I} \e + \frac{i}{3}\bigg[\bigg(Q_{I J}-\frac{3}{2}X_{I}X_{J}\bigg)BF^{J}{}_{\mu \nu \rho}\Gamma^{\mu \nu \rho}- 2\bigg(\delta^{J}{}_{I} - X^{J}X_{I}\bigg)FF_{J \mu}\Gamma^{\mu}\bigg]\e \ ,
\nonumber \\
\ee
where
\bea
\theta_{I J} &=& X_{I}\bigg(-\frac{3i}{4}\nabla_{\mu}{X_{J}}\Gamma^{\mu} + \frac{1}{12}Q_{J K}F^{K}{}_{\mu \nu}\Gamma^{\mu \nu}\bigg)
+ \frac{1}{24}C_{I J K}{\cal A}^{K} 
\nonumber \\
&+& \frac{i}{2}\chi \bigg( X_{I}V_{J} + C_{I J L}Q^{L M}V_{M} \bigg) \ .
\ee
The field equations and Bianchi identities are 
\bea
\label{5dgeineq}
E_{\mu \nu} = R_{\mu \nu}-  Q_{I J}\left(  \nabla_{\mu}X^{I}\nabla_{\nu}X^{J} +F^{I}{}_{\mu \rho}F^{J}{}_{\nu}{}^{\rho}
-\frac{1}{6}g_{\mu \nu}F^{I}{}_{\rho \lambda}F^{J \rho \lambda}\right) + \frac{2}{3} \chi^2 V(X) g_{\mu \nu} = 0 \ ,
\ee
with the scalar potential given by $V(X) = 9 V_{I} V_{J} (X^{I} X^{J} - \frac{1}{2}Q^{I J})$
\bea
\label{5dgscalareq}
FX_{I} &=& \nabla^{\mu}\left(\nabla_{\mu}{X_{I}}\right) - \nabla_{\mu}{X^{M}} \nabla^{\mu}{X^{N}} \left(\frac{1}{6}C_{I M N} - \frac{1}{2}C_{M N K} X_{I} X^{K}\right) 
\nonumber \\
&+& \frac{1}{2}F^{M}{}_{\mu \nu} F^{N \mu \nu} \left(C_{I N P} X_{M} X^{P} - \frac{1}{6}C_{I M N}-6X_{I} X_{M} X_{N}+\frac{1}{6}C_{M N J} X_{I} X^{J}\right)
\nonumber \\
&+& 3 \chi^2 V_{M} V_{N}\bigg(\frac{1}{2}C_{I J K}Q^{M J}Q^{N K} + X_{I}(Q^{M N} - 2 X^{M}X^{N})\bigg) = 0 \ ,
\ee
\bea
\label{5dgfeq}
FF_{I \mu} = \nabla^{\nu}{(Q_{I J}F^{J}{}_{\mu \nu})} -  \frac{1}{16} \epsilon_{\mu}{}^{\nu \rho \lambda \kappa} C_{I J K}F^{J}{}_{\nu \rho}F^{K}{}_{\lambda \kappa} = 0
\ee
\bea
\label{5dgbian1}
BF^{I}{}_{\mu \nu \rho} = \nabla_{[\mu}F^{I}{}_{\nu \rho]} = 0 \ .
\ee

\section{Scalar orthogonality condition}
In this section, we shall prove that if $L_{I}\partial_{a}{X^{I}} = 0$ for all 
values of $a=1, \dots, k-1$, i.e if $L_{I}$ is perpendicular to all $\partial_{a}{X^{I}}$, then it must be parallel to $X_{I}$. 
To establish the first result, it is sufficient to prove that the elements of
the set $\{ \partial_{a}{X^{I}} \, , a=1,\dots,k-1 \}$ are linearly independent.
Given this, the condition $L_{I}\partial_{a}{X^{I}} = 0$ for all 
values of $a=1, \dots, k-1$ implies that $L_I$ is orthogonal to all linearly independent
$k-1$ elements of this set, and hence must be parallel to the 1-dimensional orthogonal
complement to the set, which is parallel to $X_I$.

It remains to prove the following Lemma.
\vskip5mm

{\it Lemma:} The elements of the set $\{ \partial_{a}{X^{I}} \, , a=1,\dots,k-1 \}$ are linearly independent.

\vskip5mm

{\it Proof:} Let $N^{a}$ for $a=1,\dots,k-1$ be constants, where at least one is non-zero and suppose $N^{a}\partial_{a}{X^{I}} = 0$, then we have from (\ref{pbmetric})
\begin{eqnarray}
h_{a b}N^{a} = Q_{I J}\partial_{a}{X^{I}}\partial_{b}{X^{J}}N^{a} = 0 \ ,
\end{eqnarray}
as $h_{a b}$ is non-degenerate, this implies that $N^{a} = 0$ for all $a=1,\dots,k-1$, which is a contradiction to our assumption that not all are zero and thus the elements of the set are linearly independent.

We remark that an equivalent statement implied by the above reasoning is that 
if $L^{I}\partial_{a}{X_{I}} = 0$ for all $a=1, \dots k-1$ then $L^{I}$ must be parallel to $X^{I}$.

\section{Simplification of KSEs on ${\cal{S}}$}
In this Appendix we show how several of the KSEs on ${\cal{S}}$ are implied
by the remaining KSEs, together with the field equations and Bianchi identities. To begin,
we show that
(\ref{5dint1}), (\ref{5dint2}), (\ref{5dint5}), and (\ref{5dint8}) which contain $\tau_+$ are implied from those containing $\phi_+$, along with some of the field equations and Bianchi identities. Then,
we establish that (\ref{5dint3}) and the terms linear in $u$ in (\ref{5dint4}) and (\ref{5dint7}) from the $+$ component are implied by the field equations, Bianchi identities and the $-$ component of (\ref{5dint4}) and (\ref{5dint7}). 
A particular useful identity is obtained by considering the integrability condition of (\ref{5dint4}), which implies that
\bea
(\tilde{\nabla}_{j}\tilde{\nabla}_{i} - \tilde{\nabla}_{i}\tilde{\nabla}_{j})\phi_{\pm} &=& \bigg(\pm\frac{1}{4}\tilde{\nabla}_j(h_i)  
\mp\frac{1}{4}\tilde{\nabla}_i(h_j)
\pm \frac{i}{4}\tilde{\nabla}_j(\alpha) \Gamma_i  
\mp \frac{i}{4}\tilde{\nabla}_i(\alpha) \Gamma_j
\nonumber \\
&+& \frac{i}{2}\tilde{\nabla}_j(\tilde{F}_{i \ell})\Gamma^\ell
- \frac{i}{2}\tilde{\nabla}_i(\tilde{F}_{j \ell})\Gamma^\ell 
- \frac{i}{8}\tilde{\nabla}_j(\tilde{F}_{\ell_1 \ell_2})\Gamma_i{}^{\ell_1 \ell_2} 
\nonumber \\
&+& \frac{i}{8}\tilde{\nabla}_i(\tilde{F}_{\ell_1 \ell_2})\Gamma_j{}^{\ell_1 \ell_2} 
\mp \alpha \tilde{F}_{j \ell}\Gamma_{i}{}^{\ell} \pm \frac{1}{4}\alpha\tilde{F}_{i}{}^{\ell}\Gamma_{j}{}^{\ell}
+ \frac{1}{8}\tilde{F}_{j}{}^{\lambda}\tilde{F}_{\lambda \ell}\Gamma_{i}{}^{\ell}
\nonumber \\
&-& \frac{1}{8}\tilde{F}_{i}{}^{\lambda}\tilde{F}_{\lambda \ell}\Gamma_{j}{}^{\ell}
- \frac{3}{8}\tilde{F}_{i \ell_1}\tilde{F}_{j \ell_2}\Gamma^{\ell_1 \ell_2}
-\frac{1}{8}\alpha^2 \Gamma_{i j}+ \frac{1}{16}\tilde{F}^2 \Gamma_{i j}
\nonumber \\
&+& \frac{1}{2}\chi^2 V_{I}V_{J}X^{I}X^{J} \Gamma_{i j}
\mp \frac{i}{2}\chi V_{I}X^{I} \alpha \Gamma_{i j}
- \chi V_{I} \Gamma_{[i}\tilde{\nabla}_{j]}{(X^{I})}
\nonumber \\
&-&  \frac{3i}{2} \chi V_{I}\tilde{F}^{I}{}_{i j}
+ i \chi V_{I}X^{I} \tilde{F}_{[i | \ell |}\Gamma_{j]}{}^{\ell}
\bigg)\phi_{\pm} \ \ .
\label{5dDDphicond}
\ee
This will be used in the analysis of (\ref{5dint1}), (\ref{5dint3}), (\ref{5dint5}) and the positive chirality part of  (\ref{5dint4}) which is linear in $u$. In order to show that the conditions are redundant, we will be considering different combinations of terms which vanish
as a consequence of the independent KSEs. However, non-trivial identities are found by 
explicitly expanding out the terms in each case.

\subsection{The  condition (\ref{5dint1})}
\label{5dint1sec}
It can be shown that the algebraic condition on $\tau_+$ (\ref{5dint1}) is implied by the independent KSEs. Let us define,
\bea
\xi_1 &=& \bigg({1\over2}\Delta - {1\over8}(dh)_{ij}\Gamma^{ij} -\frac{i}{4}\beta_i \Gamma^i + \frac{3i}{2}\chi V_{I}\alpha^I \bigg)\phi_+ 
\nonumber \\
&+& 2\bigg({1\over4} h_i\Gamma^i - \frac{i}{8}(-\tilde{F}_{j k}\Gamma^{j k} + 4\alpha) + \frac{1}{2} \chi V_{I}X^{I}\bigg)\tau_+ \ ,
\ee
where $\xi_1=0$ is equal to the condition (\ref{5dint1}). 
It is then possible to show that this expression for $\xi_1$ can be re-expressed as
\bea
\label{5dint1cond}
&&\xi_1 = \bigg(-\frac{1}{4}\tilde{R} - \Gamma^{i j}\tilde{\nabla}_{i}\tilde{\nabla}_{j}\bigg)\phi_+ 
+ \mu_{I}\mathcal{A}^{I}{}_1 = 0 \ ,
\ee
where the first two terms cancel as a consequence of the definition of curvature, and
\bea
\mu_{I} = \frac{3i}{16}\Gamma^{i}\tilde{\nabla}_{i}{X_{I}} - Q_{I J}\bigg(\frac{7}{24}L^{J} + \frac{5}{48}\tilde{G}^{J}{}_{\ell_1 \ell_2}\Gamma^{\ell_1 \ell_2} \bigg) + \frac{i}{8}\chi V_{I} \ ,
\ee
the scalar curvature is can be written as
\bea
\tilde{R} = -2\Delta - \frac{1}{2}h^{2} + \frac{7}{2}\alpha^2 + \frac{5}{4}\tilde{F}^2 - \frac{2}{3}\chi^2 U + Q_{I J}\bigg(\frac{7}{3}L^{I}L^{J} + \frac{5}{6}\tilde{G}^{I\ell_1 \ell_2}\tilde{G}^{J}{}_{\ell_1 \ell_2} + \tilde{\nabla}_{i}{X^{I}}\tilde{\nabla}^{i}{X^{J}}\bigg) \ ,
\nonumber \\
\ee
and
\bea
\label{5dint1condaux}
\mathcal{A}^{I}{}_1 = \bigg[\tilde{G}^I{}_{i j}\Gamma^{i j} - 2 L^I + 2i\tilde{\nabla}_i X^I \Gamma^i - 6i\chi \bigg(Q^{I J} - \frac{2}{3}X^{I}X^{J}\bigg)V_J\bigg]\phi_+ \ .
\ee
The expression appearing in ({\ref{5dint1condaux}}) vanishes because
 $\mathcal{A}^{I}{}_1 = 0$ is equivalent to the positive chirality part of (\ref{5dint7}).
Furthermore, the expression for $\xi_1$ given in ({\ref{5dint1cond}}) also vanishes.
We also use (\ref{5dDDphicond}) to evaluate the terms in the first bracket in (\ref{5dint1cond}) and explicitly expand out the terms with $\mathcal{A}^{I}{}_1$. In order to obtain (\ref{5dint1}) from these expressions we make use of the Bianchi identities (\ref{beq}), the field equations (\ref{5dfeq1}) and (\ref{5dfeq2}). We have also made use of the $+-$ component of the Einstein equation (\ref{5dfeq4}) in order to rewrite the scalar curvature $\tilde{R}$ in terms of $\Delta$. Therefore (\ref{5dint1}) follows from (\ref{5dint4}) and (\ref{5dint7}) together with the field equations and Bianchi identities mentioned above.
\subsection{The  condition (\ref{5dint2})}
Here we will show that the algebraic condition on $\tau_+$ (\ref{5dint2}) follows from (\ref{5dint1}). It is convenient to define
\bea
\xi_2 = \bigg(\frac{1}{4}\Delta h_i \Gamma^{i} - \frac{1}{4}\partial_{i}\Delta \Gamma^{i}\bigg)\phi_+ + \bigg(-\frac{1}{8}(dh)_{ij}\Gamma^{ij} +\frac{3i}{4}\beta{}_i\Gamma^i  + \frac{3i}{2}\chi V_{I}\alpha^I\bigg) \tau_+ \ ,
\ee
where $\xi_2=0$ equals the condition (\ref{5dint2}).
One can show after a computation that this expression for $\xi_2$ can be re-expressed as
\bea
\xi_2 = -\frac{1}{4}\Gamma^{i}\tilde{\nabla}_{i}{\xi_1} + \frac{7}{16}h_{j}\Gamma^{j}\xi_1 = 0 \ ,
\ee
which vanishes because $\xi_1 = 0$ is equivalent to the condition (\ref{5dint1}). In order to obtain this, we use the Dirac operator $\Gamma^{i}\tilde{\nabla}_{i}$ to act on (\ref{5dint1}) and apply the Bianchi identities (\ref{beq}) with the field equations (\ref{5dfeq1}), (\ref{5dfeq2}) and (\ref{5dfeq5}) to eliminate the terms which contain derivatives of the fluxes, and we can also use (\ref{5dint1}) to rewrite the $dh$-terms in terms of $\Delta$. We then impose the algebraic conditions (\ref{5dint7}) and (\ref{5dint8}) to eliminate the $\tilde{\nabla}_i X^I$-terms, of which some of the remaining terms will vanish as a consequence of (\ref{5dint1}). We then obtain the condition (\ref{5dint2}) as required, therefore it follows from section \ref{5dint1sec} above that (\ref{5dint2}) is implied by (\ref{5dint4}) and (\ref{5dint7}) together with the field equations and Bianchi identities mentioned above.
\subsection{The  condition (\ref{5dint5})}
Here we will show the differential condition on $\tau_+$ (\ref{5dint5}) is not independent. Let us define
\bea
\lambda_i &=& \tilde \nabla_i \tau_{+} + \bigg( -\frac{3}{4}h_i - \frac{i}{4}\alpha\Gamma_i - \frac{i}{8}\tilde{F}_{j k}\Gamma_i{}^{j k} + \frac{i}{2}\tilde{F}_{i j}\Gamma^j - \frac{3i}{2}\chi V_{I}\tilde{A}^{I}{}_{i} - \frac{1}{2}\chi V_I X^{I}\Gamma_i\bigg )\tau_{+}
\nonumber \\
&+& \bigg(-\frac{1}{4}(dh)_{ij}\Gamma^{j} - \frac{i}{4}\beta_j \Gamma_i{}^j + \frac{i}{2}\beta_i   \bigg)\phi_{+} \ , 
\ee
where $\lambda_i=0$ is equivalent to the condition (\ref{5dint5}). We can re-express this expression for
$\lambda_i$ as
\bea
\label{5dint5cond}
\lambda_ i = \bigg(-\frac{1}{4}\tilde{R}_{i j}\Gamma^{j} + \frac{1}{2}\Gamma^{j}(\tilde{\nabla}_{j}\tilde{\nabla}_{i} - \tilde{\nabla}_{i}\tilde{\nabla}_{j}) \bigg)\phi_+  + \frac{1}{2}\Lambda_{i, I}{\cal A}^{I}{}_{1} = 0~,
\ee
where the first terms again cancel from the definition of curvature, and
\bea
\Lambda_{i, I} = \frac{3i}{8} \tilde{\nabla}_{i}X_{I} +  Q_{I J}\bigg( \frac{1}{24}\tilde{G}^{J}{}_{\ell_1 \ell_2}\Gamma_{i}{}^{\ell_1 \ell_2} - \frac{1}{6}\tilde{G}^{J}{}_{i j}\Gamma^{j} - \frac{1}{12}L^{J}\Gamma_{i}\bigg) + \frac{i}{4}\chi V_{I} \Gamma_{i} \ .
\ee
This vanishes as $\mathcal{A}^{I}{}_1 = 0$ is equivalent to the positive chirality  component of (\ref{5dint7}). The identity (\ref{5dint5cond}) is derived by making use of (\ref{5dDDphicond}), and explicitly expanding out the $\mathcal{A}^{I}{}_1$ terms. We can also evaluate (\ref{5dint5}) by substituting in (\ref{5dint6}) to eliminate $\tau_+$, and use (\ref{5dint4}) to evaluate the supercovariant derivative of $\phi_+$.  Then, on adding this to (\ref{5dint5cond}), one obtains a condition which vanishes identically on making use of the Einstein equation (\ref{5dfeq4}). Therefore it follows that (\ref{5dint5}) is implied by the positive chirality component of (\ref{5dint4}), (\ref{5dint6}) and (\ref{5dint7}), the Bianchi identities (\ref{beq}) and the gauge field equations (\ref{5dfeq1}) and (\ref{5dfeq2}).
\subsection{The  condition (\ref{5dint8})}
Here we will show that the algebraic condition containing $\tau_+$ (\ref{5dint8}) follows from the independent KSEs. We define 
\bea
\mathcal{A}^{I}{}_{2} &=& \bigg[\tilde{G}^I{}_{i j}\Gamma^{i j} + 2 L^I - 2i\tilde{\nabla}_i X^I \Gamma^i - 6i\chi \bigg(Q^{I J} - \frac{2}{3}X^{I}X^{J}\bigg)V_J \bigg]\tau_{+} + 2 M^I{}_i\Gamma^i \phi_{+} \ ,
\nonumber \\
\ee
and also set
\bea
\mathcal{A}_{I, 2} &=& Q_{I J}\mathcal{A}^{J}{}_{2} \ ,
\ee
where $\mathcal{A}^{I}{}_{2}=0$ equals the expression in (\ref{5dint8}). 
The expression for $\mathcal{A}_{I, 2}$ can be rewritten as
\bea
\label{nnaux1}
\mathcal{A}_{I, 2} &=& -\frac{1}{2}\Gamma^{i}\tilde{\nabla}_{i}{({\cal A}_{I, 1})}   
+ \Phi_{I J}{\cal A}^{J}{}_{1} \ ,
\ee
where,
\bea
\Phi_{I J} &=& \bigg(-\frac{3}{4}Q_{J K}X_{I} - \frac{1}{8}C_{I J K}\bigg)\Gamma^{\ell}\tilde{\nabla}_{\ell}{X^{K}} + \frac{i}{2}\bigg(\frac{1}{4}Q_{J K}X_{I} + \frac{1}{8}C_{I J K}\bigg)\bigg(\tilde{G}^{K}{}_{\ell_1 \ell_2}\Gamma^{\ell_1 \ell_2}
- 2 L^{K}\bigg)
\nonumber \\
&+& Q_{I J}\bigg(\frac{i}{16}\tilde{F}_{\ell_1 \ell_2}\Gamma^{\ell_1 \ell_2} - \frac{i}{8}\alpha + \frac{3}{8}h_{\ell}\Gamma^{\ell} + \frac{3i}{4}\chi V_{K}\tilde{A}^{K}{}_{\ell}\Gamma^{\ell} - \frac{3}{4}\chi V_{K}X^{K}\bigg) 
\nonumber \\
&+& \chi\bigg(-\frac{3}{8}C_{I J K}Q^{K M} - \frac{3}{4} X_{I}\delta^{M}{}_{J}\bigg)V_{M} \ ,
\ee
and $\mathcal{A}_{I, 1} = Q_{I J}\mathcal{A}^{J}{}_{1}$.
In evaluating the above conditions, we have made use of the $+$ component of (\ref{5dint4}) in order to evaluate the covariant derivative in the above expression. In addition we have made use of the Bianchi identities (\ref{beq}) and the field equations (\ref{5dfeq1}), (\ref{5dfeq2}) and (\ref{5dfeq5}).
It follows from ({\ref{nnaux1}}) that $\mathcal{A}_{I, 2}=0$ as a consequence of the condition $\mathcal{A}_{I, 1}=0$, which as we have already noted is equivalent to the positive chirality part of ({\ref{5dint7}}).
\subsection{The  condition (\ref{5dint3})}
In order to show that (\ref{5dint3}) is implied by the independent KSEs, we define
\bea
\kappa &=& \bigg(-\frac{1}{2}\Delta - \frac{1}{8}(dh)_{ij}\Gamma^{ij} -\frac{3i}{4}\beta{}_i \Gamma^i + \frac{3i}{2}\chi V_{I}\alpha^I 
\nonumber \\
&+& 2\big(-{1\over4} h_i\Gamma^i - \frac{i}{8}(\tilde{F}_{j k}\Gamma^{j k} + 4\alpha) - \frac{1}{2}\chi V_{I}X^{I}\big) \Theta_{-} \bigg)\phi_{-} \ ,
\ee
where $\kappa$ equals the condition (\ref{5dint3}). Again, this expression can be rewritten as
\bea
\kappa = \bigg(\frac{1}{4}\tilde{R} + \Gamma^{i j}\tilde{\nabla}_{i}\tilde{\nabla}_{j}\bigg)\eta_{-} - \mu_{I}\mathcal{B}^{I}{}_1 = 0 \ ,
\ee
where we use the (\ref{5dDDphicond}) to evaluate the terms in the first bracket, and
\bea
\mu_{I} = \frac{3i}{16}\Gamma^{i}\tilde{\nabla}_{i}{X_{I}} - Q_{I J}\bigg(-\frac{7}{24}L^{J} + \frac{5}{48}\tilde{G}^{J}{}_{\ell_1 \ell_2}\Gamma^{\ell_1 \ell_2} \bigg) + \frac{i}{8}\chi V_{I} \ .
\ee
 The expression above vanishes identically since the negative chirality component of (\ref{5dint7}) is equivalent to $\mathcal{B}^{I}{}_1 = 0$. In order to obtain (\ref{5dint3}) from these expressions we make use of the Bianchi identities (\ref{beq}) and the field equations (\ref{5dfeq1}),(\ref{5dfeq2}) and (\ref{5dfeq5}). Therefore (\ref{5dint3}) follows from (\ref{5dint4}) and (\ref{5dint7}) together with the field equations and Bianchi identities mentioned above.
\subsection{The positive chirality part of (\ref{5dint4}) linear in $u$}
Since $\phi_+ = \eta_+ + u\Gamma_{+}\Theta_{-}\eta_-$, we must consider the part of the positive chirality component of (\ref{5dint4}) which is linear in $u$. We begin by defining
\bea
\mathcal{B}_{I, 1} &=& \bigg[\tilde{G}^I{}_{i j}\Gamma^{i j} + 2 L^I + 2i\tilde{\nabla}_i X^I \Gamma^i - 6i\chi \bigg(Q^{I J} - \frac{2}{3}X^{I}X^{J}\bigg)V_J\bigg]\eta_{-} \ .
\ee
We then determine that $\mathcal{B}_{I, 1}$ satisfies the following expression
\bea
\label{5dint4cond}
\bigg(\frac{1}{2}\Gamma^{j}(\tilde{\nabla}_{j}\tilde{\nabla}_{i} - \tilde{\nabla}_{i}\tilde{\nabla}_{j}) - \frac{1}{4}\tilde{R}_{i j}\Gamma^{j}  \bigg)\eta_{-} + \frac{1}{2}\Lambda_{i, I}{\cal B}^{I}{}_{1} = 0 ~,
\ee
where $\mathcal{B}_{I, 1} = Q_{I J}\mathcal{B}^{J}{}_{1}$, and
\bea
\Lambda_{i, I} = \frac{3i}{8} \tilde{\nabla}_{i}X_{I} +  Q_{I J}\bigg( \frac{1}{24}\tilde{G}^{J}{}_{\ell_1 \ell_2}\Gamma_{i}{}^{\ell_1 \ell_2} - \frac{1}{6}\tilde{G}^{J}{}_{i j}\Gamma^{j} + \frac{1}{12}L^{J}\Gamma_{i}\bigg) + \frac{i}{4}\chi V_{I} \Gamma_{i} \ .
\ee
We note that $\mathcal{B}_{I, 1}=0$ is equivalent to the negative chirality component of (\ref{5dint7}). 
Next, we use (\ref{5dDDphicond}) to evaluate the terms in the first bracket in (\ref{5dint4cond}) and explicitly expand out the terms with $\mathcal{B}^{I}{}_1$. The resulting expression corresponds to the expression obtained by expanding out the $u$-dependent part of the positive chirality component of (\ref{5dint4}) by using the 
negative chirality component of (\ref{5dint4}) to evaluate the covariant derivative. We have made use of the Bianchi identities (\ref{beq}) and the gauge field equations (\ref{5dfeq1}) and (\ref{5dfeq2}).
\subsection{The positive chirality part of condition (\ref{5dint7}) linear in $u$}
Again, as $\phi_+ = \eta_+ + u\Gamma_{+}\Theta_{-}\eta_-$, we must consider the part of the positive chirality component of (\ref{5dint7}) which is linear in $u$. 
One finds that the $u$-dependent part of (\ref{5dint7}) is proportional to
\bea
-\frac{1}{2}\Gamma^{i}\tilde{\nabla}_{i}{({\cal B}_{I, 1})}   
+ \Phi_{I J}{\cal B}^{J}{}_{1} \ ,
\ee
where
\bea
\Phi_{I J} &=& \bigg(-\frac{3}{4}Q_{J K}X_{I} - \frac{1}{8}C_{I J K}\bigg)\Gamma^{\ell}\tilde{\nabla}_{\ell}{X^{K}} + \frac{i}{2}\bigg(\frac{1}{4}Q_{J K}X_{I} + \frac{1}{8}C_{I J K}\bigg)\bigg(\tilde{G}^{K}{}_{\ell_1 \ell_2}\Gamma^{\ell_1 \ell_2}
+ 2 L^{K}\bigg)
\nonumber \\
&+& Q_{I J}\bigg(\frac{i}{16}\tilde{F}_{\ell_1 \ell_2}\Gamma^{\ell_1 \ell_2} + \frac{i}{8}\alpha + \frac{1}{8}h_{\ell}\Gamma^{\ell}+ \frac{3i}{4}\chi V_{K}\tilde{A}^{K}{}_{\ell}\Gamma^{\ell} - \frac{3}{4}\chi V_{K}X^{K}\bigg) 
\nonumber \\
&+& \chi\bigg(-\frac{3}{8}C_{I J K}Q^{K M} - \frac{3}{4} X_{I}\delta^{M}{}_{J}\bigg)V_{M}
\ee
and where we use the (\ref{5dDDphicond}) to evaluate the terms in the first bracket. In addition we have made use of the Bianchi identities (\ref{beq}) and the field equations (\ref{5dfeq1}), (\ref{5dfeq2}) and (\ref{5dfeq5}).

\section{Global analysis: Lichnerowicz theorems}
\label{5dmaxpex}
To establish the Lichnerowicz type theorems, we first calculate the Laplacian of $\parallel \eta_\pm \parallel^2$. Here we will assume throughout that  ${\mathscr D}^{(\pm)}\eta_\pm=0$,
\bea
\label{lapla}
\tilde{\nabla}^i \tilde{\nabla}_i ||\eta_{\pm}||^2 = 2{\rm Re } \langle\eta_\pm,\tilde{\nabla}^i \tilde{\nabla}_i\eta_\pm\rangle + 2 {\rm Re } \langle\tilde{\nabla}^i \eta_\pm, \tilde{\nabla}_i \eta_\pm\rangle \ .
\ee
To evaluate this expression note that
\bea
\tilde{\nabla}^i \tilde{\nabla}_i \eta_\pm &=& \Gamma^{i}\tilde{\nabla}_{i}(\Gamma^{j}\tilde{\nabla}_j \eta_\pm) -\Gamma^{i j}\tilde{\nabla}_i \tilde{\nabla}_j \eta_\pm
\nonumber \\
&=& \Gamma^{i}\tilde{\nabla}_{i}(\Gamma^{j}\tilde{\nabla}_j \eta_\pm) + \frac{1}{4}\tilde{R}\eta_\pm
\nonumber \\
&=& \Gamma^{i}\tilde{\nabla}_{i}(-\Psi^{(\pm)}\eta_\pm - q_I\mathcal{A}^{I, (\pm)}\eta_{\pm}) + \frac{1}{4}\tilde{R} \eta_\pm \ .
\ee
Therefore the first term in (\ref{lapla}) can be written as,
\bea
\label{lap1}
{\rm Re } \langle\eta_\pm,\tilde{\nabla}^i \tilde{\nabla}_i\eta_\pm \rangle &=& \frac{1}{4}\tilde{R}\parallel \eta_\pm \parallel^2
+ {\rm Re } \langle\eta_\pm, \Gamma^{i}\tilde{\nabla}_i(-\Psi^{(\pm)} - q_I\mathcal{A}^{I, (\pm)})\eta_\pm\rangle
\nonumber \\
&+& {\rm Re } \langle\eta_\pm, \Gamma^{i}(-\Psi^{(\pm)} - q_I\mathcal{A}^{I, (\pm)})\tilde{\nabla}_i \eta_\pm \rangle~.
\ee
For the second term in (\ref{lapla}) we write,
\bea
\label{lap2}
{\rm Re } \langle\tilde{\nabla}^i \eta_\pm, \tilde{\nabla}_i \eta_\pm\rangle &=& {\rm Re } \langle{\hat\nabla^{(\pm)i}} \eta_\pm, {\hat\nabla^{(\pm)}_{i}} \eta_\pm\rangle - 2{\rm Re } \langle\eta_\pm, (\Psi^{(\pm)i} + \kappa_I\Gamma^{i}\mathcal{A}^{I, (\pm)})^{\dagger} \tilde{\nabla}_i \eta_\pm \rangle
\nonumber \\
&-& {\rm Re } \langle\eta_\pm, (\Psi^{(\pm)i} + \kappa_{I} \Gamma^{i}\mathcal{A}^{I, (\pm)})^{\dagger} (\Psi^{(\pm)}_i + \kappa_J \Gamma_{i} \, \mathcal{A}^{J, (\pm)}) \eta_\pm \rangle
\nonumber \\
&=& \parallel {\hat\nabla^{(\pm)}}\eta_{\pm} \parallel^2 - 2{\rm Re } \langle \eta_{\pm}, \Psi^{(\pm)i\dagger}\tilde{\nabla}_{i}\eta_{\pm}\rangle
- 2\kappa_{I} {\rm Re } \langle \eta_{\pm}, \mathcal{A}^{I, (\pm)\dagger}\Gamma^{i}\tilde{\nabla}_{i}\eta_{\pm}\rangle
\nonumber \\
&-&  {\rm Re } \langle \eta_\pm, (\Psi^{(\pm)i\dagger}\Psi^{(\pm)}_i + 2\kappa_{I} \mathcal{A}^{I, (\pm)\dagger}\Psi^{(\pm)} + 3\kappa_{I} \kappa_{J}\mathcal{A}^{I,(\pm)\dagger}\mathcal{A}^{J, (\pm)})\eta_\pm \rangle
\nonumber \\
&=& \parallel {\hat\nabla^{(\pm)}}\eta_{\pm} \parallel^2 - 2{\rm Re } \langle \eta_{\pm}, \Psi^{(\pm)i\dagger}\tilde{\nabla}_{i}\eta_{\pm}\rangle - {\rm Re } \langle \eta_\pm , \Psi^{(\pm)i\dagger}\Psi^{(\pm)}_i \eta_\pm \rangle
\nonumber \\
&+& (2\kappa_{I} q_{J} - 3\kappa_{I}\kappa_{J}){\rm Re } \langle \mathcal{A}^{I,(\pm)}\eta_\pm, \mathcal{A}^{J,(\pm)}\eta_\pm \rangle \ .
\ee
Therefore using (\ref{lap1}) and (\ref{lap2}) with (\ref{lapla}) we have,
\bea
\label{5dextralap1b}
\frac{1}{2}\tilde{\nabla}^i \tilde{\nabla}_i ||\eta_{\pm}||^2 &=& \parallel {\hat\nabla^{(\pm)}}\eta_{\pm} \parallel^2 + \, (2\kappa_I q_J - 3\kappa_I \kappa_J){\rm Re } \langle \mathcal{A}^{I,(\pm)}\eta_\pm, \mathcal{A}^{J,(\pm)}\eta_\pm \rangle \ 
\nonumber \\
&+& {\rm Re } \langle \eta_\pm, \bigg(\frac{1}{4}\tilde{R} + \Gamma^{i}\tilde{\nabla}_i(-\Psi^{(\pm)} - q_{I}\mathcal{A}^{I, (\pm)}) - \Psi^{(\pm)i\dagger}\Psi^{(\pm)}_i \bigg) \eta_\pm \rangle
\nonumber \\
&+& {\rm Re } \langle \eta_\pm, \bigg( \Gamma^{i}(-\Psi^{(\pm)} - q_{I}\mathcal{A}^{I, (\pm)}) - 2\Psi^{(\pm)i\dagger}\bigg)\tilde{\nabla}_i \eta_\pm \rangle \ .
\ee
In order to simplify the expression for the Laplacian, we shall attempt to rewrite the third line in ({\ref{5dextralap1b}}) as
\bea
\label{5dbilin}
{\rm Re } \langle \eta_\pm, \bigg( \Gamma^{i}(-\Psi^{(\pm)} - q_{I}\mathcal{A}^{I, (\pm)}) - 2\Psi^{(\pm)i\dagger}\bigg)\tilde{\nabla}_i \eta_\pm \rangle &=& {\rm Re } \langle \eta_\pm, \mathcal{F}^{(\pm)}\Gamma^{i}\tilde{\nabla}_i \eta_\pm \rangle 
\nonumber \\
&+& W^{(\pm)i}\tilde{\nabla}_i \parallel \eta_\pm \parallel^2~,
\ee
where $\mathcal{F}^{(\pm)}$ is linear in the fields and $W^{(\pm)i}$ is a vector.\footnote{This expression is useful since we can eliminate the differential term $\Gamma^{i}\tilde{\nabla}_{i}{\eta_{\pm}}$ by using the modified horizon Dirac equation (\ref{5ddgdirac}) and write it in terms of an algebraic condition on $\eta_{\pm}$. The second term gives a
vector which is consistent when you apply the maximum principle and integration by parts arguments in order to establish the generalised Lichnerowicz theorems.} After a computation, one finds that this is only possible for $q_{I} = 0$, thus we obtain
\bea
\Gamma^{i}(-\Psi^{(\pm)}) - 2\Psi^{(\pm)i\dagger} = \pm h^i + \bigg(\mp\frac{1}{4}h_{j}\Gamma^{j} \pm \frac{i}{4}\alpha + \frac{i}{8}\tilde{F}_{\ell_1 \ell_2}\Gamma^{\ell_1 \ell_2}- \frac{3i}{2} \chi V_{I}\tilde{A}^{I}{}_{\ell}\Gamma^{\ell}
- \frac{5}{2}\chi V_{I}X^{I}
\bigg)\Gamma^{i} \ ,
\nonumber \\
\ee
where
\bea
W^{(\pm)i} = \pm\frac{1}{2}h^i \ ,
\ee
\bea
\mathcal{F}^{(\pm)} = \mp\frac{1}{4}h_{j}\Gamma^{j} \pm \frac{i}{4}\alpha + \frac{i}{8}\tilde{F}_{\ell_1 \ell_2}\Gamma^{\ell_1 \ell_2}
- \frac{3i}{2} \chi V_{I}\tilde{A}^{I}{}_{\ell}\Gamma^{\ell}
- \frac{5}{2}\chi V_{I}X^{I} \ .
\ee
We remark that  $\dagger$ is the adjoint with respect to the $Spin_{c}(3)$-invariant inner product ${\rm Re } \langle \phantom{i},\phantom{i} \rangle$.\footnote{Where $Spin_{c}(3) = Spin(3) \times U(1)$ and in order to compute the adjoints above we note that the $Spin_{c}(3)$-invariant inner product is positive definite and symmetric.} We also have the following identities 
\bea
\label{5dhermiden1}
{\rm Re } \langle \eta_+, \Gamma^{\ell_1 \ell_2} \eta_+ \rangle = {\rm Re } \langle \eta_+, \Gamma^{\ell_1 \ell_2 \ell_3} \eta_+ \rangle = 0 \ ,
\ee
and
\bea
\label{5dhermiden2}
{\rm Re } \langle \eta_+, i\Gamma^{\ell} \eta_+ \rangle = 0 .
\ee
It follows that
\bea
\label{5dlaplacian}
\frac{1}{2}\tilde{\nabla}^i \tilde{\nabla}_i ||\eta_{\pm}||^2 &=& \parallel {\hat{\nabla}^{(\pm)}}\eta_{\pm} \parallel^2 - ~3\kappa_{I}\kappa_{J}{\rm Re } \langle \mathcal{A}^{I,(\pm)}\eta_\pm, \mathcal{A}^{J,(\pm)}\eta_\pm \rangle ~+ W^{(\pm)i}\tilde{\nabla}_{i}\parallel \eta_\pm \parallel^2
\nonumber \\
&+& {\rm Re } \langle \eta_\pm, \bigg(\frac{1}{4}\tilde{R} + \Gamma^{i}\tilde{\nabla}_i(-\Psi^{(\pm)}) - \Psi^{(\pm)i\dagger}\Psi^{(\pm)}_i  + \mathcal{F}^{(\pm)}(-\Psi^{(\pm)})\bigg) \eta_\pm \rangle \ .
\nonumber \\
\ee
It is also useful to evaluate ${\tilde{R}}$ using (\ref{5dfeq4}); we obtain
\bea
\tilde{R} &=& -\tilde{\nabla}^{i}(h_i) + \frac{1}{2}h^2 + \frac{3}{2}\alpha^2
+ \frac{3}{4}\tilde{F}^2 - 2\chi^2 U + Q_{I J}\bigg(\tilde{\nabla}^{i}{X^I} \tilde{\nabla}_{i}{X^J} + L^{I}L^{J} + \frac{1}{2}\tilde{G}^{I}{}_{\ell_1 \ell_2}\tilde{G}^{J \ell_1 \ell_2}\bigg) \ .
\nonumber \\
\ee
One obtains, upon using the field equations and Bianchi identities,
\bea
\label{5dquad}
\bigg(\frac{1}{4}\tilde{R} &+& \Gamma^{i}\tilde{\nabla}_i(-\Psi^{(\pm)})  - \Psi^{(\pm)i\dagger}\Psi^{(\pm)}_i  + \mathcal{F}^{(\pm)}(-\Psi^{(\pm)}) \bigg)\eta_\pm
\nonumber \\
&=& \bigg[ \frac{3i}{2} \chi V_{I}\tilde{\nabla}^{\ell}{(\tilde{A}^{I}{}_{\ell})}
\mp \frac{3i}{4} \chi V_{I}\tilde{A}^{I}{}_{\ell}h^{\ell} \mp \frac{9i}{4} \chi V_{I}X^{I} \alpha 
\nonumber \\
&+& \big(\pm \frac{1}{4}\tilde{\nabla}_{\ell_1}(h_{\ell_2}) \mp \frac{3}{16}\alpha \tilde{F}_{\ell_1 \ell_2}\big)\Gamma^{\ell_2 \ell_2} 
\nonumber \\
&+& i\big(\pm \frac{3}{4}\tilde{\nabla}_{\ell}(\alpha) + \frac{3}{4}\tilde{\nabla}^{j}(\tilde{F}_{j \ell}) - \frac{1}{8}h_{\ell}\alpha \mp \frac{1}{4}h^{j}\tilde{F}_{j \ell} - \frac{3}{2} \chi^2 V_{J}X^{J}V_{I}\tilde{A}^{I}{}_{\ell}\big)\Gamma^{\ell} 
\nonumber \\
&+& \frac{3}{8} \chi V_{I}\tilde{A}^{I}{}_{\ell_1}\tilde{F}_{\ell_2 \ell_3}\Gamma^{\ell_1 \ell_2 \ell_3}\bigg]\eta_{\pm} 
\nonumber \\
&+&  \bigg(\frac{1}{8}Q_{I J}\tilde{G}^{I \ell_1 \ell_2}\tilde{G}^{J}{}_{\ell_1 \ell_2}
+ \frac{1}{4}Q_{I J}L^{I}L^{J} 
+ \frac{9}{4}\chi^2  V_I V_J Q^{I J} - \frac{3}{2}\chi^2 V_I V_J X^{I}X^{J}
\nonumber \\
&+& \frac{1}{4}Q_{I J}\tilde{\nabla}_{\ell}{X^{I}}\tilde{\nabla}^{\ell}{X^{J}}+ \frac{3i}{8}\tilde{G}^{I}{}_{\ell_1 \ell_2}\tilde{\nabla}_{\ell_3}{X_{I}}\Gamma^{\ell_1 \ell_2 \ell_3}
-\frac{3}{2} \chi  V_{I} \tilde{\nabla}_{\ell}X^{I}\Gamma^{\ell}
\nonumber \\
&+& \frac{3i}{4} \chi V_{I} \tilde{G}^{I}{}_{\ell_1 \ell_2}\Gamma^{\ell_1 \ell_2}
\bigg)\eta_\pm
\nonumber \\
&-& {1 \over 4} \big(1 \mp 1\big) {\tilde{\nabla}}^i (h_i)  \eta_\pm  \ .
\ee
One can show that the third line in (\ref{5dquad}) can be written in terms of the Algebraic KSE (\ref{5dalg2pm}), in particular we find,
\bea
\frac{1}{16}Q_{I J}{\cal A}^{I, (\pm)\dagger}{\cal A}^{J, (\pm)}\eta_\pm &=& \bigg(\frac{1}{8}Q_{I J}\tilde{G}^{I \ell_1 \ell_2}\tilde{G}^{J}{}_{\ell_1 \ell_2}
+ \frac{1}{4}Q_{I J}L^{I}L^{J} + \frac{9}{4}\chi^2  V_I V_J Q^{I J} 
\nonumber \\
&-& \frac{3}{2}\chi^2 V_I V_J X^{I}X^{J} + \frac{1}{4}Q_{I J}\tilde{\nabla}_{\ell}{X^{I}}\tilde{\nabla}^{\ell}{X^{J}} -\frac{3}{2} \chi  V_{I} \tilde{\nabla}_{\ell}X^{I}\Gamma^{\ell}
\nonumber \\
&+& \frac{3i}{8}\tilde{G}^{I}{}_{\ell_1 \ell_2}\tilde{\nabla}_{\ell_3}{X_{I}}\Gamma^{\ell_1 \ell_2 \ell_3}
+ \frac{3i}{4} \chi V_{I} \tilde{G}^{I}{}_{\ell_1 \ell_2}\Gamma^{\ell_1 \ell_2}
\bigg)\eta_\pm \ .
\ee
Note using (\ref{5dhermiden1}) and (\ref{5dhermiden2}) all the terms on the RHS of the above expression, with the exception of the final two lines, vanish in the second line of (\ref{5dlaplacian}) since all these terms in ({\ref{5dquad}}) are anti-Hermitian.
Also, for $\eta_+$ the final line in ({\ref{5dquad}}) also vanishes and thus there is no contribution to the Laplacian of $\parallel \eta_+ \parallel^2$ in (\ref{5dlaplacian}). For $\eta_{-}$ the final line in ({\ref{5dquad}}) does give an extra term in the Laplacian of $\parallel \eta_- \parallel^2$ in (\ref{5dlaplacian}). For this reason, the analysis of the conditions imposed by the global properties of ${\cal{S}}$ is different in these two cases and thus we will consider the Laplacians of $\parallel \eta_\pm \parallel^2$ separately.

For the Laplacian of $\parallel \eta_+ \parallel^2$, we obtain from ({\ref{5dlaplacian}}):
\bea
\label{5dl1}
{\tilde{\nabla}}^{i}{\tilde{\nabla}}_{i}\parallel\eta_+\parallel^2 - \, h^i {\tilde{\nabla}}_{i}\parallel\eta_+\parallel^2 &=& 2\parallel{\hat{\nabla}^{(+)}}\eta_{+}\parallel^2 
\nonumber \\
&+&  \bigg(\frac{1}{16}Q_{I J} - 3\kappa_{I}\kappa_{J}\bigg){\rm Re } \langle {\cal A}^{I, (+)} \eta_+, {\cal A}^{J, (+)} \eta_+ \rangle  .
\nonumber \\
\ee
The maximum principle thus implies that $\eta_+$ are Killing spinors on ${\cal{S}}$, i.e. 
\bea
{{\nabla}^{(+)}}\eta_{+}=0, \quad {\cal A}^{I, (+)}\eta_{+} = 0 \ ,
\ee
and moreover $\parallel\eta_+\parallel=\mathrm{const}$. 
The Laplacian of $\parallel \eta_- \parallel^2$
is calculated from ({\ref{5dlaplacian}}), on taking account of the contribution to the second line of
({\ref{5dlaplacian}}) from the final line of ({\ref{5dquad}}). One
obtains

The Laplacian of $\parallel \eta_- \parallel^2$
is calculated from ({\ref{5dlaplacian}}), on taking account of the contribution to the second line of
({\ref{5dlaplacian}}) from the final line of ({\ref{5dquad}}). One
obtains
\bea
\label{5dl2}
{\tilde{\nabla}}^{i} \bigg(\tilde{\nabla}_{i} \parallel \eta_- \parallel^2 + \parallel \eta_- \parallel^2  h_{i}\bigg)
&=& 2 \parallel{\hat{\nabla}^{(-)}}\eta_{-}\parallel^2 
\nonumber \\
&+&~  \bigg(\frac{1}{16}Q_{I J} - 3\kappa_{I}\kappa_{J}\bigg){\rm Re } \langle {\cal A}^{I,(-)} \eta_-, {\cal A}^{J,(-)}\eta_- \rangle ~.
\nonumber \\
\ee
On integrating this over ${\cal{S}}$ and assuming that ${\cal{S}}$ is compact and without boundary, the LHS vanishes since it is a total derivative and one finds that $\eta_{-}$ are Killing spinors on ${\cal{S}}$, i.e 
\bea
{{\nabla}^{(-)}}\eta_{-}=0, \qquad {\cal A}^{I, (-)}\eta_{-} = 0 \ .
\ee
This establishes the Lichnerowicz type theorems for both positive and negative chirality spinors $\eta_\pm$ which are in the kernels of the horizon Dirac operators ${{\mathscr D}}^{(\pm)}$: i.e.
\bea
\{ \ {{\nabla}^{(\pm)}}\eta_{\pm}=0, \quad {\rm and} \quad {\cal A}^{I, (\pm)}\eta_{\pm} = 0 \ \}
\qquad \Longleftrightarrow \qquad {{\mathscr D}}^{(\pm)} \eta_\pm = 0 \ .
\ee

	\chapter{${\mathfrak{sl}}(2,\mathbb{R})$ Symmetry and Spinor Bilinears}
In this Appendix we present some formulae which are generic in the analysis
of the $\mathfrak{sl}(2,\mathbb{R})$ symmetry
\bea
\epsilon= \eta_++ u \Gamma_+\Theta_-\eta_-+ \eta_-+r \Gamma_-\Theta_+\eta_++ru \Gamma_-\Theta_+\Gamma_+\Theta_-\eta_-~.
\ee
Since the $\eta_-$ and $\eta_+$ Killing spinors appear in pairs, let us choose a $\eta_-$ Killing spinor.  Then from the results
of the previous chapters, horizons with non-trivial fluxes also admit $\eta_+=\Gamma_+\Theta_-\eta_-$ as a Killing spinors. Using $\eta_-$ and $\eta_+=\Gamma_+\Theta_-\eta_-$, one can construct two linearly independent Killing spinors on the  spacetime as
\bea
\epsilon_1=\eta_-+u\eta_++ru \Gamma_-\Theta_+\eta_+~,~~~\epsilon_2=\eta_++r\Gamma_-\Theta_+\eta_+~.
\ee
Now consider the spinor bilinear,
\bea
\label{bil}
K(\zeta_1, \zeta_2) &=& \langle(\Gamma_+-\Gamma_-) \zeta_1, \Gamma_a\zeta_2\rangle\, e^a \ .
\ee
In order to evaluate this bilinear explicitly in our analysis,
it is useful to note the following identities:
\bea
\label{iden}
\Gamma_+ \epsilon_1 &=& \Gamma_+\eta_- + 2ur\Theta_+ \eta_+ \ ,
\nonumber \\
\Gamma_- \epsilon_1 &=& u \Gamma_- \eta_+ \ ,
\nonumber \\
\Gamma_i \epsilon_1 &=& \Gamma_{i}\eta_{-} + u\Gamma_{i}\eta_{+} - u r \Gamma_{-}\Gamma_{i}\Theta_+\eta_+ \ ,
\nonumber \\
(\Gamma_+ - \Gamma_-)\epsilon_1 &=& \Gamma_+ + 2ur\Theta_+ \eta_+ - u \Gamma_- \eta_+ \ ,
\ee
and
\bea
\label{iden2}
\Gamma_+ \epsilon_2 &=& 2r\Theta_+\eta_+ \ ,
\nonumber \\
\Gamma_- \epsilon_2 &=& \Gamma_- \eta_+ \ ,
\nonumber \\
\Gamma_i \epsilon_2 &=& \Gamma_{i}\eta_{+} - r \Gamma_{-}\Gamma_{i}\Theta_+\eta_+ \ ,
\nonumber \\
(\Gamma_+ - \Gamma_-)\epsilon_2 &=& 2r\Theta_+\eta_+ - \Gamma_- \eta_+ \ .
\ee
By expanding the above in (\ref{bil}) and using the identities (\ref{iden}) and (\ref{iden2}), one can compute the 1-form bilinears of the Killing spinors $\epsilon_1$ and $\epsilon_2$
\begin{eqnarray}
K_1(\epsilon_1, \epsilon_2)&=& (2r \langle\Gamma_+\eta_-, \Theta_+\eta_+\rangle+  u^2 r \Delta \parallel \eta_+\parallel^2) \,{\bf{e}}^+-2u \parallel\eta_+\parallel^2\, {\bf{e}}^-+ V_i {\bf{e}}^i~,
\cr
K_2(\epsilon_2, \epsilon_2)&=& r^2 \Delta\parallel\eta_+\parallel^2 \,{\bf{e}}^+-2 \parallel\eta_+\parallel^2 {\bf{e}}^-~,
\cr
K_3(\epsilon_1, \epsilon_1)&=&(2\parallel\eta_-\parallel^2+4r u \langle\Gamma_+\eta_-, \Theta_+\eta_+\rangle+ r^2 u^2 \Delta \parallel\eta_+\parallel^2) {\bf{e}}^+
-2u^2 \parallel\eta_+\parallel^2 {\bf{e}}^-+2u V_i {\bf{e}}^i~,
\label{b1forms}
\nonumber \\
\end{eqnarray}
where we have set
\begin{eqnarray}
\label{vii}
V_i =  \langle \Gamma_+ \eta_- , \Gamma_i \eta_+ \rangle~.
\end{eqnarray}
Note that given a 1-form
\bea
X = X_{a}e^{a} = X_{+}e^{+} + X_{-}e^{-} + X_{i}e^{i} \ .
\ee
One can rewrite the corresponding vector field as,
\bea
\label{vec}
\tilde{X} = \tilde{X}^{a}\partial_{a} = \tilde{X}^{+}\partial_{+} + \tilde{X}^{-}\partial_{-}  + \tilde{X}^{i}\partial_{i} 
\ee
Note in this case with the given metric one has $\tilde{X}^{+} = X_{-}$ and $\tilde{X}^{-} = X_{+}$. One can also express the frame derivatives in terms of co-ordinates as
\bea
\partial_+ &=& \partial_u + \frac{1}{2}r^2\Delta\partial_r
\nonumber \\
\partial_- &=& \partial_r
\nonumber \\
\partial_i &=& \tilde{\partial}_i - rh_{i}\partial_r \ .
\ee
Now the vector field (\ref{vec}) can be expressed as
\bea
\tilde{X} = X_{-}\partial_{u} + \bigg(\frac{1}{2}r^2\Delta X_{-} + X_{+} - rX^{i}h_{i}\bigg)\partial_r + X^{i}\tilde{\partial}_i \ .
\ee
In particular, the components of $\mathcal{L}_X g$ with respect to the basis
({\ref{basis1}}) are then given by:
\bea
(\mathcal{L}_X g)_{+ +} &=& 2\partial_{u}X_{+} + r^2 \Delta \partial_{r}X_{+} - 2r\Delta X_+ + r^2(\Delta h_{i} - \tilde{\nabla}_{i} \Delta)X^{i} \ ,
\nonumber \\
(\mathcal{L}_X g)_{- -} &=& \partial_{r}X_{-} \ ,
\nonumber \\
(\mathcal{L}_X g)_{+ -} &=& \partial_{u}X_{-} + \frac{1}{2}r^2 \Delta \partial_{r}X_{-} + \partial_{r}X_+
+ r\Delta X_{-} - h_{i}X^{i} \ ,
\nonumber \\
(\mathcal{L}_X g)_{+ i} &=& \partial_{u}X_{i} + \frac{1}{2}r^2 \Delta \partial_{r}X_{i} + \tilde{\partial}_{i}X_+
- rh_{i}\partial_{r}X_+ 
\nonumber \\
&-& \frac{1}{2}r^2(\Delta h_{i} - \tilde{\nabla}_{i} \Delta)X_{-} + h_{i}X_{+} - r(dh)_{i j}X^{j} \ ,
\nonumber \\
(\mathcal{L}_X g)_{- i} &=& \partial_{r}X_{i} + \tilde{\partial}_{i}X_{-} - rh_{i}\partial_r X_{-} \ ,
\nonumber \\
(\mathcal{L}_X g)_{i j} &=& \tilde{\nabla}_{i}X_{j} + \tilde{\nabla}_{j}X_{i} \ .
\ee
	% appendices come here
	
	\addcontentsline{toc}{chapter}{Bibliography}
	\bibliographystyle{plain}
	\bibliography{bibliography/bibliography.bib}

\end{document}